\documentclass[b5paper,12pt,notitlepage,twoside,openright]{report_custom}
\usepackage[palatino]{quotchap}
\usepackage{fancyhdr}
\usepackage{bm}       
\usepackage{fourier} 

\usepackage{amsmath,amssymb,amsfonts,array,marvosym,dsfont}
\usepackage[pdftex]{graphicx}
\usepackage{verbatim}
\usepackage{enumerate}
\usepackage{subfigure}
\usepackage[english]{babel}
\usepackage{bbm}
\usepackage{slashed}
\usepackage{multirow}
\usepackage{fancybox}
\usepackage{pifont}

\fancypagestyle{plain}{%
  \fancyhf{}                          

}

\usepackage{hyperref}
\hypersetup{pdfborder={0 0 0},
	colorlinks=true,
	linkcolor=black,
	citecolor=black,
	urlcolor=black}
\newcommand{\av}[1]{\langle #1 \rangle}

\newcommand{\eps}{\varepsilon}
\newcommand{\e}{\epsilon}
\newcommand{\lT}{\lambda}
\newcommand{\g}{\gamma}
\newcommand{\kp}{\kappa}

\DeclareRobustCommand{\stirling}{\genfrac\{\}{0pt}{}}

\definecolor{rosso}{cmyk}{0,1,1,0.4}

\newcommand{\FigPath}{./graf}

\begin{document}
\pagenumbering{roman}

\begin{titlepage}
\begin{center}
\begin{center}
\includegraphics[width=.25\textwidth]{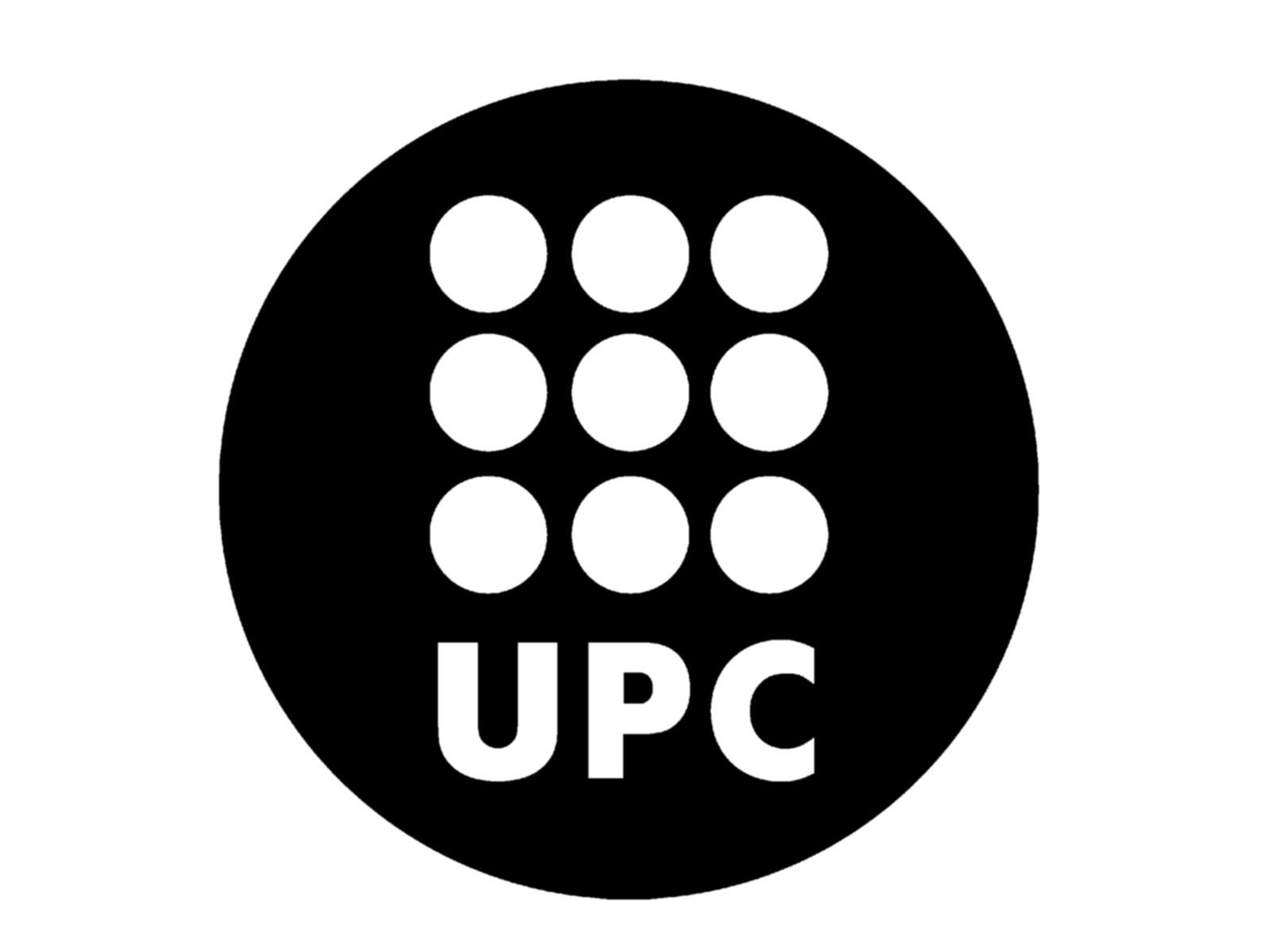}
\end{center}
\vspace{0.2cm}
\textsc{\large Universitat Polit\`ecnica de Catalunya }\\ 
\vspace{0.5cm}
\textsc{\large PhD Thesis}\\ \vspace{2 cm}
\textsc{\large  \bf Time-varying networks approach to social dynamics:\\ \vspace{.4 cm} 
From individual to collective behavior}\\
\vspace{3 cm}
\begin{minipage}{0.4\linewidth}
\flushleft  { Candidate}\\ \vspace{0.2cm} \textbf{ Michele Starnini} \\
\end{minipage}
\hspace{.01\linewidth}
\begin{minipage}{0.5\linewidth}
\flushright { Supervisor}\\ \vspace{0.2cm}
{ \bf Prof. Romualdo Pastor-Satorras} \\
\end{minipage}
\vspace{2 cm}
\textsc{\Large }\\ {October 2014}\\
\end{center}

\newpage
\thispagestyle{empty} \phantom{}
\end{titlepage}


\chapter*{\centering \begin{normalsize} \bf{Abstract}\end{normalsize}}
\begin{quotation}
\noindent

In this thesis we contribute to the understanding of the pivotal role of 
the temporal dimension in networked social systems, 
previously neglected and now uncovered by the data revolution
recently blossomed in this field.
To this aim, we first introduce the time-varying networks formalism and 
analyze some empirical data of social dynamics,
extensively used in the rest of the thesis.
We discuss the structural and temporal properties of the human contact networks, 
such as heterogeneity and burstiness of social interactions,
and we present a simple model, 
rooted on social attractiveness,
able to reproduce them.
We then explore the behavior of dynamical processes running on top of temporal networks,
constituted by empirical face-to-face interactions,
addressing in detail the fundamental cases of random walks and epidemic spreading.
We also develop an analytic approach able to compute the
 structural and percolation properties of activity driven model,
 aimed to describe a wide class of social interactions, 
driven by the activity of the individuals involved. 
Our contribution in the rapidly evolving framework of social time-varying networks
opens interesting perspectives for future work,
 such as the study of the impact of the temporal dimension on multi-layered systems.

\end{quotation}
\clearpage

\tableofcontents


\setcounter{secnumdepth}{-1}
\chapter{Invitation}

\pagenumbering{arabic}

Network science is, and will remain in the future, a crucial tool to understand the behavior of physical as well as social, biological and technological systems.
The network representation is indeed appropriate to study the emergence of the global, system-scale, properties of systems as diverse as the Internet or the brain.
In the traditional framework of network science, graphs have long been studied as static entities. Most real networks, however, are dynamic structures in which connections appear, disappear, or are rewired on various timescales. Social relationships and interactions, in particular, are inherently dynamically evolving. Very recently, the interest towards the temporal dimension of the network description has blossomed, giving rise to the emergent field of temporal or time-varying networks.
This boost in network science has been made possible by the social data revolution, with the collection of high resolution data sets from mobile devices, communication, and pervasive technologies. The ever increasing adoption of mobile technologies, indeed, allows to sense human behavior at unprecedented levels of detail and scale, providing to social networks a longitudinal dimension that calls for a renewed effort in analysis and modeling. In the future, the digital traces of human activities promise to transform the way we measure, model and reason on social aggregates and socio-technical systems.
This rapidly evolving framework brings forth new exciting challenges. The social data revolution joined with the network science approach will allow us to leverage the study of social dynamics on a different scale. We are on the way to the next paradigm shift in the understanding and modeling of the social dynamics, from the individual behavior to the social aggregate, and my thesis will dive directly into the core of this endeavor.

This thesis is organized as follows. In Chapter \ref{chap:intro} 
we introduce the time-varying networks formalism
and we present and analyze several empirical data sets of social dynamics,
 extensively used in the rest of the thesis.
The main body of the exposition is then split into two parts. 
Part \ref{part:modeling} will deal with models of social dynamics, 
focusing on two different kinds of social interactions:
 face-to-face contact networks
and  activity driven networks. 
In Chapter \ref{chap:f2f} we present a simple model
able to reproduce the main statistical regularities of empirical face-to-face interactions.
Chapter \ref{chap:activitydriven} is instead devoted to
 the analytic study of the topological properties and
the percolation processes unfolding on the activity driven networks model.
Part \ref{part:dynproc} will uncover the behavior of dynamical processes 
running on top of empirical, temporal networks. 
We will focus on two diffusive processes:
Chapter \ref{chap:RW} is devoted to the paradigmatic case of random walks,
while in Chapter \ref{chap:epidemic} we consider epidemic spreading,
with particular attention to the impact of immunization strategies on the spreading outbreak.
A detailed summary and discussion is reported at the end of each Chapter,
while conclusive remarks and perspectives for future work 
are drawn in Chapter \ref{chap:concl}. 

During my PhD studies I have been mainly focused on the topics 
exposed in the previous paragraph, 
and main results obtained have been published in scientific papers 
\cite{PhysRevE.85.056115, PhysRevLett.110.168701, citeulike:13344055, Starnini201389, starnini_topological_2013, 2013arXiv1312.5259S}.
In my PhD program I also studied 
the emergence of collective phenomena in social systems,
such as synchronization of Kuramoto oscillators, 
emergence of cooperation in social dilemmas and 
achievement of consensus in the voter model \cite{1742-5468-2012-10-P10027}.
Since these studies have been performed on static networks,
I decided not to include them in the present thesis. 
The emergence of collective phenomena in time-varying networks is,
however, a promising hot topic I plan to explore in future work.

\setcounter{secnumdepth}{2}

\chapter{From empirical data to time-varying networks}
\label{chap:intro}




Modern network science allows to represent and rationalize the
properties and behavior of complex systems that can be represented in
terms of a graph \cite{Newman2010,Dorogobook2010}.  
This approach has been proven very powerful, 
providing a unified framework to understand
the structure and function of networked systems \cite{Newman2010} and
to unravel the coupling of a complex topology with dynamical processes
developing on top of it
\cite{dorogovtsev07:_critic_phenom,BBV}.  
Until recently, a large majority of works in the field of network
science has been concerned with the study of \textit{static} networks,
 in which all the connections that appeared at least once coexist. This is the case, for
example, in the seminal works on scientific collaboration networks
\cite{newmancitations01}, or on movie costarring networks
\cite{Barabasi:1999}. In particular, dynamical processes have mainly
been studied on static complex networks \cite{BBV, dorogovtsev07:_critic_phenom, citeulike:13334906}. 
Presently, however, a lot of attention is being devoted to the
\textit{temporal} dimension of networked systems.
Indeed, many real networks are not static,
but are actually dynamical structures, 
in which connections appear,
disappear, or are rewired on various timescales \cite{temporalnetworksbook}.
Such temporal evolution is an intrinsic feature of many
natural and artificial networks, and can have profound consequences
for the dynamical processes taking place upon them. 
A static approximation is still valid when such time scales
are sufficiently large, such as in the case of the Internet
\cite{romuvespibook}. 
In other cases, however, this approximation is incorrect.
 This is particularly evident in the case in social
interactions networks \cite{wass94}, 
in which social relationships are represented by a
succession of contact or communication events, continuously created or
terminated between pairs of individuals. In this sense, the social
networks previously considered in the literature
\cite{newmancitations01,Barabasi:1999,liljeros2001web} represent a
projection or temporal integration of time-varying graphs, in which
all the links that have appeared at least once in the time window of the network evolution 
 are present in the projection.

The so-called ``\emph{data revolution}" has spurred the interest in the temporal dimension of social networks,
leading to the development of new tools and concepts embodied in the
rising field of temporal or time-varying networks
\cite{temporalnetworksbook}. 
The recent availability of large digital databases and the deployment
of new experimental infrastructures, indeed,
such as mobile phone communications \cite{Onnela:2007}, face-to-face interactions 
\cite{10.1371/journal.pone.0011596} or large
scientific collaboration databases \cite{2012arXiv1203.5351P},
 have made possible the real-time
tracking of social interactions in groups of individuals and the
reconstruction of the corresponding temporal networks
\cite{Oliveira:2005fk,barabasi2005origin,gonzalez2008understanding,10.1371/journal.pone.0011596}.
Specially noteworthy in this context is the data on face-to-face interactions
recorded by the SocioPatterns collaboration \cite{sociopatterns},
which performed a fine-grained measurement of human interactions in
closed gatherings of individuals such as schools, museums or
conferences.
The newly gathered empirical data poses new fundamental questions
regarding the properties of temporal networks, questions which have
been addressed through the formulation of theoretical models, aimed at
explaining both the temporal patterns observed and their effects on
the corresponding integrated networks
\cite{PhysRevE.83.056109,journals/corr/abs-1106-0288,citeulike:7974615,PhysRevLett.110.168701}.

Empirical analyses have revealed
rich and complex patterns of dynamic evolution
\cite{Hui:2005,PhysRevE.71.046119,Onnela:2007,Gautreau:2009,10.1371/journal.pone.0011596,Tang:2010,Bajardi:2011,Stehle:2011nx,Miritello:2011,Karsai:2011,temporalnetworksbook},
pointing out the need to characterize and model them
\cite{scherrer:inria-00327254,Hill:2009,Gautreau:2009,PhysRevE.81.035101,PhysRevE.83.056109}.
At the same time, researchers have started to study how the temporal evolution of the network substrate impacts 
the behavior of dynamical processes such as 
epidemic spreading \cite{Rocha:2010,Isella:2011,Stehle:2011nx,Karsai:2011,Miritello:2011,dynnetkaski2011},
synchronization \cite{albert2011sync}, percolation \cite{Parshani:2010,Bajardi:2011}
and social consensus \cite{consensus_temporal_nrets_2012}.
The alternation of available edges, and their rate of appearance, indeed,
can have a crucial effect on dynamical processes running on top of
temporal networks
\cite{Starnini201389,PhysRevLett.98.158702,hoffmann_generalized_2012, perra_random_2012}.
A key result of these efforts has been the observation of the ``\emph{bursty}'' dynamics of human interactions,
 revealed by the heavy tailed form of the 
distribution of gap times between consecutive social interactions
\cite{Oliveira:2005fk,Onnela:2007,Hui:2005,PhysRevE.71.046119,Tang:2010,10.1371/journal.pone.0011596},
contradicting traditional frameworks positing Poisson
distributed processes.
The origin of inhomogeneous human dynamics can be traced back to individual decision making and
 various kinds of correlations with one's social environment, 
while reinforcement dynamics \cite{PhysRevE.83.056109} , circadian and weekly fluctuations \cite{Jo_NJP2012} and 
 decision-based queuing process \cite{barabasi2005origin}
have been proposed as explanations for the observed bursty behavior of social systems.
Burstiness, however, turned out to be ubiquitous in nature, appearing in different phenomena, 
ranging from earthquakes \cite{Corral-2004-PRL} to sunspots \cite{Wheatland1998} 
and neuronal activity \cite{journals/biosystems/KemuriyamaOSMTKN10},
leading to the observation of a universal correlated bursty behavior,
which can be interpreted by memory effects \cite{Karsai:2012aa}.

The heterogeneous and bursty nature of social interactions 
and its impact on the processes taking place among individuals
makes evident the need of no longer neglecting the temporal dimension of social networks.
In this Chapter we present some empirical data of social dynamics,
and we introduce the time-varying networks formalism used to analyze them.
First, in Section \ref{sec:intro_timevar} 
we will give  some basic definitions and concepts useful to understand 
the behavior of temporally evolving systems,
extensively used in the following Chapters.
Then, in Section \ref{sec:intro_empirical} we will present and analyze 
several data sets of different social interactions.
We will discuss in details the fundamental properties of face-to-face contact networks,
as recorded by the SocioPatterns collaboration \cite{sociopatterns},
and characterize the main quantities that identify them.
This analysis will be useful in the next Chapters,
since the SocioPatterns data will be the benchmark for testing the behavior of the model presented in Chapter \ref{chap:f2f},
as well as the substrate for the diffusion of the dynamical processes considered in part \ref{part:dynproc}.
We will also show the temporal properties of the scientific collaboration networks \cite{newmancitations01}
and introduce the concept of activity potential, 
at the core of the definition of the activity driven model,
which constitutes the object of study of Chapter \ref{chap:activitydriven}.

\section{Time-varying networks formalism}
\label{sec:intro_timevar}

Network theory \cite{newman-review} traditionally maps complex interacting systems into graphs,
by representing the elements acting in the system as \emph{nodes},
and pairwise interactions between them as \emph{edges}.
In mathematical terms, a \emph{graph} is defined as a pair of sets $\mathcal{G} = (\mathcal{V}, \mathcal{E})$,
where $\mathcal{V} = \{ i, j, k, \ldots \}$ is a set of nodes
and   $\mathcal{E} = \{ (i,j), (j,k), \ldots \}$ is a set of edges
 connecting nodes among them.
We usually denote by $N$ the number of nodes and by $E$ the number of edges.
A graph can be \emph{directed} or \emph{undirected}, 
where the latter case implies the presence of a link going from node $i$ to node $j$ 
if and only if the reverted edge, 
from $j$ to $i$, exists in the graph.
A \emph{weighted} graph is defined by assigning to each edge $(i,j)$ a weight $w_{ij}$.
Weighted graphs are useful to encode additional information in the network representation.
A graph is \emph{connected} if all possible pairs of nodes are connected by at least one path.
A \emph{giant connected component}, useful to monitor in percolation processes, 
is the largest connected part of the graph, usually comprehensive of the majority of the nodes.    
A graph is usually represented by an \emph{adjacency matrix}, a $N \times N$ matrix defined such that 
\begin{equation}
a_{ij} =  \left\{
\begin{array}{rl}
1 & \textrm{if} \quad (i,j) \in \mathcal{E} \\
0 & \textrm{otherwise} \\
\end{array}
\right.
\end{equation}
For undirected graphs, the adjacency matrix is symmetric, $a_{ij} = a_{ji}$.

Network formalism can be easily extended in order to include temporally evolving graphs,
best known as temporal or time-varying networks  \cite{moody2002importance,temporalnetworksbook}.
In temporal networks, the nodes are defined by a static
collection of elements, and the edges represent pairwise
interactions, which appear and disappear over time.  
 The interactions can be aggregated  over a time window $\Delta t_0$, 
 corresponding to some characteristic time scale of the dynamics considered, 
 which can be tracked back to technical or natural constraints,
 or can be arbitrary defined depending on the context under study.
  For example, the time scale in scientific collaboration networks \cite{newmancitations01}
is given by the interval between consecutive edition of the journal considered, 
while in physical proximity networks \cite{10.1371/journal.pone.0011596}
is often established by experimental setup.
The choice of the aggregation window $\Delta t_0$ is an important issue which can
affect deeply the structure of the resulting time-varying networks \cite{TimeWindow2012}
and have non-trivial consequences in the study of dynamical processes 
taking place on top of them \cite{ribeiro2013quantifying}.
However, this procedure is standard in the time-varying networks fields, 
and represents a tractable and good approximation as far as 
the aggregation window is not too large \cite{ribeiro2013quantifying}.
Although the temporal networks formalism is general and suitable for 
the representation of any time-varying complex system,
in the present thesis we will deal exclusively with social interactions,
therefore, in the following, we will explicitly consider time-varying \emph{social} networks,
whose nodes and edges represent individuals and their interactions, respectively.

\subsection{Basics on temporal networks}

In order to build a time-varying network representation,
all the interactions established within the time window $\Delta t_0$ are considered as simultaneous 
and contribute to constitute a ``instantaneous'' contact network,
generally formed by isolated nodes and small groups of interacting individuals. 
Thus, temporal networks can be represented as a sequence of $T$ 
instantaneous graphs  $\mathcal{G}_t$, with $t = \{1, \ldots, T\}$, 
 with a number $N$ of different interacting individuals and 
 a total duration of $T$ elementary time steps, each one of fixed length $\Delta t_0$.
An exact representation of the temporal network is given in terms 
of a \emph{characteristic function} (or temporal adjacency matrix
\cite{Newman2010}) $\chi(i,j,t)$, taking the value $1$ when agents $i$
and $j$ are connected at time $t$, and zero otherwise.

Coarse-grained information about the structure of temporal networks
can be obtained by projecting them onto aggregated static networks,
either binary or weighted. 
This corresponds integrating in time up to $T$ the interactions between the agents, 
equivalent to taking the limit  $\Delta t_0 \rightarrow \infty$.
The binary projected network informs of the
total number of contacts of any given actor, while its weighted
version carries additional information on the total time spent in
interactions by each actor
\cite{Onnela:2007,temporalnetworksbook,Isella:2011,Stehle:2011}. 
The aggregated binary network is defined by an adjacency matrix of the
form
\begin{equation}
  \label{eq:9}
  a_{ij} = \Theta\left( \sum_{t} \chi(i,j,t) \right),
\end{equation}
where $\Theta(x)$ is the Heaviside theta function defined by
$\Theta(x) =1$ if $x>0$ and $\Theta(x)=0$ if $x\leq0$. In this
representation, the degree of vertex $i$, $k_i = \sum_j a_{ij}$,
represents the number of different agents with whom agent $i$ has
interacted. The associated weighted network, on the other hand, has
weights of the form
\begin{equation}
  \label{eq:10}
  \omega_{ij} = \sum_{t} \chi(i,j,t).
\end{equation}
Here, $\omega_{ij}$ represents the number of interactions between
agents $i$ and $j$. 
The strength of node $i$, $s_i = \sum_j \omega_{ij}$, 
represents the average number of interactions of agent $i$ at each time step.

The aggregated representation is an useful benchmark to point out the effect of
temporal correlations \cite{PhysRevE.85.056115},
and it allows to identify interesting properties of the system. 
For example, the correlation between 
 degree and strength of a node 
represents the tendency of an individual
to spend on average more or less time in interactions with the others, 
depending on the number of different peer he has seen,
as we will see in the next Section \ref{sec:intro_empirical}.
More in general, the strength of links and individuals helps not only to understand the structure of a social network, 
but also the dynamics of a wide range of phenomena involving human behavior,
 such as the formation of communities and the spreading of information and social influence \cite{Hill:2010fk,Onnela:2007,watts2007twenty}.

\subsection{Time-respecting paths}
\label{subsec:intro_paths}

The temporal dimension of any time-varying graph has a deep influence on
the dynamical processes taking place upon such structures. In the
fundamental example of opinion (or epidemic) spreading, for example, the
time at which the links connecting an informed (or infected) individual
to his neighbors appear determines whether the information (or
infection) will or will not be transmitted. In the same way, it is
possible that a process initiated by individual $i$ will reach
individual $j$ through an intermediate agent $k$ through the path $i
\rightarrow k \rightarrow j$ even though a direct connection between $i$
and $j$ is established later on. This information is lost in the time
aggregated representation of the network, where any two neighboring
nodes are equivalent \cite{temporalnetworksbook}.
 At a basic topological level, projected networks disregard the fact that dynamics on temporal
networks are in general restricted to follow \emph{time respecting  paths}
\cite{PhysRevE.71.046119,Kostakos:2009,Isella:2011,journals/corr/abs-1106-2134,Bajardi:2011,temporalnetworksbook}.
 In general, time respecting paths
\cite{PhysRevE.71.046119} determine the set of possible causal
interactions between the agents of the graph, and the state of any node
$i$ depends on the state of any other vertex $j$ through the collective
dynamics determining their causal relationship.  
  Therefore, not all the network is  available for
propagating a dynamics that starts at any given node, but only those
nodes belonging to its set of influence \cite{PhysRevE.71.046119},
defined as the set of nodes that can be reached from a given one,
following time respecting paths.  
Formally, if a contact between nodes $i$ and $j$ took place at
times ${\cal T}_{ij} \equiv
\{t_{ij}^{(1)},t_{ij}^{(2)},\cdots,t_{ij}^{(n)} \}$, it cannot be used
in the course of a dynamical processes at any time $t \not\in {\cal
  T}_{ij}$. 
Moreover, an important role can also
be played by the bursty nature of dynamical and social processes,
where the appearance and disappearance of links do not follow a
Poisson processes, but show instead long tails in the distribution of
link presence and absence durations, as well as long range
correlations in the times of successive link occurrences
\cite{barabasi2005origin,10.1371/journal.pone.0011596,Gautreau:2009,Bajardi:2011}.

For each (ordered) pair of nodes $(i,j)$, time-respecting paths from
$i$ to $j$ can either exist or not; moreover, the concept of shortest path on
static networks (i.e., the path with the minimum number of links
between two nodes) yields several possible generalizations in a
temporal network:
\begin{itemize}
\item the {\em fastest} path is the one that allows to go from $i$ to
  $j$, starting from the dataset initial time, in the minimum possible
  time, independently of the number of intermediate steps;

\item the {\em shortest} time-respecting path between $i$ and $j$ is
  the one that corresponds to the smallest number of intermediate
  steps, independently of the time spent between the start from $i$
  and the arrival to $j$.
\end{itemize}

For each node pair $(i,j)$, we denote by $l_{ij}^f$,
$l_{ij}^{s,temp}$, $l_{ij}^{s,stat}$ the lengths (in terms of the
number of hops) respectively of the fastest path, the shortest
time-respecting path, and the shortest path on the aggregated network,
and by $\Delta t_{ij}^f$ and $\Delta t_{ij}^s$ the duration of the
fastest and shortest time-respecting paths, where we take as initial
time the first appearance of $i$ in the dataset.  As already noted in
other works \cite{Kossinets:2008:SIP:1401890.1401945,Isella:2011}, $l_{ij}^f$ can be much
larger than $l_{ij}^{s,stat}$. Moreover, it is clear that $l_{ij}^f
\ge l_{ij}^{s,temp} \ge l_{ij}^{s,stat}$; from the duration point of
view, on the contrary, $\Delta t_{ij}^f \le \Delta t_{ij}^s$.

We therefore define the following quantities:
\begin{itemize}
\item $l_e $: fraction of the $N(N-1)$ ordered pairs of nodes for
  which a time-respecting path exists;
\item $\langle l_s \rangle $: average length (in terms of number of
  hops along network links) of the shortest time-respecting paths;
 
\item $ \langle \Delta t_s \rangle$: average duration of the shortest
  time-respecting paths;

\item $\langle l_f \rangle $: average length of the fastest
  time-respecting paths;

\item $\langle \Delta t_f \rangle$: average duration of the fastest
  time-respecting paths;

\item $\langle l_{s,stat} \rangle$: average shortest path length in
  the binary (static) projected network;
\end{itemize}

Note that these averaged quantities are not sufficient to determine the navigability of the temporal network, 
since the probability distribution of the 
 shortest and fastest time respecting path length and duration   
 may be broad tailed, indicating that a part of the network is difficult to reach.
A temporal network, indeed, may be topologically well connected and at the same time difficult to navigate or search. 
Spreading and searching processes need to follow paths 
whose properties are determined by the temporal dynamics of the network, 
and that might be either very long or very slow.

\section{Empirical data of social dynamics}
\label{sec:intro_empirical}

The recent availability of large amounts of data has indeed fostered the
quantitative understanding of many phenomena that 
had previously been considered only from a qualitative point of view
\cite{barabasi2010bursts,jackson2010social}. 
Examples range from human mobility patterns \cite{Brockmann:2006:Nature:16437114}
and human behavior in economic areas
\cite{radicchi2012rationality,preis2013quantifying}, 
to the analysis of political trends
\cite{adamic2005political,carpenter2004friends,lazer2011networks}. 
Together with the World-Wide Web, a wide array of technologies have also contributed to
this data deluge, such as mobile phones or GPS devices
\cite{eagle2009inferring,takhteyev2012geography,mocanu2013twitter},
radio-frequency identification devices \cite{10.1371/journal.pone.0011596}, 
or expressly designed online experiments \cite{centola2010spread}. 
The understanding of social networks  has clearly benefited from this trend \cite{jackson2010social}.
Different large social networks, such as mobile phone  \cite{Onnela:2007} 
or email  \cite{bird2006mining} communication networks, 
have been characterized in detail, 
while the rise of online social networks has provided an ideal playground 
for researchers in the social sciences \cite{huberman2009social,kwak2010twitter,ellison2007benefits}. 
The availability of data, finally, has allowed to test the validity of 
the different models of social networks that have mainly been published 
within the physics-oriented complex networks literature,
 bridging the gap between mathematical speculations and the social sciences \cite{toivonen2009comparative}.

Among the different kinds of social networks, 
a prominent position is occupied by the so-called \emph{face-to-face contact networks}, 
which represent a pivotal substrate for 
the transmission of ideas \cite{nohria2000face},
the creation of social bonds \cite{storper2004buzz}, 
and the spreading of infectious diseases \cite{liljeros2001web,salathe2010high}. 
The uniqueness of these networks stems from the fact that face-to-face conversation is considered 
the ``gold standard'' \cite{nardi2002place} of communication \cite{clark1991grounding,kiesler1984social}, 
and although it can be costly \cite{hollan1992beyond}, 
the benefits it contributes to workplace efficiency or 
in sustaining social relationships are so-far unsurpassed by 
the economic convenience of other forms of communication \cite{nardi2002place}. 
It is because face-to-face interactions bring about 
the richest information transfer \cite{doherty1997face}, for example,  
that in our era of new technological advancements 
business travel has kept increasing so steadily \cite{nardi2002place}.
 In light of all this, it is not surprising that face-to-face interaction networks have long been 
 the focus of a major attention \cite{bales1950interaction,bion2013experiences,arrow2000small},
  but the lack of fine-grained and time-resolved data represented 
  a serious obstacle to the quantitative comprehension of the dynamics of human contacts. 

\subsection{Face-to-face interactions networks}
\label{subsec:intro_sociopatterns}

Recently, the so-called data revolution pervaded also the study of human face-to-face contact networks  \cite{sociopatterns,stopczynski2014measuring}. 
Several techniques and methods, with different spatial and/or temporal resolution, 
have been used for monitoring physical proximity interactions.
Mobile phone traces \cite{gonzalez2008understanding, eagle2009inferring}
allow to monitor social relationship of a large number of individuals, 
but cannot assess face-to-face contacts, unless a specific software is provided.
Bluetooth and Wi-Fi networks \cite{scherrer:inria-00327254, Kostakos06instrumentingthe} 
have a spatial resolution of a few meters and can only guarantee the spatial proximity or co-location of individuals,
 being in general not a good proxy for a social interaction between them.
Finally, the MIT Reality Mining project \cite{eagle2006reality,LepriSRKFPSP12}
 collected rich multi-channel data on face-to-face interactions,
by deploying specifically designed ``sociometric badges".

\begin{figure}[tb]
\begin{center}
\includegraphics*[width=0.9\textwidth]{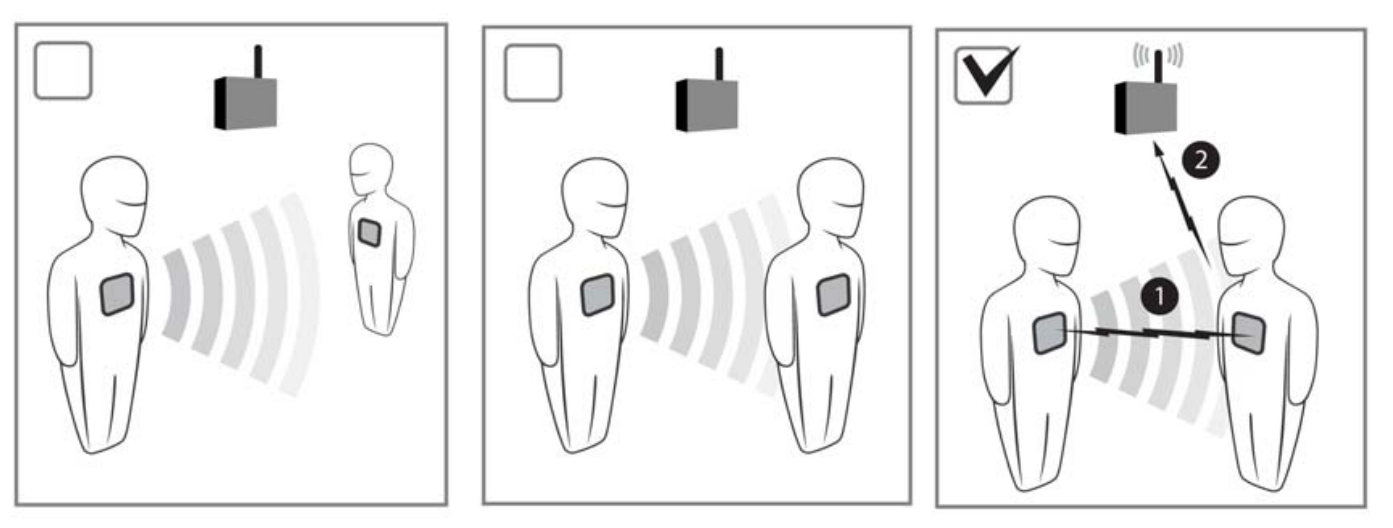}
\end{center}
\caption{ Schematic illustration of the RFID sensor system. 
	RFID tags are worn as badges by the individuals participating to the deployments.
	 A face-to-face contact is detected when two persons are close and facing each other. 
	 The interaction signal is then sent to the antenna.
	 Figure courtesy of SocioPatterns. }
  \label{fig:intro_rfid}
\end{figure}

Here we focus on the SocioPatterns collaboration \cite{sociopatterns},
which realized an experimental setup 
 merging scalability and resolution,
by means of inexpensive and unobtrusive active RFID devices.
This approach can provide data from deployments at social gatherings involving 
from few tens to several hundreds individuals,
and the spatio-temporal resolution of the devices can be tuned to probe different interaction scales,
 from co-presence in a room, to face-to-face proximity of individuals.
In the deployments of the SocioPatterns infrastructure, each individual wears
a badge equipped with an active radio-frequency identification (RFID)
device. These devices engage in bidirectional radio-communication at
very low power when they are close enough, and relay the information
about the proximity of other devices to RFID readers installed in the
environment. 
The devices properties are tuned so that face-to-face proximity (1-2 meters) of individuals 
wearing the tags on their chests can be assessed with a temporal resolution of $20$ seconds.
According to his temporal coarse-graining, two persons are considered to be ``in contact"
during an interval of 20 seconds if and only if their RFID devices 
have exchanged at least one packet during that interval. 
A schematic illustration of the sensing mechanism is shown in Fig. \ref{fig:intro_rfid}.

The empirical data collected by the SocioPatterns deployments are
naturally described in terms of temporal networks \cite{moody2002importance,temporalnetworksbook}, 
whose nodes are defined by individuals, 
and whose links represent pairwise interactions, 
which appear and disappear over time. 
As discussed in Section \ref{sec:intro_timevar}, 
in order to build temporal networks we need to aggregate 
all the interactions occurring within a time window  $\Delta t_0$,
which here is given by the temporal resolution of the deployment,
so that $\Delta t_0 = 20 $ seconds represents 
the elementary time step considered.

\begin{table}[tb]
\begin{center}
    \begin{tabular}{|c||c|c|c||c|c|c|c|c|}
     \hline
    Dataset & $N$ & $T$ & $\av{k}$ & $\overline{p}$ & $\overline{f}$ & $\overline{n}$ &$\Delta t_c$ & $\av{s}$ \\ \hline 
      25c3   &  569 & 7450 & 185 & 0.215 & 256 & 90.6 & 2.82 &6695\\
      eswc    &  173 & 4703 & 50 & 0.059 & 6.81    & 2.82 & 2.41 &370 \\
      ht         &    113 & 5093 & 39 & 0.060 & 4.06   & 1.91 & 2.13 & 366 \\
  PrimSchool  & 242 & 3100 & 69 & 0.235 & 40.8  & 25 & 1.63 &  1045\\
      sfhh           &  416 & 3834 & 54 & 0.075 & 9.15 & 27.2 & 2.96  & 502 \\
  HighSchool &  126   & 5609 & 27  & 0.069  & 5.08  & 1.95   & 2.61  & 453  \\ 
      hosp           &  84 & 20338 & 30 & 0.0485 & 2.37& 0.885 & 2.54  & 1145 \\
      \hline
      \end{tabular}
  \caption{Some average properties of the datasets under consideration.} \label{tab:summary}
\end{center}
\end{table}

In the following we present and analyze several data sets 
gathered in different social contexts: 
the European Semantic Web Conference (``eswc''), 
the 25th Chaos Communication Congress (``25c3'')
\footnote{In this particular case, the proximity detection range
  extended to 4-5 meters and packet exchange between devices was not
  necessarily linked to face-to-face proximity.}, 
    the 2009 ACM Hypertext conference (``ht"), 
a geriatric ward of a hospital in Lyon (``hosp"),  
  a primary school (``PrimSchool''),
  the 2009 congress of the
Soci\'et\'e Francaise d'Hygi\`ene Hospitali\`ere (``sfhh'')
 and  a high school (``HighSchool"). 
  A description of the corresponding contexts and various
analyses of the corresponding data sets can be found in
Refs~\cite{10.1371/journal.pone.0011596,percol,Isella:2011,Stehle:2011,Panisson:2012}.

In Table~\ref{tab:summary} we summarize the main average properties of
the datasets we are considering, that are of interest also in the context
of dynamical processes on temporal networks. 
In particular, we focus on:
\begin{itemize}
\item $N$: number of different individuals engaged in interactions;
\item $T$: total duration of the contact sequence, in units of the
  elementary time interval $\Delta t_0 = 20$ seconds;
\item $\av{k}= \sum_i k_i / N$: average degree of nodes in the
  projected binary network, aggregated over the whole data set;
\item $\overline{p}= \sum_t p(t)/T$: average number of individuals
  $p(t)$ interacting at each time step;
\item $\overline{f}= \sum_t E(t)/T= \sum_{ijt}\chi(i,j,t)/2T$: mean
  frequency of the interactions, defined as the average number of
  edges $E(t)$ of the instantaneous network at time $t$;
\item $\overline{n}= \sum_t n(t)$/2T: average number of new
  conversations $n(t)$ starting at each time step;
\item $\av{\Delta t_c}$: average duration of a contact.
\item $\av{s}= \sum_i s_i /N$: average strength of nodes in the
  projected weighted network, defined as the mean number of
  interactions per agent, averaged over all agents.
\end{itemize}

Table~\ref{tab:summary} shows the heterogeneity of the considered 
data sets, in terms of size, overall duration and contact densities. 
The contact densities, represented by the values of $\overline{p}$,
$\overline{f}$ and $\overline{n}$, are useful in order to compare and rescale 
some quantities concerning dynamical processes taking place on top of different data sets,
as we will see in part \ref{part:dynproc}. 
We note that the 25c3 data set shows a very high density
of interactions (large $\overline{p}$, $\overline{f}$ and
$\overline{n}$), due to the larger range of interaction considered (4-5 meters)
for this particular case,
 while the others are sparser. The 25c3 data set is algo the bigger in terms of size,
 thus having a larger average degree $\av{k}$.
However, even without taking into account the 25c3 data set,
all the quantities considered vary of almost an order of magnitude, 
with the exception of the average duration of contacts $\av{\Delta t_c}$, 
which is constant across the different sets.
Moreover, as also shown in the
deployments timelines in \cite{10.1371/journal.pone.0011596}, some of
the datasets show large periods of low activity, followed by bursty
peaks with a lot of contacts in few time steps, while others present
more regular interactions between elements. In this respect, it is
worth noting that we will not consider those portions of the datasets
with very low activity, in which only few couples of elements
interact, such as the beginning or ending part of conferences or the
nocturnal periods.

\begin{figure}[tb]
\begin{center}
\includegraphics*[width=0.75\textwidth]{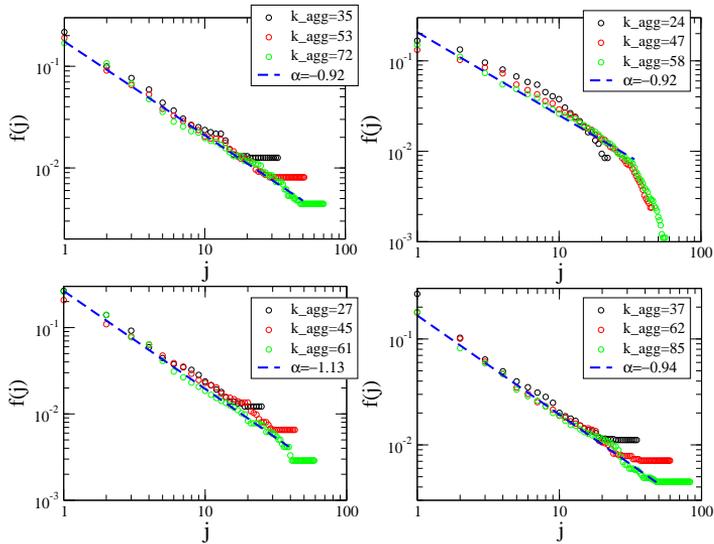}
\end{center}
\caption{ Frequency of the $j-$most met individual, $f(j)$, as a function of $j$, for 3 groups of individuals with final degree $k(T)$, for eswc (up, left), hosp (up, right), ht (down, left) and sfhh (down, right) dataset. 
    } \label{fig:zipf}
 \end{figure}

The contact patterns followed by the agents are highly heterogeneous.
To explore them, one can consider the frequency of contacts between one individual and his peers.
In this sense, one can rank the peers of each individual according to the number of times he interacts with them, 
such that for a each individual $i$, agent $j = 4$ is his forth-most-met peer.
In Fig \ref{fig:zipf} we plot the frequency distribution $f(j)$ aggregating over individuals with the same final aggregated degree $k$ (representing the total number of different agents met) of 4 different data sets.
Fig \ref{fig:zipf} shows the frequency distribution $f(j)$ for several data sets and for different final aggregated degree $k$,
finding that the probability that an individual interacts with his $j$-th most met peer 
is well approximated by the Zipf's law \cite{zipf_1949},  independent of the individual's final degree $k$:
\begin{equation}
 f(j) \sim j^{-\zeta},
 \label{eq:intro_zipf}
 \end{equation} 
 with $\zeta = 1 \pm 0.15$.
Therefore, people engage most of their interactions with few peers, 
although interacting also with many other individuals, 
met with diminished regularity.
A consequence of Eq. \eqref{eq:intro_zipf} is that the probability distribution of the frequencies of meeting different individuals, $P(f)$, turns out to be power law distributed, 
\begin{equation}
\label{eq:intro_freq}
P(f) = f ^{-\gamma}, \qquad \gamma = 1+1/\zeta \simeq 2.
\end{equation}
a behavior verified by the empirical data sets.

\begin{table}[tb]
\begin{center}
    \begin{tabular}{|c||c|c|c|c|c|c|}
     \hline
    Dataset &   $l_e$ & $\langle l_s \rangle$ & $ \langle \Delta t_s \rangle$ & $\langle l_f \rangle $ &  
$\langle \Delta t_f \rangle$ &  $\langle l_{s,stat} \rangle$ \\ \hline 
      25c3  &  0.91  & 1.67 & 1607 & 4.7  & 893 & 1.67   \\ 
      eswc  & 0.99   & 1.75 &  884 & 4.95 & 287 & 1.73   \\ 
      ht    &  0.99  & 1.67 & 1157 & 3.86 & 452 & 1.66 \\ 
 PrimSchool&  1     & 1.76 & 853  & 8.27 & 349 & 1.73  \\ 
       \hline
    \end{tabular}
  \caption{Average properties of the shortest time-respecting paths, fastest paths and shortest 
    paths in the projected network, in the datasets considered.} \label{tab:paths}
\end{center}
\end{table}

As discussed in Section \ref{subsec:intro_paths}, 
time-respecting paths are a crucial feature of any
temporal network, since they determine the set of possible causal
interactions between the actors of the graph.
Moreover, diffusion processes such as random walks or spreading
 are particularly impacted by the structure of paths between nodes,
 as we will see in part \ref{part:dynproc}.
In Table \ref{tab:paths} we report the empirical values of some quantities defined in Section \ref{subsec:intro_paths}, 
with respect to the shortest and fastest time-respecting paths between nodes, 
for some of the empirical data sets considered.
 It turns out that the great majority of pairs of
nodes are causally connected by at least one path in all
data sets. Hence, almost every node can potentially be influenced by
any other actor during the time evolution, i.e., the set of sources
and the set of influence of the great majority of the elements are
almost complete (of size $N$) in all of the considered datasets.

\begin{figure}[tb]
\begin{center}
\includegraphics*[width=0.48\textwidth]{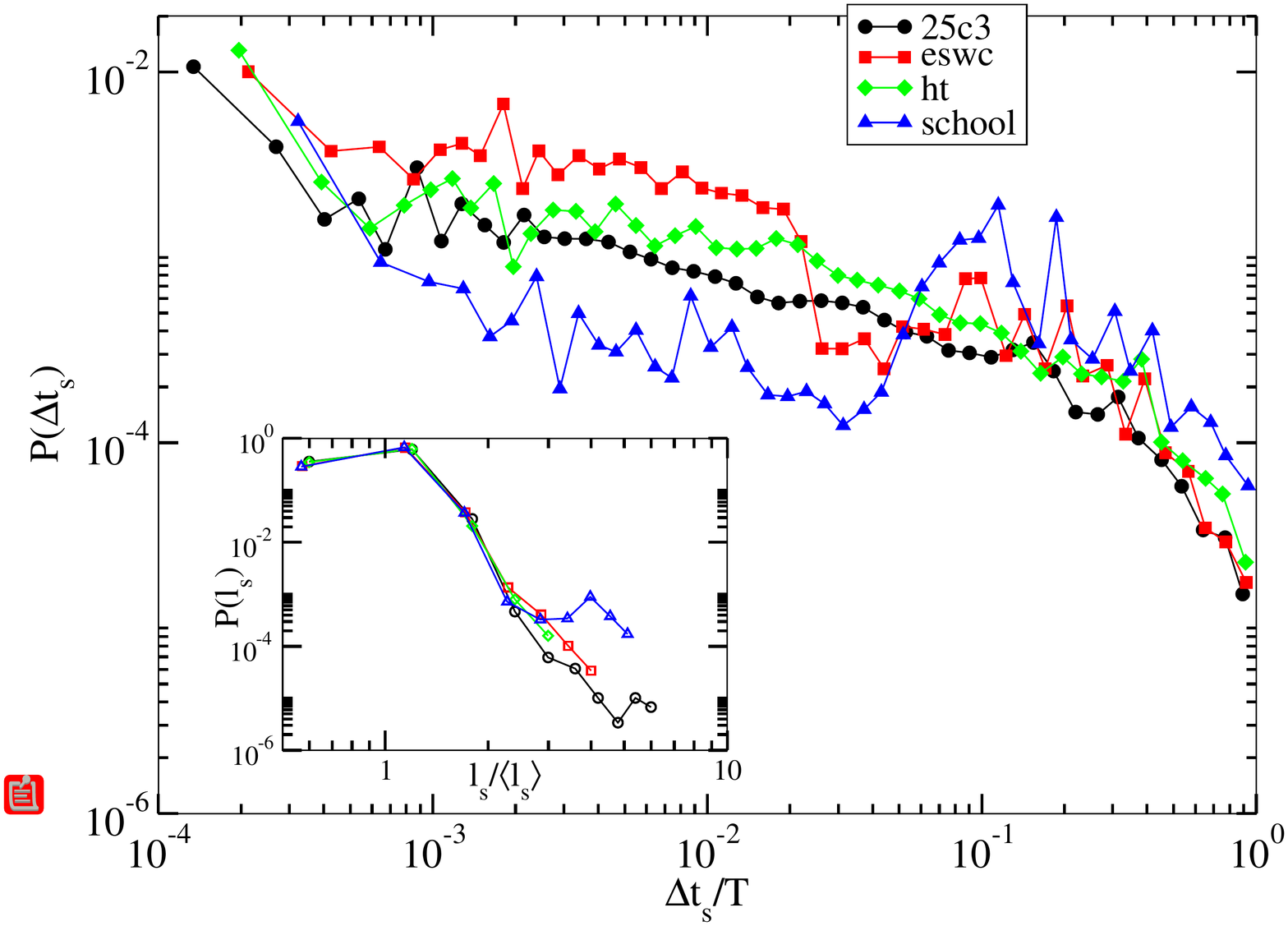}
\includegraphics*[width=0.48\textwidth]{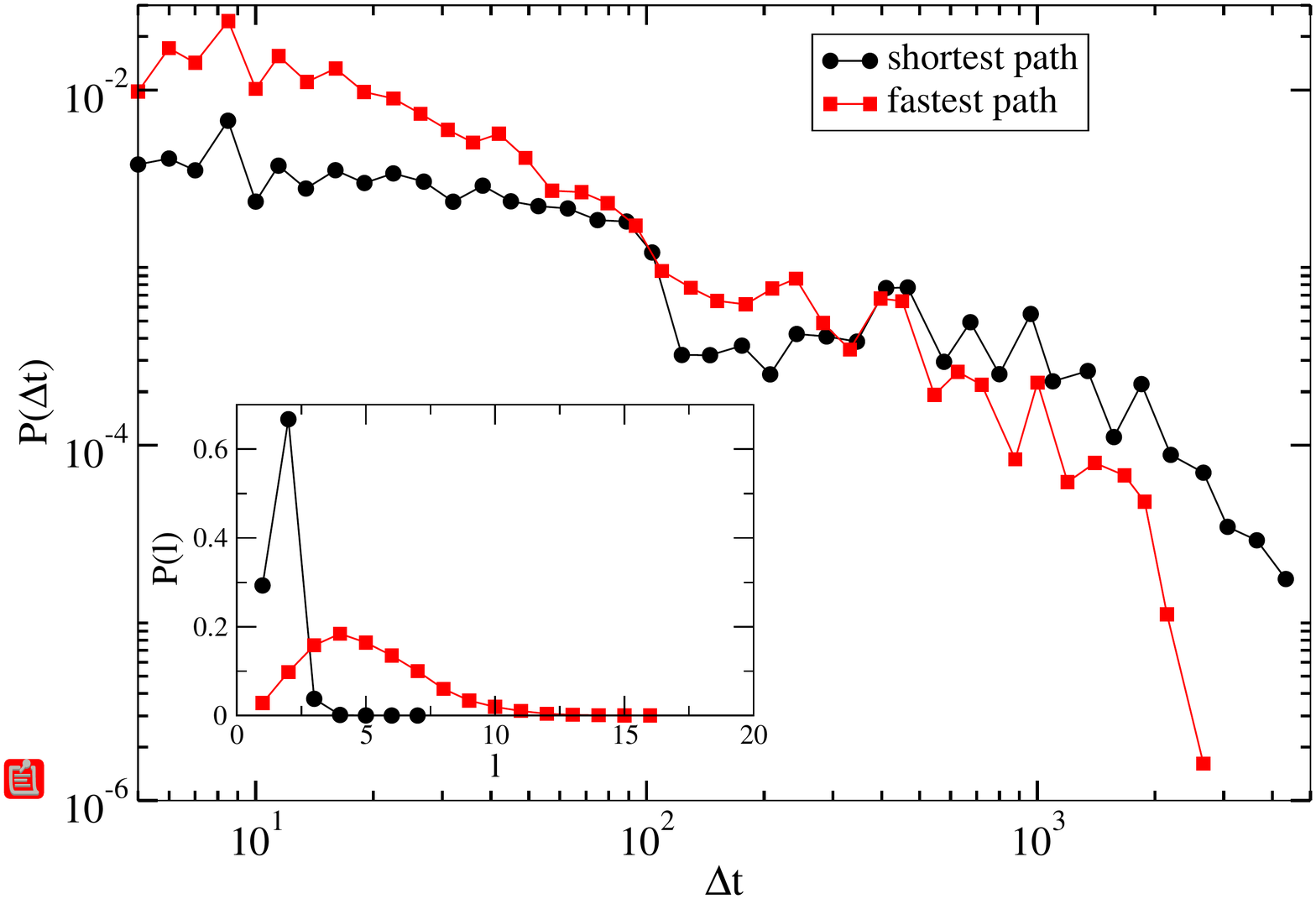}
\end{center}
\caption{ Left: Distribution of the temporal duration of the shortest
  time-respecting paths, normalized by its maximum value $T$. Inset:
  probability distribution $P(l_s)$ of the shortest path length measured
  over time-respecting paths, and normalized with its mean value
  $\langle l_s \rangle$. Note that the different datasets collapse.
  Right: Probability distribution of the duration of the shortest
  $P(\Delta t_s)$ and fastest $P(\Delta t_f)$ time-respecting paths,
  for the eswc dataset.  Inset: Probability distribution of the
  shortest $P(l_s)$ and fastest $P(l_f)$ path length for the same
  dataset.  }
  \label{fig:topodyn}
\end{figure}

In Fig. \ref{fig:topodyn} we show the distributions of the lengths,
$P(l_s)$, and durations, $P(\Delta t_s)$, of the shortest
time-respecting path for different datasets.  In the same Figure we
choose one dataset to compare the $P(l_s)$ and the $P(\Delta t_s)$
distributions with the distributions of the lengths, $P(l_f)$, and
durations, $P(\Delta t_f)$, of the fastest path.  The $P(l_s)$
distribution is short tailed and peaked on $l=2$, with a small average
value $\langle l_s \rangle $, even considering the relatively small
sizes $N$ of the datasets, and it is very similar to the projected
network one $\langle l_{s,stat} \rangle $ (see Table~\ref{tab:paths}).
The $P(l_f)$ distribution, on the contrary, shows a smooth behavior,
with an average value $\langle l_f \rangle $ several times bigger than
the shortest path one, $\langle l_s \rangle $, as expected
\cite{Kossinets:2008:SIP:1401890.1401945,Isella:2011}. 
 Note that, despite the important
differences in the datasets characteristics, the $P(l_s)$
distributions (as well as $P(l_f)$, although not shown) collapse, once
rescaled.  On the other hand, the $P(\Delta t_s)$ and $P(\Delta t_f)$
distributions show the same broad-tailed behavior, but the average
duration $\langle \Delta t_s \rangle$ of the shortest paths is much
longer than the average duration $\langle \Delta t_f \rangle$ of the
fastest paths, and of the same order of magnitude than the total
duration of the contact sequence $T$.
Thus, a temporal network may be topologically
well connected and at the same time difficult to navigate or
search. Indeed spreading and searching processes need to follow paths
whose properties are determined by the temporal dynamics of the
network, and that might be either very long or very slow.


\begin{figure}[tb]
\begin{center}
\includegraphics*[width=0.48\textwidth]{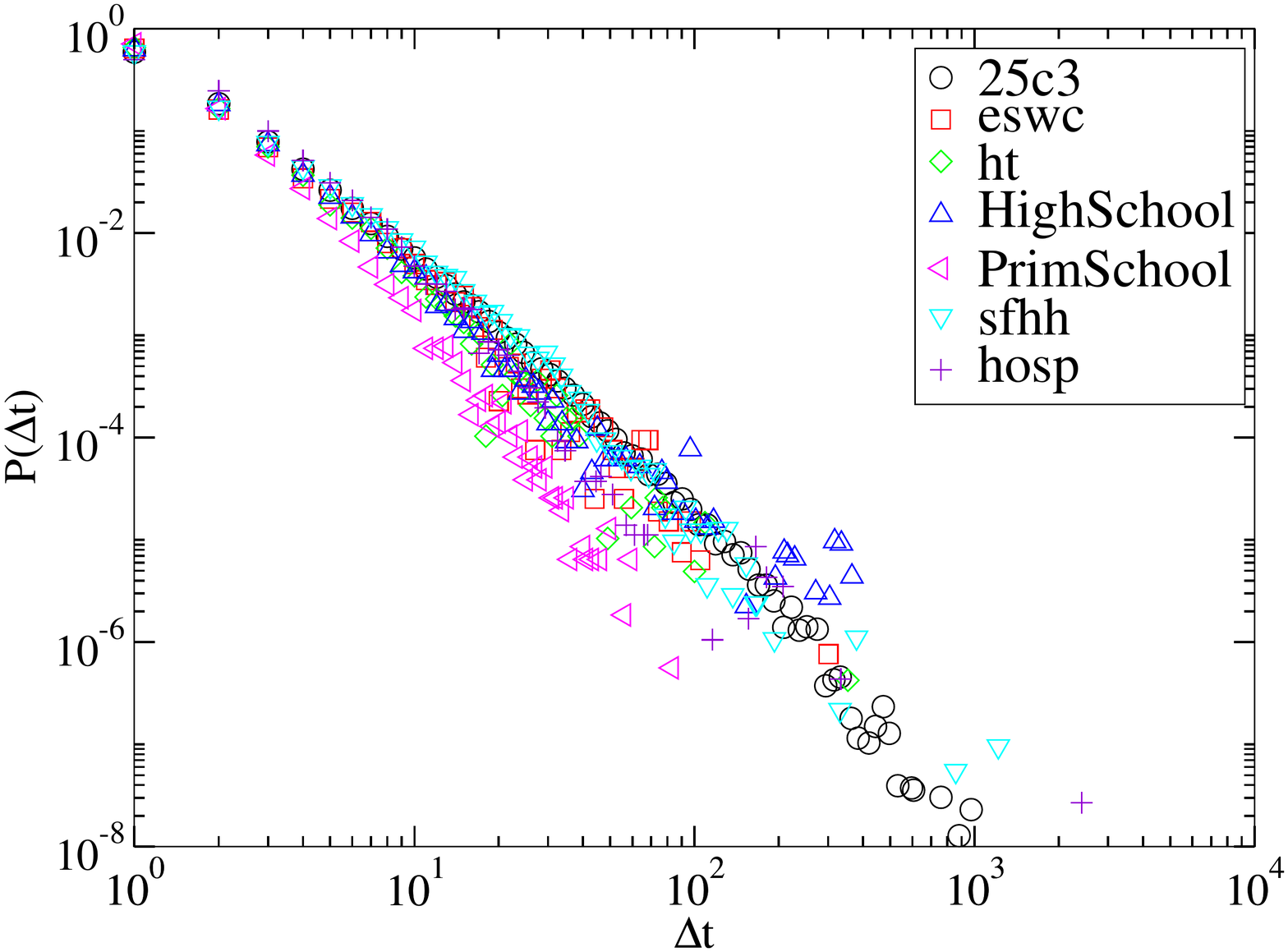}
\includegraphics*[width=0.48\textwidth]{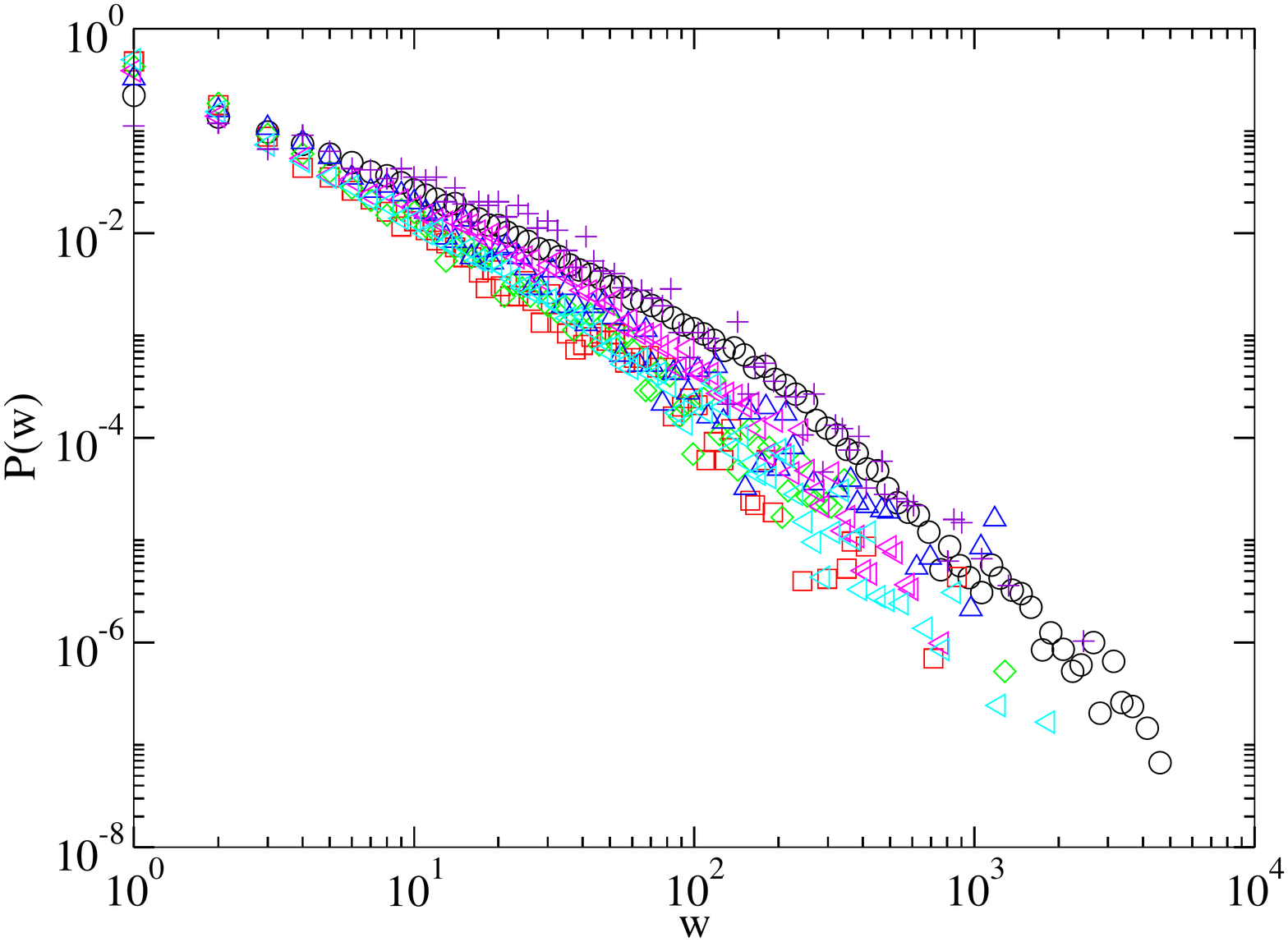}
\end{center}
\caption{
 Probability distribution of the duration $\Delta t$ of 
 the contacts between individuals,  $P(\Delta t)$, (left) 
 and the weight $w_{ij}$ representing the cumulative time spent in interaction by 
 pair of agents $i$ and $j$, $P(w)$, (right).
 } \label{fig:emp_dur}
 \end{figure}
 
The heterogeneity of the contact patterns of the
face-to-face interactions~\cite{10.1371/journal.pone.0011596} is
revealed by the study of the distribution of the duration $\Delta t$
of contacts between pairs of agents, $P(\Delta t)$.
The correspondent aggregated quantity is 
the total time spent in contact by a pair of agents,
represented in the temporal network by the weight $w$.
Fig. \ref{fig:emp_dur} show that both distributions $P(\Delta t)$ and $P(w)$ are  heavy-tailed, 
typically compatible with power-law behaviors,
with exponents $\gamma_{\Delta t} \simeq 2.5$ and  $\gamma_{w} \simeq 2.$
for $P(\Delta t)$ and $P(w)$, respectively.
This means that there are comparatively few long-lasting contacts and a multitude of brief contacts,
but the duration of the interactions does not have a characteristic time scale, 
as also found in other works based in Bluetooth technology \cite{scherer1978personality, clauset2007}.

Remarkably, despite the settings and contexts where the experiment took place 
are very diverse, different data sets display a similar behavior, 
showing a nice collapse of the different curves. 
Concerning the duration distribution $P(\Delta t)$, 
the only exception is the data set ``PrimSchool", which decays more rapidly than the others.
Specially noteworthy is the fact that also the ``25c3" data set, 
recorded with a larger detection range, displays a close behavior,
 meaning that the spatial scale of the interactions is not a discriminating signature of the observed dynamics.
The collapse in the weight distribution $P(w)$ is less striking,  
with the ``25c3" and ``hosp" data sets deviating slightly with respect to the other,
a fact due to their larger duration $T$.

 \begin{figure}[tb]
\begin{center}
\includegraphics*[width=0.48\textwidth]{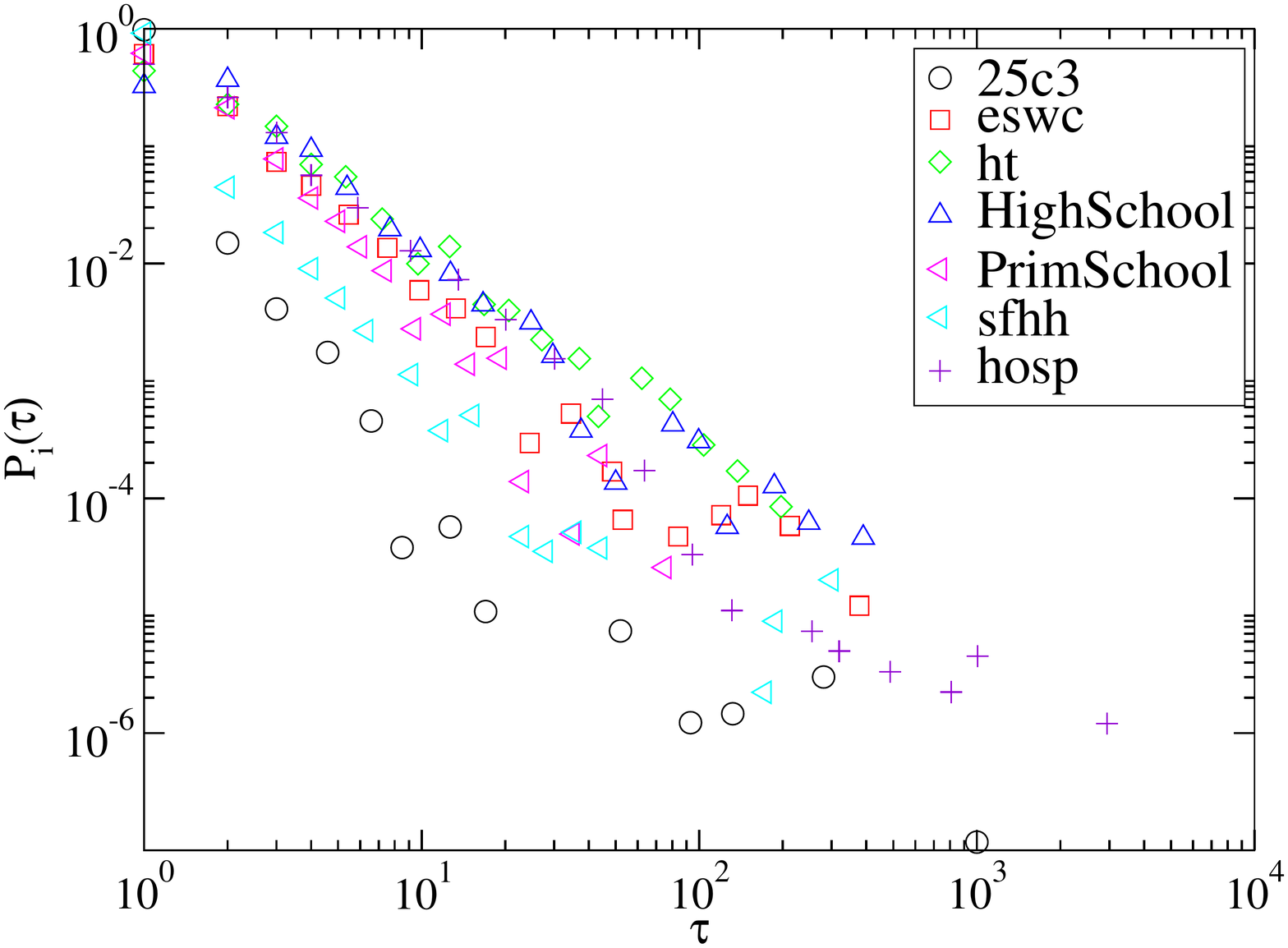}
\includegraphics*[width=0.48\textwidth]{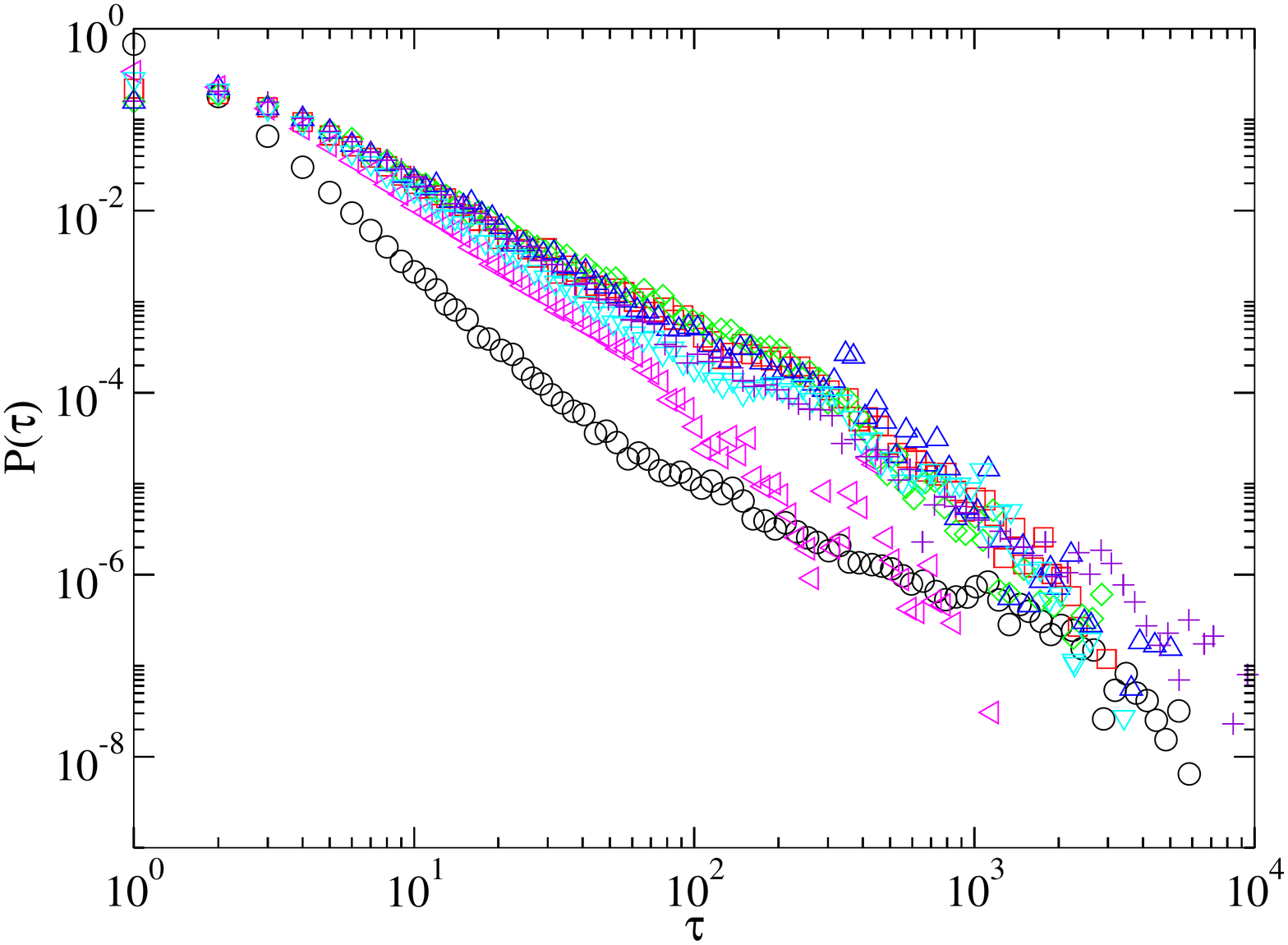}
\end{center}
\caption{
 Probability distribution of the gap times $\tau$ between consecutive interactions 
 of a single individual $i$, $P_i(\tau)$ (left),
 and aggregating over all the individuals, $P(\tau)$, (right).
 In the case of $P_i(\tau)$, we only plot the gap times
  distribution of the agent which engages in the largest number of
  conversation, but the other agents exhibit a similar behavior.
    } \label{fig:emp_burst}
 \end{figure}

 The burstiness of human interactions \cite{barabasi2005origin} is revealed by 
 the probability distribution of the interval $\tau$ between two consecutive contacts
 involving a common individual and two distinct persons, $P(\tau)$.
 In other words, if agent A starts a contact with agent B at time $t_{AB}$, 
 and later starts a different contact with agent C at $t_{AC}$, 
 the inter-contact interval is defined as $\tau = t_{AC} - t_{AB}$.
As we will see in part \ref{part:dynproc},
 measuring this quantity is relevant for the study of causal processes (concurrency) 
 that can occur on the temporal network, 
 such as diffusion processes.
 The inter-contact intervals, indeed, determine the timescale after which 
 an individual receiving some information or disease is able to propagate it to another individual. 
 Thus, the interplay between this timescale and
  the typical timescales of the spreading processes is crucial to diffusion processes.
In Fig. \ref{fig:emp_burst} we plot the inter-contact distribution of a single individual, $P_i(\tau)$, (left)
and considering all the individuals, $P(\tau)$, (right).
The broad form of both distributions indicate the absence of a characteristic timescale. 
We note that the inter-contact distribution of a single individual, $P_i(\tau)$, varies considerably depending on the data set,
while the the inter-contact distribution aggregated over all individuals, $P(\tau)$, 
shows an interesting collapse of most of the data sets, 
with the clear exceptions of the ``25c3" and ``PrimSchool" data sets.

 \begin{figure}[tb]
\begin{center}
\includegraphics*[width=0.48\textwidth]{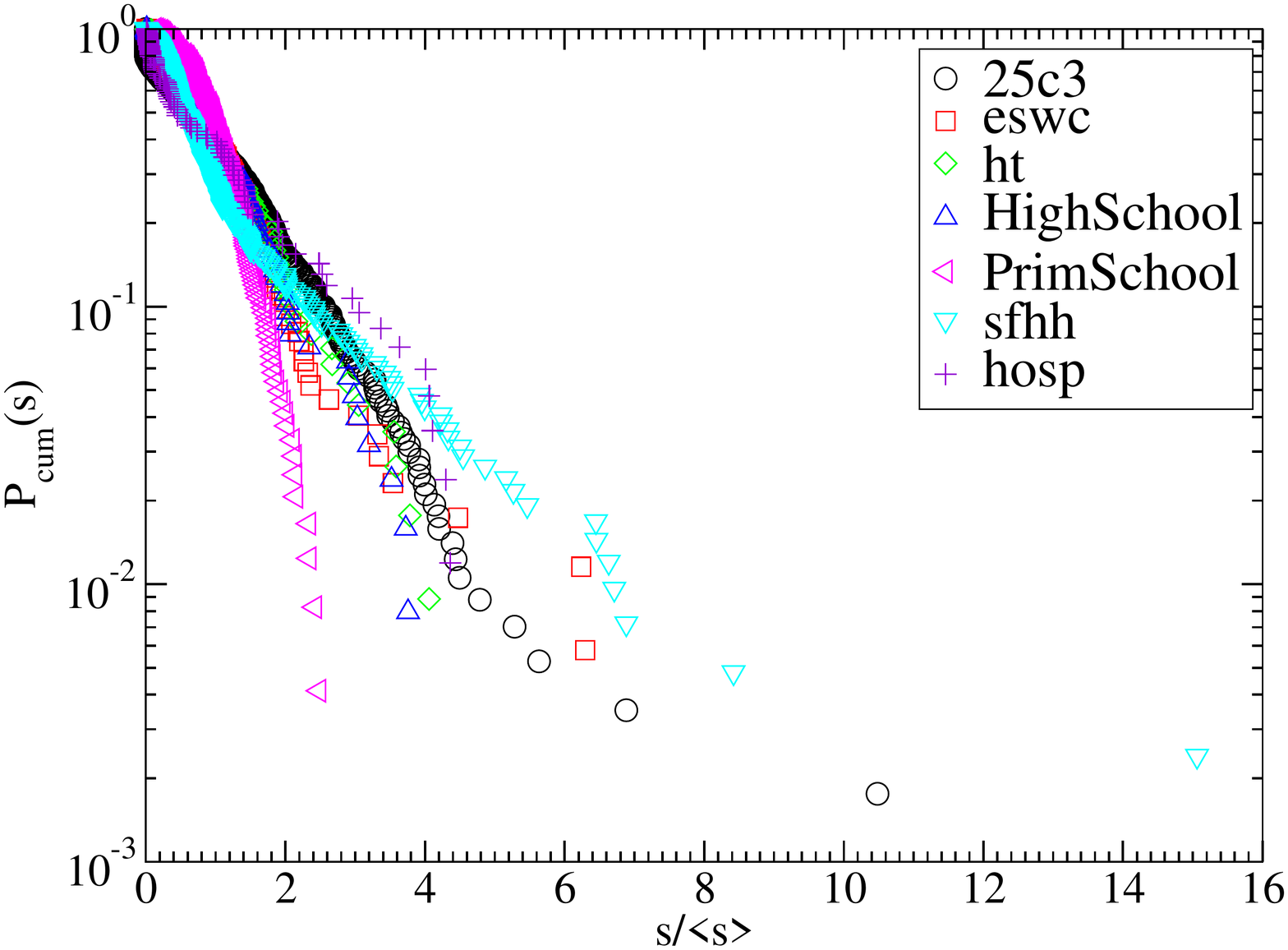}
\includegraphics*[width=0.48\textwidth]{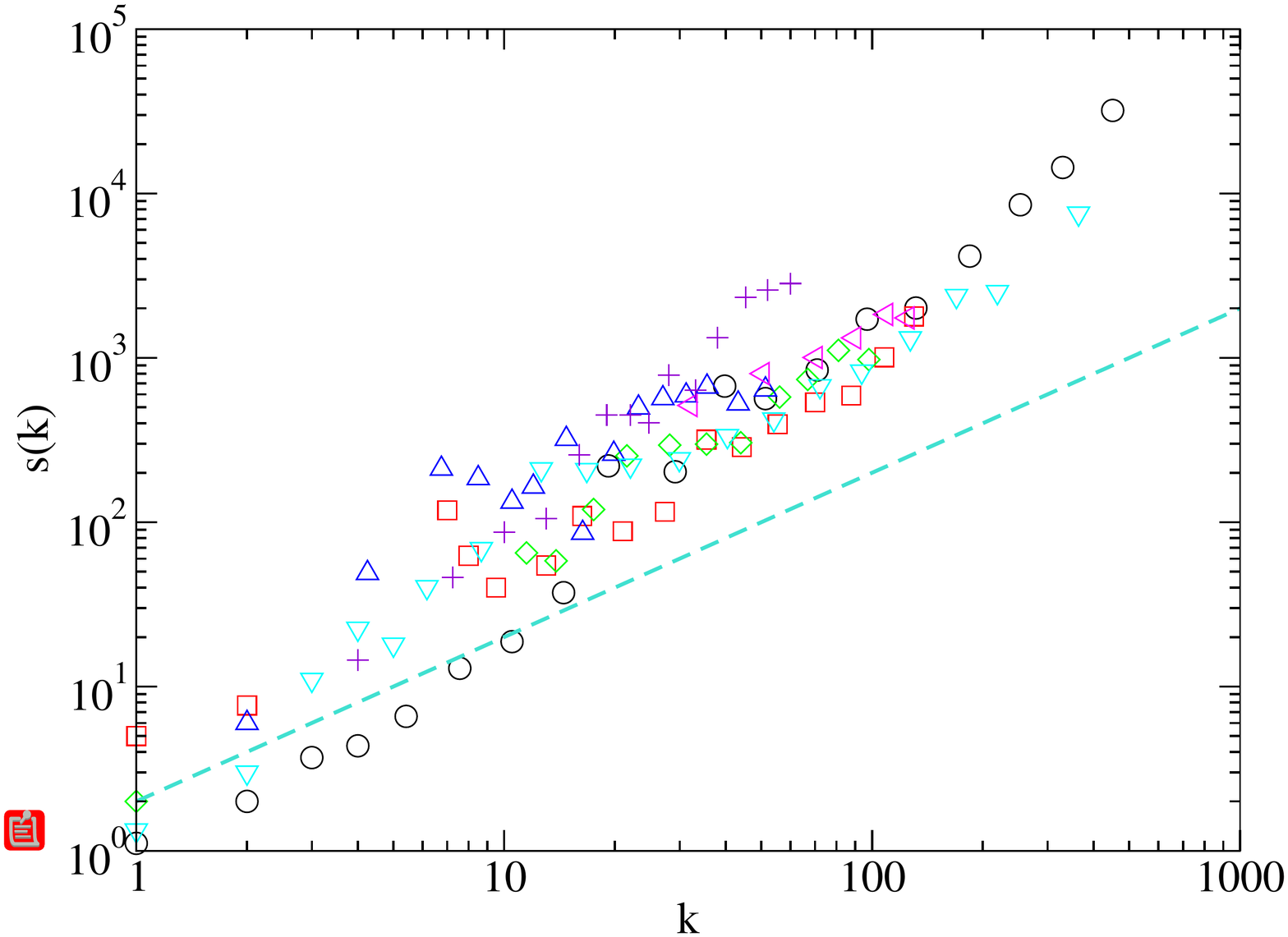}
\end{center}
\caption{  Cumulative probability distribution of the rescaled strength $s/ \av{s}$, $P_{\rm{cum}}(s)$, (left)
	and correlation between the degree and the strength of each node, $s(k)$ (right).
	In dashed line is plotted a linear correlation between strength and degree.
    } \label{fig:emp_s}
 \end{figure}

From the point of view of time-aggregated networks,
it is interesting to look and the probability distribution of the strength $s$, 
representing the total time spent in interaction by an individual.
In Fig. \ref{fig:emp_s} (left) we plot the cumulative strength distribution $P_{\rm{cum}}(s)$, 
representing the probability of finding an individual with rescaled strength greater than the $s/ \av{s}$.
One can see that the $P_{\rm{cum}}(s)$ seems to be compatible with an exponential decay,
although the small sizes of the datasets under consideration 
do not permit to establish the functional form of $P(s)$ accurately.
However, some nodes can have a very large strength, up to 5 times the average strength $\av{s}$,
in particular for the ``25c3" and ``sfhh" data sets.

Face-to-face networks can be further characterized by looking at the
correlation between the number of different contacts and the temporal
duration of those contacts. 
In temporal networks language, these correlations are estimated by
measuring the strength $s_i$ of a node $i$, representing the cumulative
time spent in interactions by individual $i$, 
as a function of its degree $k_i$, representing the total different
agents with which agent $i$ has interacted.
Fig.~\ref{fig:model_aggr} (right) shows the growth of the average strength
of nodes of degree $k$, $s(k)$, as a function of $k$ in various empirical
datasets. As one can clearly see, different data sets show a similar behavior 
and can be fitted by a power law function $s(k) \sim k^{\alpha}$ with $\alpha > 1$.
 This super-linear behavior implies that on average the nodes with high
degree are likely to spend more time in each interaction with respect
to the low-connected individuals \cite{10.1371/journal.pone.0011596}.
On the contrary, in mobile phone activity \cite{RefWorks:327} it has been reported 
a sub-linear relation between number of different contacts and time spent in interactions.
The observed phenomenon points out the presence of super-connected individuals, 
that not only engage in a large number of distinct interactions, 
but also dedicate an increasingly larger amount of time to such interactions. 
These highly social individuals may have a crucial impact in the pattern of spreading phenomena \cite{anderson92}.
 
To sum up, the empirical data of face-to-face interactions recorded by the 
SocioPatterns collaboration are naturally represented in terms of temporal networks,
and exhibit heterogeneous and bursty behavior,
 indicated by the long tailed distributions for 
 the lengths and strength of conversations, 
 as well as for the gaps separating successive interactions. 
We have underlined the importance of considering not
only the existence of time preserving paths between pairs of nodes,
but also their temporal duration: shortest paths can take much longer
than fastest paths, while fastest paths can correspond to many more
hops than shortest paths. 
Remarkably, although the data sets are collected in very diverse social contexts, 
the appropriate rescaling of the quantities considered, when necessary,
 identifies universal behaviors shared across  the different data sets considered.
These features call for a twofold effort:
On the one hand, a modeling attempt, 
able to capture the main statistical regularities exhibited by empirical  data,
and on the other hand, a study of the behavior of dynamical processes running on top of 
temporal networks constituted by the same empirical data.
These two directions will be both explored in the next Chapters.


\subsection{Scientific collaboration networks}
\label{subsec:intro_APS}

\begin{figure}[tb]
\begin{center}
\includegraphics*[width=0.48\textwidth]{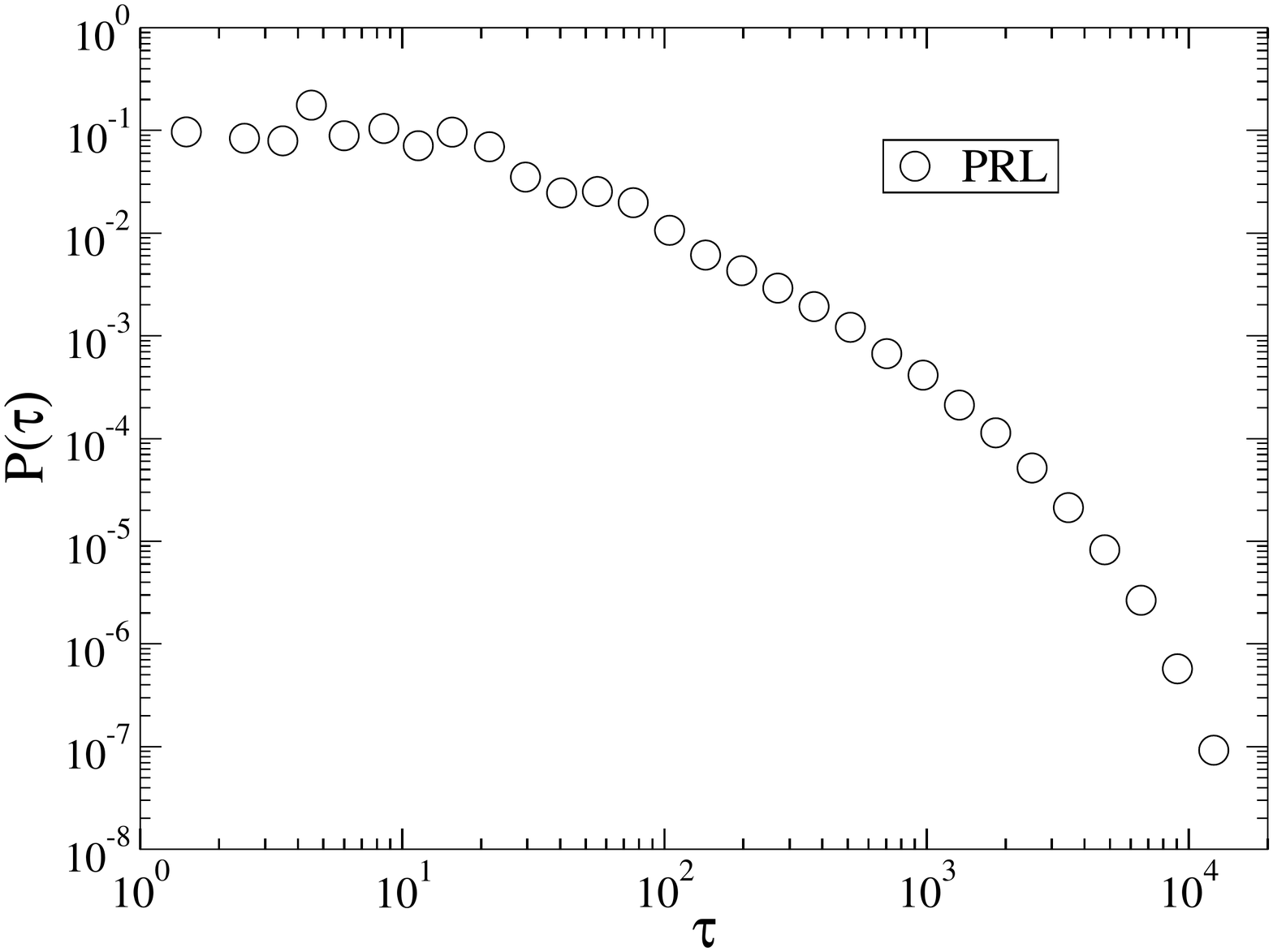}
\includegraphics*[width=0.48\textwidth]{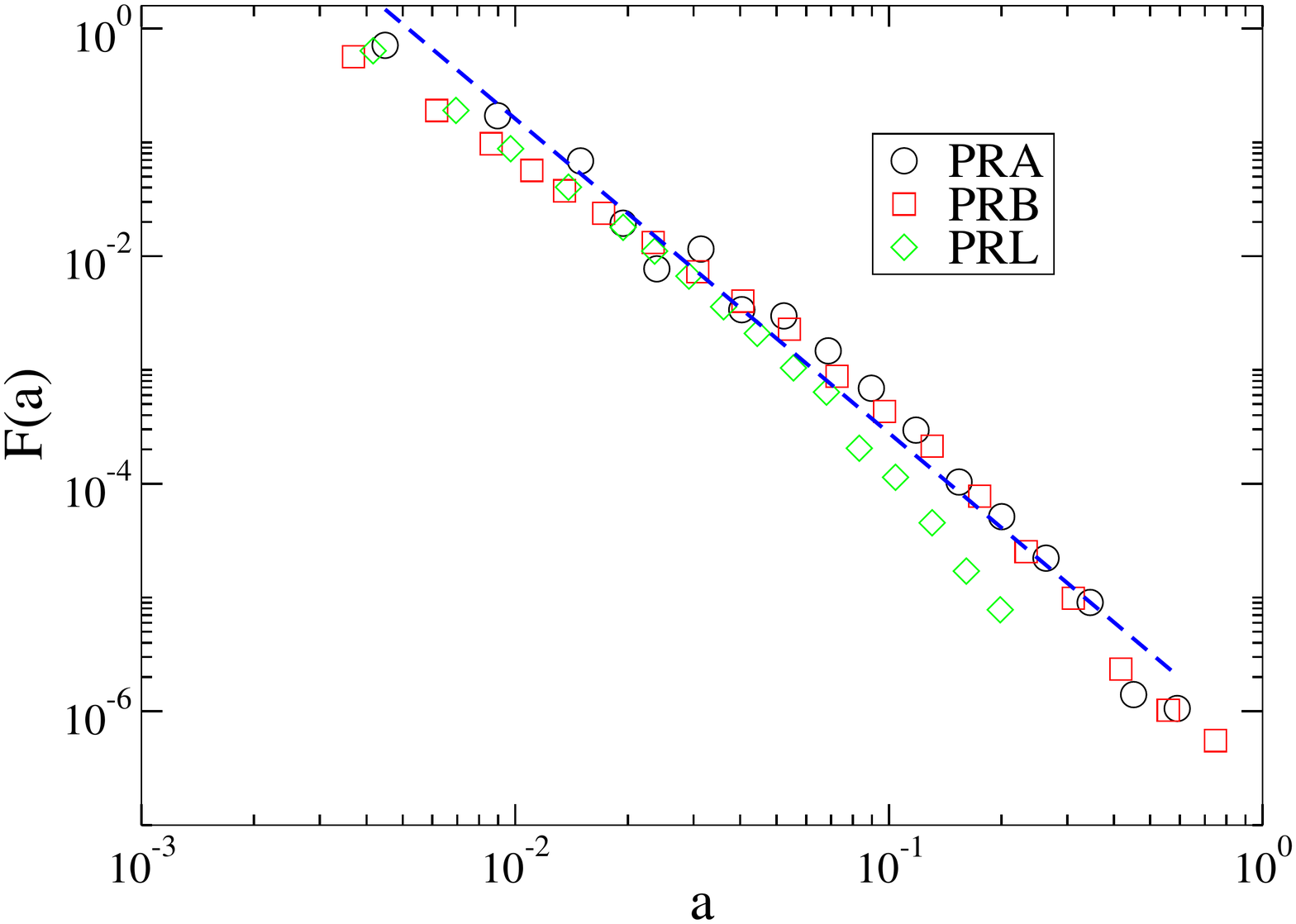}
\end{center}
\caption{ 
	 Left: Probability distribution $P(\tau)$ of the gap times $\tau$ between consecutive publications 
	  by the same author in PRL, in collaboration with one or more colleagues.
	  Right: Probability distribution of the activity of the agents, $F(a)$,
	  measured as number of papers written in unit of time,
	 in the scientific collaboration network, for different journals considered.
	 In dashed line we plot a power law distribution with exponent $\gamma = -2.7$.
	     } \label{fig:emp_APS}
 \end{figure}

Digital traces of human activity allow to grasp social behaviors 
and reconstruct the network of social interactions  
including the temporal dimension.
One of the main drawbacks of the face-to-face contact networks 
presented in the previous Section \ref{subsec:intro_sociopatterns}
is the relative small sizes of the systems considered, 
which can hardly reach one thousand individuals, due to technical and economic constraints.
Large databases of social interactions, on the contrary, 
such as mobile phone communications \cite{gonzalez2008understanding},
online social networks \cite{takhteyev2012geography}
or scientific collaboration data \cite{newmancitations01},
are cheap to collect and present the advantage of scaling up to hundreds of thousands individuals.
A scientific collaboration network, for example, 
can be easily reconstructed by using data drawn from databases of scientific publications,
such as the American Physics Society (APS) \cite{apsdata}.
In this simple network, two scientists are considered connected if they have authored a paper together. 
In the past, the fundamental work by Newman \cite{newmancitations01} showed that 
the scientific collaboration networks display scale free degree distribution and small world properties.
These networks, however, present a temporal component that can be exploited,
since the sequence of editions of a scientific journal constitutes a time-varying network, 
in which each instantaneous snapshot is formed by 
the connections between the authors who published together in the same issue of the journal.
In this case, the time scale is fixed and one unit of time corresponds to
 the interval between two consecutive issues of the journal considered,
 Physical Review Letters (PRL), for example, is weekly edited and therefore $\Delta t_0 = 1$ week.
 PRL was published for the first time in 1958, thus the time-varying network obtained from APS data
 has a total duration of more than 2700 time steps. 
The analysis presented in the previous Section, 
regarding the heterogeneity and burstiness of the temporal network and
the topological properties of the corresponding aggregated network,
can be repeated as well, leading to qualitatively similar results.
Fig. \ref{fig:emp_APS} (left), for example, shows the distribution of gap times $\tau$ 
between  two consecutive publications by the same author in PRL, $P(\tau)$.
As one can see, the inter-event time distribution is broad tailed, 
with the gap times ranging from one week up to almost twenty years.

The large sizes of the scientific collaboration networks
increase the statistical significance of the probability distribution of 
the structural and temporal properties considered,
and therefore allow to study other features of the corresponding time-varying networks.
It is possible, for example, measuring the \emph{activity} of the individuals involved in the social interactions,
representing their inclination to engage in a social act with other peers \cite{2012arXiv1203.5351P}.
The \textit{activity potential} $a_i$ of agent $i$ can be defined as the probability per unit time that 
he engages in a social interaction. In the case of the scientific collaboration networks, 
it corresponds to the number of papers written in collaboration with colleagues
in a given time window \cite{2012arXiv1203.5351P}.
In Fig. \ref{fig:emp_APS} (right) we plot the probability distribution of the activity potential, $F(a)$, 
measured in scientific collaboration networks defined by different journals, 
such as Physical Review Letters (PRL), Physical Review A (PRA), or Physical Review B (PRB),
with data drawn from the APS. 
 Fig. \ref{fig:emp_APS} (right) shows that the activity distribution $F(a)$ is broad tailed 
 and compatible with a power law form, with an exponent close to $\gamma = - 2.7$, 
 regardless of the journal considered.


The concept of activity potential is ubiquitous in social networks,
and can be applied to different kinds of social interactions. 
In  Ref. \cite{2012arXiv1203.5351P}, the authors analyzed three different empirical sets of data: 
Scientific collaborations in the journal ``Physical Review Letters", 
messages exchanged over the Twitter microblogging network, 
and the activity of actors in movies and TV series as recorded in the Internet Movie Database (IMDb). 
In the Twitter network the activity potential represents the number of messages exchanged between the users, 
while in the IMDb co-starring network two actors engage in a social act if they participate in the same movie.
It is possible to show \cite{2012arXiv1203.5351P} 
that the activity distribution is broad tailed and, importantly,
independent of the time window considered in the activity potential definition.
This observation is crucial for the definition of the \emph{activity driven} model,
 proposed by Ref. \cite{2012arXiv1203.5351P},
in which a time-varying network is built on the basis of the different activity potential of the nodes.
We will discuss in detail the activity driven model in Chapter \ref{chap:activitydriven}.



\part{Models of social dynamics}
\label{part:modeling}

Network modeling is crucial in order to identify statistical regularities 
and structural principles common to many complex systems. 
It has a long tradition in graph theory \cite{erdos59,molloy95,springerlink:10.1007/BF02294547},
with the class of growing network models \cite{Barabasi:1999,mendes99} deserving a special mention for its success, 
which has spilled over different fields \cite{mendesbook}.
Recently, the interest towards the temporal dimension of the
network description has blossomed. 
The analysis of empirical data on several types of human interactions
(corresponding in particular to phone communications or physical
proximity) has unveiled the presence of complex temporal patterns in
these systems
\cite{Hui:2005,PhysRevE.71.046119,Onnela:2007,10.1371/journal.pone.0011596,
Tang:2010,Stehle:2011nx,Miritello:2011,Karsai:2011,temporalnetworksbook},
as discussed in Chapter \ref{chap:intro}.
These findings have raised 
the necessity to outrun 
the traditional network modeling paradigm, 
rooted in the representation of the aggregated network's topology,
regardless of the instantaneous dynamics responsible of its shape. 
Efforts in temporal networks modeling range from
 social interactions \cite{PhysRevLett.110.168701, PhysRevE.81.035101} 
and mobility \cite{scherrer:inria-00327254}   to
air transportation \cite{Gautreau:2009}, 
and are based in mechanisms such as dynamic centrality \cite{Hill:2009}, 
 reinforcement dynamics \cite{PhysRevE.83.056109}
or memory \cite{Karsai:2014aa}.

The modeling effort presented in this part of the Thesis has twofold nature:
First and foremost, it is aimed to reproduce the fundamental properties of the empirical data, 
which have a deep impact on the dynamical processes taking place on top of them.
Secondly, it calls for developing a theoretical framework,
 in order to find analytic expression of the main quantity involved in the model, when possible. 
In particular, in this part we will focus on two models of social dynamics,
concerning different fields of social interactions.
On the one hand, in Chapter \ref{chap:f2f} we will devote our attention to \emph{human contact networks}.
We present and analyze a model, based in the social attractiveness of individuals,
capable to reproduce most of the main properties showed by face-to-face interactions data. 
On the other hand, in Chapter \ref{chap:activitydriven} we will focus on \emph{activity driven networks}.
The activity driven model is aimed to bridge the gap between 
the well-known topological properties of real social networks
and the microscopic mechanisms yielding the observed topology.
Its simplicity allows some analytic treatment, 
and we will derive both the topological properties and the percolation properties 
of the time-integrated networks.

\chapter{Human contact networks}
\label{chap:f2f}


Uncovering the patterns of human mobility \cite{gonzalez2008understanding} and social
interactions \cite{jackson2010social} is pivotal to decipher the
dynamics and evolution of social networks \cite{Newman2010}, with wide
practical applications ranging from traffic forecasting to epidemic
containment.  Recent technological advancements have made possible the
real-time tracking of social interactions in groups of individuals, at
several temporal and spatial scales. This effort has produced large
amounts of empirical data on human dynamics, concerning letter
exchanges \cite{Oliveira:2005fk}, email exchanges \cite{barabasi2005origin},
mobile phone communications \cite{gonzalez2008understanding}, or spatial mobility
\cite{Brockmann:2006:Nature:16437114}, among others.
The bursty dynamics of human interactions has a deep impact on the
properties of the temporally evolving networks defined by the patterns
of pair-wise interactions \cite{temporalnetworksbook}, as well as on the behavior
of dynamical processes developing on top of those dynamical networks
\cite{PhysRevE.85.056115,Stehle:2011nx,Karsai:2011,Lee:2010fk,Parshani:2010,albert2011sync,PhysRevLett.98.158702}.
A better understanding of these issues calls for new models,
capable to reproduce the bursty character of social interactions and
trace back their ultimate origin, beyond considering their temporal
evolution \cite{2012arXiv1203.5351P}. Previous modeling efforts mostly
tried to connect the observed burstiness to some kind of cognitive
mechanisms ruling human mobility patterns, such as a reinforcement
dynamics \cite{PhysRevE.83.056109}, cyclic closure
\cite{journals/corr/abs-1106-0288} or preferential return rules
\cite{citeulike:7974615}, or by focusing on the relation between
activity propensity and actual interactions \cite{2012arXiv1203.5351P}.


In this Chapter we present and analyze a simple model able to replicate
most of the main statistical regularities exhibited by human
face-to-face contact networks data. 
Avoiding any \textit{a priory} hypothesis on human mobility and
dynamics, we assume that agents perform a random walk in space
\cite{hughes} and that interactions among agents are
determined by spatial proximity \cite{consensus_temporal_nrets_2012}.  
The key insight of the model is the
suggestion that 
individuals have different degrees of social appeal or
\textit{attractiveness}, 
due to their social status or the role they
play in social gatherings, as observed in many social
\cite{citeulike:7631686}, economic \cite{hierarchy} and natural
\cite{Sapolsky05} communities. 
This insight is implemented by allowing individuals, each
characterized by an intrinsic social \textit{attractiveness}, to wander
randomly in a two dimensional space---representing the simplified
location of a social gathering---until they meet someone, at which point
they have the possibility of stopping and starting a ``face-to-face''
interaction. 
Without entering into a precise definition of
attractiveness, we adopt here an operative approach: Attractive
individuals are more likely to make people stop around them, but they
are also more prone to abandon their interactions if these are initiated
by less attractive agents.
We will see that these simple assumptions, and the asymmetry of the
interactions that they imply, are sufficient to reproduce quantitatively
the most important features of the empirical data on contact
networks.

The Chapter is structured as follows: 
 Section \ref{sec:model_def} defines in detail the model,
 while in Section \ref{sec:model_prop} we compare the 
 model behavior with respect to the properties shown by empirical data.
 In Section~\ref{sec:model_robust} we show the model
robustness concerning the variation of the main parameters
involved. Finally, Section~\ref{sec:model_discussion} is devoted to
discussion, with particular attention to the crucial role of social
attractiveness in the model.

\section{A model of social attractiveness}
\label{sec:model_def}

\begin{figure}[tb]
  \includegraphics*[width=0.95\linewidth]{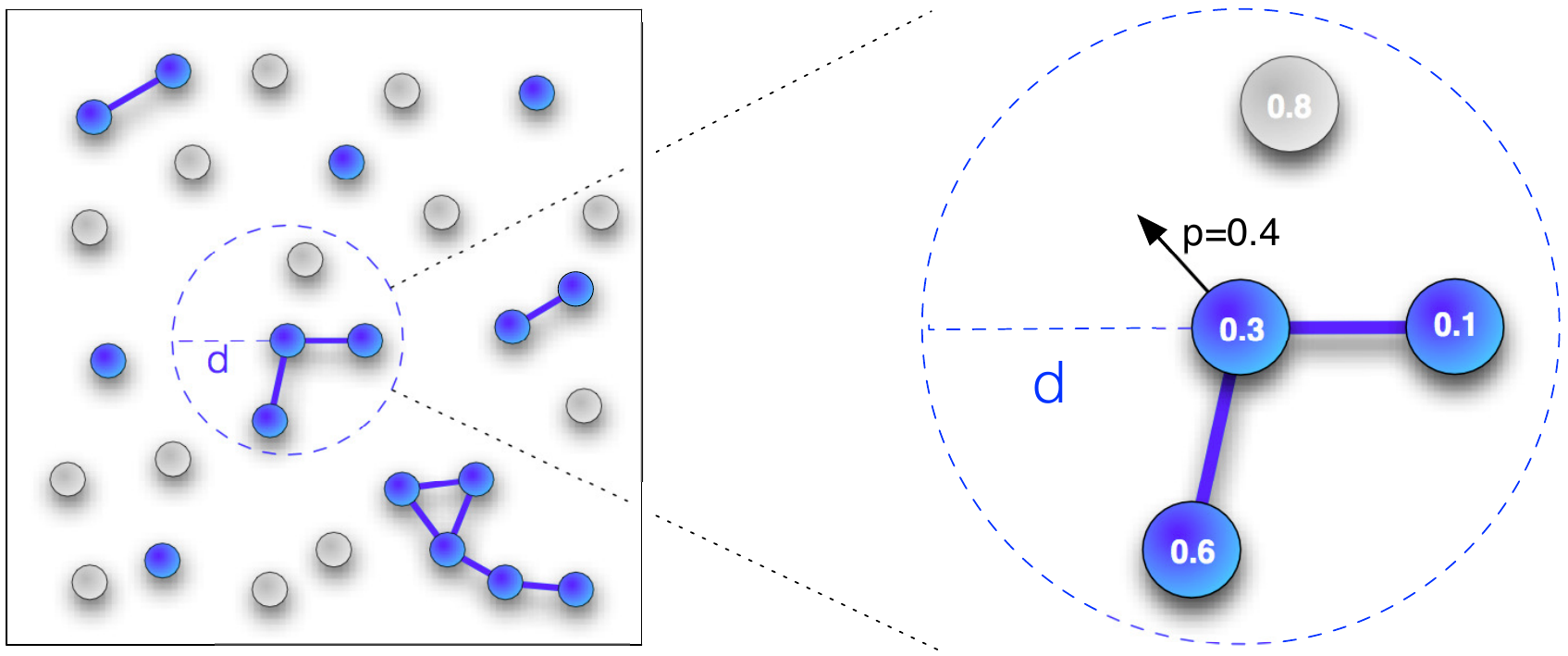}
  \caption{ Left: Blue (dark) colored agents are active,
    grey (light) agents do not move nor interact. Interacting agents,
    within a distance $d$, are connected by a link.  Right: 
    Each individual is characterized by a
    number representing her attractiveness. The probability for the
    central individual to move is $p=1.0-0.6=0.4$, since the
    attractiveness of the inactive agent is not taken into account.  }
  \label{fig:rules}
\end{figure}

The model is defined as follows (see Fig.~\ref{fig:rules}): $N$  
agents are placed in a square box of linear size
$L$ with periodic boundary conditions, corresponding to a density $\rho= N/L^2$. 
Each individual $i$ is characterized by her attractiveness or social
appeal, $a_i$ which represents her power to raise interest in the
others. The attractiveness $a_i$ of the agents is a (quenched)
variable randomly chosen from a prefixed distribution $\eta(a)$, and
bounded in the interval $a_i \in [0,1)$.  Agents perform a random walk
biased by the attractiveness of neighboring individuals. Whenever an
agent intercepts, within a distance smaller than or equal to $d$,
another individual, they start to interact. The interaction lasts as
far as the distance between them is smaller than $d$. Crucially, the more
attractive an agent $j$ is (the largest her attractiveness $a_j$), the
more interest she will raise in the other agent $i$, who will slow
down her random walk exploration accordingly. This fact is taken into
account by a walking probability $p_i(t)$ which takes the form:
\begin{equation}
\label{eq:model_rule}
p_i(t) = 1- \max_{j \in \mathcal{N}_i(t) } \{ a_j \} ,
\end{equation} 
where $\mathcal{N}_i(t)$ is the set of neighbors of agent $i$ at time
$t$, i.e. the set of agents that, at time $t$, are at a distance
smaller than or equal to $d$ from agent $i$.  Hence, the biased random
walk performed by the agents is defined as follows: At each time step
$t$, each agent $i$ performs, with probability $p_i(t)$, a step of
length $v$ along a direction given by a randomly chosen angle $\xi \in
[0, 2\pi)$. With the complementary probability $1-p_i(t)$, the agent
does not move. Thus, according to Eq.~\eqref{eq:model_rule}, if an agent $i$ is
interacting with other agents, she will keep her position in the
following time step with a probability proportional to the appeal of
his most interesting neighbor.

Furthermore, the empirical observations of SocioPatterns data show
that not all the agents involved in a social event are actually
present for its entire duration: Some agents leave the event before
the end, some join it later after the beginning, and some others leave
and come back several times.  Therefore we assume that agents can be
in an active or an inactive state. If an individual is active, she
moves in space and interacts with the other agents; otherwise she
simply rests without neither moving nor interacting.  At each time
step, one inactive agent $i$ can become active with a probability
$r_i$, while one active and isolated agent $j$ (not interacting with
other agents) can become inactive with probability $1-r_j$.  The
activation probability $r_i$ of the individual $i$ thus represents her
activeness in the social event, the largest the activity $r_i$, the
more likely agent $i$ will be involved in the event.  We choose the
activation probability $r_i$ of the agents randomly from an uniform
distribution $\zeta(r)$, bounded in $r_i \in [0,1]$, but we have
verified that the model behavior is independent of the activity
distribution functional form (even if we consider a constant activity
rate, $r_i = r $ for all agents, we obtain very similar results, see Section \ref{sec:model_robust}).
 
Within this framework, each individual performs a discrete random walk
in a 2D space, interrupted by interactions of various duration with
peers. The movement of individuals is performed in parallel in order
to implement the time resolution ($20$ seconds) at which empirical
measurements are made \cite{10.1371/journal.pone.0011596}.  
The model is Markovian, since agents do
not have memory of the previous time steps.  The full dynamics of the
system is encoded in the collision probability $p_c = \rho \pi d^2$,
the activation probability distribution $\zeta(r)$, and the
attractiveness distribution $\eta(a)$.  The latter can hardly be
accessed empirically, and is likely to be in its turn the combination
of different elements, such as prestige, status, role, etc. Moreover,
in general attractiveness is a relational variable, the same
individual exerting different interest on different agents. Avoiding
any speculations on this point, we assume the simplest case of a
uniform distribution for the attractiveness
\cite{papadopoulos2012popularity}.  Remarkably, this simple assumption
leads to a rich phenomenology, in agreement with empirical
data.

  The model has been simulated
adopting the parameters $v=d=1$, $L=100$ and $N=200$.  Different
values of the agent density $\rho$ are obtained by changing the box
size $L$. In the initial conditions, agents are placed at randomly
chosen positions, and are active with probability $1/2$. Numerical
results are averaged over $10^2$ independent runs, each one of
duration $T$ up to $T_\mathrm{max}=2 \times 10^4$ time steps.
Different choices of the parameters, if any, are specified.

Before proceeding a comment is in order. We adopt here an operational
definition of ``attractiveness" as the property of an individual to
attract the interest of other individuals, making them willing to engage
in a conversation, or to listen to what he/she is saying. Thus, we do
not enter in any speculations on what are the cultural or psychological
factors that make a person attractive in this sense, but we reckon that
many possible candidates exist, ranging from status
\cite{hollingshead1975four} to extroversion
\cite{scherer1978personality}. In light of the success of our model in
reproducing the empirical distributions (see
Section \ref{sec:model_discussion}), we consider that identifying which feature, or set of features,
the attractiveness is a proxy of represents one important direction for
future work.

\section{Modeling face-to-face interactions networks}
\label{sec:model_prop}

The empirical data of face-to-face interactions show 
several properties, universally shared across different social contexts, 
as broadly discussed in Chapter \ref{chap:intro}.
In this Section we will compare the results of the model, obtained by numerical simulations, 
against the observation from empirical data.
We focus on four data sets recorded by the SocioPatterns collaboration
representative of different social contexts:
a Lyon hospital (``Hospital"), the Hypertext 2009
conference (``Conference"), the Soci\'et\'e Francaise d'Hygi\'ene
Hospitali\'ere congress (``Congress") and a high school (`School"). 
These data sets have been extensively analyzed in Section \ref{sec:intro_empirical},
a further description can be found in
\cite{10.1371/journal.pone.0011596,Isella:2011,Stehle:2011,percol}.

In the following, we will explore human dynamics properties belonging to three
different scales. At \textit{the individual, or `microscopic', level}, we focus in
temporal properties related to the distributions of contact durations or
inter-contact times, and in structural properties related to the time
integrated representation of the contact data.  Moving beyond the analysis of
individual properties, we consider \textit{the group, or `mesoscopic', level}, represented by
groups of simultaneously interacting individuals, which typify a crucial
signature of face-to-face networks and have important consequences on
processes such as decision making and problem solving
\cite{BuchananAtom}. We measure the distribution of group
sizes as well as the distribution of duration of groups of different
size. We finally zoom one more step out and inspect the 
\textit{collective, or `macroscopic', level} looking at properties
that depend on the time interaction pattern of the whole population. We
address in particular the issue of the causality patterns of the
temporal network, as determined by the time-respecting paths between
individuals (see Section \ref{sec:intro_timevar}) and the network reachability,
defined as the minimum time for information to flow from an individual
$i$ to another individual $j$ and measured by means of a searching
process performed by a random walker
(see Chapter \ref{chap:RW} for further details). We observe that the model
 reproduces not only qualitatively, but also quantitatively, the
properties measured from empirical data at all the 
scales. 

Finally, as a check for robustness, we explore the role of the the
parameters that define the model. Particular emphasis is made on the
motion rule adopted by the individuals. While a simple random walk for
the individuals' movements is initially considered, in fact, a
consistent amount of literature suggests that L\'evy flights
\cite{viswanathan2011physics} might provide a better characterization of
human movement
\cite{Brockmann:2006:Nature:16437114,gonzalez2008understanding,rhee2011levy,baronchelli2013levy}.
We observe that the results of the model are robust with respect to
various possible alterations of the original formulation, including the
adopted rule of motion.

\subsection{Individual level dynamics}
\label{subsec:model_indiv}

Here we focus on the statistical properties of individual contacts, 
from both points of view of time-independent and time-aggregated features.

\subsubsection{Temporal correlations}

\begin{figure}[tb]
  \includegraphics*[width=0.48\linewidth]{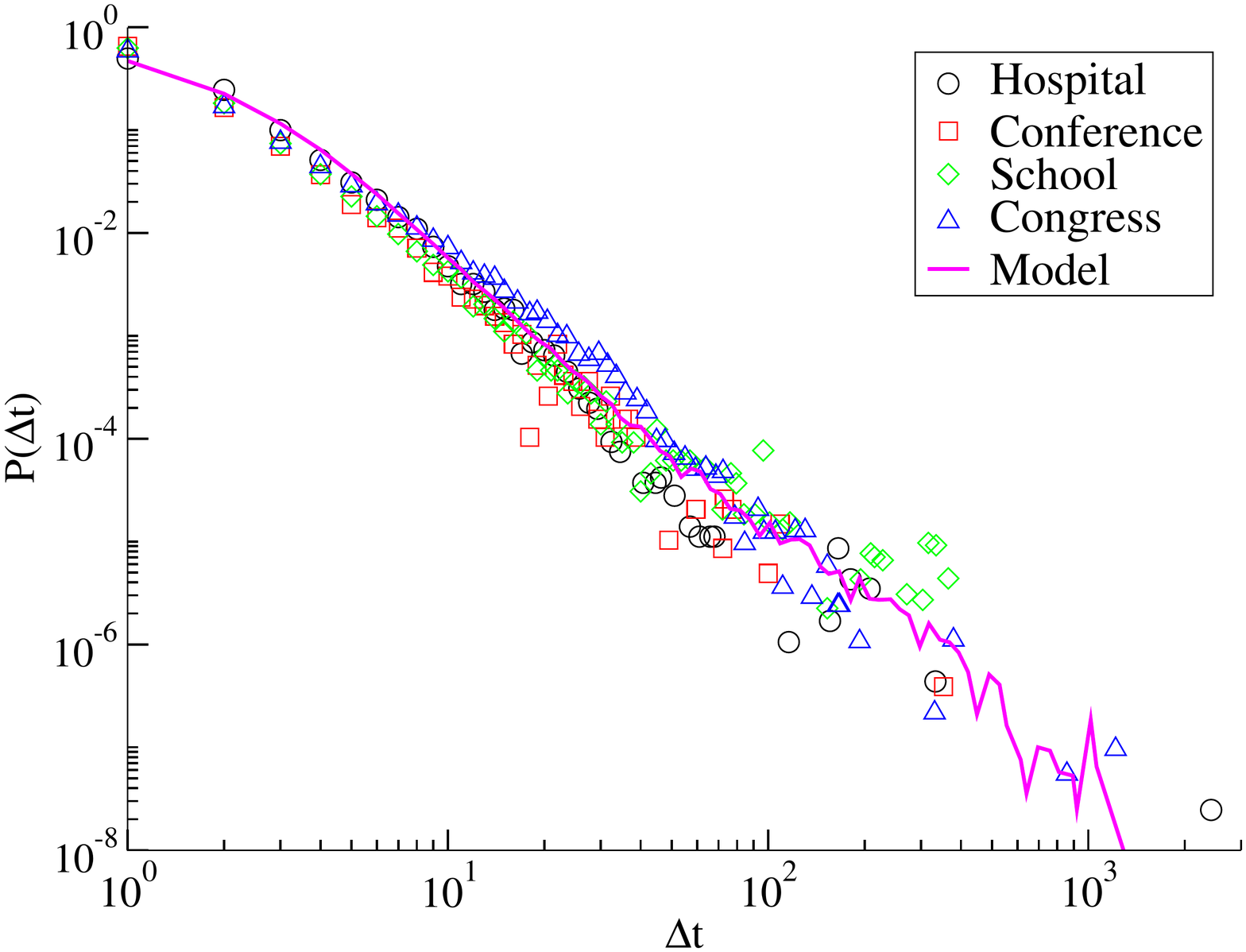}
  \includegraphics*[width=0.48\linewidth]{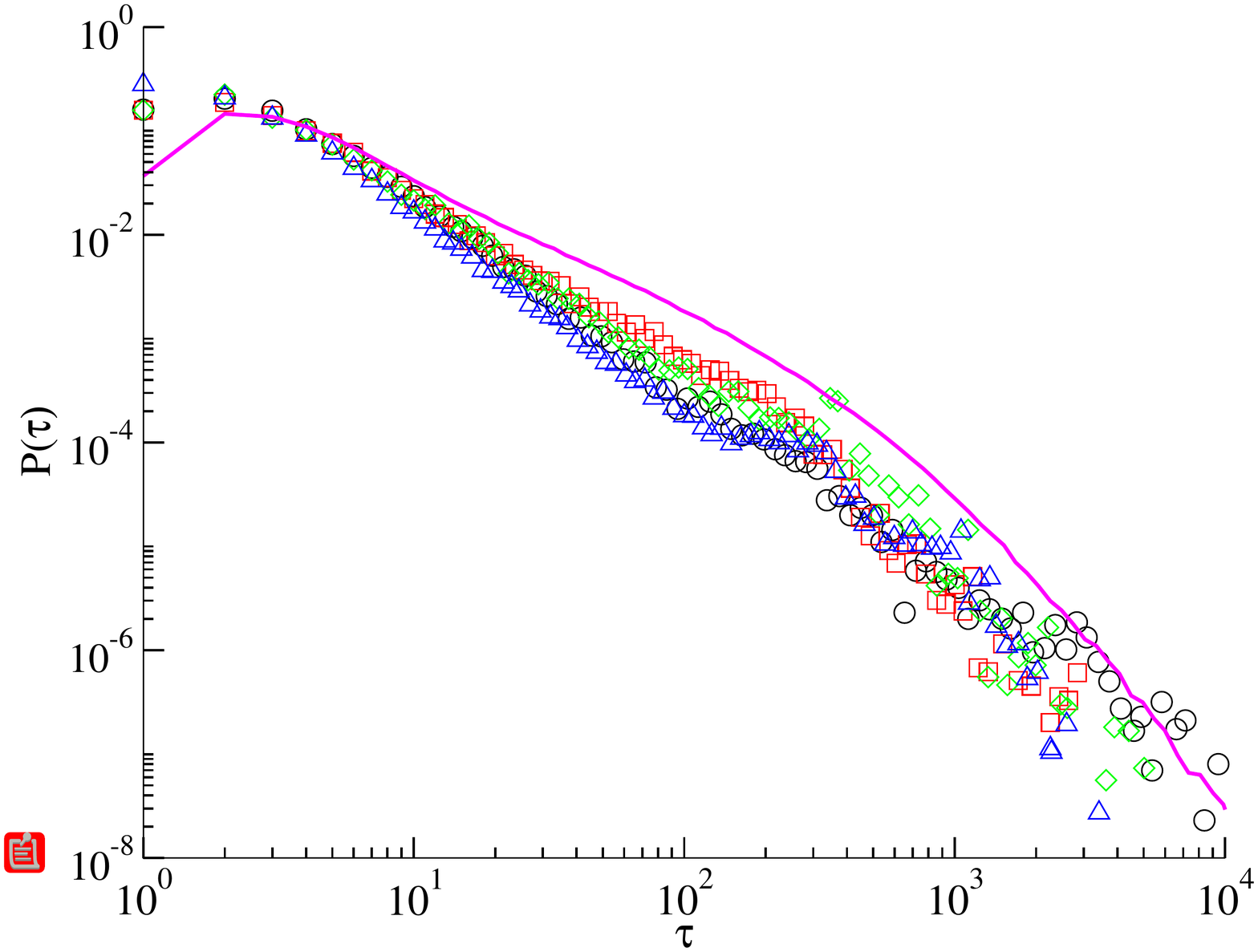}
  \caption{Distribution of the contact duration, $P(\Delta t)$, (left)
   and distribution of the time interval between consecutive contacts, $P(\tau)$, (right)
     for various datasets and for the attractiveness
     model. 
     }
    \label{fig:model_indiv} 
\end{figure}

The temporal pattern of the agents' contacts is probably the most
distinctive feature of face-to-face interaction networks
\cite{10.1371/journal.pone.0011596,PhysRevE.85.056115}. 
We therefore start by considering the distribution of the duration $\Delta t$ of
the contacts between pairs of agents, $P(\Delta t)$, and the
distribution of gap times $\tau$ between two consecutive conversations
involving a common individual, $P(\tau)$. 
In Section \ref{sec:intro_empirical} we discussed the power law form of 
both $P(\Delta t)$ and $P(\tau)$,
indicating the bursty dynamics of human interactions.
 Figure~\ref{fig:model_indiv} show both distributions
 for the various sets of empirical data along with the same distributions obtained by
simulating the model described above with density $\rho=0.02$.  In the
case of the contact duration distribution, numerical and experimental
data match almost perfectly, see
Fig.~\ref{fig:model_indiv} (left). 
It also worth highlighting the crucial role played by the heterogeneity of attractiveness $a_i$.
In fact, assuming it constant, $a_i=a$ (and neglecting excluded volume
effects between agents) our model can be mapped into a simple first
passage time problem \cite{Redner01}, leading to a distribution
$P(\Delta t) \sim (\Delta t)^{-3/2}$ with an exponential cut-off
proportional to $d^2/(1-a)$. The (non-local) convolution of the
exponential tails induced by the heterogeneous distribution of
attractiveness leads in our model to a power law form, with no
apparent cut-off, and with an exponent larger than $3/2$, in agreement
with the result observed in the SocioPatterns data.
%
%
Regarding the distribution of gap times, $P(\tau)$, the model also
generates a long-tailed form, which is compatible, although in this case 
not exactly equal, to the empirical data, see
Fig.~\ref{fig:model_indiv}(right). 

\subsubsection{Time-aggregated networks}

 \begin{figure}[tb]
   \includegraphics*[width=0.48\linewidth]{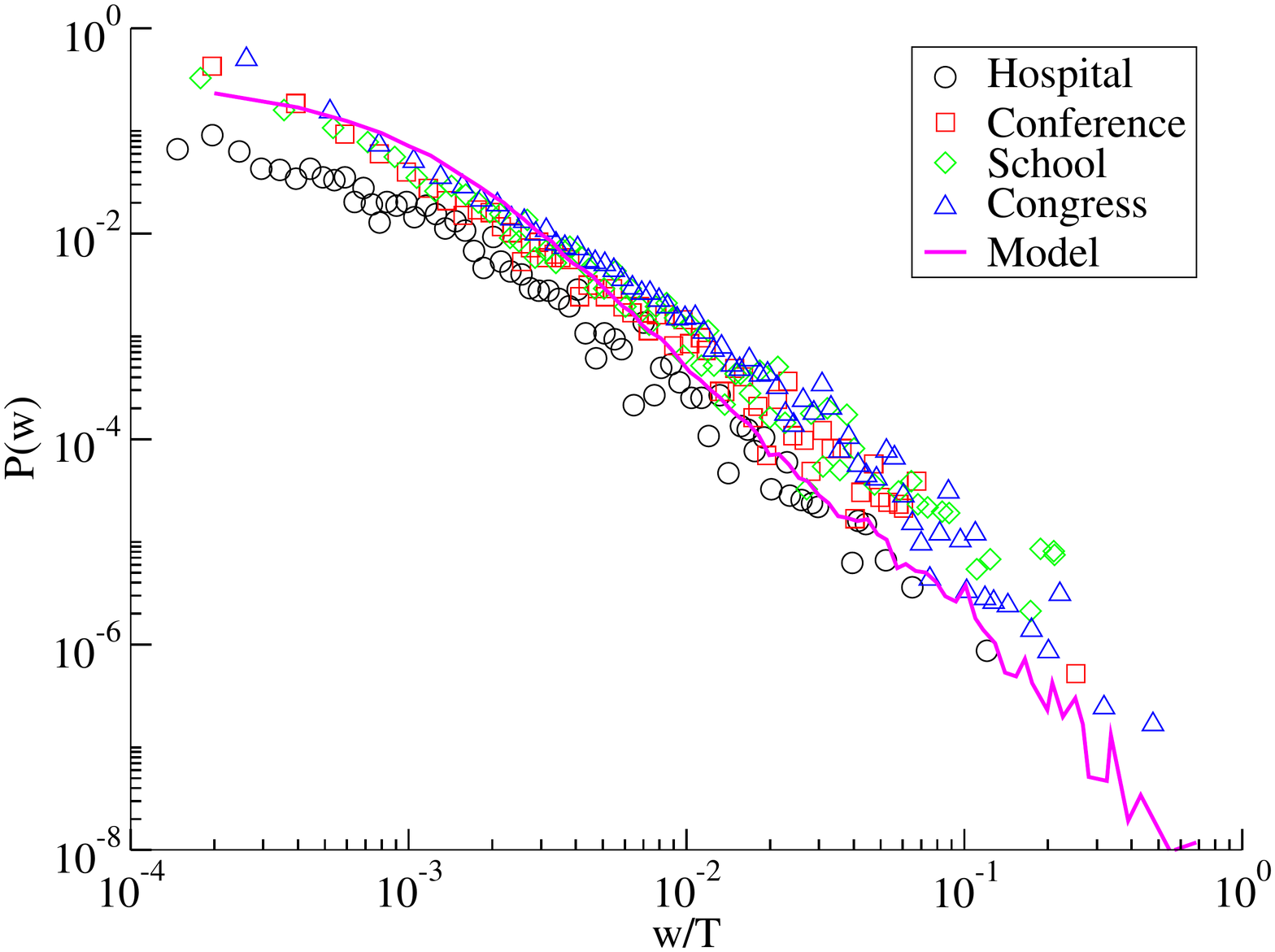}
   \includegraphics*[width=0.48\linewidth]{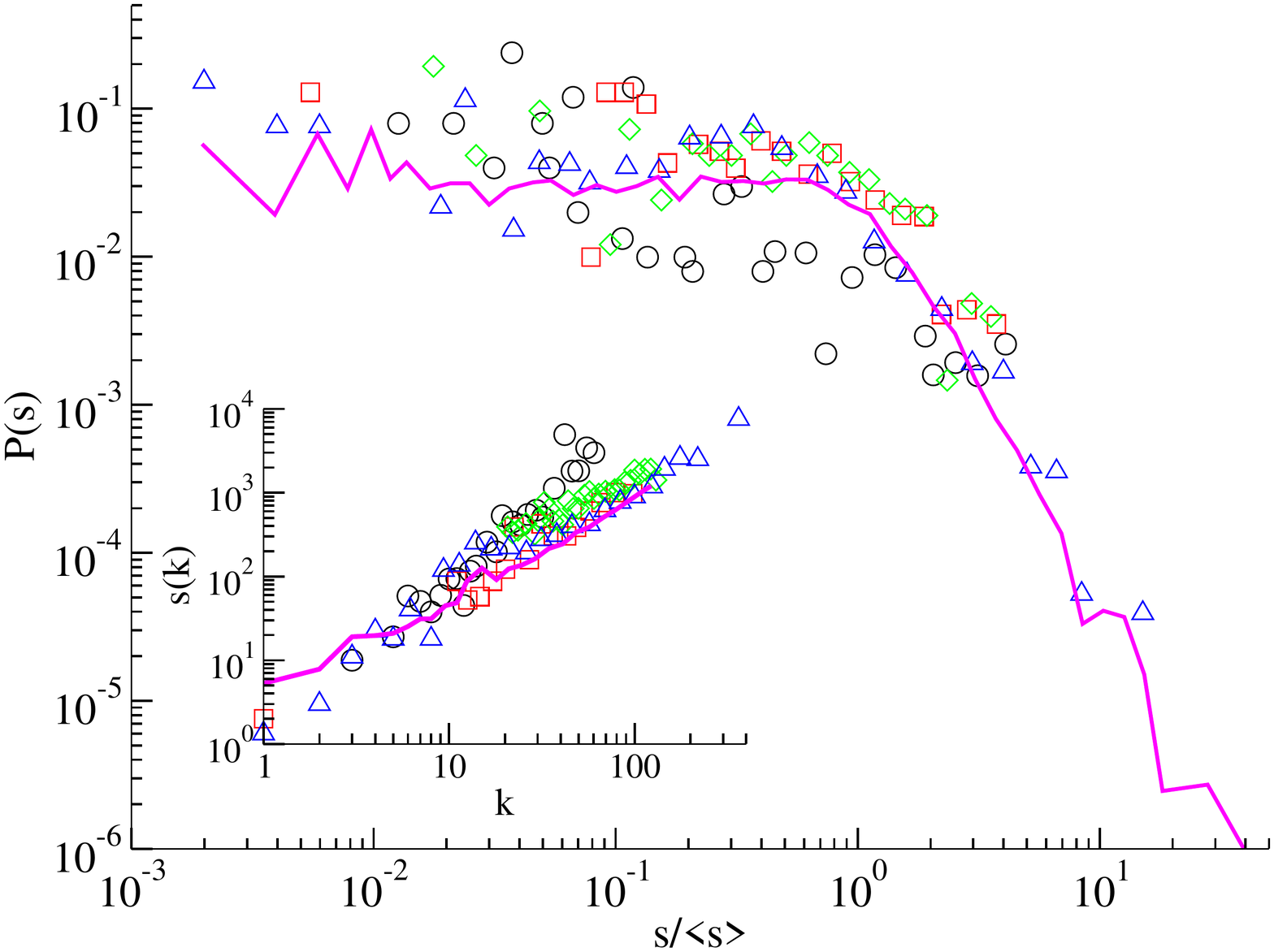}
   \caption{  
   Distribution of the weights, $P(w)$ (left) 
   and rescaled strength, $P(s)$, (right) 
   for various datasets (symbols) and for the attractiveness model (line).
   In the inset we show the superlinear correlation between the degree 
   and the strength, $s(k)$. }
     \label{fig:model_aggr} 
 \end{figure}

A different information regarding the pattern of individual interactions
is obtained by integrating the time-varying network into a aggregated, weighted network. 
As mentioned in section \ref{sec:intro_timevar}, this corresponds to considering all the interactions occurring in a given time window $\Delta t_0$ in the limit of $\Delta t_0 \rightarrow \infty$, i.e. all the interactions taking place in the dataset. 
In Fig. \ref{fig:model_aggr} we plot the distribution of weight
between pair of agents rescaled by the total duration of the contact sequence, $w_{ij}/T$, (left)
and the distribution of rescaled strength $s/\av{s}$ (right).
The heavy tailed weight distribution, $P(w)$, shows that the heterogeneity 
in the duration of individual contacts persists 
even when interactions are accumulated over longer time intervals.
The strength distribution $P(s)$, on the contrary, 
is more compatible with an exponential decay, 
as better showed in Section \ref{sec:intro_empirical}.
Again, we see that the numerical simulation of the model 
are in good agreement with empirical data.
In the inset of Fig. \ref{fig:model_aggr} (right), we show that the model is also capable to capture 
the superlinear correlations found between the degree and the strength of a node, 
$s(k)$, as discussed in Section  \ref{sec:intro_empirical}.
An exception of the good agreement is  the ``hosp''
database. The reason of the departure of this dataset with respect of
both other data set and the model could be attributed to the duration
$T$ of the corresponding sequence of contacts (see
Table~\ref{tab:summary}), which is up to four times longer than the
other data sets.  In the limit of large $T$, sporadic interactions can
lead to a fully connected integrated network, very different from the
sparser networks obtained for smaller values of $T$. These effects
extend also to the pattern of weights, which have in the ``hosp''
database a much larger average value.

\begin{figure}[tb]
 \begin{center}
   \includegraphics*[width=0.7\linewidth]{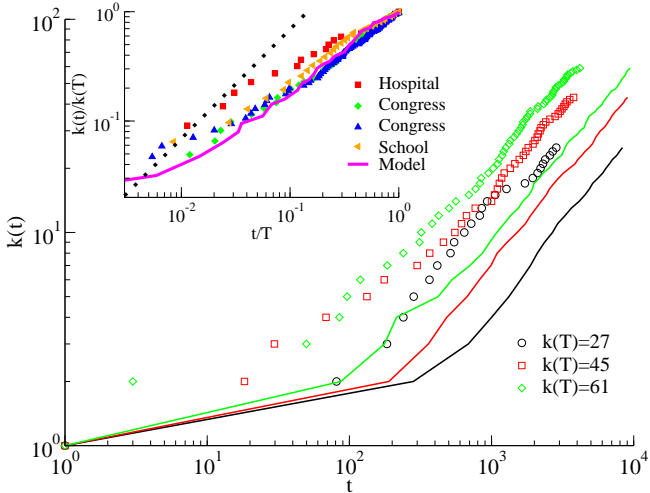}
   \caption{Main: Aggregated degree $k(t)$ versus time
     for various individuals with different final degree $k(T)$, for
     the ``Conference'' dataset (symbols) and for the network
     obtained by simulating the attractiveness model (line). Inset:
     Rescaled aggregated degree $k(t) / k(T)$ as a function of time
     $t/T$ for various datasets and for the  model. Dashed line marks a linear growth of degree in time. }\label{fig:model_k_t} 
 \end{center}
 \end{figure}

A final important feature of face-to-face interactions, also revealed
in different context involving human mobility
\cite{citeulike:7974615}, is that the tendency of an agent to interact
with new peers decreases in time. This fact translates into a
sub-linear temporal growth of the number of different contacts of a
single individuals (i.e. the aggregated degree $k_i(t)$), $k(t) \sim
t^{\mu}$, with $\mu<1$. Fig.~\ref{fig:model_k_t} shows the evolution of
$k(t)$ versus time for several agent with final aggregated degree
$k(T)$, both belonging to a single dataset (main) and for the
different datasets (inset). The sub-linear behavior of $k(t)$ is
clear, with $\mu = 0.6 \pm 0.15$ depending on the dataset. Moreover,
the shapes of the $k(t)$ functions can be collapsed in a single curve
by appropriately rescaling the data as $k(t)/k(T)$ as a function of
$t/T$, Fig.~\ref{fig:model_k_t}(inset). Fig.~\ref{fig:model_k_t} shows that,
remarkably, the attractiveness model is also capable to reproduce the
behavior of $k(t)$, up to the rescaling with total $T$ time, again
with the exception of the ``hosp'' dataset.


\subsection{Group level dynamics}
\label{subsec:model_group}

\begin{figure}[t]
\begin{center}
  \includegraphics[width=0.48\linewidth]{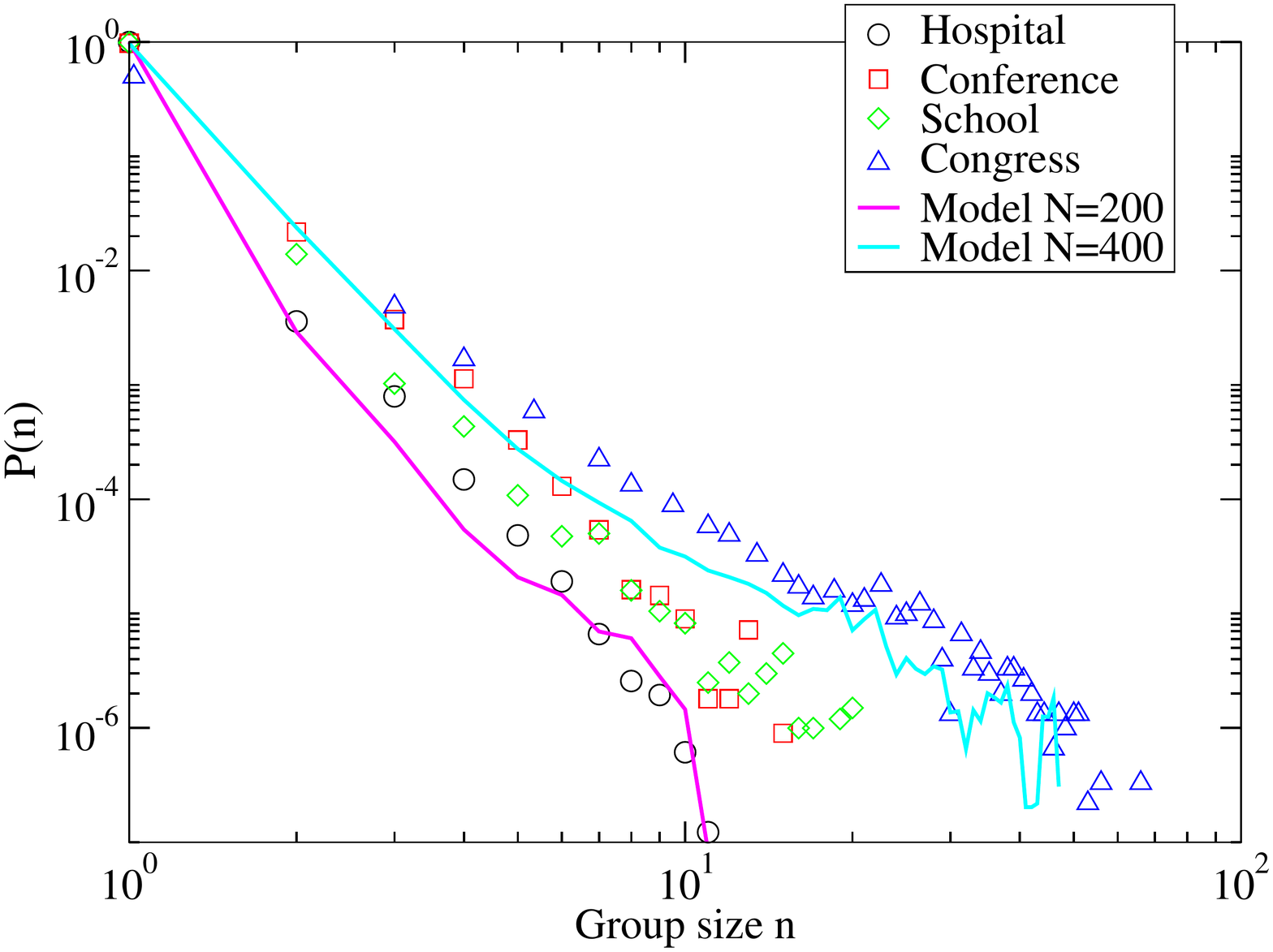}
  \includegraphics[width=0.48\linewidth]{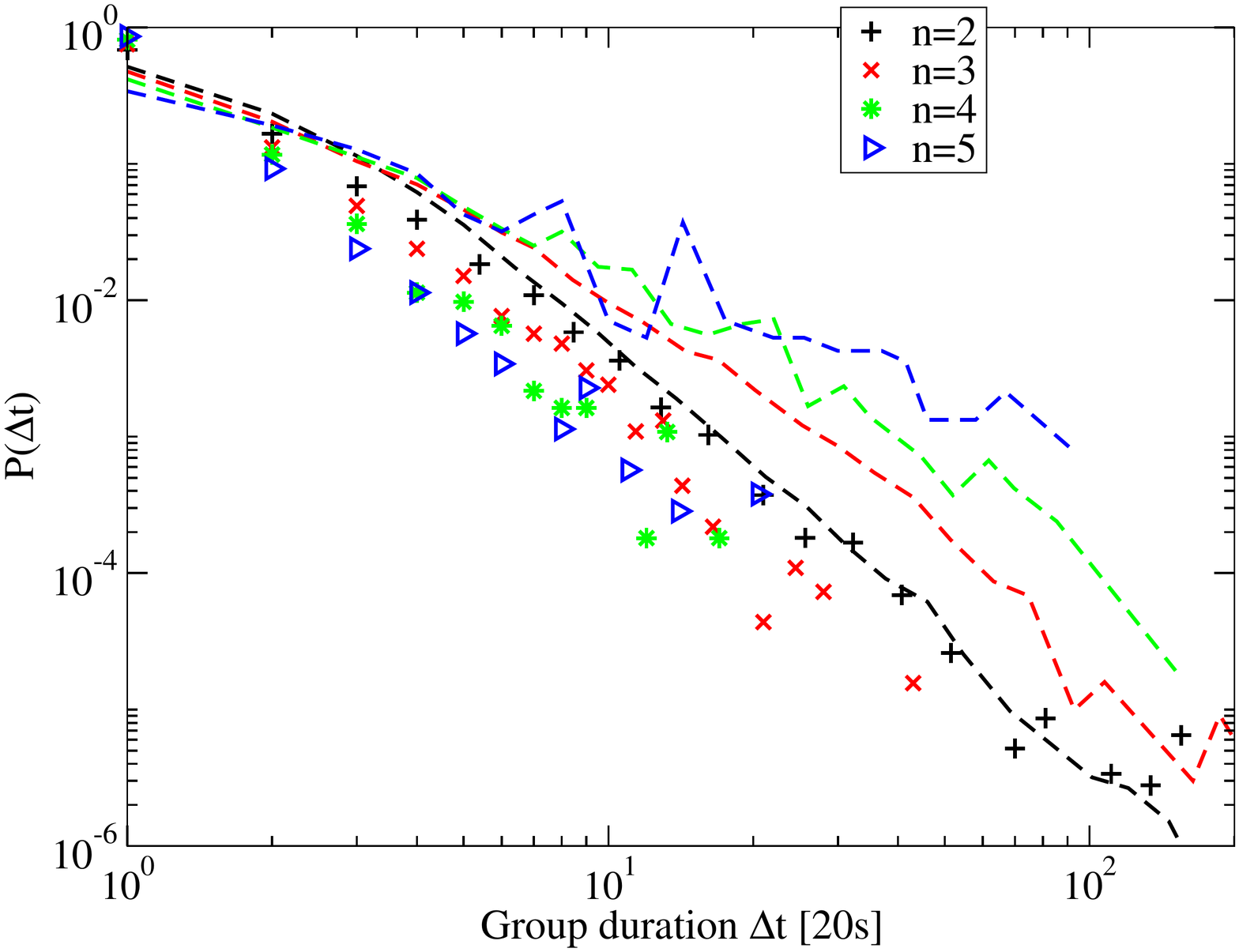}
  \caption{\label{fig:model_group} 
  Left: Group size distribution $P(n)$ for different datasets and for the model, 
  numerically simulated with different number of agents $N=200$ and $N=400$, 
  same size $L=100$ and total duration $T = 10^4$.
  Right: Lifetime distribution $P_n(\Delta t)$ of groups of different size $n$, 
  for the ``Congress" data set (symbols) 
  and for the model numerically simulated with $N=400$ and $L=100$. (dashed lines).  }
\end{center}
  \end{figure}

Another important aspect of human contact networks is the dynamics of
group formation, defined by a set of $n$ individuals interacting
simultaneously, not necessarily all with all, for a period of time
$\Delta t$. As we have noted above, such groups have a sociological
relevance in their role of catalysts for decision making and problem
solving \cite{BuchananAtom}.  In Fig. \ref{fig:model_group} (top) we plot the
probability distribution of observing a group of size $n$, $P(n)$, in
any instant of the ongoing social event, for the different empirical
data sets.  The distributions are compatible with a power law behavior,
whose exponent depends on the number of agents involved in the social
event, with larger datasets (such for example the Congress one, see
Table \ref{tab:summary}) capable of forming bigger groups with respect
to smaller data sets.  Clearly, the model predictions are in substantial
agreement with the data when we inform the model with a sensible,
data-driven, value of $N$.

In order to explore the dynamics of group formation, we define the
lifetime $\Delta t$ of a group of size $n$ as the time spent in
interaction by the same set of $n$ individuals, without any new arrival
or departure of members of the group.  In Fig. \ref{fig:model_group} (bottom)
we plot the lifetime distribution $P_n(\Delta t)$ of groups of different
sizes $n$, finding that experimental and numerical results have a
similar power-law behavior.  We note however that for empirical data the
lifetime distribution $P_n(\Delta t)$ decays faster for larger groups,
i.e. big groups are less stable than small ones, while the model outcome
follows the opposite behavior.  This means that, in the model, larger
groups are (slightly) more stable than observed in the data. This is
probably due to the fact that, the larger the group, the bigger the
probability of finding two (or more) individuals with large
attractiveness in the group, which guarantee the stability against
departures. However, an alternative explanation could be that the
RFID devices of the SocioPatterns experiment require individuals to
face each other within a given angle, making the group
definition effectively more fragile than in the model, where such directionality is absent.

\subsection{Collective level dynamics and searching efficiency}
\label{subsec:model_collective}

\begin{figure}[tb]
  \includegraphics[width=0.48\linewidth]{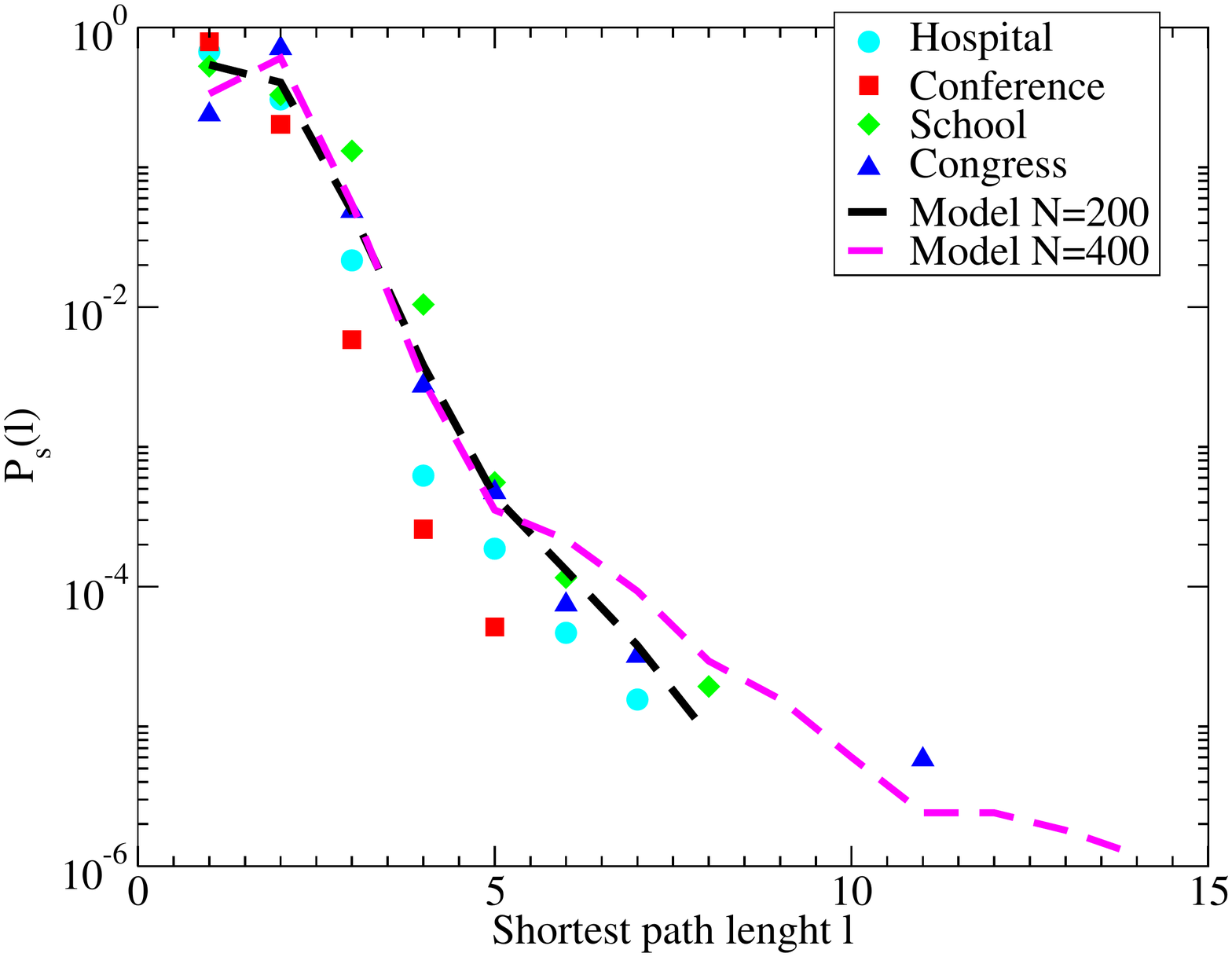}
  \includegraphics[width=0.48\linewidth]{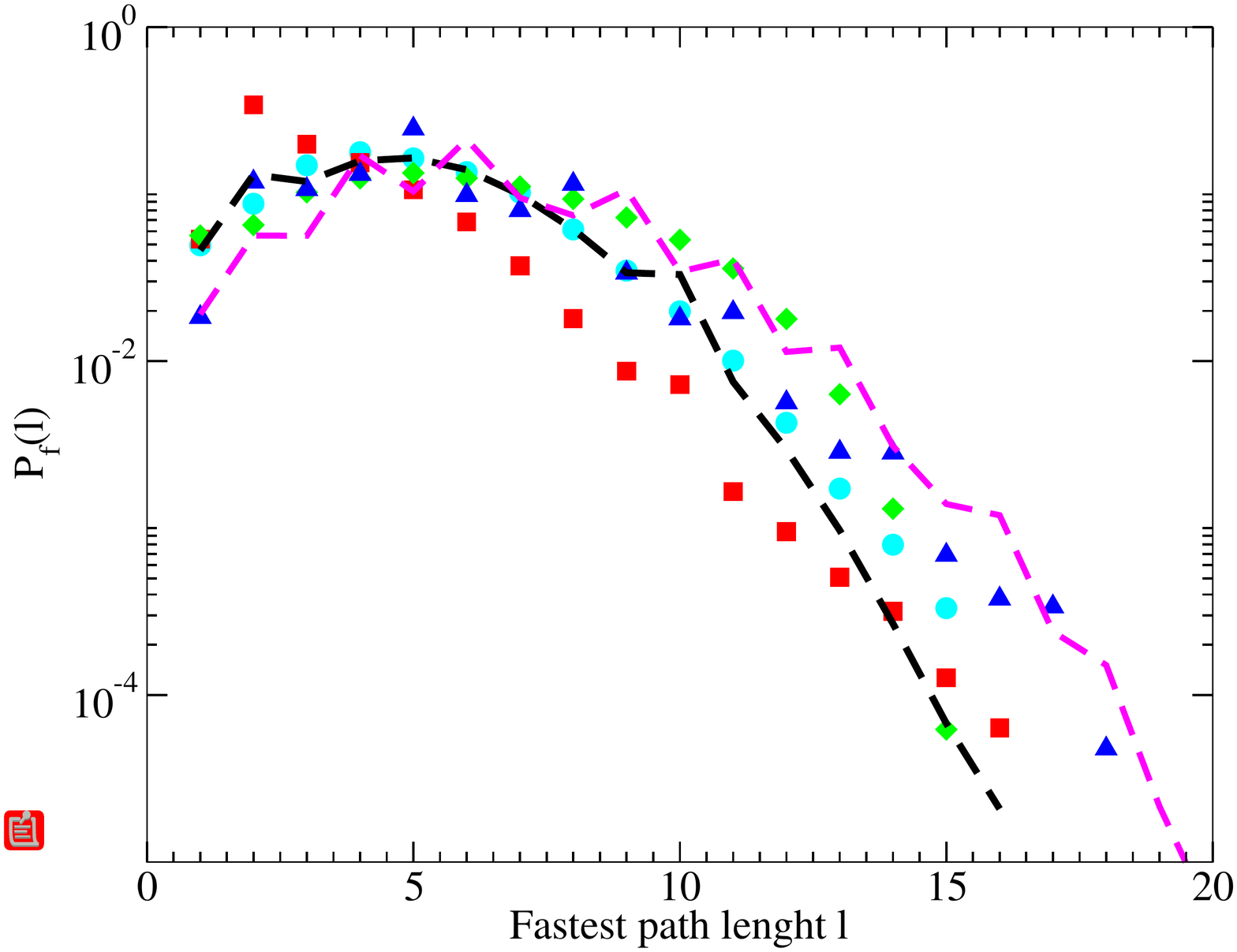}
  \caption{\label{fig:model_paths}  
  Probability distributions of the shortest, $P_s(l)$, (up) 
  and fastest, $P_f(l)$, (down) path length, 
  for the time-varying network obtained by the empirical data 
  and by the model, numerically simulated with different number of agents 
  $N=200$ and $N=400$, 
  same size $L=100$ and total duration $T=10^4$. }
\end{figure}

Time respecting paths of temporal networks have a deep impact in any collective dynamics of the individuals,
and are especially important when studying dynamical processes taking place upon such structures.
For any two vertices, we can measure the shortest time-respecting path,
$l_{ij}^{s}$, and fastest time-respecting path, $l_{ij}^{f}$, between them.
The former is defined as the path with
the smallest number of intermediate steps between nodes $i$ and $j$, and
the latter is the path which allows to reach $j$ starting from $i$
within the smallest amount of time,
as defined in detail in Section \ref{subsec:intro_paths}.
  In Fig. \ref{fig:model_paths} we plot the
probability distributions of the shortest and fastest time-respecting
path length, $P_s(l)$ and $P_f(l)$, respectively, of both empirical data
and model, finding that they show a similar behavior, decaying
exponentially, and being peaked for a small number of steps.

Given the importance of causal relationship on any spreading dynamics,
it is interesting to explicitly address the dynamical unfolding of a
diffusive process. Here we analyze the simplest example of a search
process, the random walk, which describes a walker traveling the network
and, at each time step, selecting randomly its destination among the
available neighbors of the node it occupies. The random walk represents
a fundamental reference point for the behavior of any other diffusive
dynamics on a network, when only local information is available. Indeed,
assuming that each individual knows only about the information stored in
each of its nearest neighbors, the most naive economical strategy is the
random walk search, in which the source vertex sends one message to a
randomly selected nearest neighbor \cite{PhysRevE.64.046135}.  If that
individual has the information requested, it retrieves it; otherwise, it
sends a message to one of its nearest neighbors, until the message
arrives to its final target destination.  In this context, a quantity of
interest is the probability that the random walk actually find its
target individual $i$ at any time in the contact sequence, $P_r(i)$, or
\emph{global reachability}(see Chapter \ref{chap:RW} for further details).
This quantity is related with a realistic case of searching,
and has the advantage of not being a asymptotic property of the random walk,
thus it can be computed also on finite contact sequences.

\begin{figure}[tb]
\begin{center}
  \includegraphics[width=0.75\textwidth]{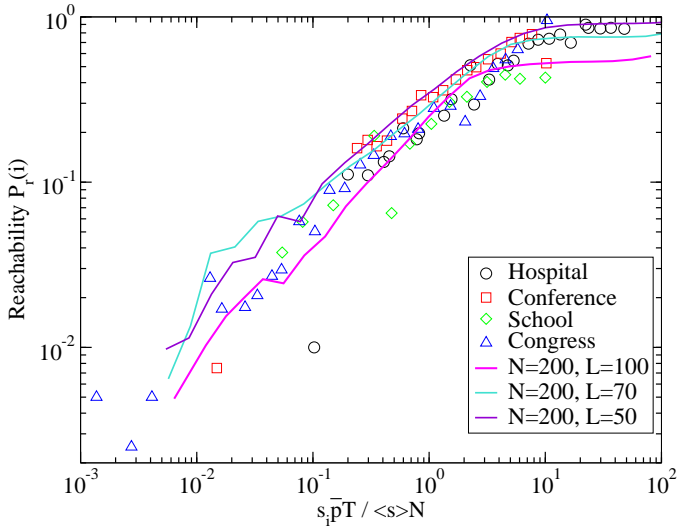}
  \caption{\label{fig:model_reach} 
  Correlation between the reachability of agent $i$,
  $P_r(i)$, and his rescaled strength, $ s_i \bar{p}T / \av{s}N$. 
  The empirical data sets considered (symbols) and the model (lines), 
  numerically simulated with different density $\rho$, follow a close behavior. 
  We averaged the reachability of each individual over at least $10^2$ different runs,
  starting with different source node.  }
\end{center}
\end{figure}

In principle, the reachability of an individual $i$ must be correlated
with the total time spent in interactions, namely his strength $s_i$,
but it also depends on the features of the considered social event, such
as the density of the interaction $\bar{p}$, the total duration $T$ and
possibly other event-specific characteristic (see Table
\ref{tab:summary} for information of the different data sets
considered).  On the basis of a simple mean field argument, it is possible to show (see Chapter \ref{chap:RW})
 that the probability of node $i$ to be
reached by the random walk, $P_r(i)$, is correlated with its relative
strength $s_i/\av{s}N$, times the average number of interacting
individuals at each time step, $\bar{p}T$.  In Fig \ref{fig:model_reach}
we plot the reachability $P_r(i)$ against the rescaled strength $ s_i
\bar{p}T / \av{s}N$, showing that very different empirical data sets
collapse into a similar functional form.  Remarkably, the model is able
to capture such behavior, with a variability, also found in the data,
which depends on the density $\rho$ of the agents involved.  As noted
for the group dynamics, a larger density corresponds to a higher
reachability of the individuals.  We note that the empirical data are
reproduced by the model for the same range of density considered in the
previous analysis.

\section{Model robustness}
\label{sec:model_robust}

\begin{figure}[tb]
\begin{center}
  \includegraphics[width=0.48\linewidth]{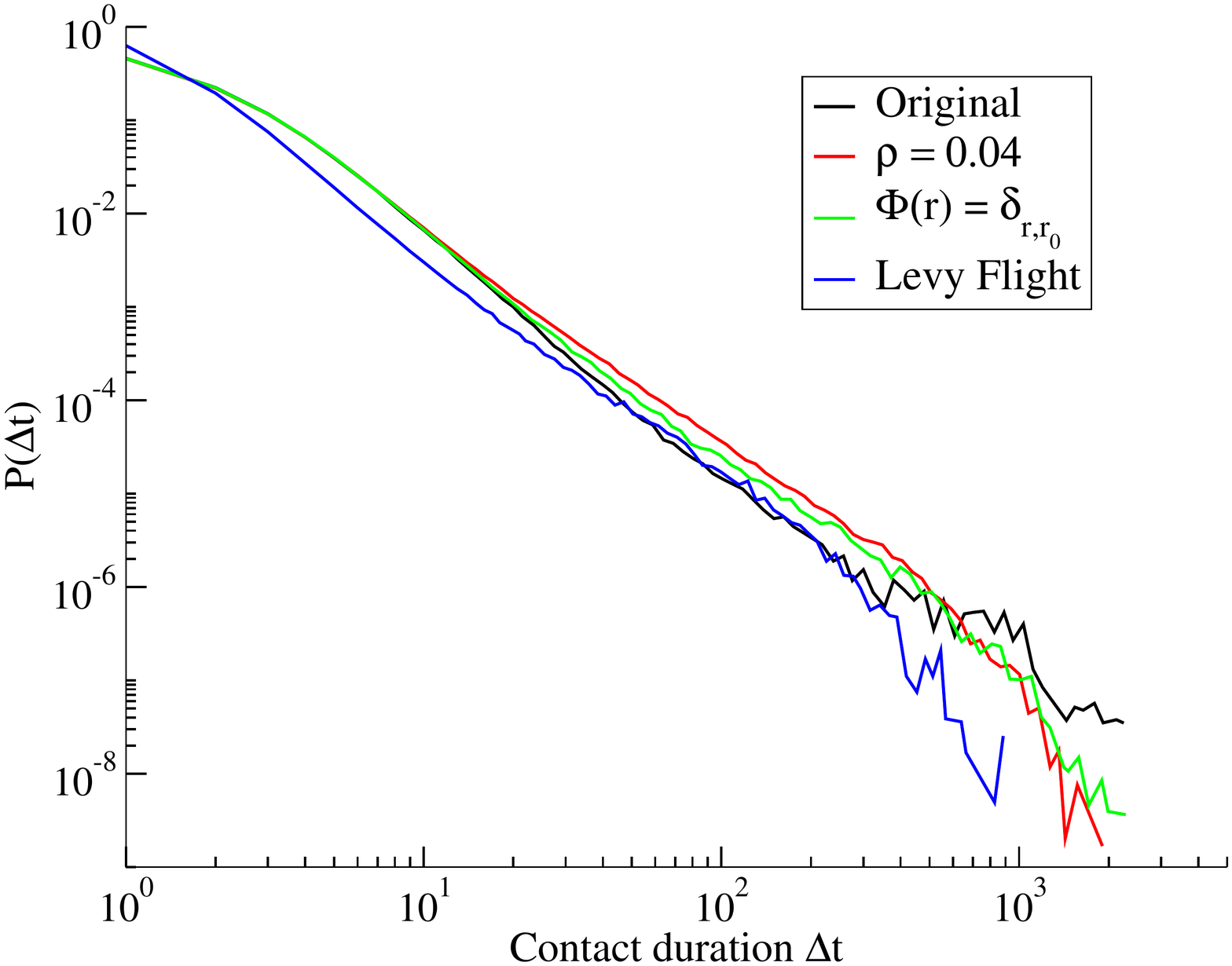}
  \includegraphics[width=0.48\linewidth]{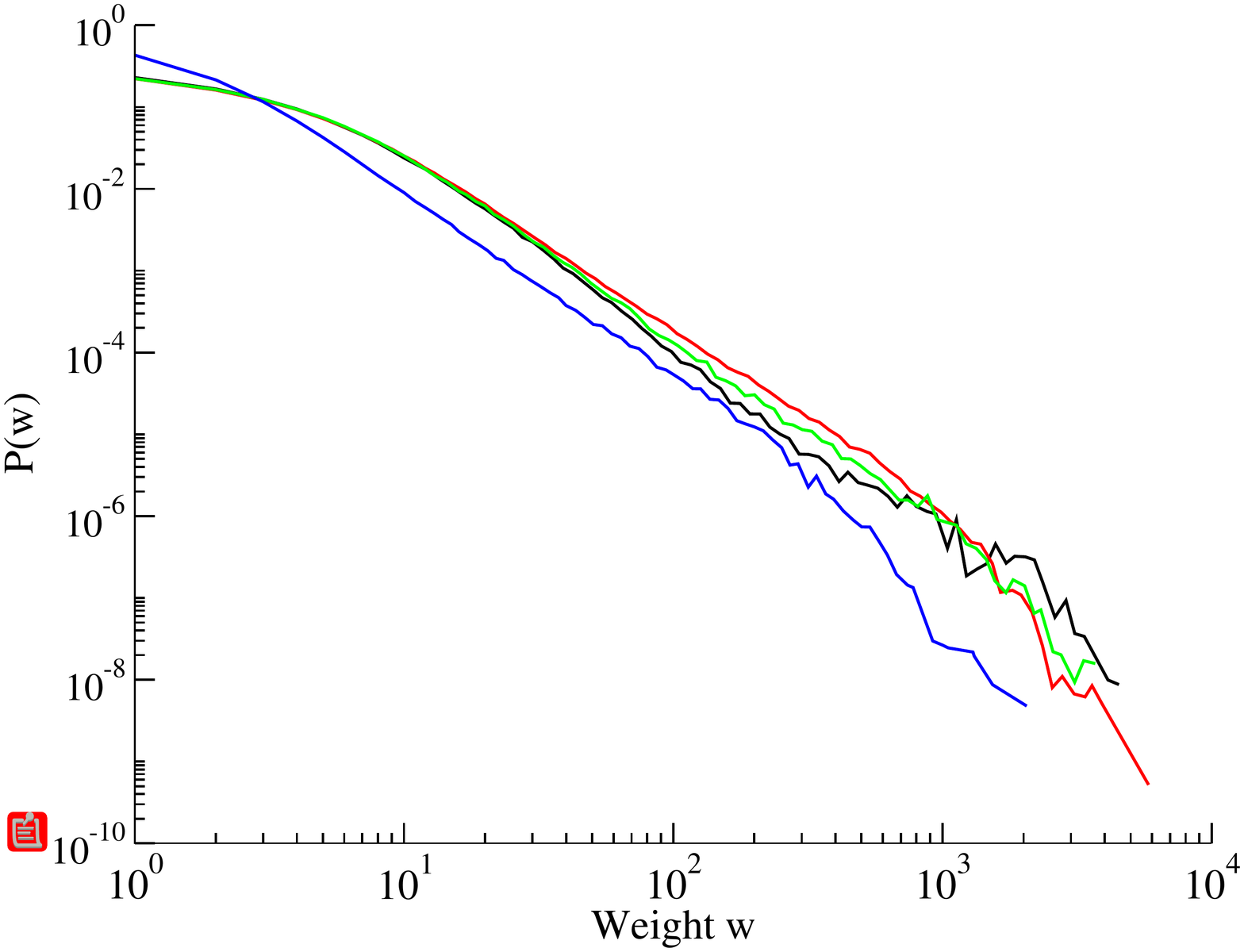}
  \caption{\label{fig:model_robust}  
  Contact duration (left) and weigh (right) probability distributions
  obtained by simulating the model in its original definition, 
  with a different density $\rho$ given by $N=400$ individuals,
  with constant activation probability $\Phi(r)=\delta_{r, r_0}$ with $r_0 = 0.5$,
  and with a L\'evy flight motion dynamics, obtained by using
  Eq. \ref{eq:model_lf} with $\gamma = 2.5$ for extracting the step length.  }
\end{center}
\end{figure}

The model discussed above depends on different numerical and functional
parameters, namely the individual density $\rho$, the attractiveness
distribution $\eta(a)$ and the activation probability distribution
$\phi(r)$.  As we have seen, some properties of the model, especially
those related to group and collective level dynamics, do indeed depend
of the density $\rho$ (or the number of individuals $N$), in such a way
that the model is able to reproduce empirical data only when fed with a
value of $N$ corresponding to the data set under consideration.  The
model properties relevant to the individual level dynamics however, such
as the contact duration and weight distributions, $P(\Delta t)$ and
$P(w)$, do not change in a reasonable range of density.
In Fig. \ref{fig:model_robust} one can observe that the functional form of
these distributions is robust with respect to changes of the individual
density, supporting the natural notion that individual level dynamics is
mainly determined by close contacts of pairs of individuals, and rather
independent of eventual multiple contacts, which become rarer for
smaller densities.

We have also explored the dependence of the model on the activation
probability distribution and the walking probability. In particular,
instead of a uniform activation probability distribution, we have
considered a constant distribution 
\begin{equation}
  \phi(r) = \delta_{r, r_0},
  \label{eq:model_distr_r}
\end{equation}
where $\delta_{r, r'}$ is the Kronecker symbol. 
 As we can see from Fig. \ref{fig:model_robust} the output of the model is 
robust with respect to changes of this functional parameter.

Finally, in the definition of the model we have adopted the simplest
motion dynamics for individuals, namely an isotropic random walk in
which the distance $v$ covered by the agents at each step is constant
and arbitrarily fixed to $v=1$. However, it has been noted for long that
a L\'evy flight turns out to provide a better characterization of human
or animal movement and foraging \cite{viswanathan2011physics}. In this
case, the random walk is still isotropic, but now the distance covered
in each step is a random variable, extracted from a probability
distribution
\begin{equation}
\label{eq:model_lf}
\mathcal{L}(v) \simeq v ^{-\gamma},  
\end{equation}
with a long tailed form.  In Fig. \ref{fig:model_robust} we show that adopting
a L\'evy flight motion dynamics gives rise to outcomes in very good
agreement with the original definition of the model.  We note, however,
that the step length probability distribution $\mathcal{L}(v)$ has a
natural cutoff given by the size of the box where the agents move,
$v<L$, reducing the degree of heterogeneity that the walk can cover.

\section{Summary and Discussion}
\label{sec:model_discussion}

Understanding the temporal and structural properties of human contact
networks has important consequences for social sciences, cognitive
sciences, and epidemiology.  The interest in this area has been fueled
by the recent availability of large amounts of empirical data, obtained
from expressly designed experimental setups.  The universal features
observed in these empirical studies prompt for the design of general
models, capable of accounting for the observed statistical regularities.

In this Chapter 
we have introduced a simple model of mobile
agents that naturally reproduces the social context described by the
Sociopatterns deployments, where several individuals move in a closed
environment
and interact between them
when situated within a small distance (the exchange range of RFID
devices). The main ingredients of the model are: (i) Agents perform a
biased random walk in two-dimensional space; (ii) their interactions
are ruled by an heterogeneous attractiveness parameter,
Eq.~\eqref{eq:model_rule}; and (iii) not all agents are simultaneously active
in the system. 
Without any data-driven mechanism, the model 
is able to quantitatively capture most
of the properties of the pattern of interactions between agents, 
at different scales represented by the individual, group and collective dynamics.
Importantly, the match between the model and the empirical
results is independent of the numerical and functional form of the
diverse parameters defining the model.  

However, the attractiveness
distribution $\eta(a)$ used in the model definition deserves a more
detailed discussion.  Its functional form is hard to access empirically,
and it is likely to be in its turn the combination of different
elements, such as prestige, status, role, etc.  Moreover, even though in
general attractiveness is a relational variable -- the same individual
exerting different interest on different agents -- we have assumed the
simplest case of a uniform distribution for the attractiveness.  For
this reason it is important to stress some facts that support our
decision, and to investigate the effect of the attractiveness
distribution on the model's outcome.

The choice of a uniform $\eta(a)$ is dictated by the maximum entropy principle, 
according to which the best guess for a unknown but bounded
distribution (as the attractiveness distribution has to be, if we want
it to represent a probability) is precisely the uniform distribution \cite{Jaynes:1957}.
However, we can also explore the relation between the
attractiveness and some other variables that can be accessed empirically.  
In particular, the attractiveness of one individual
and the strength of the corresponding node of the integrated network are expected to
be (non-trivially) related, since the more attractive an individual is,
the longer the other agents will try to engage him in interactions.   
Fig. \ref{fig:model_aggr} shows that the strength distribution $P(s)$ of
the time-integrated network obtained from the empirical data, which
follows approximately an exponential behavior, is well fitted by the model. 
Thus, if we hypothesize that the attractiveness and strength
probability distributions are related as $P(s) ds \sim \eta(a) da$, with
$\eta(a)$ uniform in [0,1], it follows that the strength of an
individual should depend on his attractiveness as
 \begin{equation}
 \label{eq:s_a}
s(a) \sim - \log (1-a).
\end{equation}
We find that this relation is fulfilled by the model (data not shown),
showing that in the model the time spent in interactions by the individuals is
directly related with their degree of attractiveness. 
Therefore the guess of a heterogeneous but uniform $\eta(a)$ leads to a exponential
decay of the $P(s)$ for the model, in accordance with experimental data,
and providing grounds to justify this choice of attractiveness distribution.
Moreover, the simple relation expressed by Eq. \ref{eq:s_a} 
may suggest a way to validate the model, once some reliable measure of attractiveness will be available.

Finally, it is worth highlighting that the form of the attractiveness
distribution $\eta(a)$ is crucial for the model outcome.  
Therefore, it is interesting to explore the effect of different functional forms of 
$\eta(a)$, for example incorporating a higher degree of heterogeneity,
such as in the case of a power-law distribution.  We note, however, that
the form of Eq. (\ref{eq:model_rule}) imposes to the $a_i$ variable to be a
probability, with the consequent constraint of being bounded, $a_i \in [0,1]$.  
A power law distribution defined over a bounded support presents 
the necessity of imposing a lower bound to prevent divergence close to $0$.  
To avoid the insertion of a free parameter in the model, 
one can redefine the motion rule of Eq. (\ref{eq:model_rule}) as
\begin{equation}
\label{eq:mr_inverse}
p^{\rm{inv}}_i(t) = \frac{1}{\max_{j \in \mathcal{N}_i(t) } \{ a'_j \} },
\end{equation} 
with the new attractiveness variable $a'_i$ unbounded,  $a'_i \in [1, \infty)$.
If we impose the walking probability of Eqs. (\ref{eq:model_rule}) and
(\ref{eq:mr_inverse}) to be the same, and we use the relation $\eta(a)
da = \zeta(a') da'$, we find that the new attractiveness distribution
$\zeta(a')$ has the form of a power law, $\zeta(a') = (\gamma -1 )
a'^{-\gamma}$, with exponent $\gamma = 2$.  Therefore, assuming a motion
rule of the form of Eq. (\ref{eq:mr_inverse}), a power law
attractiveness distribution will give rise to the same model results, as
confirmed by numerical simulations.

On the same line of argument, it would be interesting to relate the
agents' activation probability, $r_i$, with some empirically accessible
properties of the individuals. Unfortunately, finding the activation
probability distribution $\phi(r)$ is a hard task with the information
contained in the available datasets.  In the face-to-face interaction
deployment, indeed, a non-interacting but active individual is
indistinguishable from an inactive individual who is temporary not
involved into the event.  Thus, simply measuring probability to be not
involved in a conversation does not inform on the $\phi(r)$, but instead
considers something more related with the burstiness of the individual
activity.  In any case, however, the model behavior is independent of
the functional form the activation distribution, so that this point is
less crucial.

In summary, we showed that a simple model is able to account for the main statistical properties of
human contact networks at different scales. 
The proposed framework represents an important step forward in the understanding of
face-to-face temporal networks. Confronted with other modeling
efforts of SocioPatterns data \cite{PhysRevE.83.056109},  
our model is not based on any cognitive
assumption (reinforcement dynamics in Ref.~\cite{PhysRevE.83.056109}) 
and furthermore it leads to a good
agreement with experimental data without any fine tuning of internal
parameters.  
It thus opens new interesting directions for future work,
including the study of dynamical processes taking place on
face-to-face networks and possible extensions of the model to more
general settings.
Moreover, since the concept of social attractiveness is crucial in the model definition,
our finding also prompt for further empirical research, based on more detailed and extensive
experimental setups, which can shed light on the role of this attractiveness. 
Such research would help to further refine and validate
the model considered here, and could potentially provide new insights
for the social and cognitive sciences.

\chapter[Activity driven networks]{Activity driven networks}
\label{chap:activitydriven}

Research in the field of network science has mainly focused on a twofold objective: 
A data-driven effort to characterize the topological properties of real networks
\cite{barabasi02,Dorogovtsev:2002,caldarelli2007sfn}, and a posterior
modeling effort, aimed at understanding the microscopic mechanisms
yielding the observed topological properties
\cite{barabasi02,Newman2010}.
Great advancements in these tasks have been made possible through the recent data revolution,
with the availability of large digital databases and the deployment
of new experimental infrastructures.
Nevertheless, despite its importance, 
the modeling effort in the time-varying network field is still in his infancy
\cite{PhysRevE.83.056109,journals/corr/abs-1106-0288,citeulike:7974615,PhysRevLett.110.168701},
while the recent availability of time resolved data allows to deepen the study of 
the relation between the temporal patterns observed and their effects on
the corresponding integrated networks.

In this Chapter we will focus on the \textit{activity
  driven} network model recently introduced by Perra \textit{et
  al.}  \cite{2012arXiv1203.5351P}, aimed to capture the
relation between the dynamical properties of time-varying networks
and the topological properties of their corresponding aggregated social networks.
The key element in the definition of this model is the observation
that the formation of social interactions is driven by the
\textit{activity} of individuals, urging them to interact with their
peers, and by the empirical fact that different individuals show
different levels of social activity.  
Based in the concept of \textit{activity potential}, defined as the
probability per unit time that an individual engages in a social
activity, Ref.~\cite{2012arXiv1203.5351P} proposed an activity driven
social network model, in which individuals start interactions, that
span for a fixed length of time $\Delta t$, with probability
proportional to their activity potential.  The model output is thus
given by a sequence of graphs, depending on the distribution $F(a)$ of
the activity potential, which are updated every time interval $\Delta
t$.

In the landscape of temporal network models surged in the last years, 
the activity driven model
has the peculiarity of allowing some analytic treatment.
Its simplicity, indeed, suggests a analogy with a class of hidden variables models 
\cite{PhysRevLett.89.258702, PhysRevE.68.036112, Soderberg:2002fk},
which opens the path toward the analytic derivation of several quantities of interest.
Through the mapping to the hidden variable model, 
it is possible to compute analytic expressions for the topological properties of the time-integrated networks,
as a function of the integration time $T$ and the activity potential functional form $F(a)$.
Moreover, the activity driven model is an ideal framework
to study analytically another problem of interests in the temporal networks field,
namely the connectivity properties of the time-integrated networks.
The effect of the integration time $T$ in the structural properties of the aggregated
network, indeed, is an issue which has been recently shown to have relevant
consequences for dynamical processes \cite{ribeiro2013quantifying}.
The analysis presented in this Chapter is thus twofold. 
On the one hand, we focus on the topological properties of the aggregated network 
constructed by integrating the activity driven network up to a given time $T$.
On the other hand, through the expressions found for the aggregated network as a function of
the integration time $T$, we study the percolation properties of 
the time-integrated networks, 
as revealed by the percolation threshold and size of the giant connected component in time.
The formalism proposed allows to draw interesting insights 
regarding the dynamical processes running on the top of the activity driven networks, 
and can be extended to generalizations of the activity driven model.

The Chapter is organized as follow: 
First, Section \ref{sec:activity-model} defines the activity driven model and reviews briefly the hidden variables formalism.
Then in Section \ref{sec:integrated_activity} we derive the topological properties of the time-integrated network 
and in Section \ref{sec:perc_activity} we compute the temporal percolation properties of the activity driven networks.
Finally, in Section \ref{sec:activity-extensions} we discuss possible extensions of the activity driven model,
while Section \ref{sec:concl_act} is devoted to conclusions.


\section{The activity driven network model}
\label{sec:activity-model}


In Section \ref{subsec:intro_APS} we showed that 
the activity potential measured in different social contexts
is broad tailed and independent of the time window considered,
as revealed by the empirical analysis presented in Ref \cite{2012arXiv1203.5351P}.
This observation leads the authors to the definition of a simple process model 
for the generation of random dynamic graphs,
using the activity distribution to drive the formation of a time-varying network.
Importantly, the model allows to study the dynamical processes,
such as random walks \cite{perra_random_2012} or epidemic spreading \cite{PhysRevLett.112.118702}, 
unfolding on the activity driven networks
without relying on any time-scale separation approximation.



\subsection{Model definition}

The activity driven network model is defined in terms of $N$ individuals
$i$ (agents), each one of them characterized by her activity potential
$a_i$, defined as the probability that she engages in a social
act/connection with other agents per unit time. The activity of the
agents is a (quenched) random variable, extracted from the activity
potential distribution $F(a)$, which can take a priori any form.  The
model is defined by means of a synchronous update scheme, time being
measured in units of the life span of each connection $\Delta t$. It
proceeds by creating a succession of instantaneous networks
$\mathcal{G}_t$, $t=0, \Delta t, 2 \Delta t, \ldots, n \Delta t,
\ldots$ At a given time $t$, all previous edges are deleted and we
start with $N$ disconnected individuals. Each one of them is checked
and becomes active with probability $a_i \Delta t$. Active agents
generate $m$ links (start $m$ social interactions) that are connected
to $m$ other agents selected uniformly at random. Finally, time is
updated as $t \to t + \Delta t$. This procedure implies that all edges
in the temporal network have the same constant time duration $\Delta
t$.  In order to avoid complications due to the differences in the
number of emitted and received connections arising from using a
synchronous approach\footnote{Indeed, in a synchronous scheme, every
  time step an agent fires at most one connection, but can receive a
  number $n$ of connections, given trivially by a binomial
  distribution.}, here we consider a probabilistic recipe for the
instantaneous network construction: Each microscopic time step $\Delta
t$, we choose $N$ agents, uniformly at random, and check sequentially
each one of them for activation and eventual link emission.  We avoid
self and multiple connections.

To simplify the analytical calculations performed below, in the
following we choose $\Delta t= m =1$. Both quantities can be however
restored by a simple rescaling of the activity potential and the
integration time $T$. We notice that imposing $\Delta t=1$ implies
restricting the activity potential to be probability, and thus to be
limited in the interval $a \in [0,1]$.

\subsection{Hidden variables formalism: A short review}
\label{subsec:hidden_variables}

The class of network models with hidden variables was introduced in
Ref.~\cite{PhysRevE.68.036112} (see also
\cite{PhysRevLett.89.258702,Soderberg:2002fk}) as a generalization of
the random network Gilbert model \cite{Dorogobook2010}, in which the
probability of connecting two vertices is not constant, but depends on
some intrinsic properties of the respective vertices, their so-called
hidden variables. This class of models is defined as follows: Starting
from a set of $N$ disconnected vertices and a general hidden variable
$h$, we construct an undirected network with no self-edges nor
multiple connections, by applying these two rules:
\begin{enumerate}
\item To each vertex $i$, a variable $h_i$ is assigned, drawn at
  random from a probability distribution $\rho(h)$.
\item For each pair of vertices $i$ and $j$, $i \neq j$, with hidden
  variables $h_i$ and $h_j$, respectively, an edge is created with
  probability $r(h_i, h_j)$, the connection probability, which is a
  symmetric function bounded by $0 \leq r(h, h') \leq 1$.
\end{enumerate}
Each model in the class is fully defined by the functions $\rho(h)$
and $r(h, h')$, and all its topological properties can be derived as a
function of these two parameters. These topological properties are
encoded in the propagator $g(k|h)$, defined as the conditional
probability that a vertex with hidden variable $h$ ends up connected
to $k$ other vertices. The propagator is a normalized function,
$\sum_k g(k|h) =1$, whose generating function $\hat{g}(z|h) = \sum_k
z^k g(k|h)$ fulfills the general equation \cite{PhysRevE.68.036112}
\begin{equation}
  \ln   \hat{g}(z|h) = N \sum_{h'} \rho(h') \ln \left[ 1-(1-z) r(h, h')
  \right].
  \label{eq:hidden_gen_fun} 
\end{equation}
From this propagator, expressions for the topological
properties of the model can be readily obtained
\cite{PhysRevE.68.036112}:
\begin{itemize}

\item Degree distribution:
  \begin{equation}
    P(k)=\sum_h g(k|h)\rho(h).
    \label{eq:hidden_pk}
  \end{equation}

\item Moments of the degree distribution:
  \begin{equation}
  \langle k^n \rangle = \sum_{k,h} k^n g(k|h) \rho(h).
    \label{eq:hidden_moments}
  \end{equation}
  
\item Degree correlations, which can be measured by the average degree of the
  neighbors of the vertices of degree $k$, $\bar{k}^{nn}(k)$ \cite{alexei}:
  \begin{equation}
    \bar{k}^{nn}(k) = 1+ \frac{1}{P(k)} \sum_h \rho(h) g(k|h)
    \bar{k}^{nn}(h),
    \label{eq:hidden_knnk}
  \end{equation}
 where we have defined
  \begin{equation}
    \bar{k}^{nn}(h) = \frac{N}{\bar{k}(h)}  \sum_{h'}  \rho(h') \bar{k}(h')
    r(h,h'),
    \label{eq:hidden_knnh}
  \end{equation}
  and the average degree of the vertices with hidden variable $h$,
  \begin{equation}
    \bar{k}(h) = N \sum_{h'} \rho(h') r(h,h').
    \label{eq:hidden_kh}
  \end{equation}

\item Average clustering coefficient $\av{c}$, defined
  as the probability that two vertices are connected, provided that
  they share a common neighbor \cite{watts98}
  \begin{equation}
    \label{eq:hidden_av_c}
    \av{c} = \sum_h \rho(h) \bar{c}(h),
  \end{equation}
 where we have defined
  \begin{equation}
    \label{eq:hidden_clus_h}
    \bar{c}(h) = \sum_{h', h''} p(h'|h) r(h', h'') p(h''|h),
  \end{equation}
  and
  \begin{equation}
    \label{eq:hidden_phh}
    p(h'|h) = \frac{N \rho(h') r(h, h')}{ \bar{k}(h)}.
  \end{equation}
  Additionally, one can define the clustering spectrum, as measured by
  the average clustering coefficient of the vertices of degree $k$,
  $\bar{c}(k)$ \cite{alexei,ravasz_hierarchical_2003}
  \begin{equation}
    \label{eq:hidden_ck}
    \bar{c}(k) = \frac{1}{P(k)}  \sum_h \rho(h) g(k|h) \bar{c}(h),
  \end{equation}

\end{itemize}


\section{Time-integrated activity driven networks}
\label{sec:integrated_activity}

\begin{figure}[tb]
\begin{center}
  \includegraphics[width=0.9\textwidth]{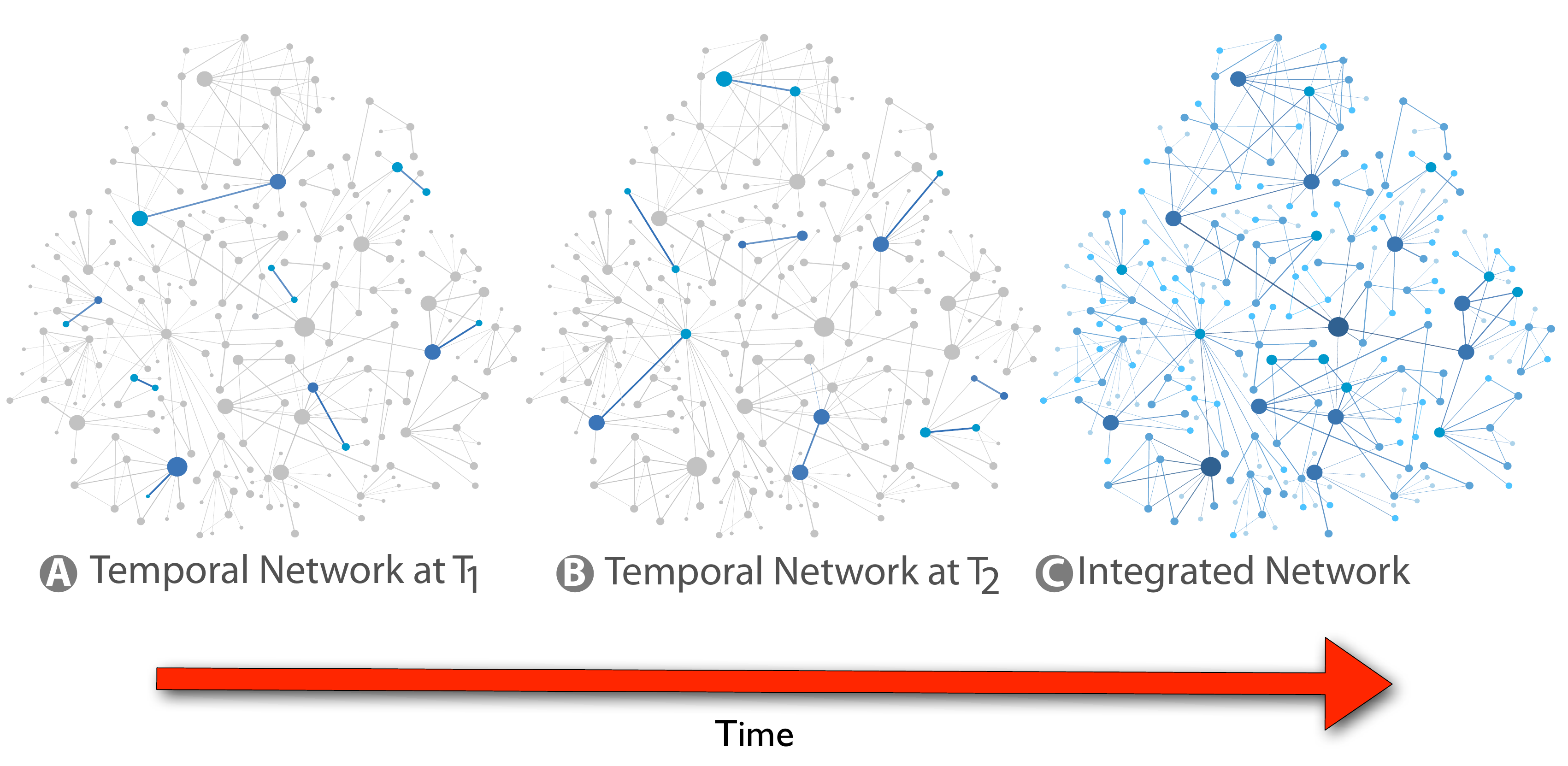}
  \caption{  Schematic representation of activity driven model. 
  (A), (B): Temporal network at two different time steps $T_1$ and $T_2$,
  represented by the graphs $\mathcal{G}_{T_1}$ and $\mathcal{G}_{T_2}$.
(C): Integrated network over a certain period of time. 
The size and color of each node describes
its activity, while the width and color of each link describes the
weight. Figure courtesy of Nicola Perra.  }
  \label{fig:act_driven_model}
\end{center}
\end{figure}

In this Section we address the study of the aggregated network generated by the
activity driven model, obtained by integrating the time-varying network up
to a given time $T$. 
The activity-driven network model, indeed, generates a time series of
instantaneous sparse graphs $\mathcal{G}_t$, each one with an average degree $\av{k}_t \simeq 2 \av{a}$, where $\av{a} = \sum_a a F(a)$. 
The integrated network at time $T$ is constructed by performing the union of 
the instantaneous networks, i.e. $\mathcal{G}_T = \cup_{t=0}^T \mathcal{G}_t$.
In this integrated network, nodes $i$ and $j$ will be joined by an edge 
if a connection has been established between $i$ and $j$ in any of the
instantaneous networks at $0 \leq t \leq T$. 
In Fig. \ref{fig:act_driven_model} we show a schematic representation of the activity driven model,
 with two snapshot at different time steps $T_1$ and $T_2$ of the time-varying network, 
 and the integrated static network at time $T \gg T_1, T_2$.

The topological properties of the integrated activity driven
network were related in Ref.~\cite{2012arXiv1203.5351P} at the level
of the degree distribution, which, by means of approximate arguments,
was shown to be proportional to the activity potential distribution
$F(a)$. However, despite the interest of the model, expressions for
the rest of topological observables have been still lacking, a fact
that hampers its possible validation as a generator of realistic
integrated social networks, as well as the identification of the
particular role that integration time has on the behavior of dynamical
processes running on top of the temporal network
\cite{perra_random_2012,ribeiro2013quantifying}.

By considering a mapping of the integrated network to 
the hidden variables model \cite{PhysRevE.68.036112} reviewed 
in Section \ref{sec:activity-model},
we are able to obtain analytic expressions of several topological properties of the integrated network,
depending on the activity potential distribution and the considered
time $T$.  We obtain a set of expressions for the degree distribution,
degree-degree correlations and clustering coefficient of the
aggregated network that are exact in the limit of large network size
$N \to \infty$ and finite time $T$, and which are amenable to analytic
asymptotic expansions in this same limit. The expressions obtained,
confirmed by numerical simulations, corroborate the basic assumption
of the activity driven model linking social activity with network
topology.

\subsection{Mapping the integrated network to a hidden variables model}
\label{subsec:map_integr_act}

The key point to map the
integrated network to a hidden variables model resides in computing
the probability $\Pi_T(i,j)$ that two vertices $i$ and $j$ become
eventually joined at time $T$. This probability is given by
$\Pi_T(i,j) =1- Q_T(i,j)$, where $Q_T(i,j)$, the probability that no
connection has ever been created between agents $i$ and $j$ up to time
$T$, can be calculated as follows: At time $T$, an agent $i$ will have
become active $z$ times with probability $P_T(z)$.  Given the
definition of the model, at time $T$ we have selected $T N$ agents to
check for activation. The number of times $z$ that agent $i$ has
become active will be given by the binomial distribution
\begin{equation}
  \label{eq:activity_binom}
  P_T(z) = \binom{T N }{z} \left(
    \frac{a_i}{N}\right)^{z} 
  \left(1- \frac{a_i}{N}\right)^{TN  -z},
\end{equation}
and analogously for agent $j$.  Now, vertices $i$ and $j$ will be
connected in the integrated network if at least one of the links
generated from $i$ reaches $j$, or vice-versa. Since every time that
she becomes active, an agent creates a connection targeted to a
randomly chosen peer, the probability $Q_T(i,j)$ is given by
\begin{equation}
\label{eq:activity_prob_q}
  Q_T(i,j)  =  \sum_{z_i,z_j}  P_T(z_i)  P_T(z_i)
  \left(1-\frac{1}{N}\right)^{ z_i + z_j} =
  \left[ \left( 1 - \frac{a_i}{N^2}
      \right)
    \left( 1 - \frac{a_j}{N^2}
       \right)
  \right]^{T N }, 
\end{equation}
where we have performed the summation using the probability
distribution in Eq.~\eqref{eq:activity_binom}. We see now that the probability
that agents $i$ and $j$ are connected in the integrated network at
time $T$ depends only on their respective activity potentials $a_i$
and $a_j$, which are random variables with distribution $F(a)$.  The
mapping to a hidden variables network is thus transparent:
\begin{itemize}
\item Hidden variable: $h \to a$.
\item Distribution of hidden variables: $\rho(h) \to F(a)$.
\item Connection probability: $r(h, h') \to \Pi_T(a, a')$.
\end{itemize}

At very large times, the integrated network emerging from the activity
driven model will trivially tend to a fully connected
network. Interesting topology will thus be restricted to the limit of
small $T$ compared with the network size $N$. In this limit,
Eq.~(\ref{eq:activity_prob_q}) can be simplified, yielding
\begin{equation}
\label{eq:activity_conn_prb}
  \Pi_T(a,a')  =  1-Q_T(a,a') \simeq1 - \left[ 1 - \frac{(a + a')}{N^2}
  \right]^{TN} \simeq 1- \exp\left[ -
    \lT (a +a') \right],
\end{equation}
where we have neglected terms of order $\mathcal{O}(N^{-2})$ and
defined the parameter
\begin{equation}
  \lambda =  \frac{T}{N}.
\end{equation}
An explicit calculation of the connection probability for a factor
$m>1$ and a time interval $\Delta t \neq 1$ can be easily performed;
in the limit of large $N$ and constant $\lambda$, the only change
ensuing is a rescaling of time, $T \to Tm$, the value of $\Delta t$
becoming canceled in the process of taking the limit $ \lambda\to0$.

\subsection{Topological properties}
\label{subsec:topol_prop_act}

Here we will apply the formalism presented in
Sec.~\ref{subsec:hidden_variables} to provide analytic expressions
characterizing the topology of the integrated network resulting from
the activity driven model.  For the sake of concreteness, we will
focus in the following activity potential distributions, in the
continuous $a$ limit:
\begin{itemize}
\item Constant activity:
  \begin{displaymath}
    F(a) = \delta_{a, a_0} , \;\; \mathrm{with} \; 0< a_0 < 1.
  \end{displaymath}
\item Homogeneous activity:
  \begin{equation}
    F(a) = 1/a_{\max}, \;\; \mathrm{with} \;  0 \leq a \leq a_{\max}
    \leq 1. 
  \end{equation}
\item Power-law distributed activity:  
  \begin{equation}
     F(a) = (\gamma-1) \eps^{\gamma-1} a^{-\gamma}, \;\; \mathrm{with} \; a \in
    [\eps,1].
  \end{equation}
\end{itemize}
In the last case, where we consider $\gamma>2$, in accordance with
experimental evidence \cite{2012arXiv1203.5351P}, we have introduced a
lower cut-off $0<\eps \ll 1$ in order to avoid dangerous divergences
in the vicinity of zero.

\subsubsection{Degree distribution}
\label{subsubsec:deg_distr_act}

In order to compute the degree distribution we have to solve and
invert the generating function equation Eq.~\eqref{eq:hidden_gen_fun}, an almost
impossible task to perform exactly, except in the case of very simple
forms of the activity potential distribution. So, in the case of
constant activity, $F(a) = \delta_{a, a_0}$, we have
\begin{equation}
   \hat{g}(z|a_0) = \left[ z \Pi_T(a_0,a_0) + (1-\Pi_T(a_0,a_0)) \right]^N,
\end{equation}
which corresponds to the generating function of a binomial
distribution \cite{Wilf:2006:GEN:1204575}. Therefore, in the limit of
large $N$ and constant $\lT$, the degree distribution takes the
Poisson form
\begin{equation}
  \label{eq:activity_cnst_pk}
  P_T(k) = e^{-\mu} \frac{\mu^k}{k!}
\end{equation}
with parameter $\mu = N \left(1- e^{-2 \lT a_0}\right)$, which, for
fixed $T$ and large $N$, can be approximated as $\mu \simeq 2 T
a_0$. 

\begin{figure}[tb]
  \includegraphics[width=0.48\textwidth]{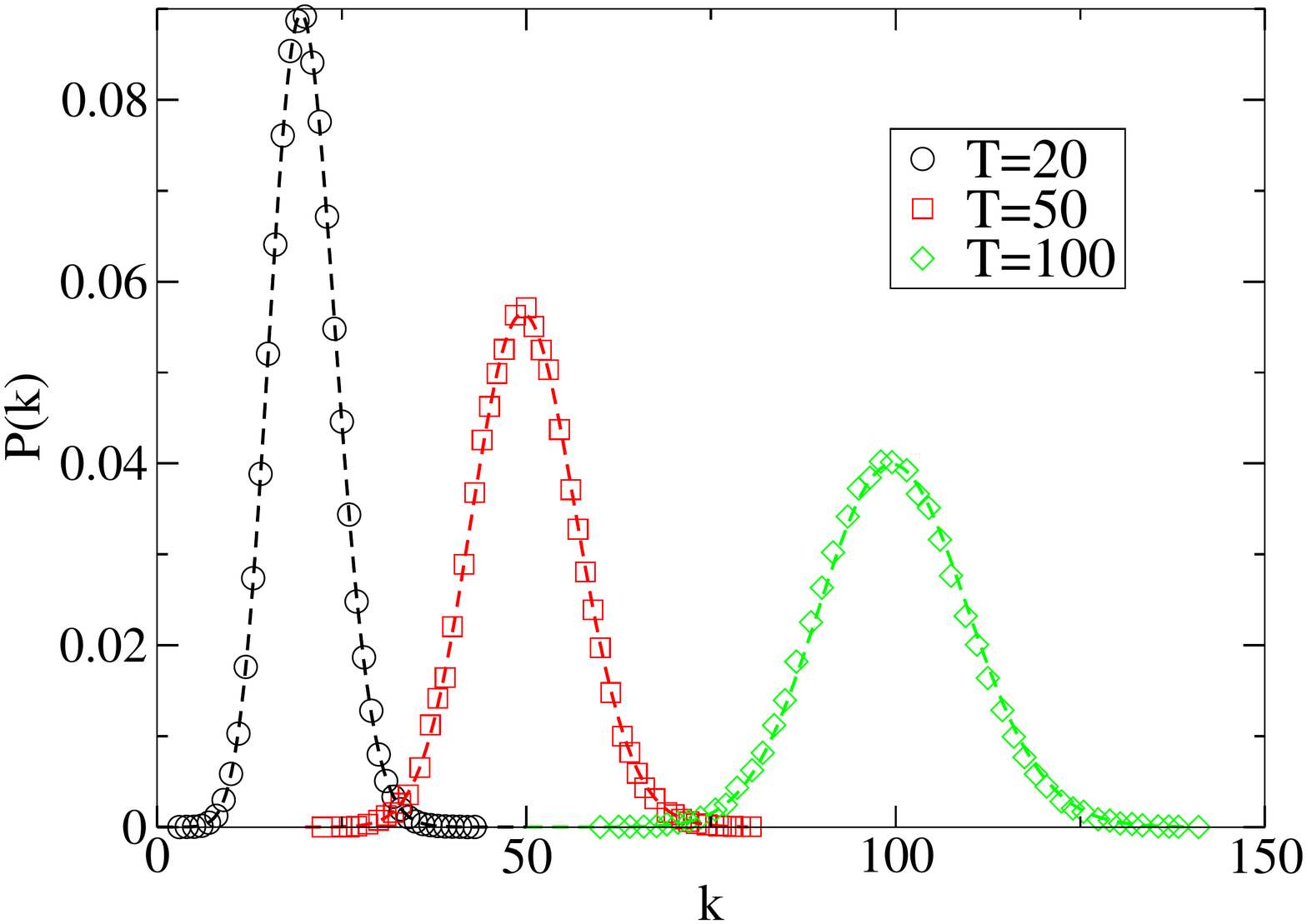}
    \includegraphics[width=0.48\textwidth]{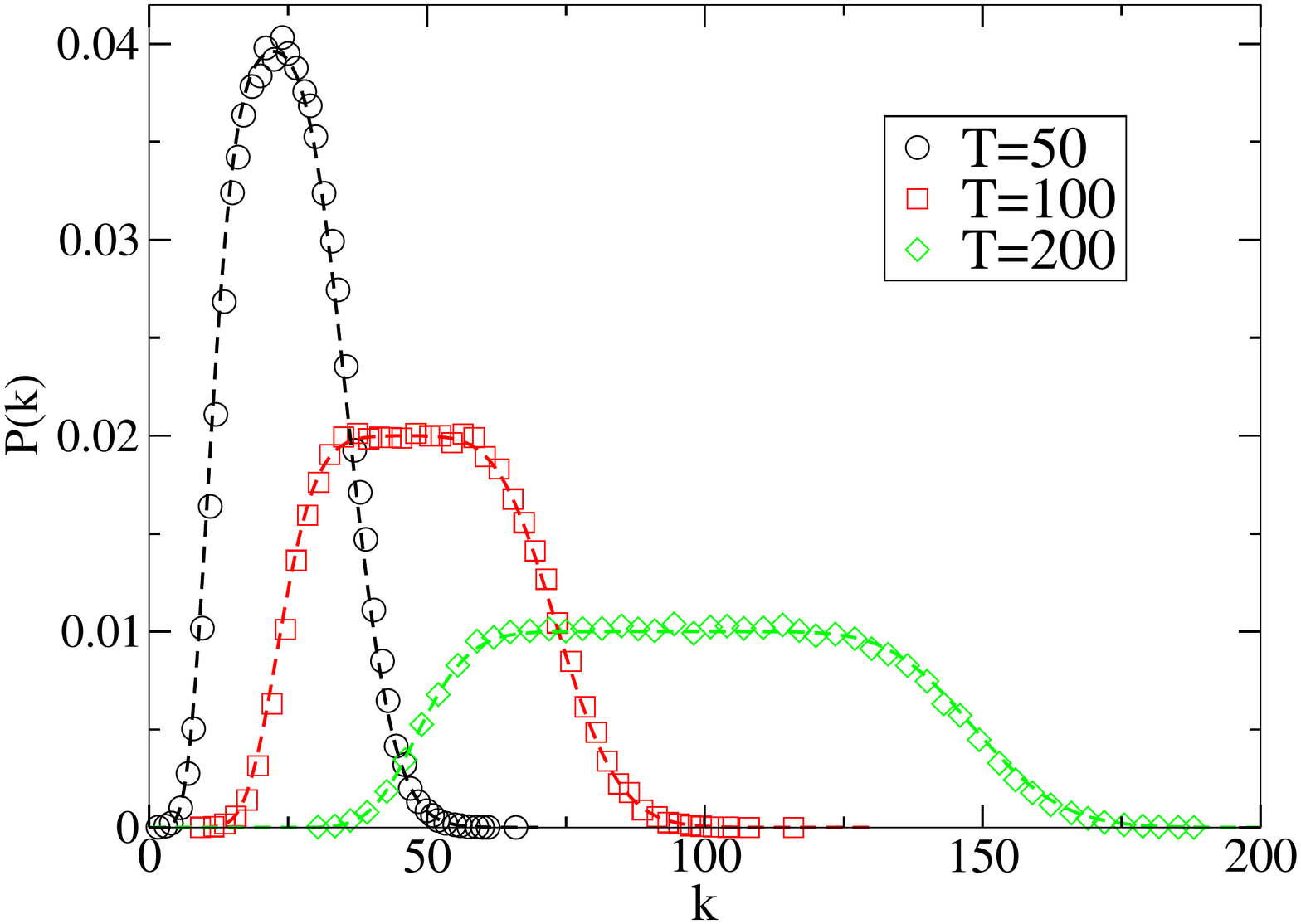}
  \caption{   \label{fig:activity_hom_k_distr}
 	Degree distribution $P_T(k)$ for
    integrated networks corresponding to different values of $T$, with
    constant activity distribution $F(a) = \delta_{a,a_0}$ with $a_0=0.5$ (left),
    and uniform activity distribution $F(a) = a_{\rm{max}}^{-1}$, with $ a_{\rm{max}} = 0.5$.
      The behavior predicted by Eq. \eqref{eq:activity_cnst_pk}, on the left, 
      and  by Eq. \eqref{eq:activity_pk_hom}, on the right,  
      is showed as dashed lines.     }
\end{figure}

For a nontrivial activity distribution $F(a)$, we must resort to
approximations. We therefore focus in the interesting limit of small
$\lT$, which corresponds to fixed $T$ and large $N$, which is the one
yielding a non-trivial topology.  In this limit, we can approximate
the connection probability as
\begin{equation}
  \label{eq:activity_conn_prob_aprox}
  \Pi_T(a,a') \simeq \lT(a +a').
\end{equation}
Introducing this expression into Eq.~\eqref{eq:hidden_gen_fun} and performing a new
expansion at first order in $\lambda$, we obtain
\cite{PhysRevE.68.036112}
\begin{equation}
   \ln   \hat{g}(z|a) 
   \simeq (1-z) \lT N \sum_{a'} F(a') (a+a') 
 =  (1-z) \lT N (a+\av{a}).  
\end{equation}
The generating function of the propagator is a pure exponential, which
indicates that the propagator itself is a Poisson distribution
\cite{Wilf:2006:GEN:1204575}, i.e.
\begin{equation}
  \label{eq:activity_prop}
  g_T(k|a) = e^{-T (a+\av{a})} \frac{\left[ T (a+\av{a})
    \right]^k}{\Gamma(k+1)},
\end{equation}
where $\Gamma(x)$ is the Gamma (factorial) function \cite{abramovitz}.
From Eq.~\eqref{eq:hidden_pk} we obtain the general expression for the degree
distribution
\begin{equation}
  \label{eq:activity_pk_general}
  P_T(k) =  \frac{T^k}{\Gamma(k+1)} \sum_a
    F(a) \left[a+\av{a} \right]^k e^{-T (a+\av{a})}.
\end{equation}
In the case of a homogeneous activity distribution, $F(a)=
{a_{\max}}^{-1}$, for which $\av{a} = a_{\max}/ 2$, we can integrate
directly Eq.~\eqref{eq:activity_pk_general}, to obtain
\begin{equation}
  \label{eq:activity_pk_hom}
   P_T(k) =  \frac{\Gamma\left(k+1, T \av{a} \right) -
     \Gamma\left(k+1, 3T \av{a} \right)}{2T\av{a} \; \Gamma(k+1)}.
\end{equation}
where $\Gamma(x,z)$ is the incomplete Gamma function
\cite{abramovitz}.  
In Fig. \ref{fig:activity_hom_k_distr} we plot the degree distribution  $P_T(k)$ of 
the aggregated network at different values of $T$, for constant and homogeneous activity potential, 
$F(a) = \delta_{a, a_0}$ and $F(a) = a_{\max}^{-1}$, respectively,
showing the perfect agreement between numerical results and theoretical prediction 
given by Eqs. \eqref{eq:activity_cnst_pk} and  \eqref{eq:activity_pk_hom}.

More complex forms of the activity distribution do not easily yield to
an exact integration, and more approximations must be performed. In
particular, the asymptotic form of the degree distribution can be
obtained by performing a steepest descent approximation. Thus, we can
write
\begin{equation}
  P_T(k) =  \frac{1}{\Gamma(k+1)}  \int F(a) e^{\phi_T(a)} da,
\end{equation}
where we have defined
\begin{equation}
  \phi_T(a) = k \ln[T(a + \av{a})] - T(a+\av{a}).
\end{equation}
The function $\phi_T(a)$ has a sharp maximum around $a_M = \frac{k}{T} -
\av{a}$. Performing a Taylor expansion up to second order, we can
write $\phi_T(a) \simeq \phi(a_M) - \frac{T^2}{2 k}
\left[a-a_M\right]^2$, with $ \phi(a_M) = k \ln(k) - k$. Now, for $T^2
/ k \gg 1$, the function $e^{- \frac{T^2}{2 k} \left[a-a_M\right]^2}$
is strongly peaked around the maximum $a_M$; therefore we can
substitute the activity potential by its value at the maximum, to
obtain
\begin{equation}
\label{eq:activity_pk_step_desc_aprox}
  P_T(k) \simeq   \frac{e^{\phi(a_M)}F(a_M)
   }{\Gamma(k+1)}  \int_{-\infty}^{\infty} e^{- \frac{T^2}{2 k}
     \left[a- a_M\right]^2} da 
   = \frac{ \sqrt{2 \pi k} k^k e^{-k}
   }{T \Gamma(k+1)} F\left(\frac{k}{T} - \av{a}\right),
\end{equation}
where we have extended the integration limits to plus and minus
infinity. In the large $k$ limit, we can use Stirling's approximation,
$\Gamma(k+1) \sim \sqrt{2 \pi k} k^k e^{-k}$, to obtain the asymptotic
form
\begin{equation}
  \label{eq:activity_pk_aprox}
  P_T(k) \sim \frac{1}{T} F\left(\frac{k}{T} -
    \av{a}\right). 
\end{equation}
In this expression we recover, using more rigorous arguments, the
asymptotic form of the integrated degree distribution obtained in
Ref.~\cite{2012arXiv1203.5351P}.
The limits of validity of this expression are however now transparent, 
being explicitly $N \gg T \gg 1$ and $T^2 \gg k \gg 1$.

\begin{figure}[tb]
  \includegraphics[width=0.48\textwidth]{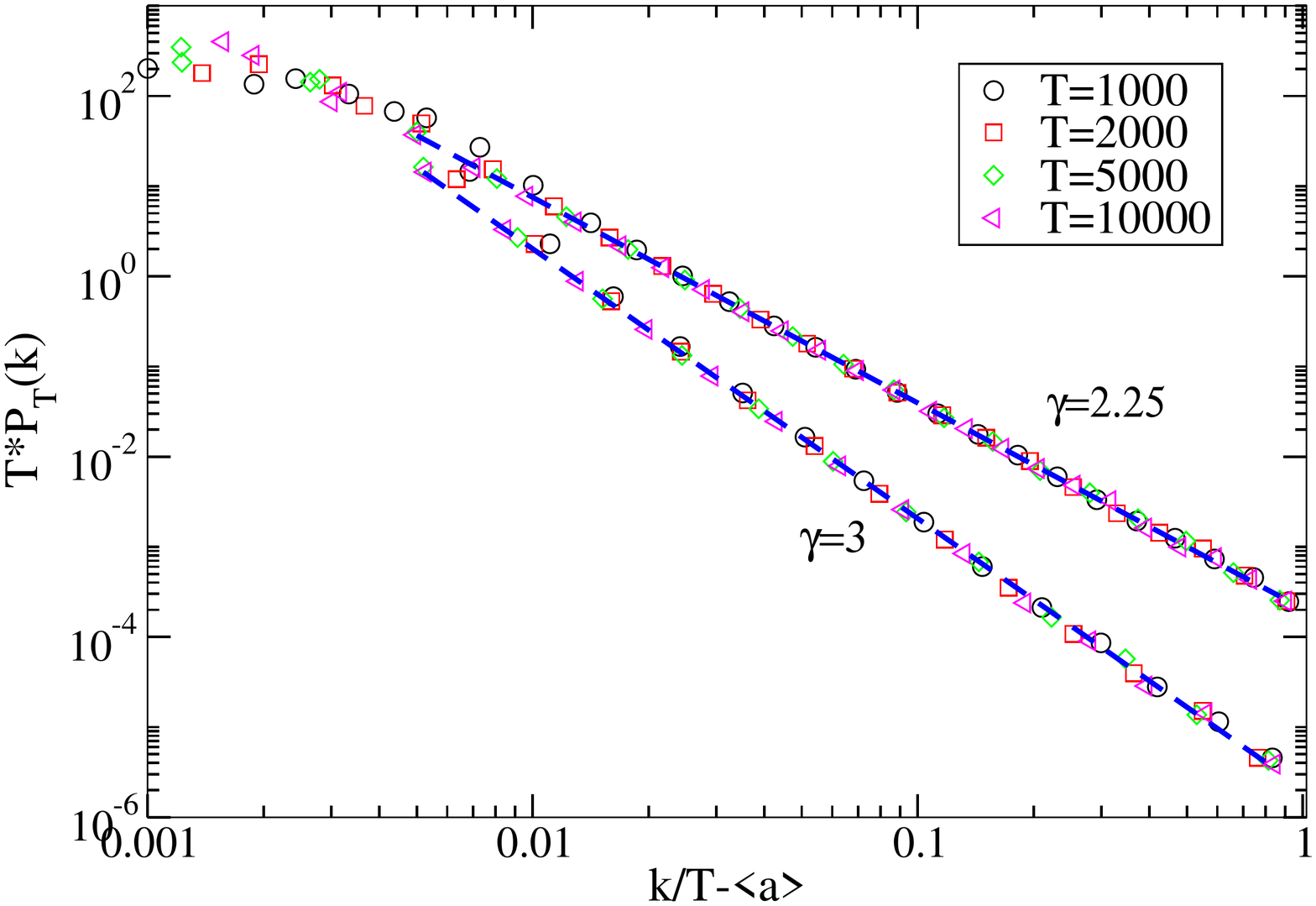}
    \includegraphics[width=0.48\textwidth]{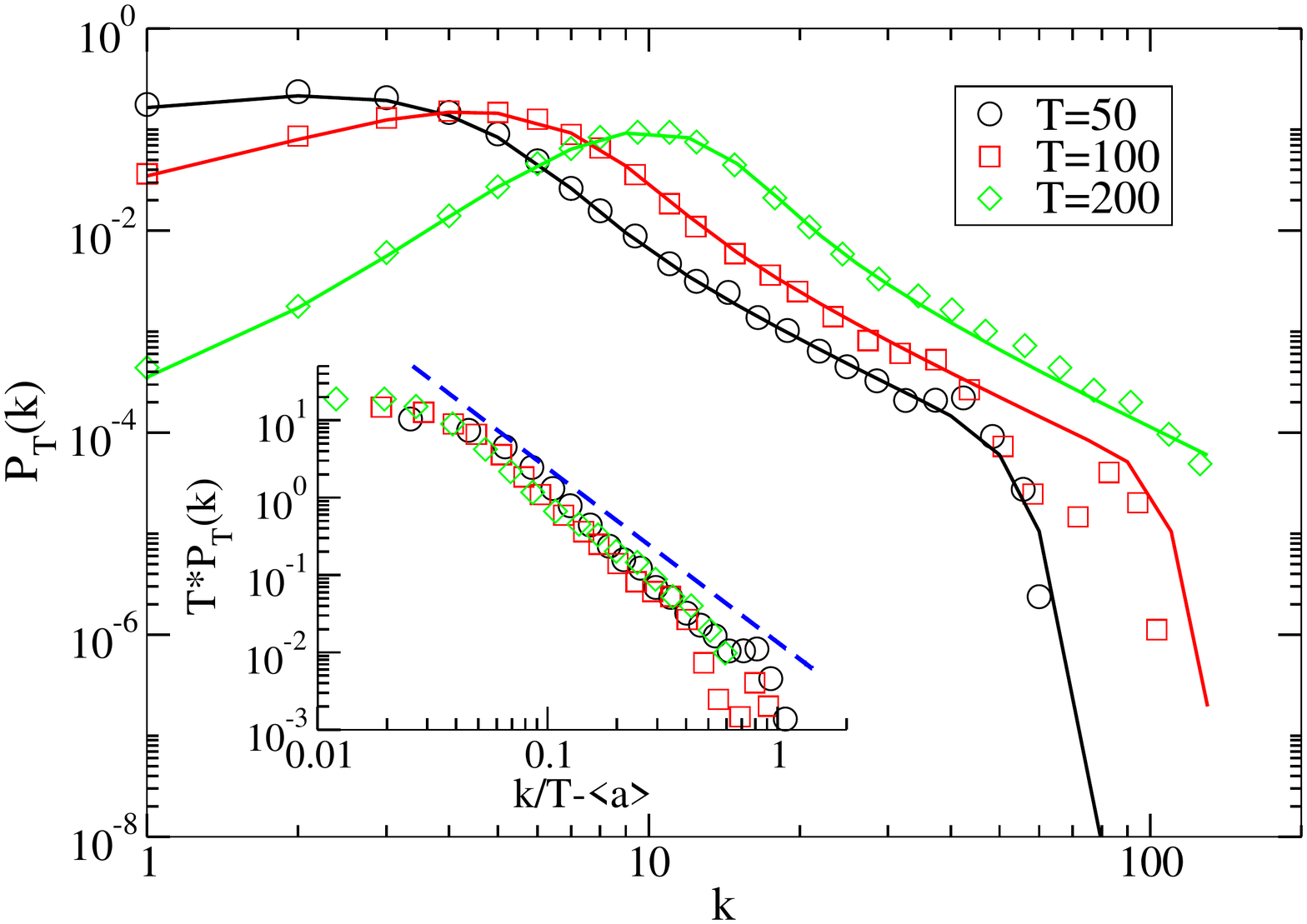}
  \caption{   \label{fig:activity_k_distr}
 	Left: Rescaled degree distribution $P_T(k)$ for
    integrated networks corresponding to different values of $T$, with
    power law activity distribution with exponents $\gamma = 3$ and
    $2.25$. Network size $N=10^6$.  The behavior predicted by
    Eq.~(\ref{eq:activity_pk_aprox}) is represented as dashed lines.
	Right: Degree distribution $P_T(k)$ for integrated
    networks corresponding to different values of $T$, with power law
    activity distribution with exponent $\gamma = 2.25$. Network size $N=10^3$.
    The result of a numerical integration of Eq.~(\ref{eq:activity_pk_general})
    is showed as continuous lines.  Inset: Rescaled $P_T(k)$ shown
    against Eq.~(\ref{eq:activity_pk_aprox}), dashed in blue.    
        }
\end{figure}

For the case of constant activity, $F(a) = \delta_{a, a_0}$, the
asymptotic form of the degree distribution is ${P_T(k) \sim \delta_{k,
    T a_0} / T}$, while the exact form is a Poisson distribution
centered at $2 T a_0$.  For a uniform activity, on the other hand, the
asymptotic prediction is a flat distribution, while the exact
expression can be quite different, in particular for large and small
values of $k$, see Eq.~(\ref{eq:activity_pk_hom}).  For the case of a power-law
distributed activity, in Fig.~\ref{fig:activity_k_distr} we plot the degree
distribution $P_T(k)$ of the aggregated network at different values of
$T$ for networks of size $N=10^6$ and two different values of
$\gamma$.  As we can see, for such large networks sizes and values of
$\lambda \sim 10^{-2} - 10^{-3}$, the asymptotic expression
Eq.~\eqref{eq:activity_pk_aprox} represents a very good approximation to the model
behavior.  In Fig.~\ref{fig:activity_k_distr} we plot the degree distribution
for a smaller network size $N=10^3$. As one can see, a numerical
integration of Eq.~(\ref{eq:activity_pk_general}) recovers exactly the behavior
of $P_T(k)$ even for small values of $k$.  With such small network
size, however, the asymptotic prediction of Eq.~\eqref{eq:activity_pk_aprox} is less
good, as shown in the inset of Fig.~\ref{fig:activity_k_distr}.

\subsubsection{Moments of the degree distribution}
\label{subsubsec:moments_deg_distr_act}

The moments $\langle k^n \rangle_T$ of the degree distribution 
present a very simple form given by Eq. \eqref{eq:hidden_moments}.
 Since the propagator $g_T(k|a)$ has the form of a Poisson distribution, given by Eq. \eqref{eq:activity_prop}, 
 $\langle k^n \rangle_T$ are a direct combination of the moments $\langle a^n \rangle$ of the activity distribution, 
 and simply read as
\begin{equation}
  \label{eq:activity_moments}
  \av{k^n}_T = \sum_{m=1}^n \stirling{n}{m}  T^m \kp_m,
\end{equation}
where $\stirling{n}{m} $ are the Stirling numbers of the second kind
\cite{gradshteyn2007} and
\begin{equation}
  \kp_m = \sum_a F(a)(a+\av{a})^m = \sum_{i=0}^m \binom{m}{i} \av{a^i}
  \av{a}^{m-i}.
\end{equation}

\subsubsection{Degree correlations}
\label{subsubsec:deg_corr_act}

We start from Eq.~\eqref{eq:hidden_kh}, which takes the form, as a function
of time
\begin{equation}
\label{eq:activity_ka}
\bar{k}_T(a) = N[1-e^{-\lT a} \Psi(\lT)],
\end{equation}
where $\Psi(\lambda)$ is the Laplace transform 
\begin{equation}
   \Psi(\lambda) \equiv  \sum_{a} F(a) e^{- \lambda a}.
\end{equation}
We can now use Eq.~\eqref{eq:hidden_knnh}, which leads to the exact expression
 \begin{equation}
  \label{eq:activity_knna}
   \bar{k}^{nn}_T(a) =
  N \left\{ 1 -  \Psi(\lT)\frac{\Psi(\lT)-\Psi(2 \lT) e^{-\lT
        a}}{1-\Psi(\lT) e^{-\lT a}}  
   \right\}.
\end{equation}
In order to obtain an explicit expression for $\bar{k}^{nn}_T(k)$ we
must perform the integral in Eq.~\eqref{eq:hidden_knnk}. In the case of a
constant activity potential, $F(a) = \delta_{a,a_0}$, we have $P_T(k)
= g(k|a_0)$. Since in this case $\Psi(\lT) = e^{-\lT a_0}$, we have
\begin{equation}
  \bar{k}^{nn}_T(k) = 1 + N \left[ 1 - e^{-2\lT a_0}  \right] \simeq 1 +
  2 T a_0,
\end{equation}
where the last expression corresponds to the limit of small $\lT$.
This function is independent of $k$, indicating that the integrated
network corresponding to constant activity potential has no degree
correlations. 

For more complex forms of $F(a)$, we resort to an expansion in powers
of $\lT$ to obtain an approximate expression, which at lowest order
takes the form
\begin{equation}
\label{eq:activity_knna_aprox}
  \bar{k}^{nn}_T(a) \simeq \frac{\lT N }{a+\av{a}} \left[ \av{a^2} +
  \av{a}^2 + 2a \av{a} \right].
\end{equation}
Inserting this expression into Eq.~\eqref{eq:hidden_knnk}, and considering the
Poisson form of the propagator Eq.~\eqref{eq:activity_prop}, we can write
\begin{eqnarray}
  \bar{k}^{nn}_T(k) &\simeq & 1 + \frac{T^2 (\av{a^2} + \av{a}^2)}{k P(k)}  \int da F(a) g(k-1|a) + \nonumber \\ \nonumber
                          &           &  \frac{2T^2 \av{a}}{k P(k)}  \int da a F(a) g(k-1|a) \\  \nonumber
  &\simeq& 1 +  T^2 \frac{P(k-1)}{k P(k)} \left[ \sigma_a^2 + 2\av{a} \left(\frac{k}{T} \right) \right],  
\end{eqnarray}
where in the last expression we have performed the steepest descent
approximation used to obtain Eq.~\eqref{eq:activity_pk_step_desc_aprox}, and $\sigma_a^2 =
\av{a^2} - \av{a}^2$ is the variance of the activity potential
$F(a)$. In the limit of large $k$, where $P(k-1) / P(k) \sim 1$, we
have the general form for the degree correlations
\begin{equation}
  \label{eq:activity_knnk_aprox}
  \frac{ \bar{k}^{nn}_T(k) -1} {T}    \simeq  2\av{a}  + \sigma_a^2
  \left( \frac{ k }{ T} \right)^{-1 } .
\end{equation}
This expression recovers in a natural way the exact result for
constant activity potential, where $\sigma_a^2 = 0$.  From
Eq.~(\ref{eq:activity_knnk_aprox}) we conclude that, in general, for an non-constant
activity distribution, the integrated networks resulting from the
activity driven model show disassortative mixing by degree
\cite{PhysRevLett.89.208701}, with a $\bar{k}^{nn}_T(k)$ function
decreasing as a function of $k$. This disassortative behavior, which
can be however quite mild in the case of small variance $\sigma_a$, as
in the case of a power law distributed activity with small $\eps$, is
in any case at odds with the assortative form observed for degree
correlations in real social networks \cite{Newman2010}.

\begin{figure}[tb]
  \includegraphics[width=0.49\textwidth]{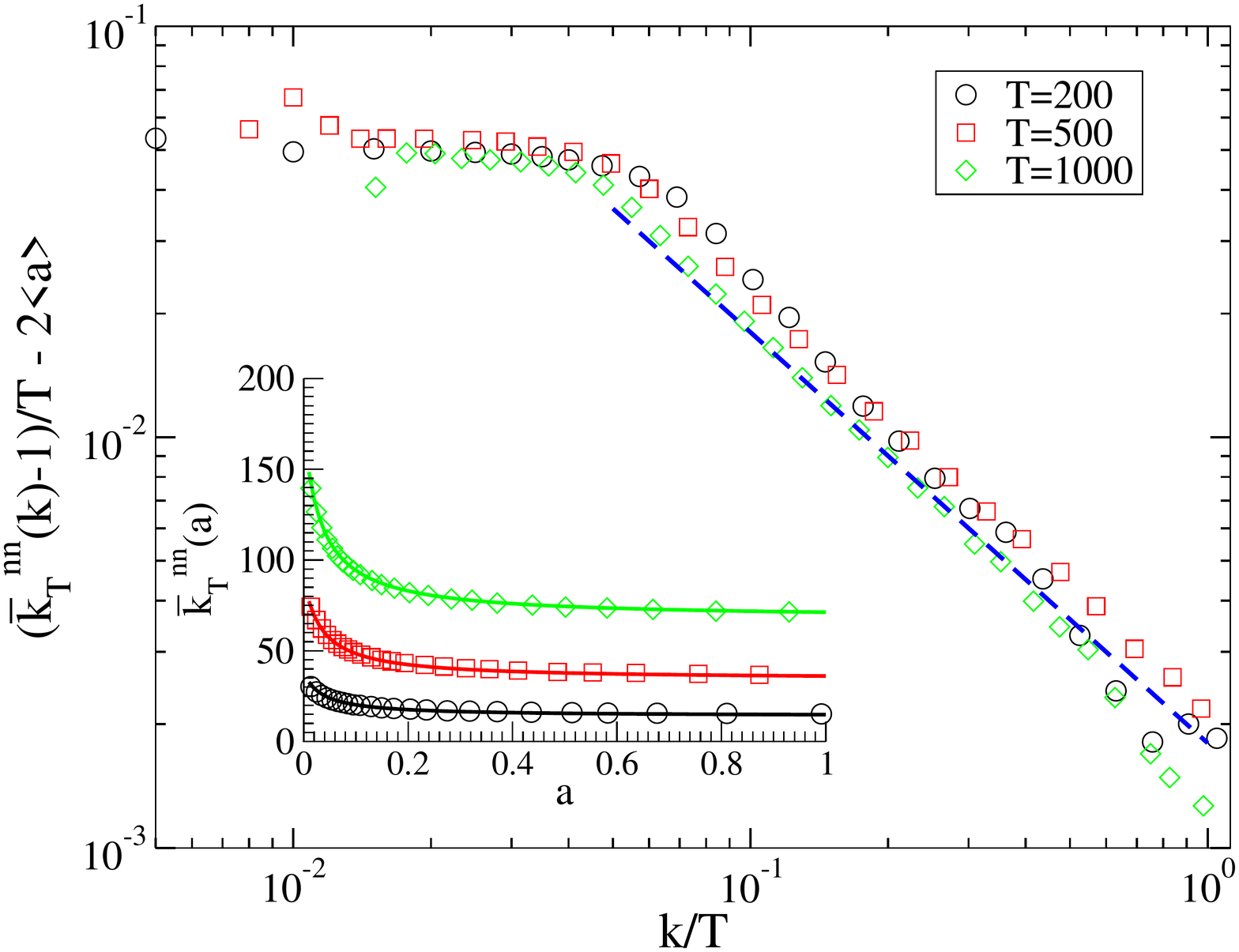}
  \includegraphics[width=0.49\textwidth]{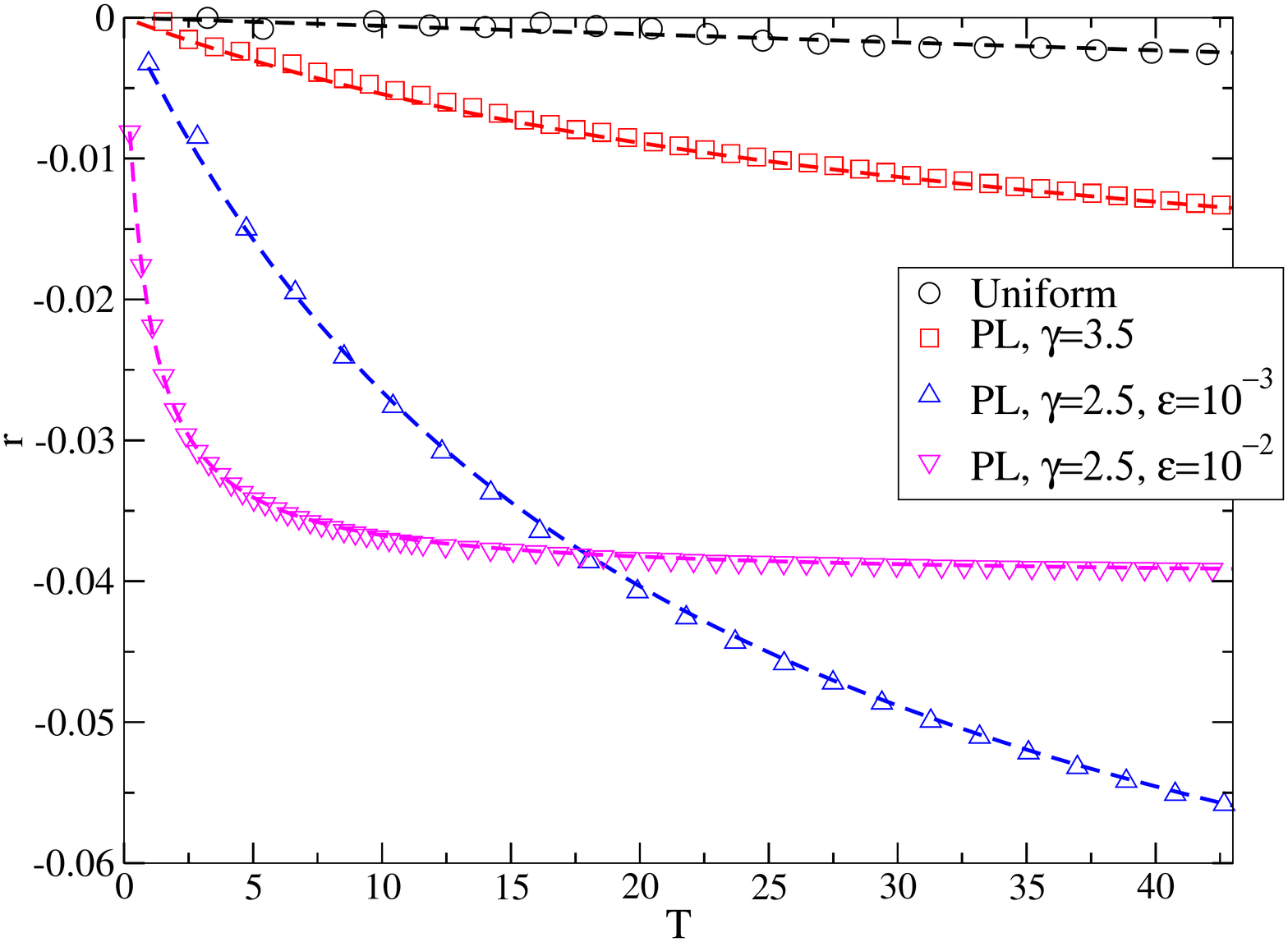}
  \caption{Left: Rescaled average degree of the nearest
    neighbors of the vertices of degree $k$, $\bar{k}^{nn}_T(k)$, for
    the integrated network with size $N=10^4$ and power law activity
    distribution with $\gamma = 2.5$, for different values of $T$.
    The prediction of Eq.~(\ref{eq:activity_knnk_aprox}) is dashed in blue.  Inset:
    Average degree of the neighbors of the vertices with activity $a$,
    $\bar{k}^{nn}_T(a)$, for the same integrated network.  The
    predictions from Eq.~(\ref{eq:activity_knna_aprox}) are shown as continuous lines. 
    Right: Assortativity $r$ as a function of the integration time $T$ 
    of the activity driven network, for different activity    distribution:
     Uniform with $a_{\textrm{max}} = 10^{-3}$,
     power law (PL) with $\gamma=3.5$,
     power law (PL) with $\gamma=2.5$,
     with $\e=10^{-2}$, and $\e=10^{-3}$.
      In dashed line we plot the
     prediction given by Eq. (\ref{eq:activity_r_th}).
     Results are averaged over $10^2$ runs, $N=10^7$.
}    
  \label{fig:activity_knnk}
\end{figure}

In Fig.~\ref{fig:activity_knnk} (left) we check the validity of Eq.~(\ref{eq:activity_knna_aprox})
and the asymptotic form Eq.~(\ref{eq:activity_knnk_aprox}) in the case of power law
distributed activity. We observe that the prediction of
Eq.~(\ref{eq:activity_knna_aprox}) recovers exactly the model behavior, also in the
case of small activity $a$ (shown in the inset).  The degree
correlation, $\bar{k}^{nn}_T(k)$, is also correctly captured by the
asymptotic form Eq.~(\ref{eq:activity_knnk_aprox}). Note however that, since the
variance $\sigma_a$ is small (of order $\eps^{\gamma-1}$ for
$\gamma<3$ and order $\eps^2$ for $\gamma>3$), the net change in the
average degree of the neighbors is quite small, and the integrated
network may be considered as approximately uncorrelated, depending on the degree of accurateness desired.

Another, global measure of degree correlations
can be defined in terms of the Pearson correlation coefficient $r$
between the degree of a node and the mean degree of its neighbors
\cite{PhysRevLett.89.208701}, taking the form
\begin{equation}
  \label{eq:activity_general_r}
  r =  \frac{ \av{k} \sum_k k^2 \bar{k}^{nn}(k) P(k) - \av{k^2}^2 }
  {\av{k} \av{k^3} - \av{k^2}^2 }.  
\end{equation} 
We can easily evaluate the sum $\sum_k k^2 \bar{k}^{nn}_T(k) P_T(k) $
by applying the hidden variable formalism presented in
Sec.~\ref{subsec:hidden_variables}. Inserting in Eq. (\ref{eq:activity_general_r})
the first moments of the degree distribution as obtained from
Eq.~(\ref{eq:activity_moments}), the coefficient $r$ in the limit of large $N$ reads
\begin{equation}
  \label{eq:activity_r_th}
  r_T = - \frac{  \left( \sigma_a^2 \right)^2}{  \dfrac{\kp_1\kp_2}{T}
    +  \kp_1\kp_3 -  \kp_2^2 },
\end{equation}
where $\sigma_a^2 = \av{a^2} - \av{a}^2$ is the variance of the activity distribution.  
Both the decreasing functional form of $\bar{k}^{nn}_T(k)$ and the
negative value of $r$ (since $\kp_1\kp_3 > \kp_2^2$ for any
probability distribution with a positive support), indicate the
presence of dissasortative correlations \cite{PhysRevLett.89.208701}
in the integrated activity driven networks, correlations whose
amplitude is modulated by $\sigma_a^2$.
In Fig. \ref{fig:activity_knnk} we show that the coefficient $r$ as measured on activity driven networks 
obtained by numerical simulations is very well fitted by the prediction of Eq. \ref{eq:activity_r_th}. 
We note in particular that scale free activity distribution with $\gamma<3$ lead to 
relevant degree correlations, as measured by the coefficient $r$.  

\begin{figure}[tb]
\begin{center}
  \includegraphics[width=0.7\textwidth]{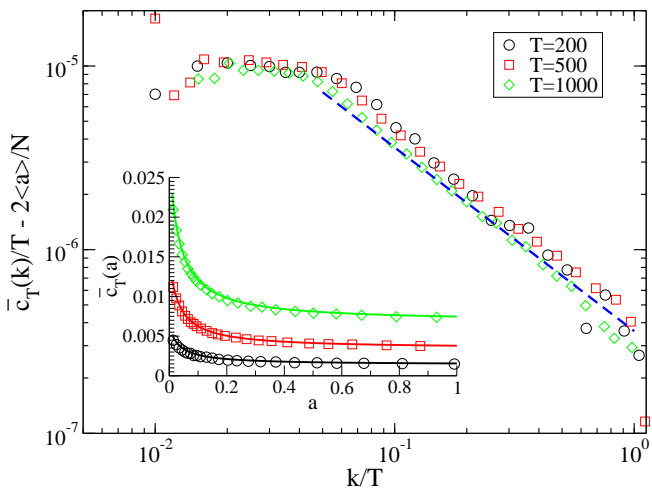}
   \caption{
     Rescaled clustering coefficient of the
    nodes of degree $k$, $c_T(k)$, of the integrated network with size
    $N=10^4$ and power law activity distribution with $\gamma = 2.5$,
    for different values of $T$.  The prediction of Eq.~(\ref{eq:activity_ck_aprox})
    is dashed in blue.  Inset: Clustering coefficient of the nodes
    with activity $a$, $c_T(a),$ of the same integrated network.  The
    predictions from Eq.~(\ref{eq:activity_clus_a_aprox}) are shown as continuous lines.  }
   \label{fig:activity_ck}
\end{center}
 \end{figure}

\subsubsection{Clustering coefficient}
\label{subsubsec:clust_coeff_act}

The expression of the clustering spectrum at time $T$, $\bar{c}_T(k)$,
takes the form, from Eq.~\eqref{eq:hidden_ck}
\begin{equation}
  \label{eq:activity_ck}
  \bar{c}_T(k) = \frac{1}{P_T(k)} \sum_a F(a) g(k|a) \bar{c}_T(a). 
\end{equation}
Using Eqs.~\eqref{eq:hidden_clus_h}, \eqref{eq:hidden_phh} and the expression for
$\bar{k}(a)$, we can write the exact form
\begin{equation}
  \label{eq:activity_clus_a}
  \bar{c}_T(a) = 1 - \left[\frac{\Psi(\lT) - e^{-\lT a} \Psi(2\lT)}{1 -
      e^{-\lT a} \Psi(\lT)} \right]^2.
\end{equation}
Again in the simplest case of a constant activity potential, $F(a) =
\delta_{a, a_0}$, we have $\bar{c}_T(a) = 1 - e^{-2 \lT a_0}$, which
leads to a clustering spectrum
\begin{equation}
  \bar{c}_T(k) \equiv \av{c}_T = 1 -  e^{-2 \lT a_0} \simeq \frac{2 T
    a_0}{N},
\end{equation}
where the last expression is valid for small $\lT$.  The clustering
spectrum is in this case constant, and equal to the average clustering
coefficient. For fixed time $T$, it is inversely proportional to the
network size, in correspondence to a purely random network. It
increases with $T$, saturating at $\av{c}_\infty =1$ for a fully
connected network in the infinite time limit.

For a general activity potential distribution, we need to perform
again an expansion in $\lT$, which in this case takes the form, at
first order in  $\lT$,
\begin{equation}
  \label{eq:activity_clus_a_aprox}
  \bar{c}(a) \simeq \frac{2 \lT}{a + \av{a}} [\av{a^2} + a \av{a}].
\end{equation}
Inserting this form into Eq.~\eqref{eq:activity_ck}, and performing the same
steepest descend approximation applied in Eq.~\eqref{eq:activity_knnk_aprox}, we obtain
\begin{equation}
\bar{c}(k) \simeq  \frac{ 2 T^2}{N}  \frac{P(k-1)}{k P(k)}  
\left[ (\av{a^2} - \av{a}^2) +  \av{a} \left(\frac{k}{T} \right) \right].
\end{equation}
In the limit of large $k$, we obtain the general form of the
clustering spectrum, valid for any activity potential,
\begin{equation}
  \label{eq:activity_ck_aprox}
  \frac{ \bar{c}(k)}{T} \simeq \frac{2 \av{a}}{N} +  \frac{2 \sigma_a^2}{N} \left( \frac{ k }{ T} \right)^{-1 }.
\end{equation}
In Fig.~\ref{fig:activity_ck} (right) we plot the clustering coefficient as a function
of the degree (main) and the activity (inset), in the case of power
law distributed activity. We observe that both Eq.~(\ref{eq:activity_clus_a_aprox}) and
Eq.~(\ref{eq:activity_ck_aprox}) recover correctly the clustering coefficient
behavior.

\section{Temporal percolation on activity driven networks}
\label{sec:perc_activity}

The connectivity properties of the time-integrated networks $\mathcal{G}_T$, 
and in particular the birth and evolution of a giant connected component,
 may have relevant consequences for dynamical processes 
 running on the top of  $\mathcal{G}_T$, \cite{ribeiro2013quantifying}. 
Indeed, at a given instant of time $t$, a temporal network can be represented by 
a single network snapshot, which is usually very sparse, 
composed by isolated edges, stars or cliques. 
As we integrate more and more of those snapshots,
the integrated network will grow, until at some time $T_p$ it will
percolate, i.e. it will possess a giant connected component with a
size proportional to the total number of individuals in the network. 
The time of the first appearance of this giant component is 
not only an important topological property of integrated networks, 
but it is also relevant for the evolution of dynamical processes running on the top of the time-varying graph,
in the sense that any process with a characteristic lifetime $\tau < T_p$
will be unable to explore a sizable fraction of the network.

In this Section we will focus on the percolation properties of the
time-integrated form of activity driven networks.
Building on the analytic expression for the topological properties of 
the integrated networks at time $T$, $\mathcal{G}_T$, 
found in the last Section \ref{sec:integrated_activity},
we compute analytic expressions for the percolation time and the size of the
giant component of the integrated network. 
An added value of our
approach is the possibility to extend the mapping of epidemic
spreading into percolation processes in static networks
\cite{newman02} into the temporal case. Thus our results can
be extended to provide the epidemic threshold and the outbreak size of
the susceptible-infected-susceptible epidemic model \cite{anderson92}.

\subsection{Generating function approach to percolation}
\label{subsec:perc_gener_func}

Percolation in random networks can be studied applying the generating
function approach developed in Ref.~\cite{Newman2001}, which
is valid assuming the networks are degree uncorrelated.  Let us define
$G_0(z)$ and $G_1(z)$ as the degree distribution and the excess degree
distribution (at time $T$) generating functions, respectively, given
by \cite{Newman2010}
\begin{equation}
  G_0(z) = \sum_k P_T(k) z^k, \quad G_1(z) =\frac{G'_0(z)}{G'_0(1)}.  
  \label{eq:activity_generdefs}
\end{equation}
The size of the giant connected component, $S$, is then given by
\begin{equation}
\label{eq:activity_S}
  S = 1 - G_0(u),
\end{equation}
where $u$, the probability that a randomly chosen vertex is not
connected to the giant component, satisfies the self-consistent
equation
\begin{equation}
  \label{eq:activity_S_self}
  u = G_1(u).
\end{equation}
The position of the percolation threshold can be simply obtained by
considering that $u=1$ is always a solution of Eq.~(\ref{eq:activity_S_self}),
corresponding to the lack of giant component. A physical solution
$u<1$, corresponding to a macroscopic giant component, can only take
place whenever $G'_1(1)>1$, which leads to the Molloy-Reed criterion
\cite{molloy95}:
\begin{equation}
  \label{eq:activity_molloy}
  \frac{\av{k^2}_T}{\av{k}_T} > 2.
\end{equation}
The moments of the degree distribution $\langle k^n \rangle$ has been computed in Section \ref{subsubsec:moments_deg_distr_act}: 
The ratio $\av{k^2}_T / \av{k}_T $ is a monotonic, growing function of
$T$, and it will fulfill the condition of Eq.~(\ref{eq:activity_molloy}) for $T>T_p^0$,
defining a percolation time
\begin{equation}
  \label{eq:activity_Tc_uncor}
  T_p^0 =  \frac{2 \av{a}}{\av{a^2} +
    3\av{a}^2}.
\end{equation}
This percolation time is independent of $N$, and thus guarantees the
fulfillment of the condition $\lT \ll 1$ assumed in the derivation of
Eq.~(\ref{eq:activity_conn_prob_aprox}).  We can obtain information on the size of
the giant component $S$ for $T>T_p^0$ from Eqs.~(\ref{eq:activity_generdefs})
and (\ref{eq:hidden_pk}), using the Poisson form of the propagator, which
allows to write the simplified expressions
\begin{eqnarray}
\label{eq:activity_genfun0}
  G_0(u) &=& \sum_a  F(a) e^{-(1-u)T(a+\av{a})}, \\
  G_1(u) &=& \frac{1}{2 \av{a}} \sum_a F(a) [a+\av{a}]
  e^{-(1-u)T(a+\av{a})}.    
\end{eqnarray}
From the self-consistent Eq.~(\ref{eq:activity_S_self}), setting $\delta =
1-u$, and solving at the lowest order in $\delta>0$, we find, close to
the transition,
\begin{equation}
  \delta \simeq \frac{2\kp_1}{\kp_3 T^2} \left( \frac{T-T_p^0}{T_p^0}
  \right), 
\end{equation} 
where $\kp_n$ has been defined in Section \ref{subsubsec:moments_deg_distr_act}.
Thus we recover the Molloy Reed criterion, Eq.~(\ref{eq:activity_Tc_uncor}), for the
onset of the giant component.  Since the derivatives of $G_0(u)$ are
finite, we can obtain the size of the giant component $S$ by expanding
Eq.~(\ref{eq:activity_genfun0}) close to $u =1$,
\begin{eqnarray}
\label{eq:activity_S_appr}
  S & \simeq& 1 - G_0(1) + \delta G_0'(1) - \frac{\delta^2}{2}   G_0''(1) \nonumber \\ 
& = &\frac{2 \kp_1^2 }{ \kp_3 T} \left( \frac{T-T_p^0}{T_p^0} \right) -
  \frac{2 \kp_2 \kp_1^2}{\kp_3^2 T^2} \left( \frac{T-T_p^0}{T_p^0}
  \right)^2 .
\end{eqnarray}
Since Eq. (\ref{eq:activity_S_appr}) is obtained from a Taylor expansion for
$\delta \ll 1$, we expect it to be valid only close to the percolation
threshold.

In order to check the validity of the analytical results developed
above, we consider the concrete case of two different forms of
activity distribution, namely a uniform activity distribution $F(a) =
a_{\textrm{max}}^{-1}$, with $a \in [0, a_{\textrm{max}}]$, and the
empirically observed case of power law activity distribution in social
networks \cite{2012arXiv1203.5351P}, $F(a) \simeq (\g -1) \e^{\g -1}
a^{-\g}$, $a \in [\e, 1]$, where $\e$ is the minimum activity in the
system.  In this last case, we note that the analytical form of the
activity distribution is valid for small $\e$ only in the limit of
large $N$. Indeed, a simple extreme value theory calculation
\cite{gumbel2004statistics} shows that, in a random sample of $N$
values $a_i$, the maximum activity scales as $\min\{ 1, \eps
N^{1/(\gamma-1)} \}$. Therefore, when performing numerical simulations
of the model, one must consider systems sizes with $N > N_c =
\e^{1-\g}$ in order to avoid additional finite-size effects.  In case
of performing simulations for system with small sizes $N<N_c$, for
example to study the finite size effects on the percolation threshold
(see next Section), we used a deterministic power law distribution to
avoid the cutoff effect on the maximum value of the activity due to a
random sampling of values.
 
\begin{figure}[tb]
\begin{center}
\includegraphics[width=0.48\textwidth]{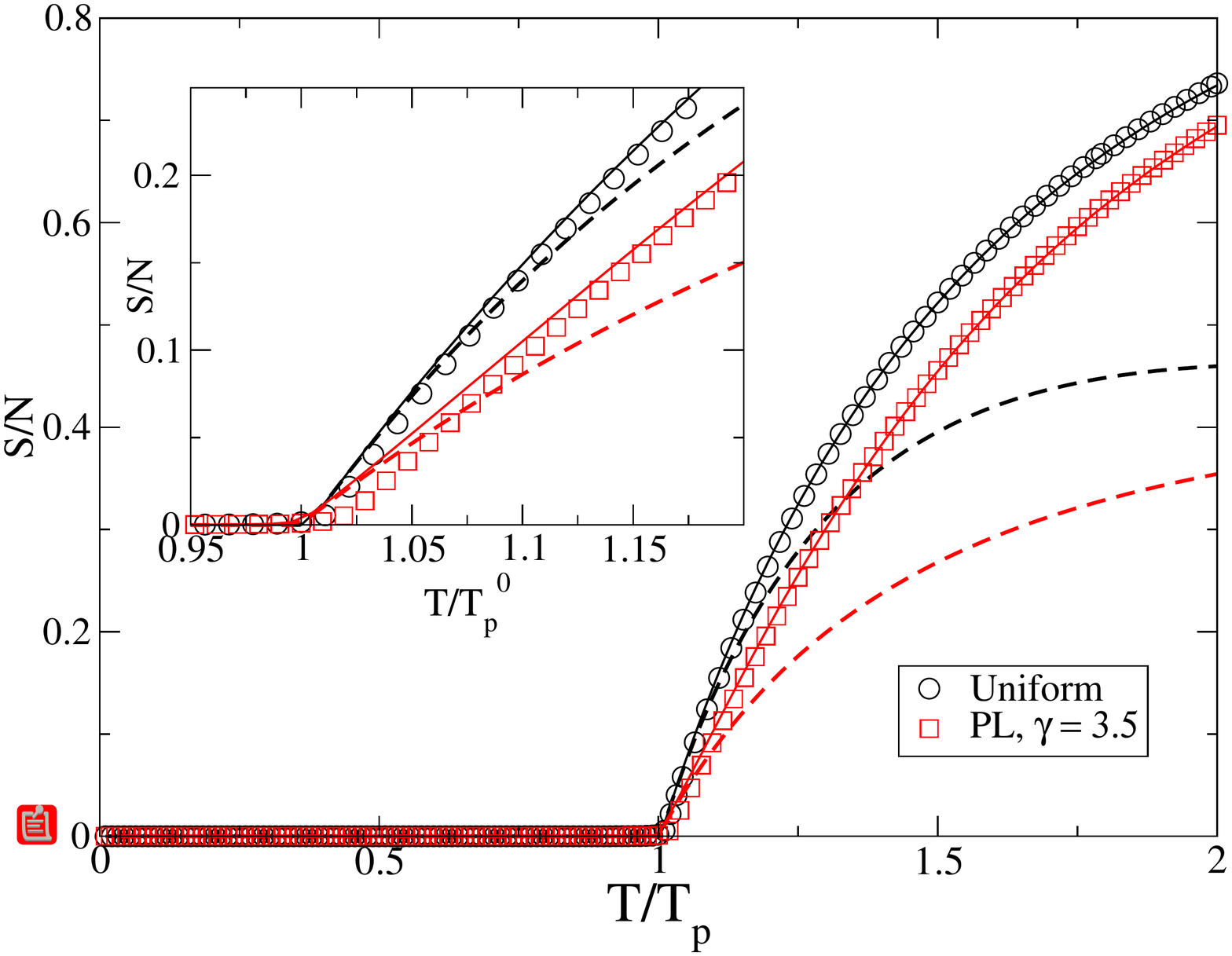}
\includegraphics[width=0.48\textwidth]{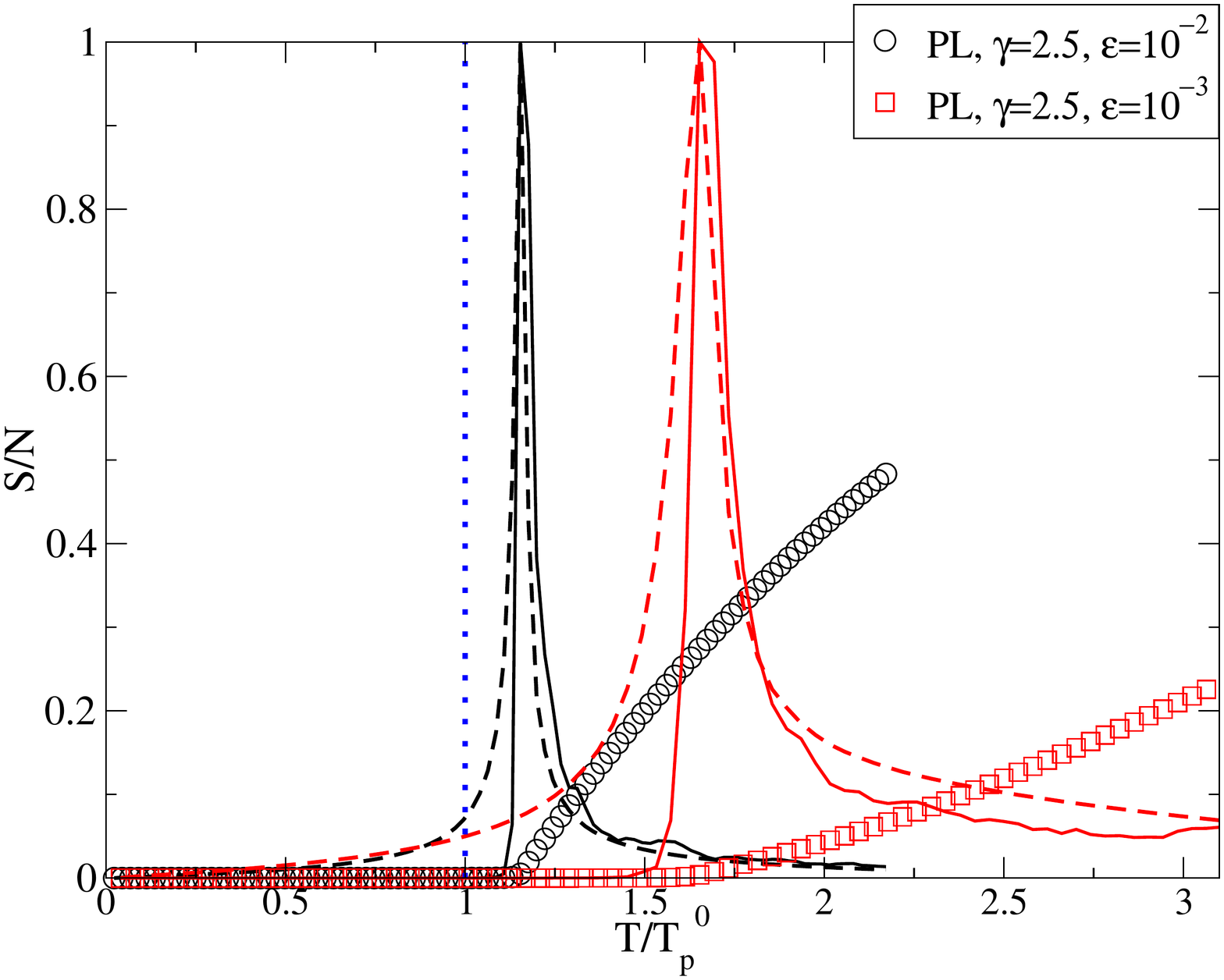}
  \caption{ Rescaled giant component size $S/N$ as a function of the
    rescaled time $T/T_p^0$ for activity driven networks.  Left: Uniform
    ($a_{\textrm{max}} = 0.01$) and power law ($\g = 3.5$, $\e =
    0.01$) activity distributions, compared with the numerical
    integration of the generating function equations (continuous line)
    and the theoretical approximation Eq.~(\ref{eq:activity_S_appr}) (dashed
    line). The inset shows the details close to the percolation threshold $T_p$. 
    Right: Power law activity distribution with $\g = 2.5$, $\e = 10^{-2}$ 
    and $\e = 10^{-3}$.  The peaks of the variance of the giant
    component size, $\sigma(S)^2$, and the susceptibility of the
    clusters size, $\chi(s)$, are plotted in continuous and dashed
    line, respectively. Percolation threshold $T_p$ is plotted in dotted line for reference.
    Results averaged over $10^2$ runs, network size $N=10^7$. }
  \label{fig:activity_S_comp}
\end{center}
\end{figure}

Fig. \ref{fig:activity_S_comp} shows the giant component size $S$ of the
activity driven network as a result of numerical simulations for both
uniform and power law activity distributions.  In
Fig.~\ref{fig:activity_S_comp} (left) we compare $S$ with the analytical
approximation Eq.~(\ref{eq:activity_S_appr}), as well as with the result of a
direct numerical integration\footnote{The numerical integration has been performed by using the function 
\emph{quad} of the package \emph{SciPy} with \emph{Python 2.7}.  }
 of Eqs.~(\ref{eq:activity_S}) and~(\ref{eq:activity_S_self}),
for uniform activity and power-law activity with exponent
$\gamma>3$. In both cases we observe an almost exact match between
numerical simulations and the numerical integration of the generating
function equations, and a very good agreement with the analytical
approximation in the vicinity of the percolation threshold.
Fig.~\ref{fig:activity_S_comp} (right), on the contrary, focuses on power-law
activity distributions with an exponent smaller than three. In this
case we additionally plot a numerical estimation of the percolation
threshold as given by the peak of both the variance of the giant
connected component size $S$, $\sigma(S)^2 = \av{S^2}-\av{S}^2$, and
the susceptibility of the clusters size, $\chi(s) = \sum_{s=2}^{S-1}
s^2 n_s$, where $n_s$ is the number of cluster of size $s$. From this
figure we can see that the numerical percolation threshold strongly
deviates in this case from the theoretical prediction
Eq. (\ref{eq:activity_Tc_uncor}), deviation that increases when the
distribution cutoff $\eps$ becomes smaller.

\subsection{Effect of degree correlations on the temporal percolation threshold}
\label{subsec:act_perc_degr_corr}

The origin of the disagreement for the case of power law activity
distribution with $\gamma < 3$ can be traced back to the effect of
degree correlations in the integrated networks generated by the
activity driven model. Indeed, as stated above, the generating
function technique makes the explicit assumption of lack of degree
correlations \cite{Newman2001}.  
However, in Section \ref{sec:integrated_activity} we showed
that the integrated activity driven network exhibits degree correlations.
The average degree of the neighbors of the vertices of degree $k$, $\bar{k}^{nn}_T(k)$ \cite{alexei}, 
is a decreasing function of $k$ (see Fig. \ref{fig:activity_knnk}), 
and the Pearson correlation coefficient $r$
between the degree of a node and the mean degree of its neighbors
\cite{PhysRevLett.89.208701} can assume non-negligible values, 
especially for networks obtained with  scale-free activity distribution with $\gamma < 3$, 
as shown in Fig. \ref{fig:activity_knnk}.

In order to take into account the effect of degree correlations let us
consider the general problem of percolation in a correlated random
network \cite{PhysRevE.78.051105}.  The effect of the degree
correlations are accounted for by the branching matrix 
\begin{equation}
  B_{kk'} =
  (k'-1)P(k'|k),
  \label{eq:activity_branch_mat}
\end{equation}
where $P(k'|k)$ is the conditional probability that a node with degree
$k$ is connected to a node with degree $k'$ \cite{alexei}.  
The percolation threshold is determined by the largest eigenvalue
$\Lambda_1$ of $B_{kk'}$ through the condition $\Lambda_1 = 1$.  
If the network is uncorrelated, the branching matrix does not depend on $k$, $B_{kk'} = \frac{k'(k'-1)P(k')}{\av{k}}$, 
and thus $\Lambda_1$ reduces
to the ratio of the first two moment of the degree distribution,
\begin{equation}
\label{eq:activity_eigen0}
\Lambda_1^0 = \sum_{k'} \frac{k'(k'-1)P(k')}{\av{k}}= \frac{\av{k^2}}{ \av{k} } -1 
\end{equation}
thus recovering the Molloy-Reed criterion Eq.~(\ref{eq:activity_molloy}).

In activity driven networks we can compute the largest eigenvalue
$\Lambda_1$ in the limit of small $\lT$ by applying the hidden
variables mapping from Sec.~\ref{subsec:map_integr_act}. In fact, the
conditional probability $P_T(k'|k)$ of the integrated network at time
$T$ can be written as \cite{PhysRevE.68.036112}
\begin{equation}
  P_T(k'|k) = \frac{N}{P_T(k)} \sum_{a,a'} g_T(k-1|a') F(a')
  \frac{\Pi_T(a',a)}{\bar{k}_T(a)}  F(a) g_T(k|a),
\end{equation}
where $\bar{k}_T(a) = N \sum_a F(a) \Pi_T(a,a')$, as follows from Eq. \eqref{eq:hidden_kh}.
From here, the branching matrix takes the form
 \begin{equation}
   B_{kk'} = (k'-1) \left[ p_{k'-1} + \frac{p_{k-1}}{k p_{k}}
    \left( k' p_{k'} - \av{k}  p_{k'-1} \right) \right], 
\end{equation}
where we write $P_T(k)$ as $p_k$ for brevity.
Assuming that the branching matrix is irreducible, and given that it is
non-negative (see Eq.~(\ref{eq:activity_branch_mat})) we can compute its largest
eigenvalue by applying Perron-Frobenius theorem \cite{gantmacher} and
looking for a principal eigenvector $v_k$ with positive components. 
Using the ansatz $v_k = 1 + \alpha p_{k-1}/k p_{k}$, we obtain that,
in order to be an eigenvector, the following conditions must be
fulfilled:
\begin{eqnarray}
  \Lambda_1 &=& \av{k}_T  +\alpha \sum_k\frac{(k-1) p_{k-1}^2}{k p_k} \nonumber \\  
  \Lambda_1 \alpha &=& \av{k^2}_T - \av{k}_T -\av{k}_T^2 + \alpha \av{k}_T
  \left(1- \sum_k\frac{(k-1) p_{k-1}^2}{k p_k} \right)\nonumber.  
\end{eqnarray}
One can see that $\sum_k(k-1) p_{k-1}^2/k
p_{k} \simeq 1$, in the limit of large $N$.  Thus we obtain the equation for $\Lambda_1$
\begin{equation}
  \Lambda_1(T)^2 - \av{k}_T \lambda_1(T) - \av{k^2}_T + \av{k}^2_T  +
  \av{k}_T = 0.
  \label{eq:activity_eq_eigenval}
\end{equation}
By using the form of the moments of the degree distribution given by
Eq.~(\ref{eq:activity_moments}), we solve Eq.~(\ref{eq:activity_eq_eigenval}). Excluding the
non-physical solution $\Lambda_1 < 0$, one finally find the largest eigenvalue 
\footnote{From Eq. (\ref{eq:activity_eigenval}) one can find 
$\alpha = \frac{\sigma_a^2}{(\sqrt{\av{a^2}} + \av{a})} T > 0$, 
confirming the validity of the proposed ansatz for $v_k$, implying $v_k >0 \forall k$.} 
of the branching matrix as
\begin{equation}
  \label{eq:activity_eigenval}
  \Lambda_1(T) = \left( \sqrt{\av{a^2}} + \av{a} \right) T. 
\end{equation} 
From here, the percolation threshold in activity driven networks
follows as 
\begin{equation}
  \label{eq:activity_perc_real}
  T_p = \frac{1}{\sqrt{\av{a^2}} + \av{a}}.
\end{equation}
 
We can understand the results of Fig.~\ref{fig:activity_S_comp} by comparing
the ratio of the exact threshold $T_p$ with the uncorrelated value $T_p^0$, 
\begin{equation}
  Q = \frac{T_p - T_p^0}{T_p^0} = \frac{\sigma_a^4}{2\av{a} \left(
      \sqrt{\av{a^2} } + \av{a} \right)^3 }. 
  \label{eq:activity_Q}
\end{equation}
In the case of a uniform activity distribution, we have $Q =
13/\sqrt{3} - 15/2 \simeq 5.5 \times 10^{-3}$, and therefore the
temporal percolation threshold is given with very good accuracy by the
uncorrelated expression. For a power-law activity distribution, the
ratio $Q$ depends simultaneously of the exponent $\gamma$ and the
minimum activity $\eps$. Thus, for $\gamma<3$, we have that $Q \sim
\eps^{(\gamma-3)/2}$, which diverges for $\eps\to0$, indicating a
strong departure from the uncorrelated threshold. For $\gamma>3$, on
the other hand, $Q$ becomes independent of $\eps$, and it goes to $0$
in the limit of large $\gamma$. In the case $\gamma=3.5$ and
$\eps=0.01$, for example, we obtain $Q \simeq 1.6 \times 10^{-2}$.
This implies an error of less than $2\%$ in the position of the
percolation threshold as given by the uncorrelated expression,
explaining the good fit observed in Fig.~\ref{fig:activity_S_comp}(a).

\begin{figure}[tb]
\begin{center}
  \includegraphics[width=0.7\textwidth]{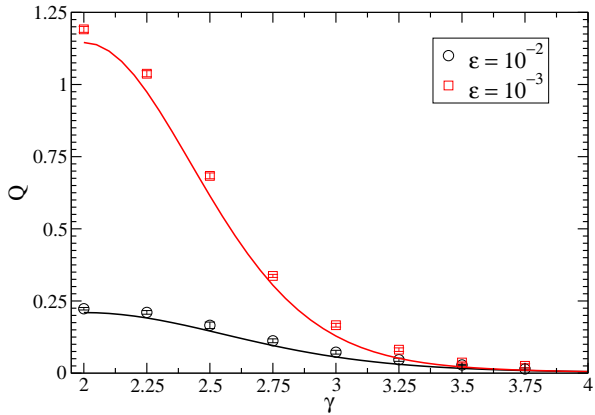}
  \caption{ Ratio $Q$, defined in Eq. (\ref{eq:activity_Q}), as a function of
    the exponent $\gamma$ of the activity potential $F(a) \sim
    a^{-\gamma}$, for $\epsilon = 10^{-2}$ and $\epsilon = 10^{-2}$.
    We compare $Q$ as obtained by estimating the percolation threshold
    $T_p$ from the peak of the variance of the giant component size,
    $\sigma(S)^2$, by means of a numerical simulation of a network
    with size $N=10^7$ (symbols), with the prediction of
    Eq. (\ref{eq:activity_Q}) (lines).
    The results of numerical simulations are averaged over $10^2$ runs.}
  \label{fig:activity_Q}
\end{center}
\end{figure}

\begin{figure}[tb]
\begin{center}
  \includegraphics[width=0.48\textwidth]{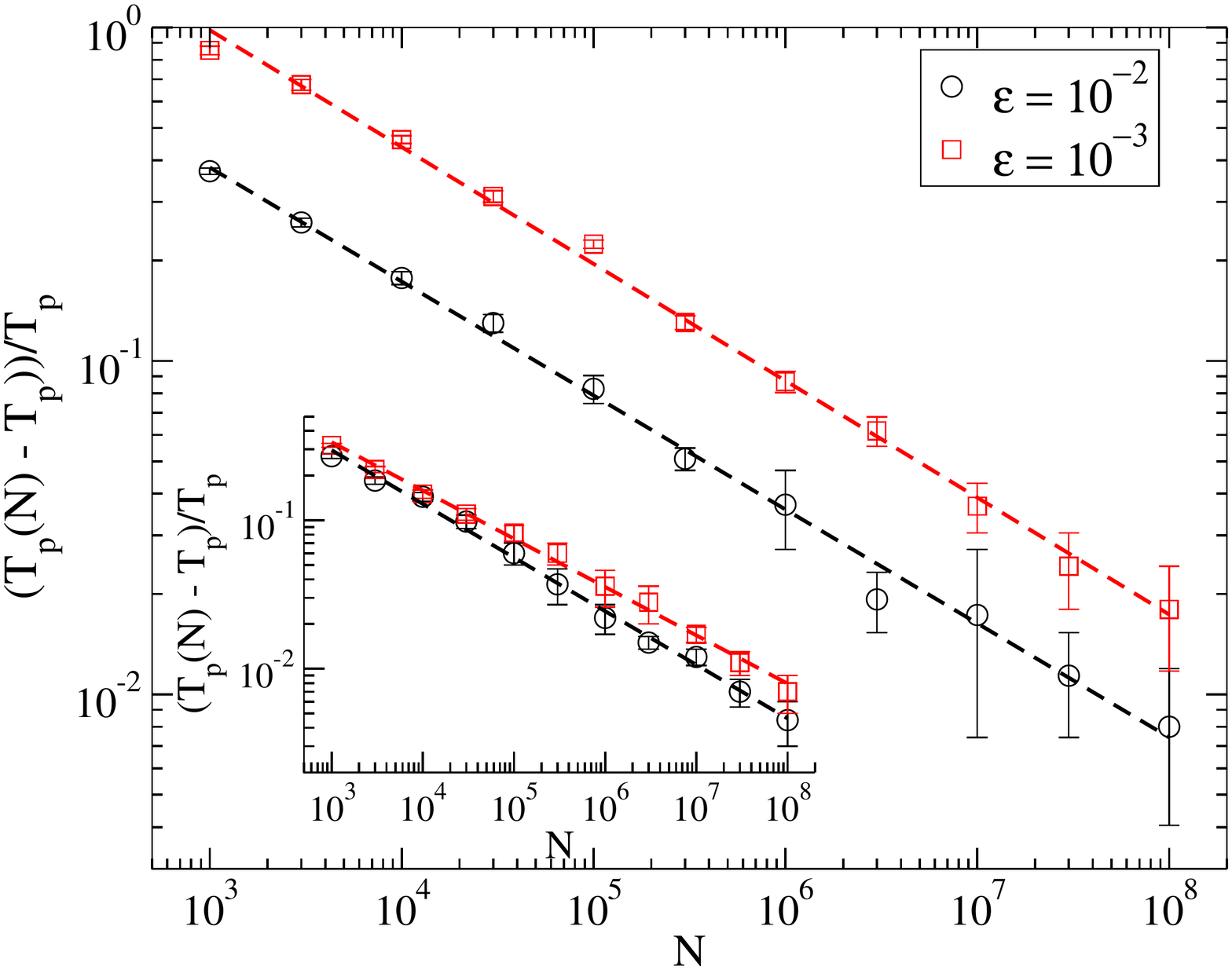}
  \includegraphics[width=0.48\textwidth]{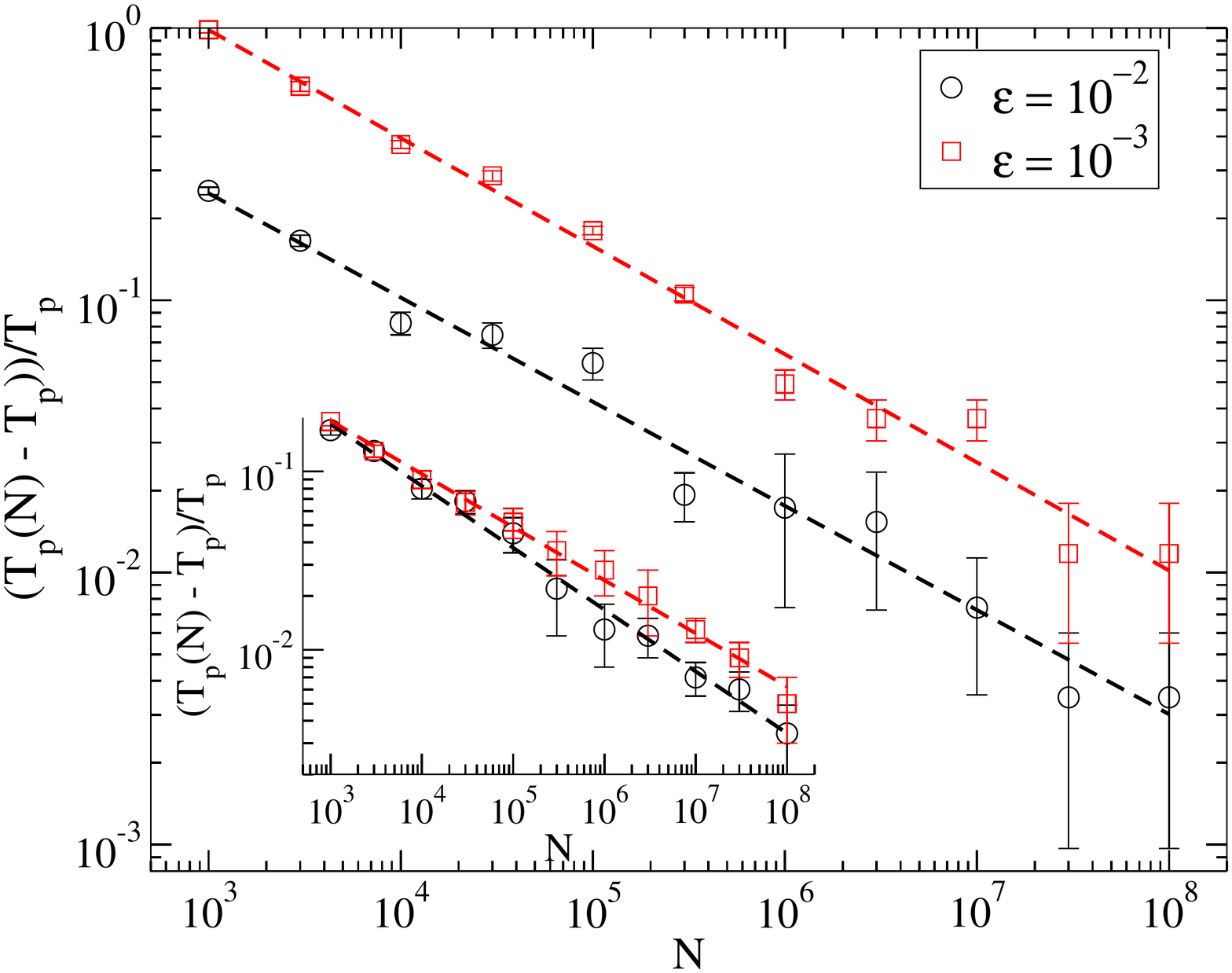}
  \caption{ Finite size scaling of the percolation threshold as
    estimated by the peak of the variance of the giant component size $\sigma(S)^2$, (left), 
    and by the peak of the susceptibility of the cluster size $\chi(s)$, (right),
    for a network with power law activity distribution with
    $\gamma = 2.5$ (main) and $\gamma = 3.5$ for different values of $\epsilon$.
    We plot $(T_p(N) - T_p)/T_p$ as a function of $N$, finding a
    scaling of the form of Eq. \ref{eq:activity_ffs}, plotted in dashed line.
    The two methods give similar exponents, 
    although the $\sigma(S)^2$ method seems to be more precise:
	$\nu_{\sigma} = 0.34 \pm 0.01$ for $\gamma = 2.5$ and
    $\nu_{\sigma} = 0.34 \pm 0.02$ for $\gamma = 3.5$ with the $\sigma(S)^2$ method,
	$\nu_{\chi} = 0.38 \pm 0.03$ for $\gamma = 2.5$ and
    $\nu_{\sigma} = 0.32 \pm 0.04$ for $\gamma = 3.5$ with the $\chi(s)$ method.
    Results are averaged over $10^2$ runs.
    }
  \label{fig:activity_fss}
  \end{center}
\end{figure}

In Fig.~\ref{fig:activity_Q} we show the ratio $Q$ as a function of the
exponent $\gamma$ of a power-law distributed activity potential for
different values of $\epsilon$, computed from numerical simulations by
evaluating the percolation threshold from the peak of the variance of the
giant component size, $\sigma(S)^2$. The numerical result is compared
with the analytical prediction given by Eq. (\ref{eq:activity_Q}).
In this Figure one can see that, although numerical and analytical
results are in quite good agreement, they still do not exactly
coincide for $\gamma < 3$.  This is due to the presence of finite size
effects, which have not been taken into account in the percolation
theory developed.  We can consider the finite size effects on the
percolation time $T_p(N)$ in a network of size $N$ by putting forward
the standard hypothesis of a scaling law of the form \cite{stauffer94}
\begin{equation}
\label{eq:activity_ffs}
T_p(N) = T_p + A N^{-\nu}.
\end{equation} 
In Fig. \ref{fig:activity_fss} we plot the rescaled numerical thresholds
$[T_p(N)-T_p]/T_p$ estimated by the peak of the variance of the giant
component size $\sigma(S)^2$ (left), and by the susceptibility of the clusters size $\chi(s)$ (right),
 as a function of the network size $N$.
We can observe that the numerical thresholds $T_p(N)$ asymptotically
tend to the theoretical prediction $T_p$ by following the scaling law
of Eq. (\ref{fig:activity_fss}), with an exponent close to $\nu \simeq 0.34
\pm 0.04$, and different values of the prefactor $A$ depending on the
values of $\gamma$ and $\epsilon$. 
The two methods used to estimate the numerical threshold give similar values for the exponent $\ni$.  
 and a slightly different prefactor $A$.

\subsection{Application to epidemic spreading}
\label{subsec:perc_appl_epidemic}

The concept of temporal percolation can be applied to gain
understanding of epidemic processes on activity driven
temporal networks \cite{2012arXiv1203.5351P}. Let us focus in the
susceptible-infected-susceptible (SIR) model \cite{anderson92}, which
is the simplest model representing a disease that confers immunity and
that is defined as follows: Individuals can be in either of three
states, namely susceptible, infected, and removed. Susceptible individuals acquire the disease by contact with
infected individuals, while infected individuals heal spontaneously
becoming removed, which cannot contract the disease anymore. On a
temporal network, the SIR model is parametrized by the rate $\mu$
(probability per unit time) at which infected individuals become
removed, and by the transmission probability $\beta$ that the
infection is propagated from an infected individual to a susceptible
individuals by means of an instantaneous contact.

We can approach the behavior of the SIR model on activity driven
networks by extending the mapping to percolation developed in
Ref.~\cite{newman02} into the temporal case. To do so, let
us consider first a modified SIR model in which individuals stay in
the infected state for a fixed amount of time $\tau$. We define the
transmissibility $\mathcal{T}_{ij}$ as the probability that the
infection is transmitted from infected individual $i$ to susceptible
individual $j$. Considering that contacts last for an amount of time
$\Delta t = 1/N$, the transmissibility can be written as
\begin{equation}
 \mathcal{T}_{ij}(\beta, \tau) = 1 - \left( 1 - \beta  p_{ij} \right)
 ^{\tau N},  
\end{equation}
where $p_{ij} = (a_i + a_j)/N^2$ is the probability that individuals
$i$ and $j$ establish a contact in any given time step $\Delta t$,
as given by Eq. \eqref{eq:activity_conn_prob_aprox}.
In the limit of large $N$, we can thus write
\begin{equation}
 \mathcal{T}_{ij}(\beta, \tau) = 1 - \exp\left(-\frac{\beta \tau
      [a_i+a_j]}{N}\right). 
\end{equation}
From here we can deduce the form of the transmissibility when healing
is not deterministic but a Poisson process with rate $\mu$. In this
case, the probability that an infected individual remains infected a
time $\tau$ is given by the exponential distribution $P(\tau) = \mu
e^{- \mu \tau}$.
Therefore, we can write \cite{newman02}
\begin{eqnarray}
\label{eq:activity_transm}
   \mathcal{T} _{ij}(\beta, \mu) &=& \int_0^\infty   \mathcal{T}_{ij}(\beta, \tau) P(\tau)\;  d\tau \nonumber \\
   &=& 1- \left(1+\frac{\beta}{\mu}\frac{a_i + a_j}{N} \right)^{-1} 
\simeq \frac{\beta}{\mu}\frac{a_i + a_j}{N}
\end{eqnarray}
in the limit of large $N$.  If we consider the process of infection as
equivalent to establishing a link between infected and susceptible
individuals and we compare this expression with
Eq.~(\ref{eq:activity_conn_prob_aprox}), we can see that the SIR process can be
mapped to the creation of the integrated network in the activity
driven model up to a time $T = \beta /\mu$.  The epidemic threshold
will be given by the existence of a finite cluster of recovered
individuals, and therefore will coincide with the temporal percolation
threshold, i.e.
\begin{equation}
  \left(\frac{\beta}{\mu}\right)_c = T_p.
\end{equation}
The temporal percolation threshold given by Eq.~(\ref{eq:activity_perc_real})
recovers the epidemic threshold obtained in
Ref.~\cite{PhysRevLett.112.118702} using a mean-field rate equation
approach\footnote{Notice that in Ref.~\cite{PhysRevLett.112.118702} the
  per capita infection rate $\beta' = 2 \av{a}\beta$ is used.}. A
particular benefit of this percolation mapping is the fact that it
makes accessible the calculation of explicit approximate forms for the
size of epidemic outbreaks, Eq.~(\ref{eq:activity_S_appr}) (valid however in
certain limits), which are not easily available in mean-field
approximations \cite{2012arXiv1203.5351P,PhysRevLett.112.118702}.
Moreover, the finite-size scaling presented in Section \ref{subsec:act_perc_degr_corr}
allows to estimate corrections to the epidemic threshold due to the size $N$ of the system considered.

\section{Extensions of the activity driven model}
\label{sec:activity-extensions}

The potency of the hidden variables formalism we have introduced above
to solve the activity driven model allows to easily extended it to
tackle the analysis of generalized models inspired in the same
principles. We can consider, indeed,
different rules for activation and reception of connections. The only
limitation to be imposed in order to properly implement the formalism 
is that connection rules must be local, i.e. involving only properties of the
emitting and receiving agent. 
As a simple example, we consider a
sort of ``inverse'' activity driven model, in which every agent $i$
becomes active with the same constant probability $a_i = a_0$ and, when
active, she sends a connection to another agent $j$, chosen at random
with probability proportional to some (quenched) random quantity $b_j$, 
i.e. with probability $b_j / \av{b}N$. 
In this case the quantity $b_j$ can represent the importance of the individual $j$
in the social context, being the agents in the system interested in engaging a social act with individual $j$.
One can easily repeat the steps of the
mapping presented in Sec.~\ref{subsec:map_integr_act}: The number of
times $z$ that agent $i$ becomes active is now
\begin{equation}
  P'_T(z) = \binom{T N }{z} \left(
    \frac{a_0}{N}\right)^{z} 
  \left(1- \frac{a_0}{N}\right)^{TN  -z},
\end{equation}
and the probability that $i$ and $j$ become connected up to
time $T$ is
\begin{eqnarray}
  Q'_T(i,j) &=& \sum_{z_i,z_j}  P'_T(z_i)  P'_T(z_j)
  \left(1-\frac{b_j}{\av{b}N}\right)^{ z_i} 
  \left(1-\frac{b_i}{\av{b}N}\right)^{ z_j}  \nonumber \\
 & = &
  \left[ \left( 1 - \frac{a_0 b_i}{\av{b}N^2}
      \right)
    \left( 1 - \frac{a_0 b_j}{\av{b}N^2}
       \right)
  \right]^{T N } \nonumber \\
  & \simeq &   \exp\left[ - \lT'(b_i + b_j)\right],
\nonumber
\end{eqnarray}
where we have defined the new parameter $\lT' = a_0T/\av{b}N$ and, in
the last step of the previous expressions, we have performed and
expansion for large $N$ a finite $\lT'$ and assumed a bounded
distribution $F(b)$ for the values $b_i$. From here, we obtain
$\Pi'_T(i,j) = 1- Q'_T(i,j)$. As one can see, this modified model can
be exactly mapped to the activity driven model (see
Eq.~(\ref{eq:activity_conn_prb}), with the simple translation $\lT \to \lT'$; all the
general expression derived above hold thus in this case, and can be worked
out, upon providing the selected expression for the distribution $F(b)$.

On different grounds, one of the most striking features of the empirical social interactions under consideration is the 
broad tailed form of the interevent time probability distribution, 
the well known phenomenon of burstiness of human dynamics, presented in Chapter \ref{chap:intro}.
In its present formulation, the activity driven model does not reproduce this empirical fact,
since one can easily check that the number of time steps $\tau$ between two consecutive fires of a node $i$
has a probability distribution of the form $\phi_i(\tau) = a_i e^{-a_i \tau}$,
a fact rooted in the Poissonian process ruling the node's activation.
The waiting time distribution $\phi_i(\tau)$ decays exponentially in time, 
with a characteristic time scale given by the inverse of the activity $a_i$ of node $i$,
at odds with the scale free form of the interevent time distribution found in many cases. 
In order to overcome this issue, one can move forward and 
consider a generalization of the activity driven model, 
in which the activation probability $a_i(t)$ of node $i$ explicitly depends also on time $t$.
The process of activation of an individual can be interpreted as a \emph{renewal process} \cite{renewal},
being $a_i(t)dt$ the probability that individual $i$ becomes active in the interval $[t, t+dt]$,
with $t$ the time occurred since the last activation event.
The waiting time distribution can be obtained by applying renewal theory  \cite{renewal},
and reads
\begin{equation}
\phi_i(\tau) = a_i(\tau) \exp \left[ - \int_0^{\tau} a_i(\tau') d\tau' \right] .
\end{equation}
The original activity driven model can be recovered by considering a constant activation probability, $a_i(t) = a_i$.
The explicit time-dependence of $a_i(t)$, on the contrary, opens the path toward non-Poissonian dynamics,
with the possibility of considering power law form of the waiting time distribution.
Importantly, the hidden variables formalism presented in Section \ref{subsec:hidden_variables} still applies,
 and it is possible to compute expressions for 
 the topological properties of the time-integrated networks as done in Section \ref{sec:integrated_activity}.
Therefore, although this conclusion arises from a preliminary study,
 it seems to be possible to incorporate in the activity driven model 
a more realistic non-Markovian activation process, 
which leads to broad tailed form of the interevent time distribution.

\section{Summary and Discussion}
\label{sec:concl_act}

The activity driven model represents an interesting approximation to
temporal networks, providing a preliminary explanation of the origin
of the degree distribution of integrated social networks, in terms of
the heterogeneity of the agents' activity, and the distribution of
this quantity. 
In this Chapter we addressed the activity driven networks under an analytic point of view.

First, we have explored the full relation between
topology and activity distribution, obtaining analytical expressions
for several topological properties of the integrated social networks
for a general activity potential, in the thermodynamic limit of large
number of agents, $N\rightarrow \infty$, and finite integration time
$T$.  To tackle this issue, we have applied the hidden variables
formalism, by mapping the aggregated network to a model in which the
probability of connecting two nodes depends on the hidden variable (in
this case represented by the activity potential) of those nodes.  Our
analysis is complemented by numerical simulations in order to check
theoretical predictions against concrete examples of activity
potential distributions.  Using this formalism, we can demonstrate
rigorously that the integrated degree distribution at time $T$ takes
the same functional form as the activity potential distribution, with
the rescaled degree $k/T-\av{a}$, in the limit of large system size $N \rightarrow \infty$.
This is however an asymptotic result, which is well fulfilled for an activity potential power-law
distributed, as empirically measured in a wide range of social
interaction settings, but fails for simple constant or homogeneous
distributions. We also show that the aggregated social networks show
 in general disassortative degree correlations, at odds with the
assortative mixing revealed in real social networks.  The clustering
coefficient is low, $\av{c} \sim T/N$, comparable with a random
network.

Secondly, we have studied the time evolution of the connectivity
properties of the integrated network. 
We have focused in particular in the onset of the giant
component in the aggregated network, defined as the largest set of
connected agents that have established at least one contact up to a
fixed time $T$. 
We have been able to provide analytic expressions for the 
percolation time $T_p$,  at which the onset of the giant component takes place, 
 depending on the details of temporal network dynamics. 
Assuming lack of degree correlations in the initial evolution of the integrated
network, the application of the generating function formalism
\cite{Newman2001} allows to obtain an explicit general form for
the temporal percolation threshold, as well as analytic asymptotic
expressions for the size of the giant component in the vicinity of the
threshold. These expressions turn out to be in good agreement with
numerical results for particular forms of the activity distribution
imposing weak degree correlations. For a skewed, power-law distributed
activity $F(a) \sim a^{-\gamma}$, the uncorrelated results are still
numerically correct for large values of $\gamma$. When $\gamma$ is
small, however, strong disagreements arise. 
Applying a percolation formalism for correlated networks
\cite{PhysRevE.78.051105}, we have been able to obtain the exact threshold $T_p$,
valid for any kind of activity potential.   
For $\gamma >3$, the correlated threshold collapses
onto the uncorrelated result, which thus provides a very good
approximation to the exact result.  For small $\gamma<3$, the
percolation threshold as obtained by numerical simulation of large
networks is in very good agreement with the analytical 
prediction.

Our study opens interesting direction for future work, 
in the first instance concerning the possible modifications of the activity driven network model, in
order to incorporate some properties of real social networks currently
missed, such as a high clustering coefficient,
assortative mixing by degree or a community structure \cite{Newman2010}.
A particularly promising line of investigation regards the possibility of 
considering an activity driven model in which the probability to fire new connections 
does not follow a Poissonian or periodic process, 
but it is driven by a non-Markovian or renewal process. 
In this case, the model would incorporate one of the most relevant features of human dynamics,
the burstiness of interactions, represented by a power law distributed interevent time \cite{barabasi2005origin}.
Through the hidden variable formalism and renewal theory, it is possible to show that
the degree distribution of the aggregated network still displays a broad tailed behavior, 
whose exact form depends on the interplay between the renewal process and the activity potential form.  
The study of the percolation properties of integrated temporal
networks also opens new interesting venues of future research,
related in particular to the properties of dynamical processes running
on top of them and to the coupling of their  different time
scales. One such application in the context of epidemic spreading is
the study of the SIR model, which we have shown can be mapped to a
temporal percolation problem in activity driven networks, thus
providing explicit forms (albeit valid in certain limits of weak
degree correlations) for the size of epidemic outbreaks in this class
of systems.

\part{Dynamical processes on empirical time-varying networks}
\label{part:dynproc}


The traditional approach to the study of dynamical processes on complex networks 
considered a time-scale separation between 
the network evolution and the dynamical process unfolding on its structure \cite{BBV}.
In most cases, the substrate is taken as a static entity, or \emph{quenched} in its connectivity pattern,
with connections frozen or evolving at a time scale
much longer than the one of the process under study \cite{BBV, dorogovtsev07:_critic_phenom}.
Researchers have also focused on the opposite limit, 
in which the process dynamics are much slower than the network evolution,
thus the interactions among individuals can be replaced by effective random couplings,
equivalent to consider the graph as \emph{annealed} \cite{Boguna09}.
While time scale separation is extremely convenient for the numerical and analytical tractability of the models, 
in many cases of interest the two timescales are comparable and cannot be decoupled 
\cite{butts:revisiting,moody2002importance,Panisson:2012}.
The time duration and the co-occurrence of links is crucial.  
This is particularly true when the substrate is represented by a social network, 
in which connections between individuals are constantly rewired. 
 Longitudinal data has traditionally
been scarce in social network analysis, but, thanks to recent
technological advances, researchers are now in a position to gather
data describing the contacts in groups of individuals at several
temporal and spatial scales and resolutions.

The analysis of empirical data on several types of human interactions
(corresponding in particular to phone communications or physical
proximity) has unveiled the presence of complex temporal patterns in
these systems, \cite{Hui:2005,PhysRevE.71.046119,Onnela:2007,10.1371/journal.pone.0011596,Tang:2010,
Stehle:2011nx,Miritello:2011,Karsai:2011,temporalnetworksbook}.
In particular, the heterogeneity and burstiness of the contact patterns 
 of human interactions~\cite{barabasi2005origin}, 
 as broadly discussed in Chapter \ref{chap:intro},
stimulated the study of the impact of a network's dynamics on the
dynamical processes taking place on top of it. 
The processes studied in this context include synchronization~\cite{albert2011sync},
percolation~\cite{Parshani:2010,Bajardi:2012}, social
consensus~\cite{consensus_temporal_nrets_2012}, or
diffusion~\cite{PhysRevE.85.056115}. 
Epidemic-like processes have
also been explored, both using realistic and toy models of propagation
processes
\cite{Rocha:2010,Isella:2011,Stehle:2011nx,Karsai:2011,Miritello:2011,dynnetkaski2011,Panisson:2012,Holme:2013,Rocha:2013,Masuda13}.
The study of simple schematic spreading processes over temporal
networks helps indeed expose several properties of their dynamical
structure: dynamical processes can in this context be conceived as
probing tools of the network's temporal structure \cite{Karsai:2011}.

In this part of the Thesis, we will uncover the behavior of dynamical processes 
running on the top of empirical temporal networks.
 To this aim, we consider as typical examples of temporal networks the
dynamical sequences of contact between individuals in various social
contexts, as recorded by the SocioPatterns project \cite{sociopatterns},
(see Chapter \ref{chap:intro}).
We will devote our attention to two simple cases of dynamical processes:
in Chapter \ref{chap:RW} we address random walks, 
while Chapter \ref{chap:epidemic} is dedicated to epidemic spreading.
Considering the pivotal role of these processes, 
we believe that our understanding 
could help to shed light on the behavior of more complex dynamics on temporally evolving networks.

\chapter{Random walks}
\label{chap:RW}

%

Random walks are a paradigm of dynamical processes,
 having a glorious tradition in statistical physics,
 and since they lie at the core of many real-world phenomena,
they have been extensively studied in many other fields,  ranging from biology to economics
  \cite{WeissRandomWalk, hughes, lovasz}.
The random walk is indeed the simplest diffusion model, and its dynamics
provides fundamental hints to understand the whole class of diffusive
processes on networks. Moreover, it has relevant applications in
such contexts as spreading dynamics (i.e. virus or opinion spreading)
and searching.  For instance, assuming that each vertex knows only
about the information stored in each of its nearest neighbors, the
most naive economical strategy is the random walk search, in which the
source vertex sends one message to a randomly selected nearest
neighbor \cite{PhysRevE.64.046135,Lv:2002,BBV}. If that vertex has the
information requested, it retrieves it; otherwise, it sends a message
to one of its nearest neighbors, until the message arrives to its
finally target destination. Thus, the random walk represents a lower
bound on the effects of searching in the absence of any information in
the network, apart form the purely local information about the
contacts at a given instant of time.

In this Chapter we will focus on the dynamics of a random walker
unfolding on a temporal network \cite{temporalnetworksbook},
 represented by the empirical contact sequence 
of face-to-face interactions \cite{sociopatterns},
presented in Chapter \ref{chap:intro}.
In our study, we introduce different randomizing strategies that
 allow us to single out the role of the different properties of the empirical networks,
 such as the heterogeneity and burstiness of the social interactions. 
 We will show that the random walk exploration is slower on temporal networks than
  it is on the aggregate projected network, even when the time is properly rescaled.
   In particular, we point out that a fundamental role is played by the temporal correlations
    between consecutive contacts present in the data. 
We also address the consequences of the intrinsically limited duration of many real world dynamical networks,
by considering the \emph{reachability} of the individuals when performing a random walk search.
Although these results refer to temporal networks represented by face-to-face social interactions, 
 our findings, especially regarding the impact of burstiness in slowing down the dynamics,
 turn out to be crucial in the study of a wide class of dynamical processes on temporal networks  
  \cite{citeulike:12739344, Lambiotte:2013aa, citeulike:13132337}. 
  
The Chapter is structured as follows. In Sec \ref{sec:rw_overv}
we review some of the fundamental results for random walks on static networks. 
In Sec. \ref{sec:rw_synth_extens} we recall the datasets used in the study and 
we introduce suitable randomization procedures, which will help later on to pinpoint the
role of the correlations in the real data, and we write down mean-field equations for
the case of maximally randomized dynamical contact networks.
In Sec. \ref{sec:rw_num_sim} we investigate the random walk
dynamics numerically, focusing on the exploration properties and on
the mean first passage times, while Sec. \ref{sec:rw_finite} is
devoted to the analysis of the impact of the finite temporal duration
of real time series. Finally, we summarize our results and comment on
some perspectives in Sec \ref{sec:rw_concl}.

\section{A short overview of random walks on static networks}
\label{sec:rw_overv}

The random walk (RW) process is defined by a
walker that, located on a given vertex $i$ at time $t$, hops to a
nearest neighbor vertex $j$ at time $t+1$.

In binary networks, defined by the adjacency matrix $a_{ij}$ (see Chapter \ref{chap:intro})
the transition probability at each time step from $i$ to $j$ is
\begin{equation}
  p_b(i \to j) = \frac{a_{ij}}{\sum_r a_{ir}} \equiv \frac{a_{ij}}{k_i},
\end{equation}
where $k_i= \sum_j a_{ij}$ is the degree of vertex $i$: the walker
hops to a nearest neighbor of $i$, chosen uniformly at random among
the $k_i$ neighbors, hence with probability $1/k_i$ (note that we
consider here undirected networks with $a_{ij}=a_{ji}$, but the
process can be considered as well on directed networks).  In
weighted networks with a weight matrix $\omega_{ij}$, the transition
probability takes instead the form
\begin{equation}
  p_w(i \to j) = \frac{w_{ij}}{\sum_r w_{ir}} \equiv \frac{w_{ij}}{s_i},
\end{equation}
where $s_i=\sum_j \omega_{ij}$ is the strength of vertex $i$
\cite{Barrat16032004}. Here the walker chooses a nearest neighbor with
probability proportional to the weight of the corresponding connecting
edge.

The basic quantity characterizing random walks in networks is the
\textit{occupation probability} $\rho_i$, defined as the steady state
probability (i.e., measured in the infinite time limit) that the
walker occupies the vertex $i$, or in other words, the steady state
probability that the walker will land on vertex $i$ after a jump from
any other vertex. Following rigorous master equation arguments, it is
possible to show that the occupation probability takes the form
\cite{PhysRevLett.92.118701,wu07:_walks}
\begin{equation}
\label{eq:rw_rho_stat}
  \rho_i^b = \frac{k_i}{\av{k}N}, \qquad 
  \rho_i^w = \frac{s_i}{\av{s}N},
\end{equation}
 in binary and weighted networks, respectively.

Other characteristic properties of the random walk, relevant to the
properties of searching in networks, are the \textit{mean
  first-passage time} (MFPT) $\tau_i$ and the \textit{coverage} $C(t)$
\cite{WeissRandomWalk,hughes,lovasz}. The MFPT of a node $i$ is
defined as the average time taken by the random walker to arrive for
the first time at $i$, starting from a random initial position in the
network. This definition gives the number of messages that have to be
exchanged, on average, in order to find vertex $i$. The coverage
$C(t)$, on the other hand, is defined as the number of different
vertices that have been visited by the walker at time $t$, averaged
for different random walks starting from different sources. The
coverage can thus be interpreted as the searching efficiency of the
network, measuring the number of different individuals that can be
reached from an arbitrary origin in a given number of time steps.

At a mean-field level, these quantities are computed as follows: let us
define $P_f(i; t)$ as the probability for the walker to arrive for the
first time at vertex $i$ in $t$ time steps. Since in the steady state
$i$ is reached in a jump with probability $\rho_i$, we have $P_f(i;t)
= \rho_i [1-\rho_i]^{t-1} $. The MFPT to vertex $i$ can thus be
estimated as the average $\tau_i = \sum_t t P_f(i;t)$, leading to
\begin{equation}
  \label{eq:rw_static_mfpt}
  \tau_i = \sum_{t=1}^\infty t \rho_i [1-\rho_i]^{t-1} \equiv \frac{1}{\rho_i}.
\end{equation}
On the other hand, we can define the {\it random walk reachability} of vertex
$i$, $P_r(i;t)$, as the probability that vertex $i$ is visited by a
random walk starting at an arbitrary origin, at any time less than or
equal to $t$. The reachability takes the form 
\begin{equation}
  \label{eq:rw_static_reach}
  P_r(i;t)=1-[1-\rho_i]^t \simeq 1-\exp(-t \rho_i),
\end{equation}
where the last expression is valid in the limit of sufficiently small
$\rho_i$. The coverage of a random walk at time $t$ will thus be given by
the sum of these probabilities, i.e.
\begin{equation}
  \label{eq:rw_static_cov}
  \frac{C(t)}{N} = \frac{1}{N} \sum_i  P_r(i;t)\equiv 1-
  \frac{1}{N} \sum_i  \exp\left(-t \rho_i\right). 
\end{equation}
For sufficiently small $\rho_i t$, the
exponential in Eq.~\eqref{eq:rw_static_cov} can be expanded to yield $C(t)\sim
t$, a linear coverage implying that at the initial stages of the walk,
a different vertex is visited at each time step, independently of the
network properties \cite{stauffer_annealed2005,almaas03:_scaling}.

It is  now important to note that the random walk process has
been defined here in a way such that the walker performs a move and
changes node at each time step, potentially exploring a new node:
except in the pathological case of a random walk starting on an
isolated node, the walker has always a way to move out of the node it
occupies. In the context of temporal networks, on the other hand, the
walker might arrive at a node $i$ that at the successive time step
becomes isolated, and therefore has to remain trapped on that node
until a new link involving $i$ occurs. In order to compare in a
meaningful way random walk processes on static and dynamical networks,
and on different dynamical networks, we consider in each dynamical
network the average probability $\overline{p}$ that a node has at
least one link. The walker is then expected to move on average once
every $\frac{1}{\overline{p}}$ time steps, so that we will consider
the properties of the random walk process on dynamical networks as a
function of the rescaled time $\overline{p} t$.
The values of the mean ratio of interacting individuals $\overline{p}$ 
for each dataset is reported in Table \ref{tab:summary}, 
as one can see, these values are far from $\overline{p} = 1$, 
which implies that the slowing down due to the temporal nature of the network
 is considerable and cannot be neglected.

\section{Synthetic extensions of empirical contact sequences}
\label{sec:rw_synth_extens}

In this study we use as substrate for the random walks dynamics 
the temporal networks representing the face-to-face interactions 
recorded by the SocioPatterns collaboration \cite{sociopatterns}.
In particular, we focus on data sets measured in several
different social contexts: the 2010 European Semantic Web Conference
(``eswc''), a geriatric ward of a hospital in Lyon (``hosp''), the
2009 ACM Hypertext conference (``ht''), and the 2009 congress of the
Soci\'et\'e Francaise d'Hygi\`ene Hospitali\`ere (``sfhh'').  
A detailed description of the properties of these data sets has been provided in Chapter \ref{chap:intro},
while a broader analysis including a discussion on experiment setup can be found in
\cite{10.1371/journal.pone.0011596,percol,Isella:2011,Stehle:2011,Panisson:2012}.

The empirical contact sequences represent the proper dynamical network
substrate upon which the properties of any dynamical process should be
studied. In many cases however, the finite duration of empirical
datasets is not sufficient to allow these processes to reach their
asymptotic state \cite{Pan:2011,Stehle:2011nx}.  This issue is
particularly important in processes that reach a steady state, such as
random walks. As discussed in Sec.~\ref{sec:rw_overv}, a walker
does not move at every time step, but only with a probability
$\overline{p}$, and the effective number of movements of a walker is
of the order $T \overline{p}$. 
For the considered empirical sequences, 
this means that the ratio between the number of hops of
the walker and the network size, $T \overline{p} / N$, assumes values
between $3.01$ for the school case and $1.60$ for the eswc case.  Typically,
for a random walk processes such small times permit to observe
transient effects only, but not a stationary behavior. Therefore we will first
explore the asymptotic properties of random
walks in synthetically extended contact sequences, and we will consider
the corresponding finite time effects in Sec.~\ref{sec:rw_finite}. 
The synthetic extensions preserve at different levels the statistical properties
observed in the real data, thus providing null models of
dynamical networks. 


Inspired by previous approaches to the synthetic extension of
empirical contact sequences
\cite{PhysRevE.71.046119,Pan:2011,Stehle:2011nx,dynnetkaski2011,temporalnetworksbook}, we
consider the following procedures:
\begin{itemize}
\item \textbf{SRep}: Sequence replication. The contact sequence is
  repeated periodically, defining a new extended characteristic
  function such that $\chi_e^{SRep}(i,j,t) = \chi(i,j,t \bmod
  T)$. This extension preserves all of the statistical properties of
  the empirical data (obviously, when properly rescaled to take into
  account the different durations of the extended and empirical time
  series), introducing only small corrections, at the topological
  level, on the distribution of time respecting paths and the
  associated sets of influence of each node. Indeed, a contact present
  at time $t$ will be again available to a dynamical process starting
  at time $t' > t$ after a time $t +T$.

\item \textbf{SRan}: Sequence randomization. The time ordering of the
  interactions is randomized, by constructing a new characteristic
  function such that, at each time step $t$,
  $\chi_e^{SRan}(i,j,t)=\chi(i,j,t')$ $\forall i$ and $\forall j$,
  where $t'$ is a time chosen uniformly at random from the set $\{1,
  2, \ldots, T\}$.  This form of extension yields at each time step an
  empirical instantaneous network of interactions, and preserves on
  average all the characteristics of the projected weighted network,
  but destroys the temporal correlations of successive contacts,
  leading to Poisson distributions for $P(\Delta t)$ and $P_i(\tau)$.

\item \textbf{SStat}: Statistically extended sequence. An intermediate
  level of randomization can be achieved by generating a synthetic
  contact sequence as follows: we consider the set of all
  conversations ${c(i,j,\Delta t)}$ in the sequence, defined as a
  series of consecutive contacts of length $\Delta t$ between the 
  pair of agents $i$ and $j$.  The new sequence is generated, at each
  time step $t$, by choosing $\overline{n}$ conversations
  ($\overline{n}$ being the average number of new conversations
  starting at each time step in the original sequence, see
  Table~\ref{tab:summary}), randomly selected from the set of
  conversations, and considering them as starting at time $t$ and
  ending at time $t+\Delta t$, where $\Delta t$ is the duration of the
  corresponding conversation. In this procedure we avoid choosing
  conversations between agents $i$ and $j$ which are already engaged
  in a contact started at a previous time $t' < t$. This extension
  preserves all the statistical properties of the empirical contact
  sequence, with the exception of the distribution of time gaps
  between consecutive conversations of a single individual,
  $P_i(\tau)$.
\end{itemize}

\begin{figure}[tb]
\begin{center}
\includegraphics*[width=0.48\textwidth]{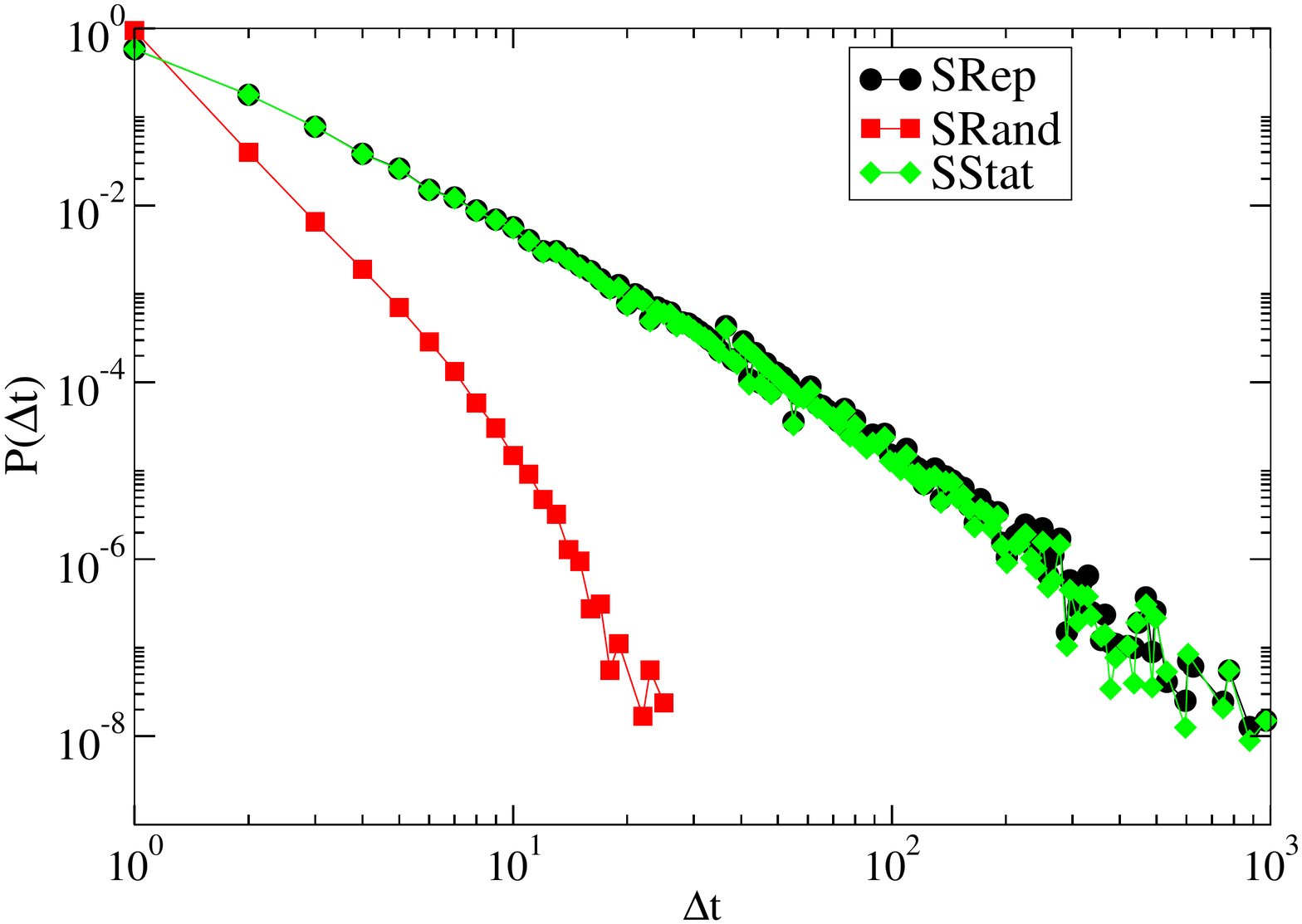}
\includegraphics*[width=0.48\textwidth]{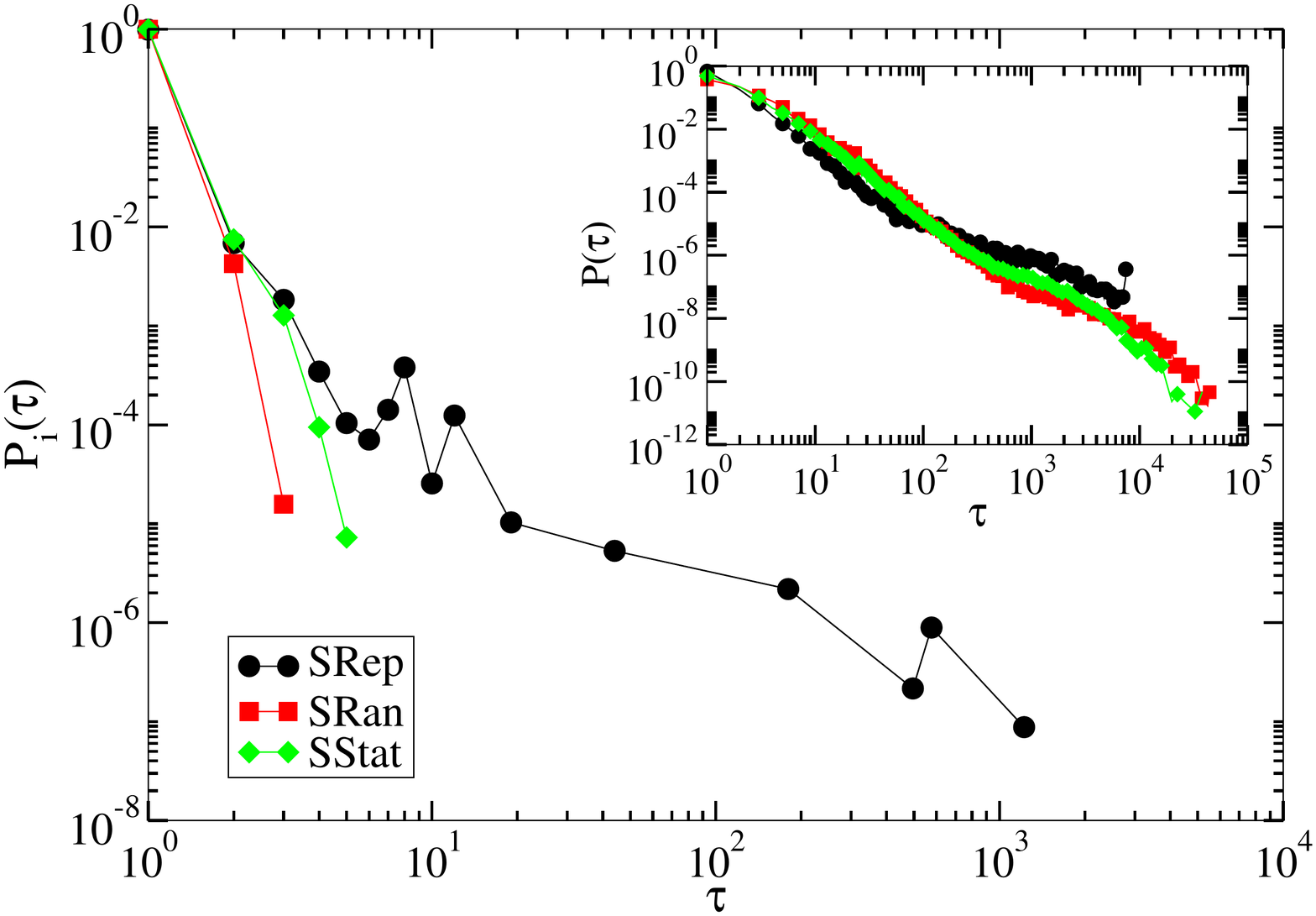}
\end{center}
\caption{Left: Probability distribution $P(\Delta t)$ for the extended contact
  sequences SRep, SRan and SStat of the 25c3 dataset.   
  Right:  Interevent time probability distribution $P_i(\tau)$ of a single
  individual $i$ (chosen as the most connected one) for the SRep,  SRan and SStat  extensions of the  25c3 dataset.  
  Inset: Interevent time probability distribution $P(\tau)$ for $N$ agents.  
  The weight   distribution $P(w)$ of the original contact sequence is preserved  for every extension.  }
  \label{fig:comp_extens}
\end{figure}

In Fig. \ref{fig:comp_extens} we plot the distribution of the duration
of contacts, $P(\Delta t)$, and the distribution of gap times between
two consecutive conversations realized by a single individual, $P
(\tau_i )$, for the extended contact sequences SRep, SRan and SStat.
One can check that the SRep extension preserves all the $P(w)$,
$P(\Delta t)$ and $P_i(\tau)$ distributions of the original contact
sequence, the SRan extension preserves only $P(w)$ and the SStat
extension preserves both the $P(w)$ and the $P(\Delta t)$ but not the
$P_i(\tau)$, as summarized in Table \ref{tab:extens}.
 Interestingly,
we note that the distribution of gap times for all agents, $P(\tau)$,
is also broadly distributed in the SRan and SStat extensions, despite the fact
that the respective individual burstiness $P_i(\tau)$ are bounded, see
Fig.~\ref{fig:comp_extens}. This fact can be easily understood by
considering that $P(\tau)$ can be written in terms of a convolution of
the individual gap distributions times the probability of starting a
conversation. In the case of SRan extension, the probability $r_i$
that an agent $i$ starts a new conversation is proportional to its
strength $s_i$, i.e. $r_i=s_i / (N \langle s \rangle) $. Therefore,
the probability that it starts a conversation $\tau$ time steps after
the last one (its gap distribution) is given by $P_i(\tau) = r_i
[1-r_i]^{\tau - 1} \simeq r_i \exp(-\tau r_i )$, for sufficiently
small $r_i$. The gap distribution for all agents $P(\tau)$ is thus
given by the convolution
\begin{equation}
  P(\tau)= \int P(s)  \dfrac{s}{N \langle s \rangle} \exp\left(-\tau
  \dfrac{s}{N \langle s \rangle} \right) ds,  
  \label{eq:rw_ptau}
\end{equation}
where $P(s)$ is the strength distribution. This distribution has an
exponential form, as shown in Section \ref{sec:intro_empirical}.
This leads, from Eq.~\eqref{eq:rw_ptau}, to a total gap
distribution $P(\tau) \sim (1+ \tau/N)^{-2}$, with a heavy
tail. Analogous arguments can be used in the case of the SStat extension.

\begin{table}[tb]
\begin{center}
    \begin{tabular}{|c||  c | c  | c |}
    \hline 
    Extension & $P(w)$ & $P(\Delta t)$ &  $P_i(\tau)$\\ \hline 
      SRep  &   \checkmark  &  \checkmark &  \checkmark   \\ 
      SRan  &  \checkmark & \ding{55} &  \ding{55}   \\
      SStat  &   \checkmark &  \checkmark & \ding{55} \\ 
      \hline 
    \end{tabular}
  \caption{Comparison of the properties of the original contact sequence preserved in the synthetic extensions.}
  \label{tab:extens}
\end{center}
\end{table}

\subsection{Analytic expressions for the extended  sequences}
\label{subsec:rw_analyit_sran}

Let us consider a random walk on the sequence of instantaneous networks at
discrete time steps, which is equivalent to a message passing strategy
in which the message is passed to a randomly chosen neighbor. The
walker present at node $i$ at time $t$ hops to one of its neighbors,
randomly chosen from the set of vertices
\begin{equation}
  \mathcal{V}_i(t) = \left\{ j \; | \; \chi(i,j,t)=1 \right\},
\end{equation}
of which there is a number 
\begin{equation}
  k_i(t) = \sum_j  \chi(i,j,t),
\end{equation}
If the node $i$ is isolated at time $t$, i.e. $\mathcal{V}_i(t) =
\varnothing$, the walker remains at node $i$. 
In any case, time is increased $t\to t+1$.

Analytical considerations analogous to those in
Sec.~\ref{sec:rw_overv} for the case of contact sequences are
hampered by the presence of time correlations between contacts. In
fact, as we have seen, the contacts between a given pair of agents are
neither fixed nor completely random, but instead show long range
temporal correlations. An exception is represented by the randomized
SRan extension, in which successive contacts are by construction
uncorrelated. Considering that the random walker is in vertex $i$ at
time $t$, at a subsequent time step it will be able to jump to a
vertex $j$ whenever a connection between $i$ and $j$ is created, and a
connection between $i$ and $j$ will be chosen with probability
proportional to the number of connections between $i$ and $j$ in the
original contact sequence, i.e. proportional to $\omega_{ij}$. That
is, a random walk on the extended SRan sequence behaves essentially as
in the corresponding weighted projected network, and therefore the
equations obtained in Sec.~\ref{sec:rw_overv}, namely
\begin{equation}
  \label{eq:rw_mf_mfpt}
  \tau_i = \frac{\av{s} N}{s_i},
\end{equation}
and
\begin{equation}
  \frac{C(t)}{N} = 1- \frac{1}{N} \sum_i  \exp\left(-t
    \frac{s_i}{\av{s} N}\right)
\end{equation}
apply. In this last expression for the coverage we can approximate the
sum by an integral, i.e.
\begin{equation}
  \label{eq:rw_mf_cov}
  \frac{C(t)}{N} = 1- \int ds P(s)  \exp\left(-t
    \frac{s}{\av{s} N}\right),
\end{equation}
being $P(s)$ the distribution of strengths.  Giving that $P(s)$ has an
exponential behavior, we can obtain from the last expression
\begin{equation}
  \label{eq:rw_aprox_cov}
   \frac{C(t)}{N} \simeq 1- \left(1+\frac{t}{N}\right)^{-1}.
\end{equation}

\section{Random walks on extended contact sequences}
\label{sec:rw_num_sim}

In this Section we present numerical results from the simulation of
random walks on the extended contact sequences described above.
Measuring the coverage $C(t)$ we set the duration of these sequences
to $50$ times the duration of the original contact sequence $T$, while
to evaluate the MFPT between two nodes $i$ and $j$, $\tau_{ij}$, we
let the RW explore the network up to a maximum time $t_{max}=10^8$, 
time steps, corresponding to $10^4 - 10^5$ times the duration of the original contact sequences.
Each result we report is averaged over at least $10^3$ independent
runs.

\begin{figure}[tb]
\begin{center}
\includegraphics*[width=0.7\textwidth]{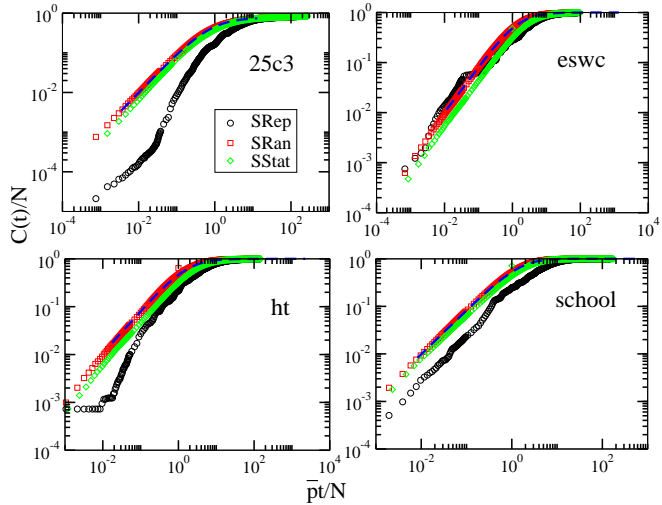}
\end{center}
\caption{ Normalized coverage $C(t)/N$ as a function of the 
  time $\overline{p} t/N$, for the SRep, SRan and SStat extension of
  empirical data. The numerical evaluation of Eq.~(\ref{eq:rw_mf_cov}) is
  shown as a dashed line. 
  The exploration of the
  empirical repeated data sets (SRep) is slower than the other
  cases, the SRan is in agreement with the theoretical
  prediction, and the SStat case shows a close (but systematically
  slower) behavior.
  }
  \label{fig:covextended}
  \end{figure}

\subsection{Network exploration}
\label{subsec:network-coverage}

The network coverage $C(t)$ describes the fraction
of nodes that the walker has discovered up to time $t$.
Figure~\ref{fig:covextended} shows the normalized coverage $C(t)/N$ as
a function of time, averaged for different walks starting from
different sources, for the dynamical networks obtained using the
 SRep, SRan and SStat prescriptions.
Time is
rescaled as $ t \to \overline{p} t$ to take into account that the walker can find 
itself on an isolated vertex, as discussed before.  While for SRep and
SRan extensions the average number of interacting nodes $\overline{p}$
is by construction the same as in the original contact sequence, for
the SStat extension we obtain numerically different values of
$\overline{p}$, which we use when rescaling time in the corresponding
simulations.

The coverage corresponding to the SRan
extension is very well fitted by a numerical integration of
Eq.~(\ref{eq:rw_mf_cov}), which predicts the coverage $C(t)/N$ obtained in the
correspondent projected weighted network. Moreover, when using the
rescaled time $\overline{p} t$, the SRan coverages for different
datasets collapse on top of each other for small times, with a linear
time dependence $C(t)/N \sim t/N$ for $t \ll N$ as expected in static
networks, showing a universal behavior. 

\begin{figure}[tb]
  \begin{center}
    \includegraphics*[width=0.5\textwidth]{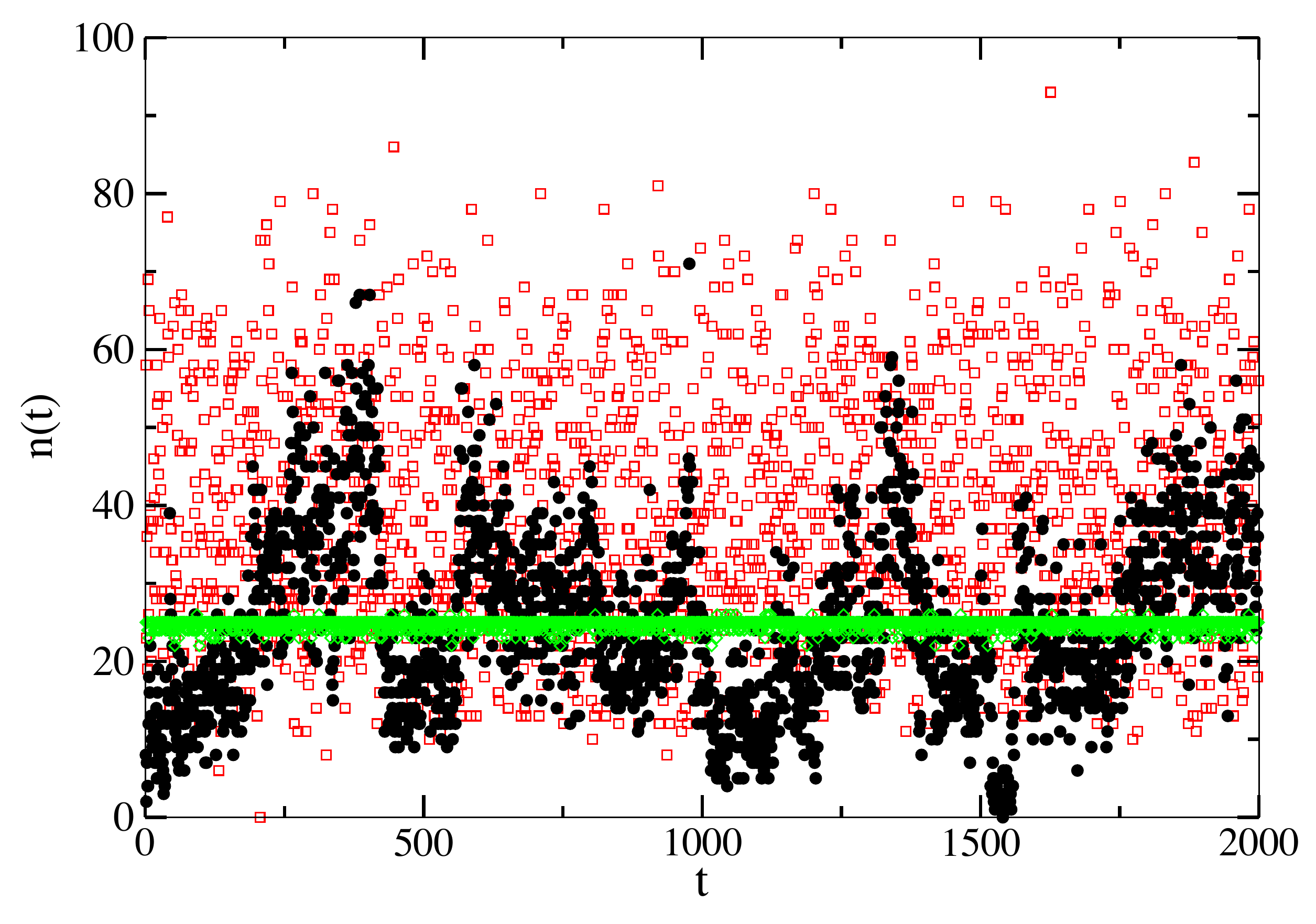}
  \end{center}
  \caption{ Number of new conversations $n(t)$ started per unit time in the SRep (black, full dots), SRan (red, empty squares) and SStat (green, diamonds) extensions of the
    school dataset. 
    }
  \label{fig:numlinksschool}
\end{figure}

The coverage obtained on the SStat extension is systematically
smaller than in the SRan case, but follows a similar evolution. On the
other hand, the RW exploration obtained with the SRep prescription is
generally slower than the other two, particularly for the 25c3 and ht
datasets. As discussed before, the original contact sequence, as well
as the SRep extension, are characterized by irregular distributions of
the interactions in time, showing periods with few interacting nodes
and correspondingly a small number $n(t)$ of new started
conversations, followed by peaks with many interactions  (see
Fig.~\ref{fig:numlinksschool}). This feature slows down the RW
exploration, because the RW may remain trapped for long times on
isolated nodes. The SRan and the SStat extensions, on the contrary,
both destroy this kind of temporal structure, balancing the periods of
low and high activity: the SRan extension randomizes the time order of
the contact sequence, and the SStat extension evens the number of
interacting nodes, with $\overline{n}$ new conversations starting at
each time step.

The similarity between the random walk processes on the SRan and SStat
dynamical networks shows that the random walk coverage is not very sensitive 
to the heterogenous durations of the
conversations, as the main difference between these two cases
is that $P(\Delta t)$ is narrow for SRan and broad for SStat. In these
cases, the observed behavior is instead well accounted for by Eq. \eqref{eq:rw_mf_cov}, 
taking into account only the weight distribution of the projected network, i.e.,
the heterogeneity between aggregated conversation durations.  Therefore, the
slower exploration properties of the SRep sequences can be
mostly attributed to the correlations between consecutive
conversations of the single individuals, as given by the individual
gap distribution $P_i(\tau)$, (see
\cite{Stehle:2011nx,Karsai:2011,dynnetkaski2011} for analogous results
in the context of epidemic spreading).

A remark is in order for the 25c3 conference. A close inspection of
Fig.~\ref{fig:covextended} shows that the RW does not reach the whole
network in any of the extensions schemes, with $C_{max} < 0.85$,
although the duration of the simulation is quite long
$\overline{p}t_{max}>10^2 N$.  The reason is that this dataset
contains a group of nodes (around $20\%$ of the total) with a very low
strength $s_i$, meaning that there are agents who are isolated for
most of the time, and whose interactions are reduced to one or two
contacts in the whole contact sequence. Given that each extension
we use preserves the $P(w)$ distribution, the discovery of these nodes
is very difficult.
The consequence is that we observe an extremely slow approach to the
asymptotic value $\lim_{t\to\infty} \frac{C(t)}{N} = 1$. Indeed, the 
mean-field calculations presented in Secs.~\ref{sec:rw_overv}
and \ref{subsec:rw_analyit_sran} suggest a power-law decay with
$(1+\bar{p}t/N)^{-1}$ for the residual coverage $1-C(t)/N$. 

\begin{figure}[tb]
\begin{center}
\includegraphics*[width=0.48\textwidth]{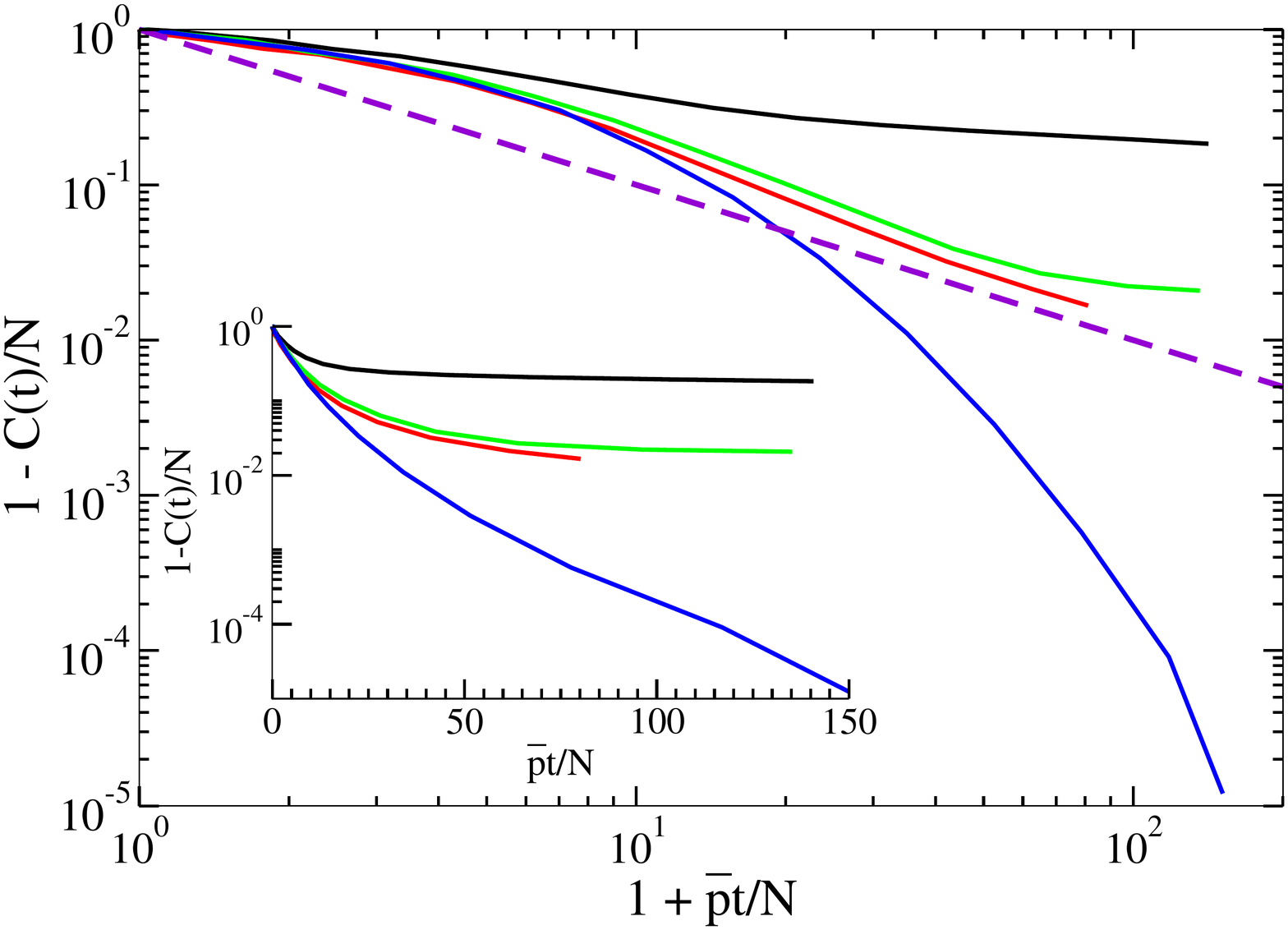}
\includegraphics*[width=0.48\textwidth]{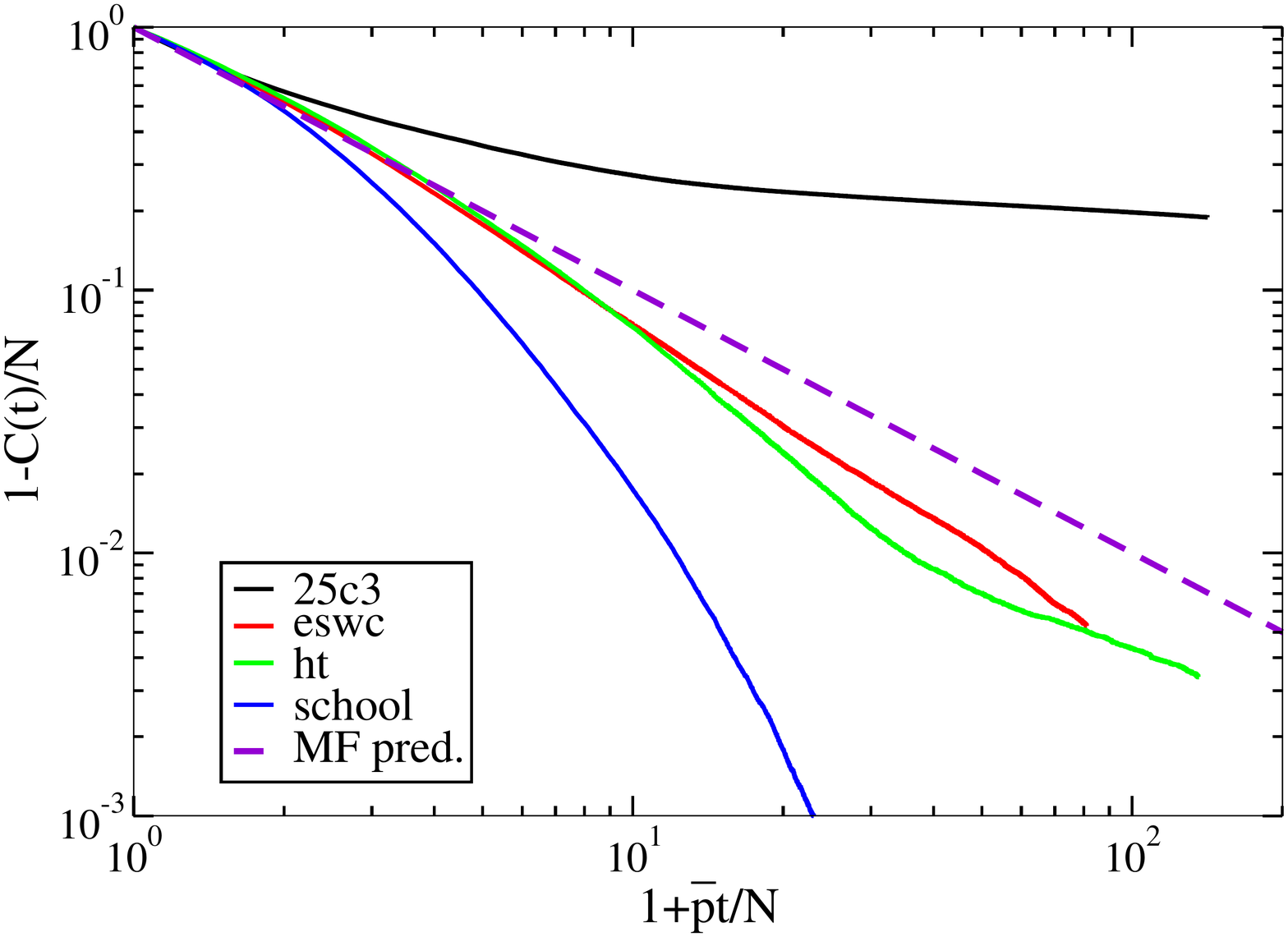}
\end{center}
\caption{ Asymptotic residual coverage $1-C(t)/N$ as a function 5of $\bar{p} t/N$ for the SRep (left) and SRan (right) extended sequences, 
for different datasets. In the inset we show the same plot in semilog scale. }
  \label{fig:residual}
\end{figure}

In Fig.~\ref{fig:residual} we plot the asymptotic coverage for large
times in the 4 datasets considered. We can see that RW on the eswc and ht dataset
conform at large times quite reasonably to the expected theoretical
prediction in Eq.~\eqref{eq:rw_aprox_cov}, both for the SRep and SRan
extensions. The 25c3 dataset shows, as discussed above, a considerable
slowing down, with a very slow decay in time. Interestingly, the school
dataset is much faster than all the rest, with a decay of the residual
coverage $1-C(t)/N$ exhibiting an approximate exponential decay. It is
noteworthy that the plots for the randomized SRan sequence do not
always obey the mean-field prediction (see lower plot in
Fig.~\ref{fig:residual}). This deviation can be attributed to the fact
that SRan extensions preserve the topological structure of the
projected weighted network, and it is known that, in some instances,
random walks on
weighted networks can deviate from the mean-field predictions
\cite{dynam_in_weigh_networ}. These deviations are particularly strong in the
case of the 25c3 dataset, where connections with a very small weight
are present.

\subsection{Mean first-passage time}
\label{subsec:mean-first-passage}

\begin{figure}[tb]
\begin{center}
\includegraphics*[width=0.8\textwidth]{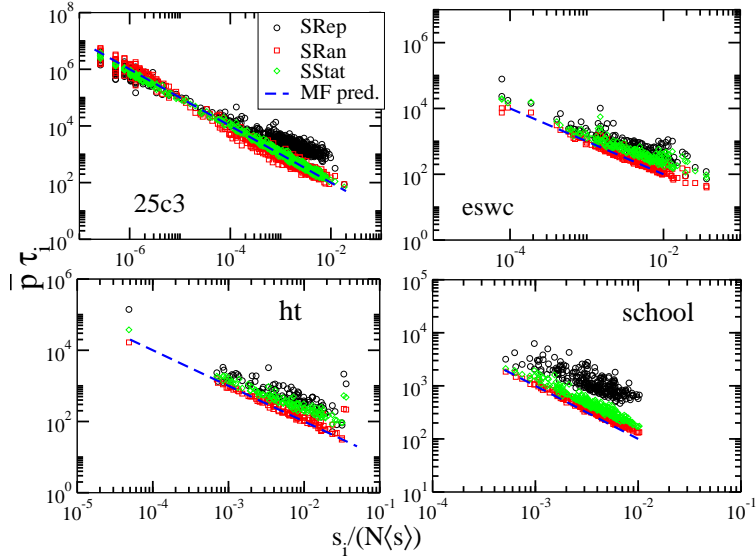}
\end{center}
\caption{ Rescaled mean first passage time $\tau_i$,
  shown against the strength $s_i$, normalized with the total strength
  $N \langle s \rangle $, for the SRep, SRan and SStat extensions of
  empirical data. The dashed line represents the prediction of
  Eq.~\eqref{eq:rw_mf_mfpt}. Each panel in the figure corresponds to one of the
  empirical datasets considered.  }
  \label{fig:mpftextended}
\end{figure}

Let us now focus on another important characteristic property of random
walk processes, namely the MFPT defined in Section
\ref{sec:rw_overv}. 
Figure \ref{fig:mpftextended} shows the correlation between the MFPT $\tau_i$
of each node, measured in units of rescaled time $\overline{p} t$, and
its normalized strength $s_i/(N \langle s \rangle )$. 

\begin{figure}[tb]
\begin{center}
\includegraphics*[width=0.75\textwidth]{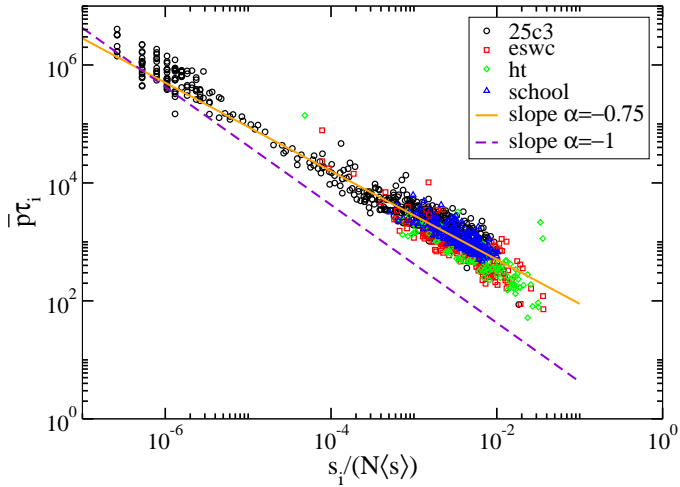}
\end{center}
  \caption{(Mean first passage time at node $i$, in units of rescaled
    time $\overline{p} t$, vs. the strength $s_i$, normalized with the
    total strength $N \langle s \rangle $, for RW processes on the
    SRep datasets extension. All data collapse close to the continuous
    line whose slope, $\alpha \simeq 0.75$, differs from the
    theoretical one, $\alpha = 1.0$, shown as a dashed line. }
  \label{fig:mpft-rp}
\end{figure}

The random walks performed on the SRan and SStat extensions are very
well fitted by the mean field theory, i.e. Eq.~\eqref{eq:rw_mf_mfpt} 
(predicting that $\tau_i$  is inversely proportional to $s_i$), for every
dataset considered; on the other hand, random walks on the extended
sequence SRep yield at the same time deviations from the mean-field
prediction and much stronger fluctuations around an average
behavior. Figure \ref{fig:mpft-rp} addresses this case in more detail,
showing that the data corresponding to RW on different datasets collapse on 
an average behavior that can be fitted by a scaling function of the form
\begin{equation}
  \label{eq:7}
  \tau_i \sim \frac{1}{\overline{p}} \times \left( \frac{s_i}{N
      \langle s \rangle 
    } \right)^{-\alpha}, 
\end{equation} 
with an exponent $\alpha \simeq 0.75$.

These results show that the MFPT, similarly to the coverage, is rather
insensitive to the distribution of the contact durations, as long as
the distribution of cumulated contact durations between individuals is
preserved (the weights of the links in the projected network). Therefore, the
deviations of the results obtained with the SRep extension of the
empirical sequences have their origin in the burstiness of the
contact patterns, as determined by the temporal correlations between
consecutive conversations. The exponent $\alpha<1$ means that the
searching process in the empirical, correlated, network is slower than
in the randomized versions, in agreement with the smaller coverage
observed in Fig.~\ref{fig:covextended}. 

The data collapse observed in Fig. \ref{fig:mpft-rp} for the SRep case leads 
to two noticeable conclusions.  First, although the various datasets studied
correspond to different contexts, with different numbers of
individuals and densities of contacts, simple rescaling procedures are
enough to compare the processes occurring on the different temporal
networks, at least for some given quantities.  Second, the MFPT at a
node is largely determined by its strength. This can indeed seem
counterintuitive as the strength is an aggregated quantity (that may
include contact events occurring at late times). However, it can be
rationalized by observing that a large strength means a large number
of contacts and therefore a large probability to be reached by the
random walker. Moreover, the fact that the strength of a node is an
aggregate view of contact events that do not occur homogeneously for
all nodes but in a bursty fashion leads to strong fluctuations around
the average behavior, which implies that nodes with the same strength can
also have rather different MFPT (Note the logarithmic scale on the
y-axis in Fig.  \ref{fig:mpft-rp}).

\section{Random walks on finite contact sequences}
\label{sec:rw_finite}

\begin{figure}[tbp]
\begin{center}
\includegraphics*[width=0.49\textwidth]{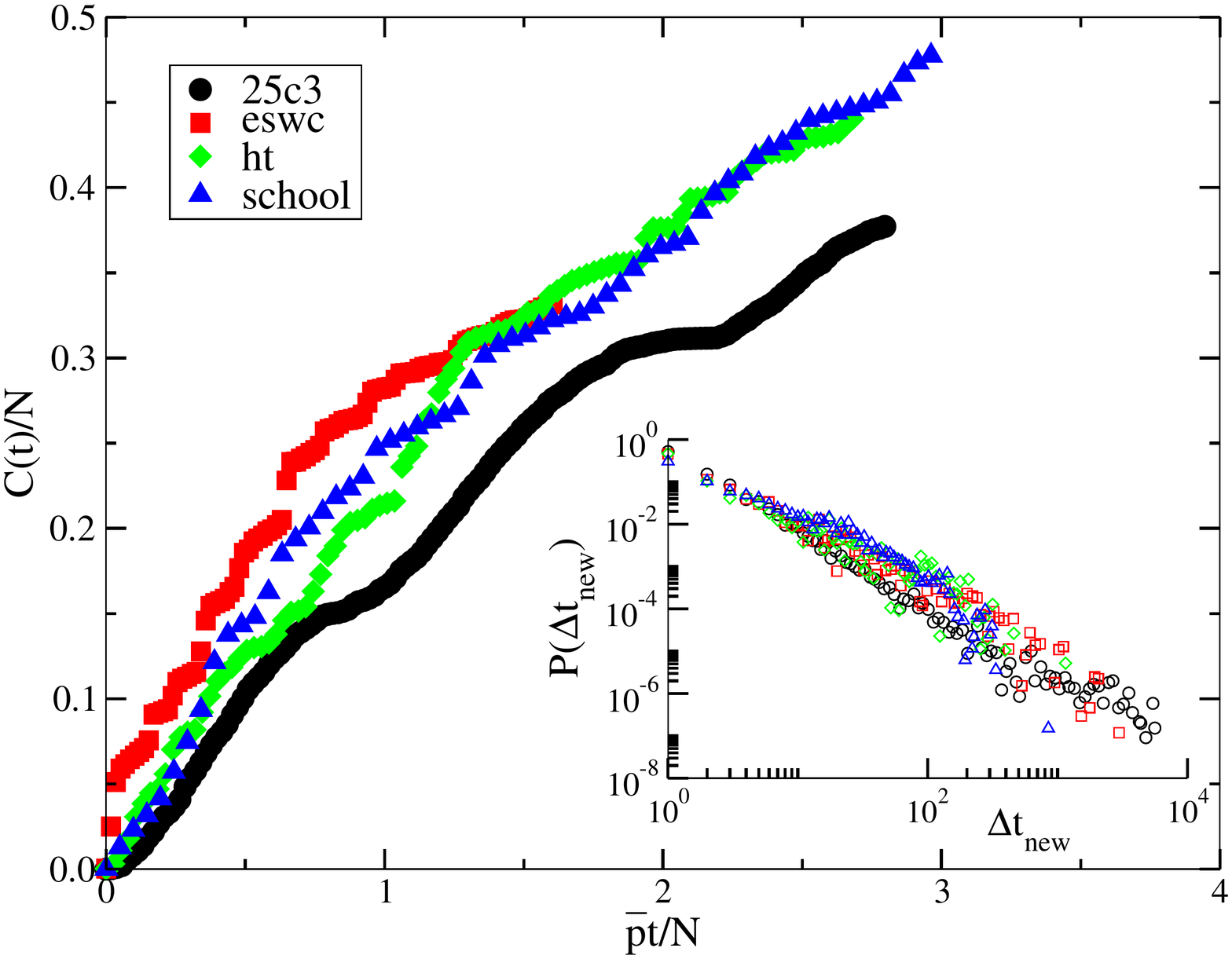}
\includegraphics*[width=0.49\textwidth]{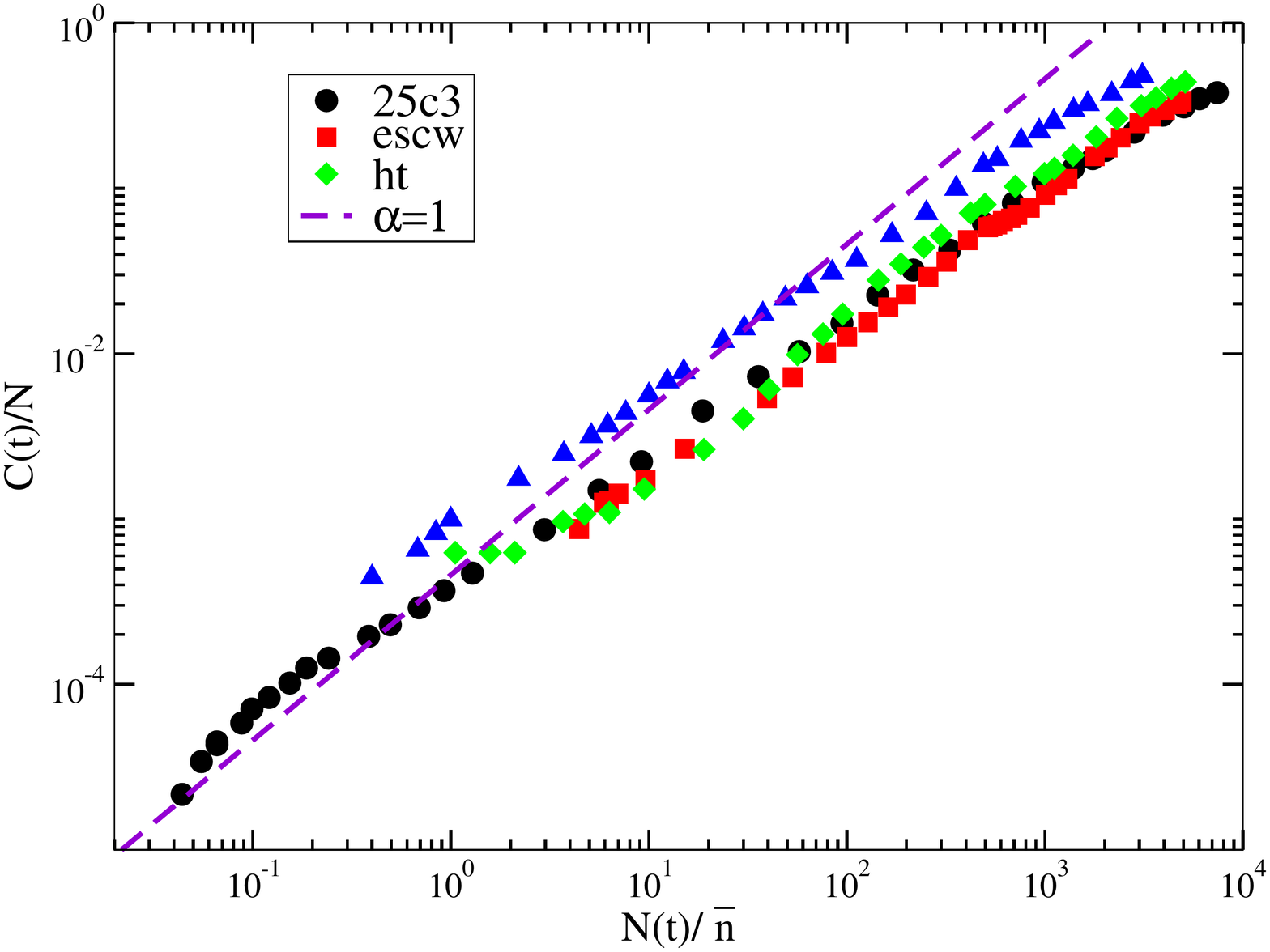}
\end{center}
  \caption{Left: Normalized coverage $C(t)/N$ as function of the 
    time $\overline{p} t / N$ for the different datasets.  
    Inset: probability distribution $P(\Delta t_{new})$ of the
    time lag $\Delta t_{new}$ between the discovery of two new vertices.
    Only the discovery of the first $5\%$ of the network is
    considered, to avoid finite size effects     \cite{PhysRevE.78.011114}.
    Right: Coverage $C(t)/N$ as a function of the number of new
  conversation realized up to time $t$, normalized for the mean number
  of new conversation per unit of time, $\overline{n}$, for different
  datasets.
	  \label{fig:cov}}
\end{figure}

The case of finite sequences is interesting
from the point of view of realistic searching processes. The limited duration of a
human gathering, for example, imposes a constraint on the length of any searching strategy. 
Fig. \ref{fig:cov} ((left) shows the normalized $C(t)/N$ coverage as a
function of the rescaled time $\overline{p} t / N$. The coverage exhibits a considerable variability in the different datasets, which do not obey the rescaling
obtained for the extended SRan and SStat sequence.  The probability distribution 
of the time lags $\Delta t_{new}$ between the discovery of two new vertices
\cite{PhysRevE.78.011114} provides further evidence
of the slowing down of diffusion in temporal networks. 
The inset of Fig. \ref{fig:cov} (left)
indeed shows broad tailed distributions $P(\Delta t_{new})$ for all the
dataset considered, differently from the exponential decay observed in
binary static networks \cite{PhysRevE.78.011114}. 

The important differences in the rescaled coverage $C( t )/N$ between
the various datasets, shown in Fig. \ref{fig:cov} (left), can be attributed to
the choice of the time scale, $\overline{p}t/N$, which corresponds to
a temporal rescaling by an average quantity.  We can argue,
indeed, that the speed with which new nodes are found by the RW is
proportional to the number of new conversations $n(t)$ started at each
time step $t$, thus in the RW exploration of the temporal network the effective
time scale is given by the integrated number of new conversations up
to time $t$, $N(t) = \int^t_0 n(t') dt'$.  In the same Fig.~\ref{fig:cov} (right)
we display the correlation between the coverage $C(t)/N$ and the number
of new conversations realized up to time $t$, $N(t)$, normalized for
the mean number of new conversations per unit of time, $\overline{n}$.
While the relation is not strictly linear, a very strong
positive correlation appears between the two quantities.

\begin{figure}[tb]
\begin{center}
\includegraphics*[width=0.46\textwidth]{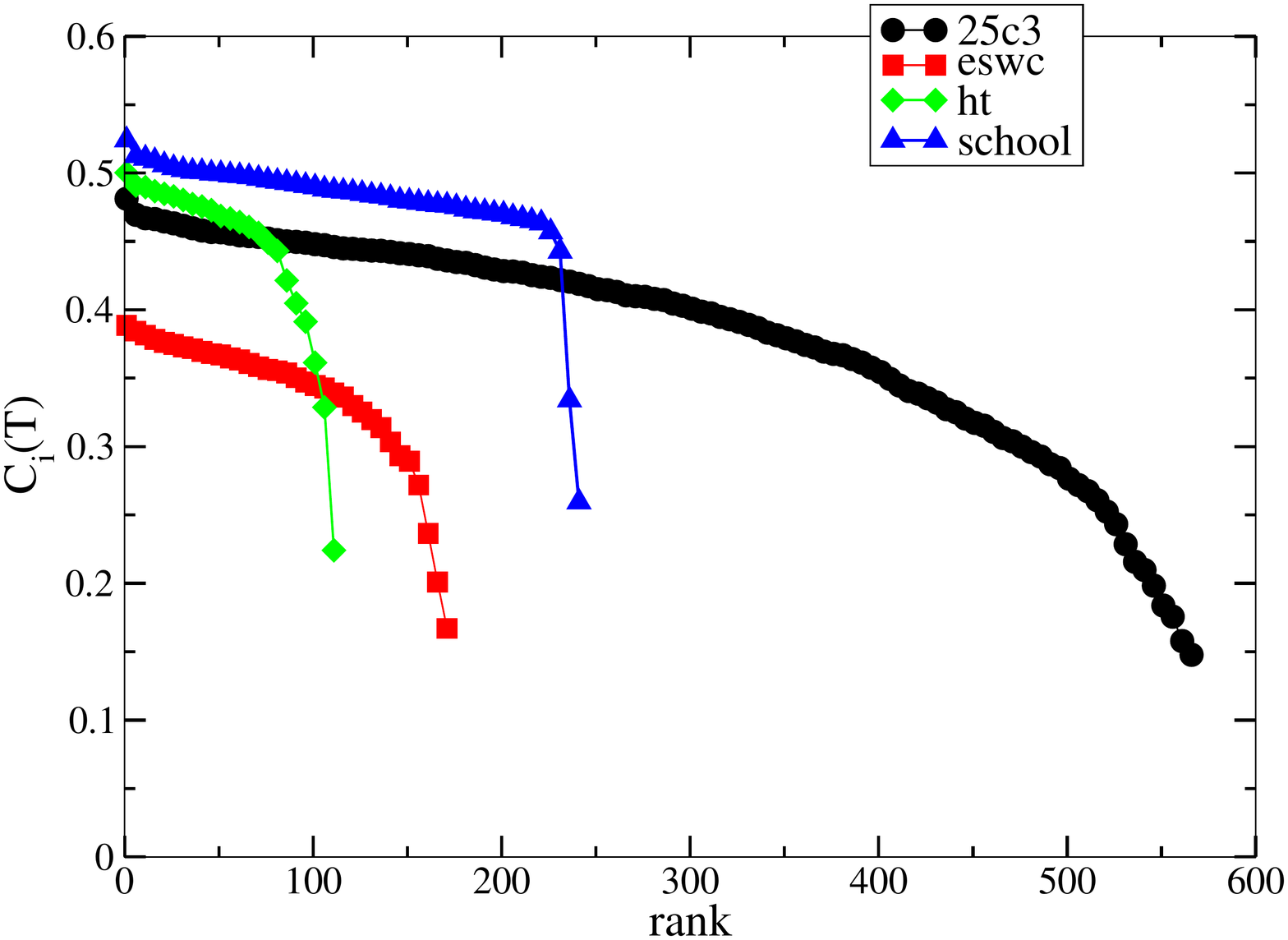}
\includegraphics*[width=0.46\textwidth]{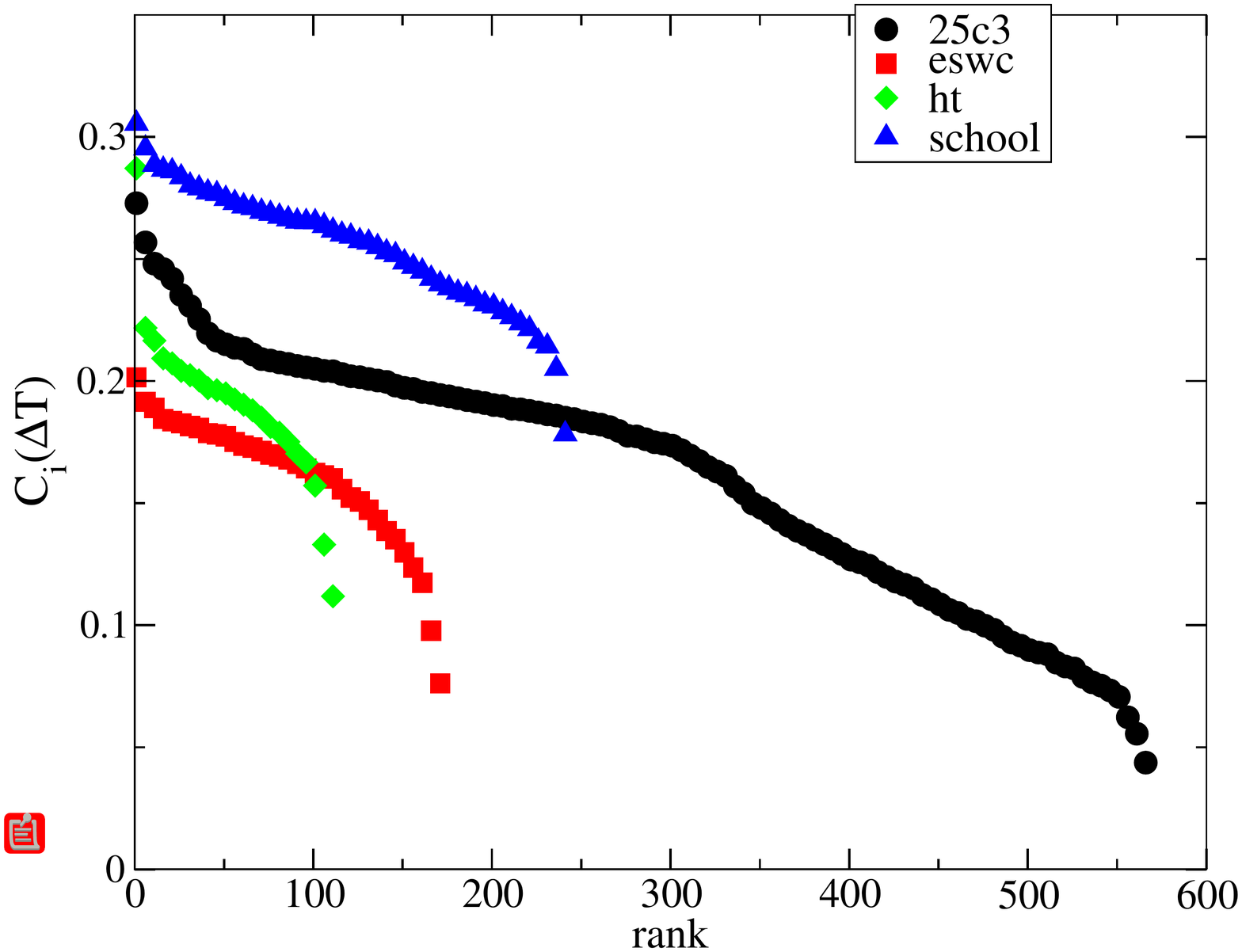}
\end{center}
  \caption{ (Left: Rank plot of the coverage $C_i$ obtained starting from node
    $i$ in the contact sequence of duration $T$, averaged over $10^3$ runs.
    Right: Rank plot of the coverage $C_i (\Delta T)$ up to a fixed time
    $\Delta T=10^3$. }
 \label{fig:rank_cov} 
\end{figure}

The complex pattern shown by the average coverage $C(t)$
originates from the lack of self-averaging in a dynamic network. Figure \ref{fig:rank_cov}
shows the rank plot of the coverage $C_i$ obtained at the end
of a RW process starting from node $i$, and averaged over $10^3$ runs. 
Clearly, not all vertices are equivalent. 
A first explanation of the variability in $C_i$ comes from the fact that
not all nodes appear simultaneously on the network at time $0$. If
$t_{0,i}$ denotes the arrival time of node $i$ in the system, a random
walk starting from $i$ is restricted to $T^r_i =T-t_{0,i}$: nodes
arriving at later times have less possibilities to explore their set
of influence, even if this set includes all nodes. To put all nodes on
equal footing and compensate for this somehow trivial difference
between nodes, we consider the coverage of random walkers starting on
the different vertices $i$ and walking for exactly $\Delta T$ time
steps (we limit of course the study to nodes with $t_{0,i} < T- \Delta
T$). Differences in the coverage $C_i(\Delta T)$ will then depend on
the intrinsic properties of the dynamic network. For a static network
indeed, either binary or weighted, the coverage $C_i(\Delta T)$ would
be independent of $i$, as random walkers on static networks lose the
memory of their initial position in a few steps, reaching very fast
the steady state behavior Eq. \eqref{eq:rw_rho_stat}.  As the inset of
Fig. \ref{fig:rank_cov} shows, important heterogeneities are instead
observed in the coverage of random walkers starting from different
nodes on the dynamic network, even if the random walk duration is the
same.

\begin{figure}[tbp]
\begin{center}
\includegraphics*[width=0.48\textwidth]{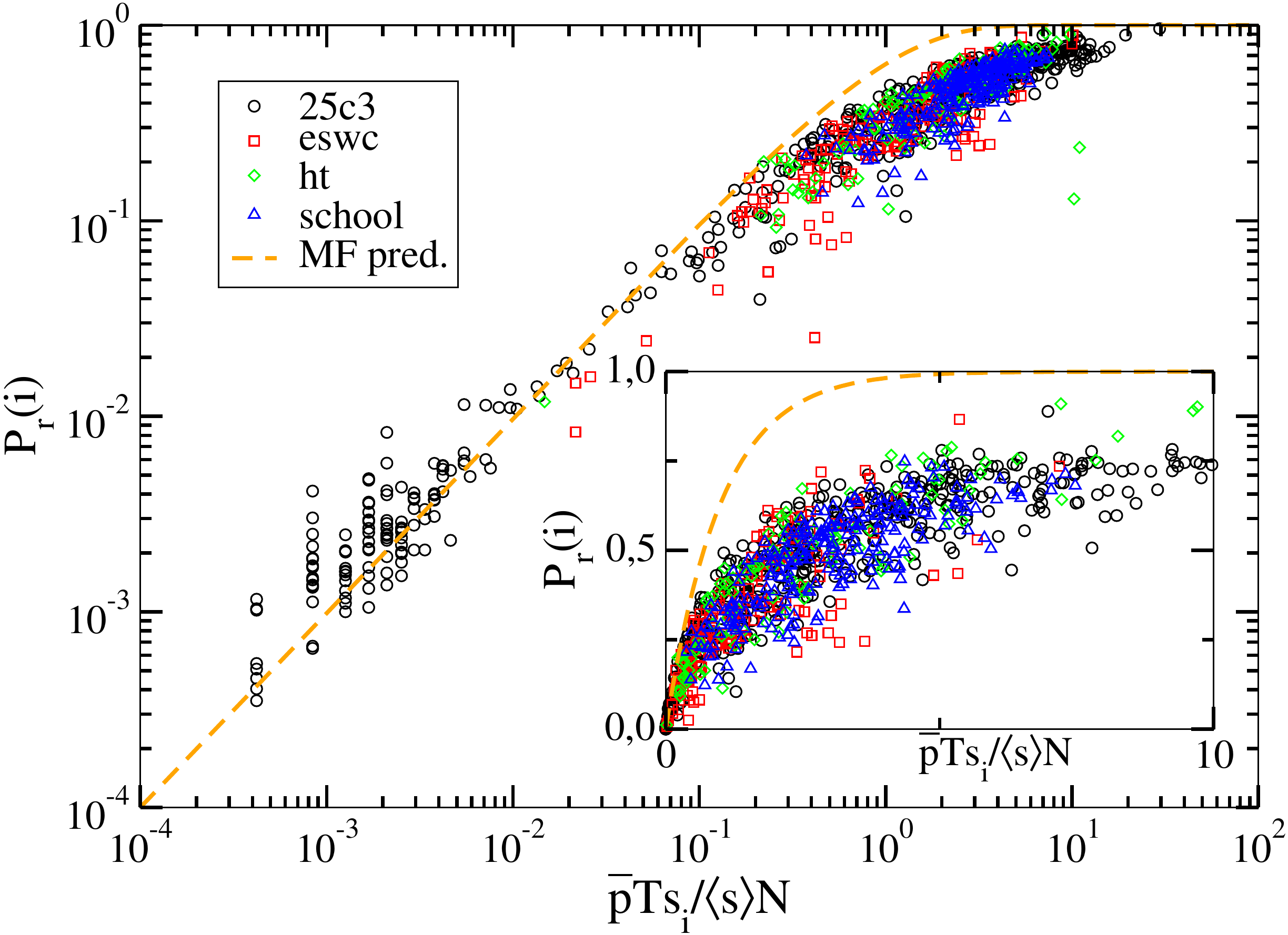}
\includegraphics*[width=0.48\textwidth]{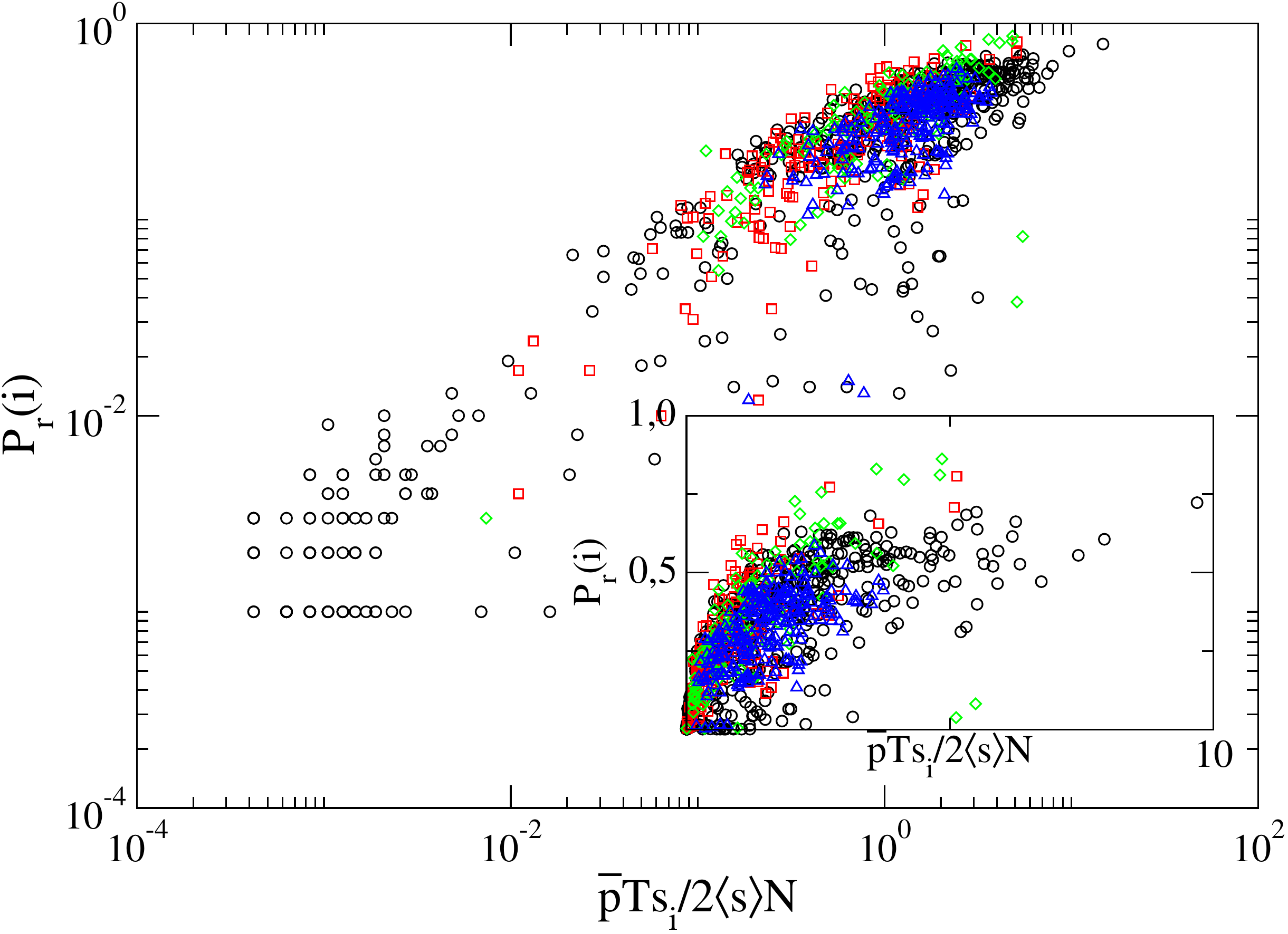}
\end{center}
  \caption{ Correlation between the probability of node $i$ to be
    reached by the RW, $P_r(i)$, and the rescaled strength
 $\overline{p} T s_i/N \langle s \rangle $ for different datasets. The
 curves obtained by different dataset collapse, 
but they do not follow the mean-field behavior predicted by of
Equation (\ref{eq:prob_reach}) (dashed line). The inset shows the same
data on a linear scale, to emphasize the deviation from mean-field.
Right: Correlation between the probability of node $i$ to be
    reached by a RW of length $T/2$, $P_r(i)$, and the rescaled strength
 $\overline{p} T s_i/N \langle s \rangle$ for different datasets,
 where $s_i$ is computed on the whole dataset of length $T$. 
 The inset shows the same data on a linear scale.
 \label{fig:reach-s} }
\end{figure}

Another interesting quantity is the probability that a
vertex $i$ is discovered by the random walker.
As discussed in Section
\ref{sec:rw_overv}, at the mean field level the probability that a
node $i$ is visited by the RW at any time less than or equal to $t$
(the random walk reachability) takes the form $P_r (i; t) = 1 - \exp[
- t \rho (i) ]$.  Thus the probability that the node $i$ is reached by
the RW at any time in the contact sequence is
\begin{equation}
\label{eq:prob_reach}
P_r (i) = 1 - \exp \left( -\frac{\overline{p} T s_i}{N \langle s \rangle } \right),
\end{equation}
where the rescaled time $\overline{p}t $ is taken into account.  
In Fig. \ref{fig:reach-s}, we plot the probability
$P_r(i)$ of node $i$ to be reached by the RW during the contact
sequence as a function of its strength $s_i$. $P_r(i)$ exhibits a
clear increasing behavior with $s_i$,  larger strength corresponding to 
larger time in contact and therefore larger probabilities to be
reached. 
Interestingly, the simple rescaling by $\overline{p}$ and $\langle s \rangle$ leads to an approximate data collapse for the RW processes on the various
dynamical networks, showing a very robust behavior. Similarly to the
case of the MFPT on extended sequences, the dynamical property
$P_r(i)$ can be in part ``predicted'' by an aggregate quantity such as $s_i$.
Strong deviations from the mean-field prediction of Eq. (\ref{eq:prob_reach}) are
however observed, with a tendency of $P_r(i)$ to saturate at large
strengths to values much smaller than the ones obtained on a static
network. Thus, although the set of sources of almost every node $i$ has size
$N$, as determined by the large fraction $l_e$ of pairs of nodes connected by at least one
time respecting path, shown in Table \ref{tab:paths} (see Section \ref{subsec:intro_sociopatterns}),
the probability for node $i$ to be effectively reached by a RW is far from being equal to $1$.

Moreover, rather strong fluctuations of $P_r(i)$ at given $s_i$ are also
observed: $s_i$ is indeed an aggregate view of contacts which are typically
inhomogeneous in time, with bursty behaviors\footnote{When
  considering RW on a contact sequence of length $T$
  randomized according to the SRan procedure instead,
  Eq. (\ref{eq:prob_reach}) is well obeyed and only small fluctuations
of $P_r(i)$ are observed at a fixed $s_i$.}. 
Figure \ref{fig:reach-s}  also shows that the
reachability computed at shorter time (here $T/2$)
displays stronger fluctuations as a function of the strength $s_i$
computed on the whole time sequence: $P_r(i)$ for shorter RW is
naturally less correlated with an aggregate view which takes into
account a more global behavior of $i$.

\section{Summary and Discussion}
\label{sec:rw_concl}

In this Chapter we have investigated the behavior of random walks on
temporal networks. In particular, we have focused on real face-to-face
interactions concerning four different datasets. 
Given the finite life-time of each network, we have considered as
substrate for the random walk process the replicated sequences in which
the same time series of contact patterns is indefinitely repeated. At
the same time, we have proposed two different randomization procedures
to investigate the effects of correlations in the real dataset. The
``sequence randomization" (SRan) destroys any temporal correlation by
randomizing the time ordering of the sequence. This allows to write
down exact mean-field equations for the random walker exploring these
networks, which turn out to be substantially equivalent to the ones
describing the exploration of the weighted projected network. The
``statistically extended sequence" (SStat), on the other hand, selects
random conversations from the original sequence, thus preserving the
statistical properties of the original time series, with the exception
of the distribution of time gaps between consecutive conversations.

We have performed numerical analysis both for the coverage and the
MFPT properties of the random walker. In both cases we have found that
the empirical sequences deviate systematically from the mean field
prediction, inducing a slowing down of the network exploration and of
the MFPT. Remarkably, the analysis of the randomized sequences has
allowed us to point out that this is due \textit{uniquely} to the
temporal correlations between consecutive conversations present in the
data, and \textit{not} to the heterogeneity of their lengths.  Finally, we have
addressed the role of the finite size of the empirical networks, which
turns out to prevent a full exploration of the random walker,
though differences exist across the four considered cases. In this
context, we have also shown that different starting nodes provide on
average different coverages of the networks, at odds to what happens
in static graphs. In the same way, the probability that the node $i$
is reached by the RW at any time in the contact sequence exhibits a
common behavior across the different time series, but it is not
described by the mean-field predictions for the aggregated network,
which predict a faster process.

In conclusion, the contribution of the analysis presented in this Chapter is two-fold. On the
one hand, we have proposed a general way to study dynamical processes
on temporally evolving networks, by the introduction of randomized
benchmarks and the definition of appropriate quantities that
characterize the network dynamics. On the other hand, for the
specific, yet fundamental, case of the RW, we have obtained
detailed results that clarify the observed dynamics, and that will
represent a reference for the understanding of more complex diffusive
dynamics occurring on dynamic networks. Our investigations also open
interesting directions for future work. For instance, it would be
interesting to investigate how random walks starting from different
nodes explore first their own neighborhood \cite{baronchelli2006ring}, 
which might lead to hints about the definition of ``temporal communities'' (see
e.g. \cite{Pons:2005} for an algorithm using RW on static networks
for the detection of static communities); various measures of
nodes centrality have also been defined in temporal networks
\cite{Braha:2009,Tang:2010proc,Lerman:2010,Pan:2011,temporalnetworksbook}, but their
computation is rather heavy, and RW processes might present
interesting alternatives, similarly to the case of static networks
\cite{Newman:2005}.

\chapter{Epidemic spreading}
\label{chap:epidemic}

The topology of the pattern of contacts between individuals plays a
fundamental role in determining the spreading patterns of epidemic
processes \cite{keeling05:_networ}. The first predictions of
classical epidemiology \cite{anderson92,Keeling07book} were
based on the homogeneous mixing hypothesis, assuming that all
individuals have the same chance to interact with other individuals in
the population. This assumption and the corresponding results were
challenged by the empirical discovery that the contacts within
populations are better described in terms of networks with a
non-trivial structure \cite{Newman2010}. Subsequent studies were
devoted to understanding the impact of network structure on the
properties of the spreading process. The main result obtained
concerned the large susceptibility to epidemic spread shown by
networks with a strongly heterogeneous connectivity pattern, as
measured by a heavy-tailed degree distribution $P(k)$ \cite{pv01a,lloyd01,newman02,refId0}.

The study of spreading patterns on networks is naturally complemented
by the formulation of immunization strategies tailored to the
specific topological  properties of each network.
Optimal strategies shed light on how the role and importance of nodes
depend on their properties, and can yield importance rankings of
nodes. In the case of static networks, this issue has been
particularly stimulated by the fact that heterogeneous networks with a
heavy-tailed degree distribution have a very large susceptibility to
epidemic processes, as represented by a vanishingly small epidemic
threshold. In such networks, the simplest strategy consisting in
randomly immunizing a fraction of the nodes is ineffective. More
complex strategies, in which nodes with the largest number of
connections are immunized, turn out to be effective
\cite{PhysRevE.65.036104} but rely on the global knowledge of the
network's topology. This issue is solved by the so-called acquaintance
immunization~\cite{PhysRevLett.91.247901}, which prescribes the
immunization of randomly chosen neighbors of randomly chosen
individuals.
In the temporal networks field, few works have addressed the issue of the design of immunization
strategies and their respective efficiency ~\cite{Lee:2010fk,Tang:2011,Takaguchi:2012,Masuda13}.  
In particular, \cite{Lee:2010fk} consider datasets describing the
contacts occurring in a population during a time interval $[0,T]$;
they define and study different strategies that use information from the
interval $[0,\Delta T]$ to decide which individuals should be
immunized in order to limit the spread during the remaining time
$[\Delta T, T]$.  
Using a large $\Delta T=75\% T$, the authors show
that some tailored strategies, based on the contact patterns,
perform better than random immunization and show
that this is related to the temporal correlations of the networks.

In this Chapter, we study the dynamics of a simple spreading process 
running on  top of empirical temporal networks.
In particular, we investigate the effect of several immunization strategies, 
including the ones considered by \cite{Lee:2010fk},
and address the issue of the length $\Delta T$ of the
``training window'', which is highly relevant in the context of
real-time, specific tailored strategies. The scenario we have in mind
is indeed the possibility to implement a real-time immunization
strategy for an ongoing social event, in which the set of individuals
to be immunized is determined by strategies based on preliminary
measurements up to a given time $\Delta T$. The immunization problem
takes thus a two-fold perspective: The specific rules (strategy) to
implement, and the interval of time over which preliminary data are
collected. Obviously, a very large $\Delta T$ will lead to more
complete information, and a more satisfactory performance for most
targeting strategies, but it incurs in the cost of a lengthy data
collection. On the other hand, a
short $\Delta T$ will be cost effective, but yield a smaller amount of
information about the observed social dynamics.

In order to investigate the role of the training window length on the
efficiency of several immunization strategies, we consider a simple
snowball susceptible-infected (SI) model of epidemic spreading or
information diffusion~\cite{anderson92}. In this model, individuals
can be either in the susceptible (S) state, indicating that they have
not been reached by the ``infection'' (or information), or they can be
in the infectious (I) state, meaning that they have been infected by
the disease (or that they have received the information) and can
further propagate it to other individuals.  Infected individuals do
not recover, i.e., once they transition to the infectious state they
remain indefinitely in that state.  Despite its simplicity, this model
has indeed proven to provide interesting insights into the temporal
structure and properties of temporal networks. Here we focus on the
dynamics of the SI model over empirical time-varying social
networks. The networks we consider describe time-resolved face-to-face
contacts of individuals in different environments and were measured by
the SocioPatterns collaboration (see Chapter\ref{chap:intro}).
 We consider the effect on the spread of an SI model of the immunization of a fraction of
nodes, chosen according to different strategies based on different
amounts of information on the contact sequence.  We find a saturation
effect in the increase of the efficiency of strategies based on nodes
characteristics when the length of the training window is
increased. The efficiency of strategies that include an element of
randomness and are based on temporally local information do not
perform as well but are largely independent on the amount of
information available.

The Chapter is organized as follows: 
 In Sec. \ref{sec:epid_models} we define the spreading model and
some quantities of interest. The immunization strategies we consider
are listed in Sec. \ref{sec:epid_imm_strat}.
Section \ref{sec:epid_num_res} contains the main numerical results,
and we discuss in Sec. \ref{subsec:epid_temp_corr} and
Sec. \ref{subsec:epid_non_det} the respective effects of temporal
correlations and of randomness effects in the spreading model. Section
\ref{sec:epid_concl} finally concludes with a discussion of our
results.

\section{Epidemic models and numerical methods}
\label{sec:epid_models}

Here we present the results of numerical simulation of the 
 susceptible-infected (SI) spreading dynamics on the
 empirical data of human face-to-face proximity. 
 The process is initiated by a single infected individual
(``seed'').  At each time step, each infected individual $i$ infects
with probability $\beta$ the susceptible individuals $j$ with whom $i$
is in contact during that time step.  The process stops either when
all nodes are infected or at the end of the temporal sequence of
contacts.

\begin{figure}[tb]
  \begin{center}
    \includegraphics[width=0.48\textwidth]{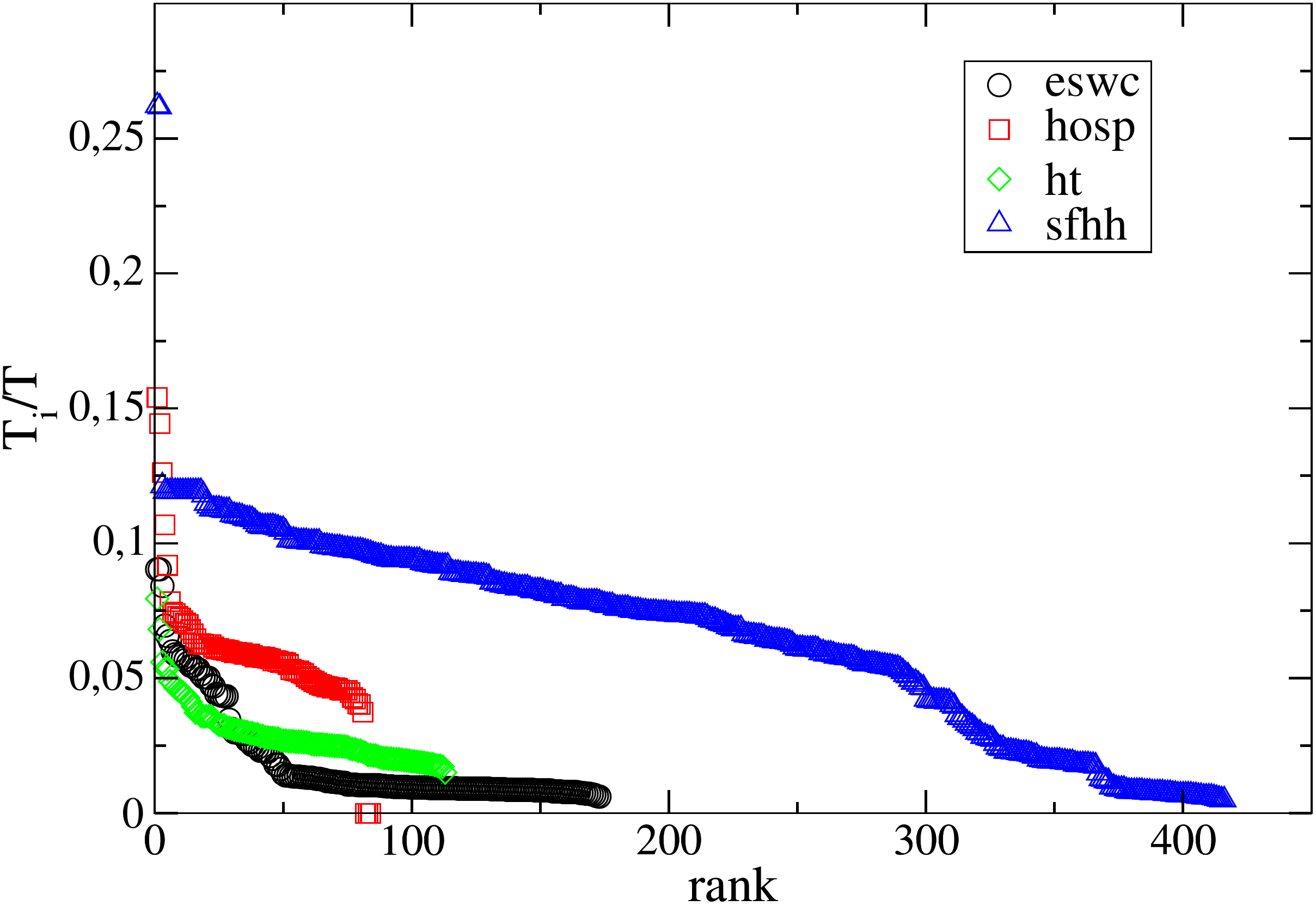}
    \includegraphics[width=0.48\textwidth]{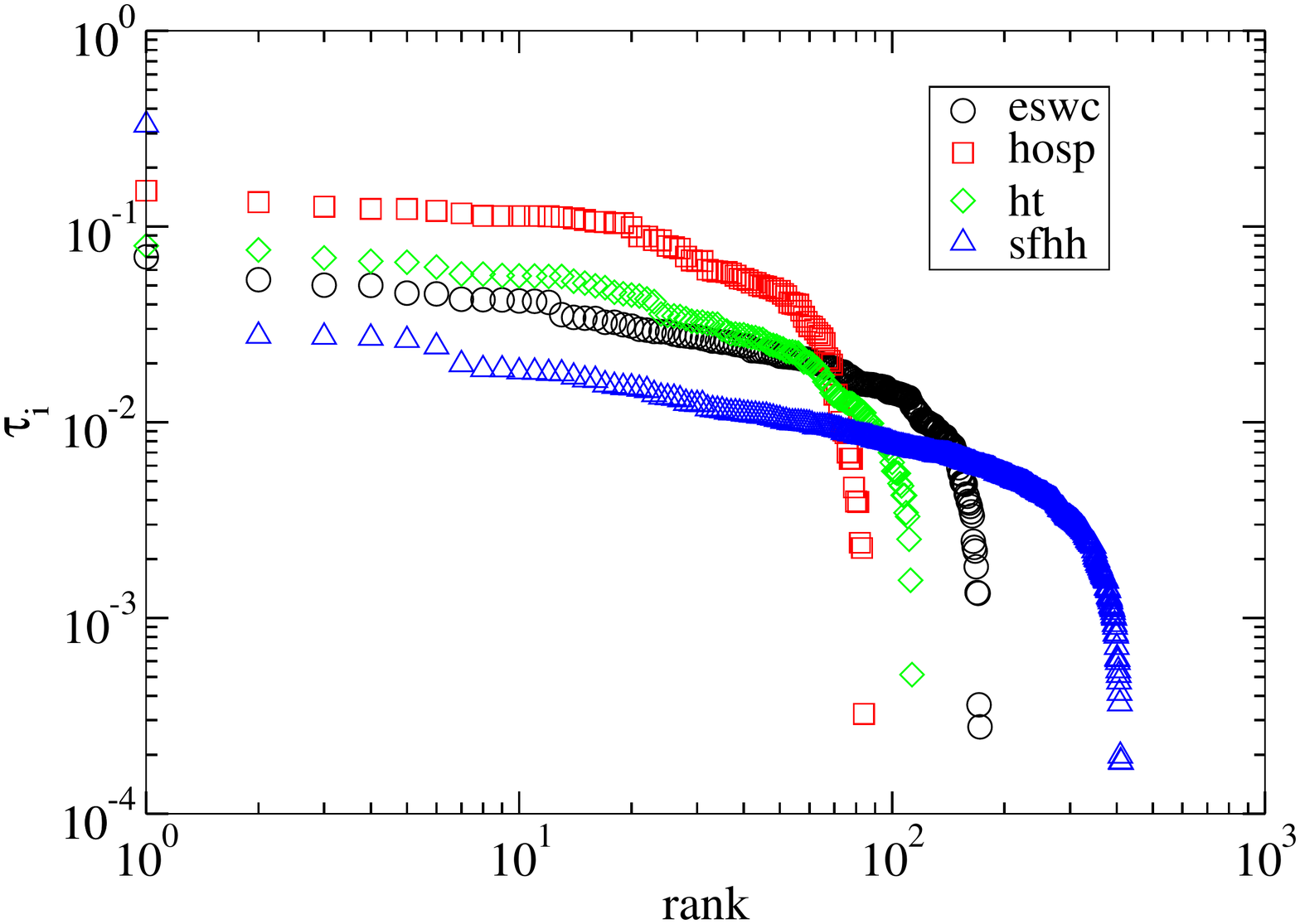}
  \end{center}
  \caption{Left: Rank plot of the half-infection times $T_i$ divided by the
    contact sequence duration $T$ for the various datasets.
    Right: Rank plot of the infection delay ratio $\tau_i$ for various
  datasets.}
  \label{fig:rank_t}
\end{figure}

Different individuals have different contact patterns and \textit{a
  priori} contribute differently to the spreading process. In order to
quantify the spreading efficiency of a given node $i$, we proceed as
follows: We consider $i$ as the seed of the SI process, all other
nodes being susceptible. We measure the half prevalence time, i.e.,
the time $t_i$ needed to reach a fraction of infected nodes equal to
$50 \%$ of the population.  Since not all nodes appear simultaneously
at $t=0$ of the contact sequence, we define the \textit{half-infection
  time} of seed node $i$ as $T_i = t_i - t_{0,i}$, where $t_{0,i}$ is
the time at which node $i$ first appears in the contact sequence. The
half-infection time $T_i$ can thus be seen as a measure of the spreading power of
node $i$: smaller $T_i$ values correspond to more efficient spreading
patterns.
We first focus on the deterministic case $\beta=1$ (the effects of
stochasticity, as given by $\beta<1$, are explored in
Sec.~\ref{subsec:epid_non_det}).  Figure~\ref{fig:rank_t} (left) shows the
rank plot of the rescaled half-infection times $T_i/T$ for various
datasets, where $T$ is the duration of the contact sequence.  We note
that $T_i$ is quite heterogeneous, ranging from $T_i < 5\% T$ up to
$T_i \simeq 20 \% T$ \footnote{We note that defining $T_i$ as the time
  needed to reach a different fraction of the population, such as
  e.g. $25\%$, leads to a similar heterogeneity.}.

Some nodes are therefore much more efficient spreaders than
others. This implies that the immunization of different nodes could
have very different impacts on the spreading process. To estimate this
impact, we define for each node $i$ the infection delay ratio $\tau_i$
as
\begin{equation}
\label{eq:epid_tau_i}
\tau_i = \left\langle \frac{ T_j^{i}- T_j }{T_j} \right\rangle_{j \neq i} ,
\end{equation}
where $T_j^{i} $ is the half-infection time obtained when node $j$ is
the seed of the spreading process and node $i$ is immunized, and the
ratio is averaged over all possible seeds $j \neq i$ and over
different starting times for the SI process (as the half-infection time
is typically much smaller than the total duration of the contact sequence,
$T_j \simeq 0.1 T$) \footnote{Note that in some cases, node $i$ is not
  present during the time window in which the SI process is simulated;
  in this case, $T_j^{i}= T_j$.}.  The infection delay ratio $\tau_i$
quantifies therefore the average impact that the immunization of node
$i$ has on SI processes unfolding over the temporal network.
Individuals with large $\tau_i$ have a high impact in delaying the infection propagation
and should be immunized, while individuals with low $\tau_i$ will have 
a marginal impact in slowing down the spreading if immunized.
Figure \ref{fig:rank_t} (right) displays a rank plot of $\tau_i$ for various
datasets. As expected, the immunization of a single node does most
often lead to a limited delay of the spreading dynamics.
Interestingly however, $\tau_i$ is broadly distributed and large
values are also observed.

The infection delay ratio of a single node $i$, $\tau_i$, can be
generalized to the case of the immunization of any set of nodes
$\mathcal{V} =\{i_1,\ldots, i_n \}$, with $n<N$.  We measure the
spreading slowing down obtained when immunizing the set $\mathcal{V}$
through the infection delay ratio
\begin{equation}
\label{eq:epid_inf_del}
\tau_{\mathcal{V}} = \left\langle \frac{ T_j ^\mathcal{V} - T_j }{T_j}
\right\rangle_{j \notin \mathcal{V}} , 
\end{equation}
where $T_j^\mathcal{V}$ is the half-infection times of node $j$ when
all the nodes of set $\mathcal{V}$ are immunized, and the average is
performed over all possible seeds $j \notin \mathcal{V}$ and over different
starting times for the SI process.

In addition to slowing down the propagation process, the immunization
of certain individuals can also block the spreading paths towards
other, non-immunized, individuals, limiting in this way the final
number of infected individuals. We measure this effect through the
\textit{average outbreak size ratio}
\begin{equation}
\label{eq:epid_outbreak_ratio}
i_\mathcal{V} = - \left\langle \frac{ I_j ^\mathcal{V} - I_j }{I_j}
\right\rangle_{j \notin \mathcal{V}} , 
\end{equation}
where $I^\mathcal{V}_j$ and $I_j$ are the final numbers of infected
individuals (outbreak size) for an SI process with seed $j$, with and
without immunization of the set $\mathcal{V}$, respectively.  The
ratio is averaged over all possible seeds $j \notin \mathcal{V}$ and
over different starting times of the SI process.

\section{Immunization strategies}
\label{sec:epid_imm_strat}

An immunization strategy is defined by the choice of the set
$\mathcal{V}$ of nodes to be immunized.  We define here different
strategies, and we compare their efficiency in section
\ref{sec:epid_num_res} by measuring $\tau_\mathcal{V}$ and
$i_\mathcal{V}$.  More precisely, for each contact sequence of
duration $T$ we consider an initial temporal window $[0, \Delta T]$
over which various node properties can be measured.  A fraction
$f$ of the nodes, chosen according to different possible rules, is
then selected and immunized (it forms the set $\mathcal{V}$). 
Finally, $\tau_\mathcal{V}$ and
$i_\mathcal{V}$ are computed by simulating the SI process with and
without immunization and averaging over starting seeds and times.  For
each selection rule, the two relevant parameters are $f$ and $\Delta
T$.  Larger fractions $f$ are naturally expected to lead to larger
$\tau_\mathcal{V}$ and $i_\mathcal{V}$.  Here we also consider the
effect of $\Delta T$, where a larger $\Delta T$ corresponds to a
larger amount of information on the contact sequence.  We investigate
whether and how more information about the contact sequence yields a
higher efficiency of the immunization strategy.

We consider the following strategies (or ``protocols"):
\begin{itemize}
\item[\textbf{K}] Degree protocol. We immunize the $f N$ individuals
  with the highest aggregated degree in $[0, \Delta T]$
  \cite{PhysRevE.65.036104}; the aggregated degree of an individual
  $i$ corresponds to the number of different other individuals with
  whom $i$ has been in contact during $[0, \Delta T]$;

\item[\textbf{BC}] Betweennness centrality protocol. We immunize the $f
  N$ individuals with the highest betweenness centrality measured on
  the aggregated network in $[0, \Delta T]$
  \cite{PhysRevE.65.056109};
  
\item[\textbf{A}] Acquaintance protocol. We choose randomly an
  individual and immunize one of his contacts in $[0,\Delta T]$, chosen at random,
  repeating the process until $f N$ individuals are immunized
  \cite{PhysRevLett.91.247901};

\item[\textbf{W}] Weight protocol. We choose randomly an individual and
  immunize his {\em most frequent} contact in $[0,\Delta T]$, repeating the process 
  until $f N$ individuals are immunized
  \cite{Lee:2010fk};

\item[\textbf{R}] Recent protocol. We choose randomly an individual and
  immunize his {\em last} contact in $[0,\Delta T]$, repeating the process
  until $f N$ individuals are immunized \cite{Lee:2010fk}.

\end{itemize}


As a benchmark, we also consider the following two strategies:
\begin{itemize}
\item[\textbf{Rn} ] Random protocol. We immunize $f N$ individuals
  chosen randomly among all nodes;

\item[\textbf{T}] $\tau$-protocol. We immunize the $f N$
  individuals with the highest $\tau_i$, where the values of $\tau_i$ are calculated
  according to Eq.~(\ref{eq:epid_tau_i}) in the interval $[0,\Delta T]$.
\end{itemize}
The \textbf{Rn} strategy uses no information about the contact
sequence and we use it as a worst case performance baseline.  The
\textbf{T} strategy makes use, through the quantity $\tau_i$, of the
most complete information available in the interval considered, 
since it takes into account the average effect of node immunization on SI
processes taking place over the contact sequence.
It could thus be
expected to yield the best performance among all strategies.

The \textbf{A}, \textbf{W}, \textbf{R} and \textbf{Rn} strategies
involve a random choice of individuals. In these cases, we
therefore average the results over $10^2$ independent runs (each run
corresponding to an independent choice of the immunized individuals).

\section{Numerical results}
\label{sec:epid_num_res}

\begin{figure}[tb]
\begin{center}
\includegraphics*[width=0.8\textwidth]{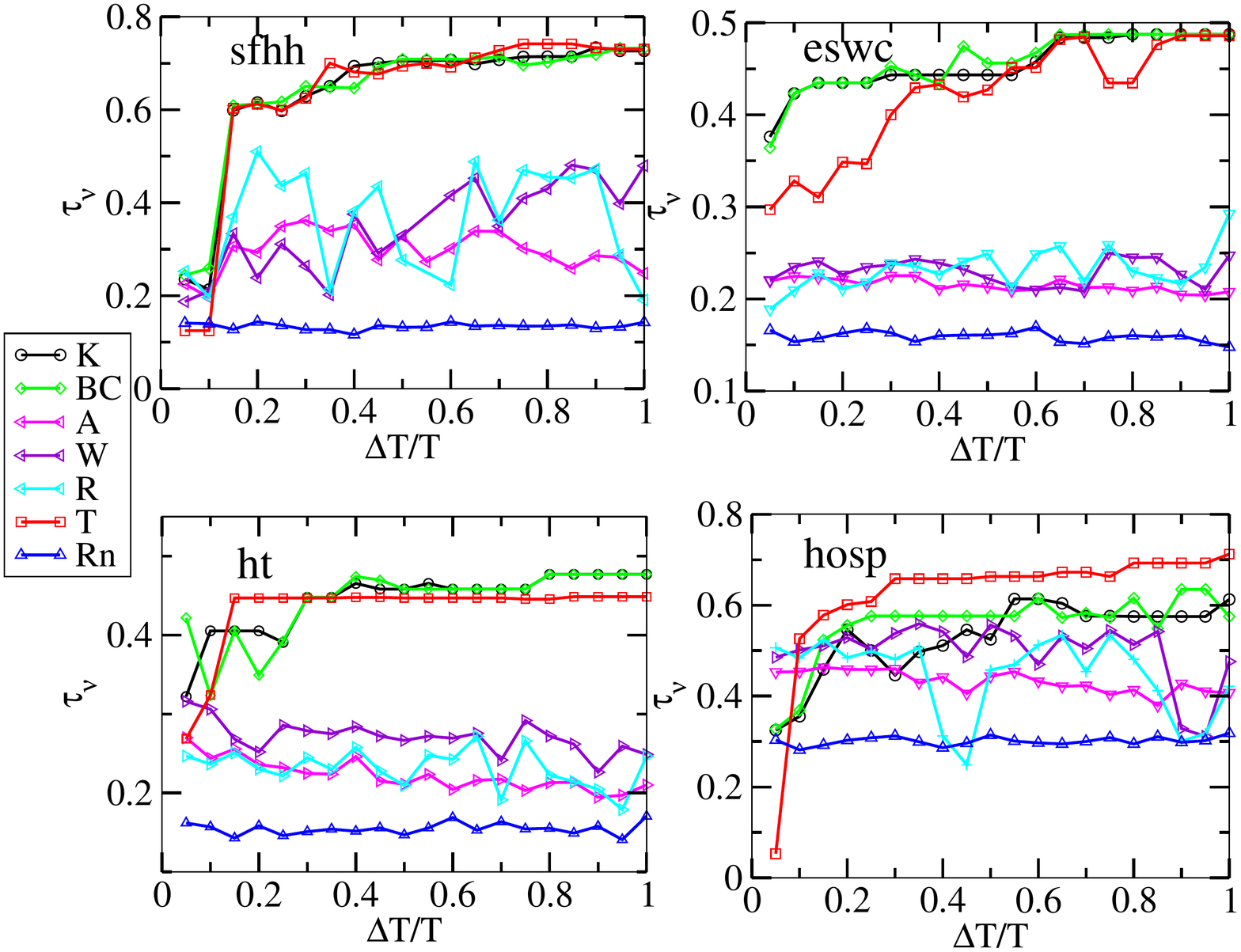}
\end{center}
\caption{Infection delay ratio $\tau_{\mathcal{V}}$ as a function of
  the training window $\Delta T$ for different immunization protocols,
  for various datasets.  The fraction of immunized individuals is
  $f=0.05$.}
  \label{fig:delay_dt}
\end{figure}

\begin{figure}[tb]
\begin{center}
\includegraphics*[width=0.8\textwidth]{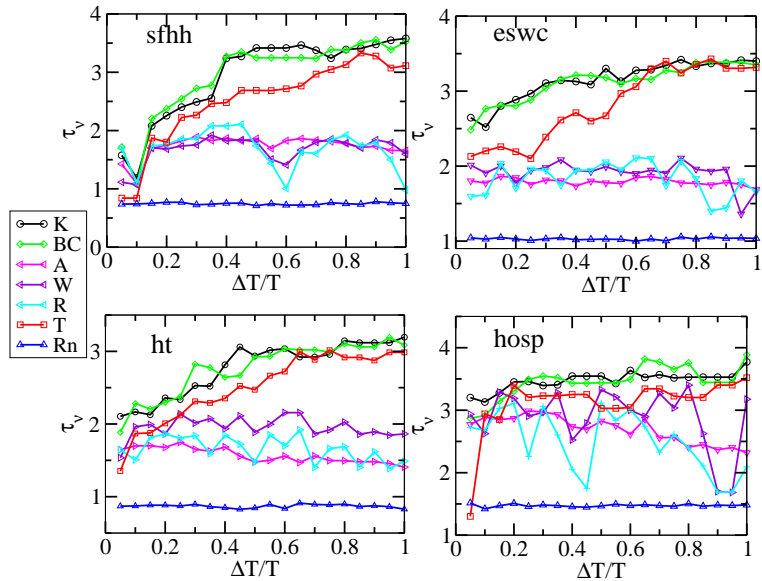}
\end{center}
\caption{Infection delay ratio $\tau_{\mathcal{V}}$ as a function of
  the training window $\Delta T$ for different immunization protocols,
  for various datasets.  The fraction of immunized individuals is
  $f=0.2$.}
  \label{fig:delay_dt_f02}
\end{figure}

We first study the role of the temporal window $\Delta T$ on the
efficiency of the various immunization strategies. To this aim, we
consider two values of the fraction of immunized individuals, $f=0.05$
and $f=0.2$, and compute the infection delay ratio $\tau_\mathcal{V}$ as a
function of $\Delta T$ for each immunization protocol and for each
dataset. Considering the limited duration of the contact sequence, 
we simulate the spreading dynamics over the whole sequence $[0,T]$,
and not only over  the remaining part $[\Delta T , T]$. 
The results, displayed in Figs.~\ref{fig:delay_dt} and
\ref{fig:delay_dt_f02}, show that an increase in the amount of
information available, as measured by an increase in $\Delta T$, does
not necessarily translate into a larger efficiency of the
immunization, as quantified by the delay of the epidemic process.  The
$\textbf{A}$, $\textbf{W}$ and $\textbf{R}$ protocols have in all
cases lower efficiencies that remain almost independent on $\Delta
T$. Moreover, and in contrast with the results of \cite{Lee:2010fk} on
a different dataset, $\textbf{W}$ and $\textbf{R}$ do not perform
better than $\textbf{A}$.  On the other hand, the immunization
efficiency of the $\textbf{K}$, $\textbf{BC}$ and $\textbf{T}$
protocols increases at small $\Delta T$ and reaches larger values for
all the datasets. As expected, the \textbf{Rn} protocol, which does not use any
information, fares the worst. For $f=0.05$, all protocols yield an
infection delay ratio that is largely independent of $\Delta T$ for
large enough training windows $\Delta T \gtrsim 0.2 T$.  For $f=0.2$,
the increase of $\tau_\mathcal{V}$ is more gradual but tends to saturate for
$\Delta T \gtrsim 0.4 T$ as well.  In all cases, a limited knowledge
of the contact time series is therefore sufficient to estimate which
nodes have to be immunized in order to delay the spreading dynamics,
especially for small $f$, i.e., in case of limited resources.
Interestingly, in some cases, the $\textbf{K}$ and $\textbf{BC}$
protocols lead to a larger delay of the spread than the $\textbf{T}$
protocol, despite the fact that the latter is explicitly designed to 
identify the nodes which yield the maximal (individual) infection
delay ratio. This could be ascribed to correlations between the
activity patterns of nodes, leading to a non-linear dependence on $f$
of the immunization efficiency (in particular, the list of nodes to
immunize is built using the list of degrees, betweenness centralities,
and $\tau_i$ values computed on the original network, without
recomputing the rankings each time a node is immunized, i.e., effectively removed from the network).

\begin{figure}[tb]
\begin{center}
\includegraphics*[width=0.8\textwidth]{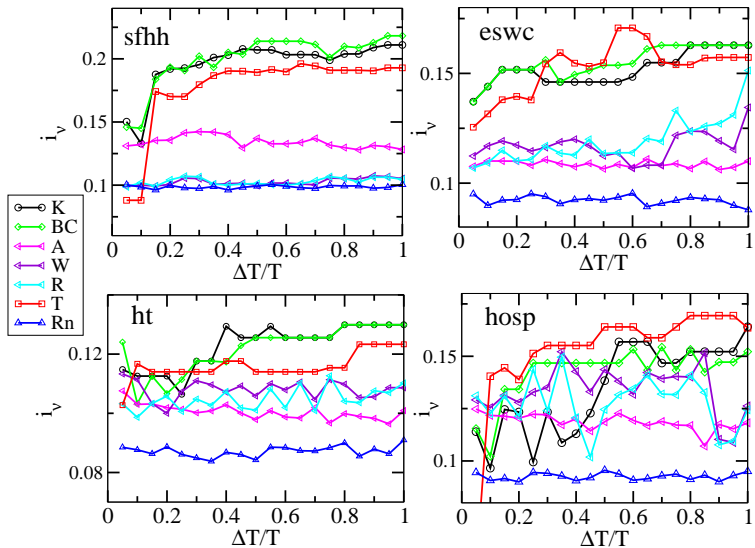}
\end{center}
\caption{Average outbreak ratio $i_{\mathcal{V}}$ as a function of the temporal
  window $\Delta T$ for different immunization protocols, for various
  datasets.  Here the fraction of immunized individuals is $f=0.05$.}
  \label{fig:infect_dt}
\end{figure} 

Figure \ref{fig:infect_dt} reports the outbreak ratio
$i_{\mathcal{V}}$ as a function of the temporal window $\Delta T$ for
different immunization protocols. Results similar to the case of the
infection delay ratio are recovered: the reduction in outbreak size,
as quantified by the average outbreak size ratio defined in
Eq. (\ref{eq:epid_outbreak_ratio}), reaches larger values for the degree, betweenness
centrality and $\textbf{T}$ protocols than for the $\textbf{A}$,
$\textbf{W}$ and $\textbf{R}$ protocols.

\begin{figure}[tb]
\begin{center}
\includegraphics*[width=0.8\textwidth]{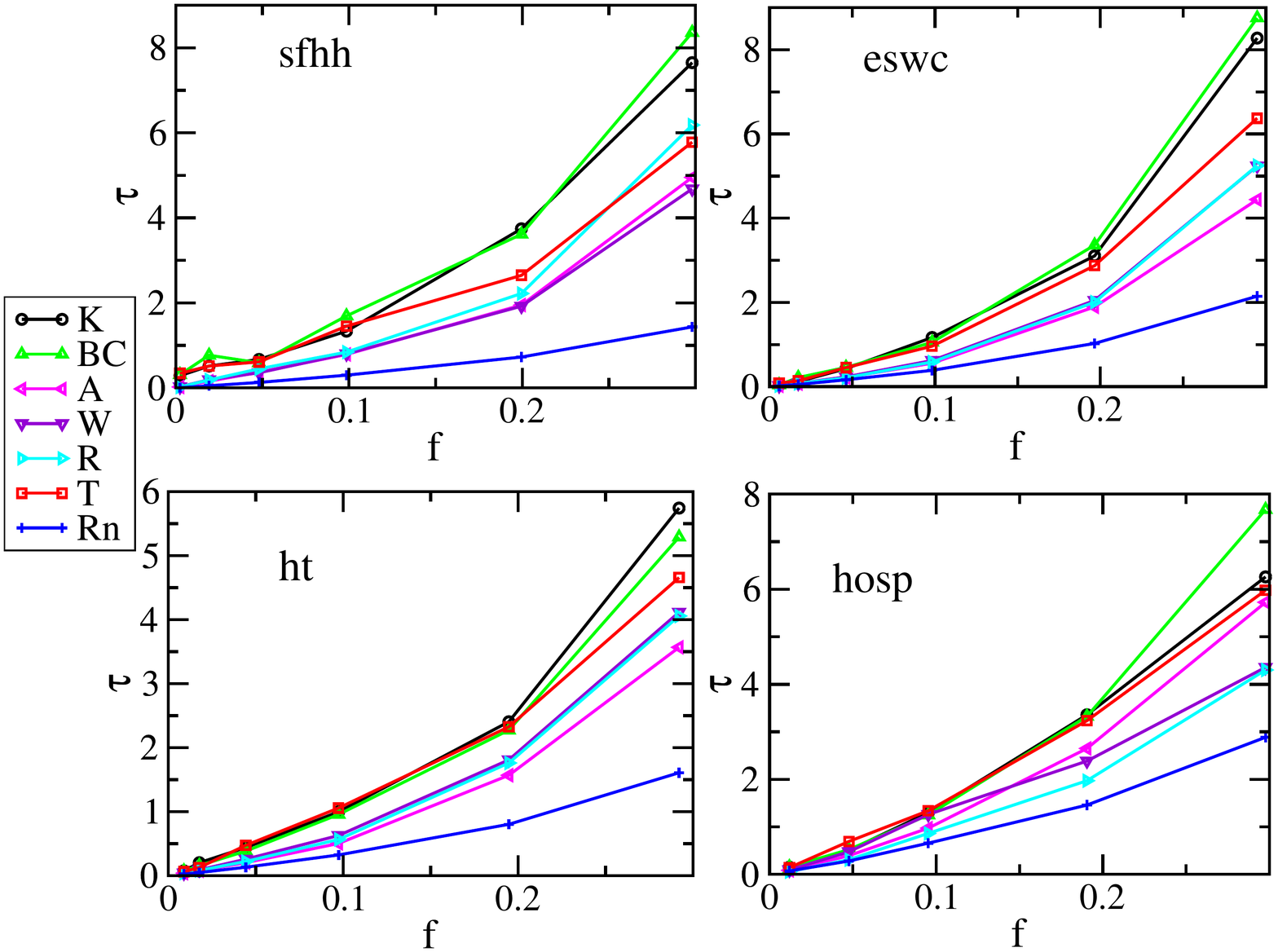}
\end{center}
\caption{Infection delay ratio $\tau_{\mathcal{V}}$ as a function of
  the fraction $f$ of immunized individuals, for different immunization
  protocols, for various datasets, and a fixed $\Delta T=0.4 T$. }
  \label{fig:delay_f}
\end{figure}

\begin{figure}[tb]
\begin{center}
\includegraphics*[width=0.8\textwidth]{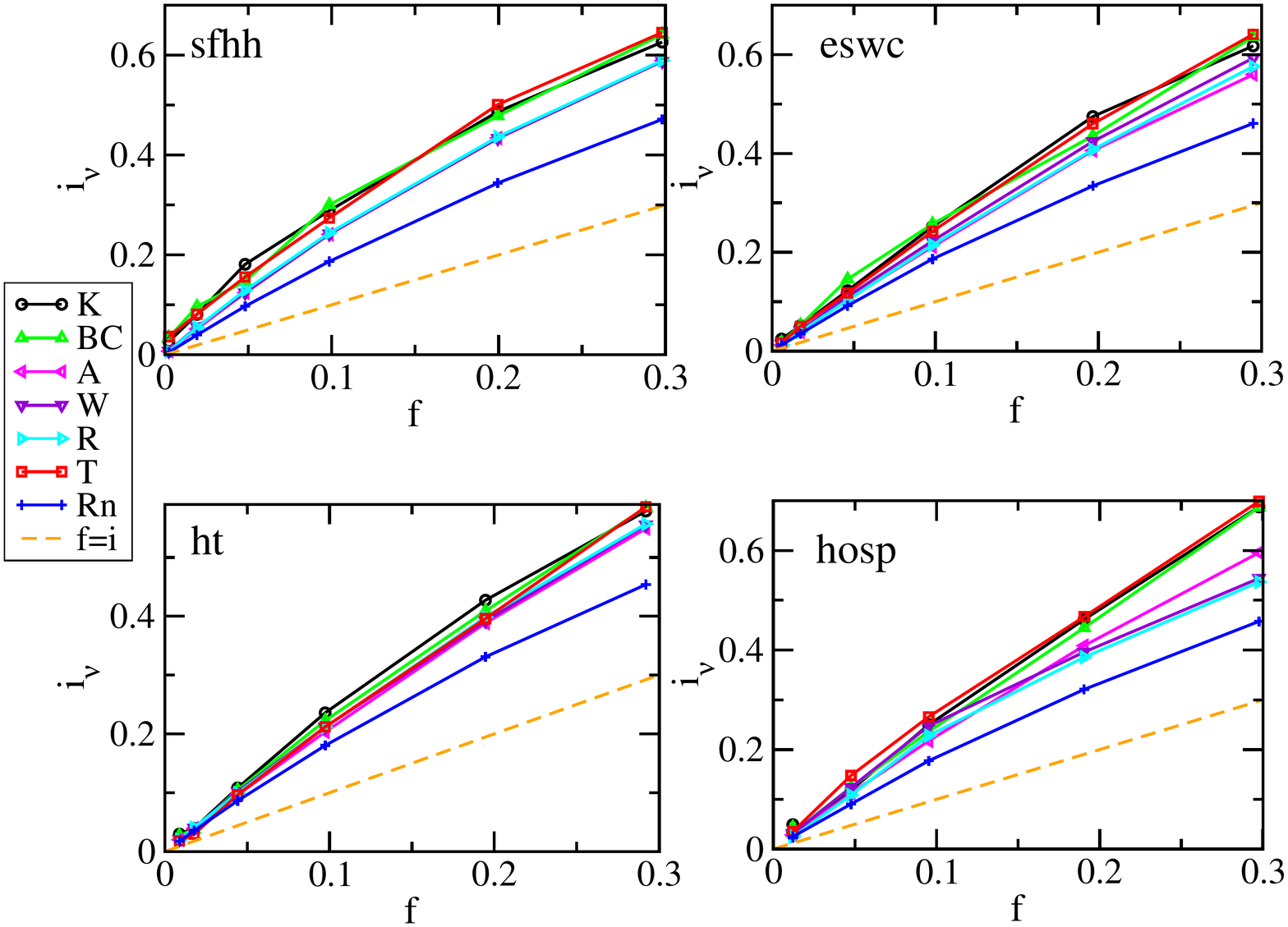}
\end{center}
\caption{Average outbreak ratio $i_{\mathcal{V}}$ as a function of the
  fraction $f$ of immunized individuals, for different immunization
  protocols, for various datasets. The dashed line represents the
  fraction $f$ of immunized individuals. Here $\Delta T=0.4 T$. }
  \label{fig:infect_f}
\end{figure}

We finally investigate the robustness of our results when the fraction
of immunized individuals varies. To this aim, we use a fixed length
$\Delta T=0.4 T$ for the training window and we plot the infection
delay ratio $\tau_{\mathcal{V}}$ and the average outbreak size ratio
$i_{\mathcal{V}}$ as a function of $f$, respectively, in
Figs. \ref{fig:delay_f} and \ref{fig:infect_f}.  The results show that
the ranking of the strategies given by these two quantities is indeed
robust with respect to variations in the fraction of immunized
individuals.  In particular, the $\textbf{K}$ and $\textbf{BC}$
protocols perform much better than the $\textbf{W}$ and $\textbf{R}$
protocols for at least one of the efficiency indicators.
 
\subsection{Effects of temporal correlations}
\label{subsec:epid_temp_corr}

Real time-varying networks are characterized by the presence of bursty
behavior and temporal correlations, which impact the unfolding of
dynamical processes
\cite{PhysRevE.71.046119,Kostakos:2009,Isella:2011,journals/corr/abs-1106-2134,Bajardi:2012,temporalnetworksbook,PhysRevE.85.056115}.
For instance, if a contact between vertices $i$ and $j$ takes place
only at the (discrete) times ${\cal T}_{ij} \equiv
\{t_{ij}^{(1)},t_{ij}^{(2)},\cdots,t_{ij}^{(n)} \}$, it cannot be used
in the course of a dynamical processes at any time $t \not\in {\cal
  T}_{ij}$.  A propagation process initiated at a given seed might
therefore not be able to reach all the other nodes, but only those
belonging to the seed's set of influence \cite{PhysRevE.71.046119},
i.e., those that can be reached from the seed by a time respecting
path.

In order to investigate the role of temporal correlations, we consider
a reshuffled version of the data in which correlations between
consecutive interactions among individuals are removed.
To this aim, we consider the list of events $(i,j,t)$ describing a contact between $i$
and $j$ at time $t$ and reshuffle at random their timestamps to build
a synthetic uncorrelated new contact sequence, 
as done in the study of the random walks process with the ``SRan" randomization.
We then apply the same
immunization protocols to this uncorrelated temporal network.

\begin{figure}[tb]
  \begin{center}
    \includegraphics[width=0.8\textwidth]{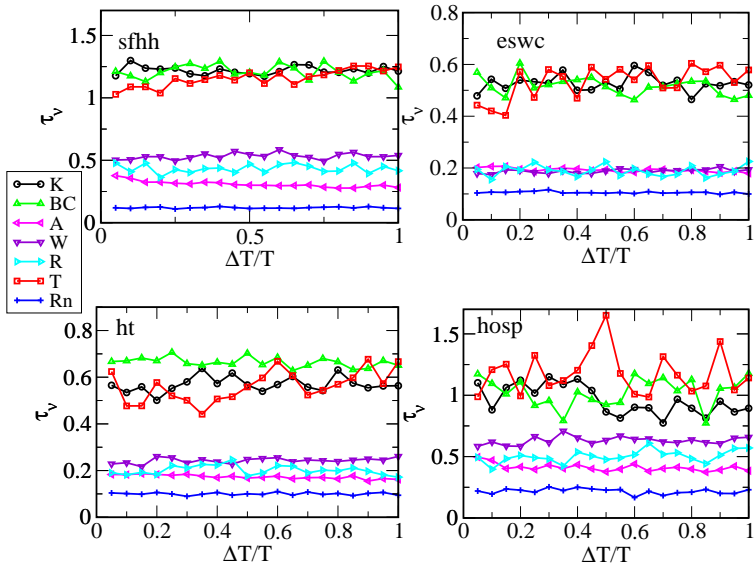}
  \end{center}
  \caption{Infection delay ratio $\tau_{\mathcal{V}}$ as a function of
    the training window length $\Delta T$, computed on one instance of
    a randomized dataset in which the timestamps of the contacts have
    been reshuffled, for different immunization protocols and various
    datasets, with $f=0.05$.  }
  \label{fig:delay_sran}
\end{figure}

Figure \ref{fig:delay_sran} displays the corresponding results for the
infection delay ratio $\tau_{\mathcal{V}}$ computed for SI spreading
simulations performed on a randomized dataset (similar results are
obtained for the average outbreak size ratio $i_{\mathcal{V}}$).  We
have checked that our results hold across different realizations of
the randomization procedure.  The efficiency of the protocol is then
largely independent of the training window length.  As the contact sequence is
random and uncorrelated, all temporal windows are statistically
equivalent, and no new information is added by increasing $\Delta T$:
in particular, as nodes appear in the randomly reshuffled sequence
with a constant probability that depends on their empirical activity,
the ranking of nodes according for instance to their aggregated degree
remains very stable as $\Delta T$ changes, so that a very small
$\Delta T$ is enough to reach a stable ranking. Nevertheless,
the efficiency ranking of the different protocols is unchanged: the
degree, betweenness centrality, and $\textbf{T}$ protocols outperform
the other immunization strategies.  Moreover, the efficiency levels
reached are higher than for the original contact sequence: the
correlations present in the data limit the immunization efficiency in
the case of the present datasets. Note that studies of the role of
temporal correlations on the speeding or slowing down of spreading
processes have led to contrasting results, as discussed by
\cite{Masuda13}, possibly because of the different models and dataset
properties considered.

\subsection{Non-deterministic spreading}
\label{subsec:epid_non_det}

\begin{figure}[t]
  \begin{center}
    \includegraphics[width=0.8\textwidth]{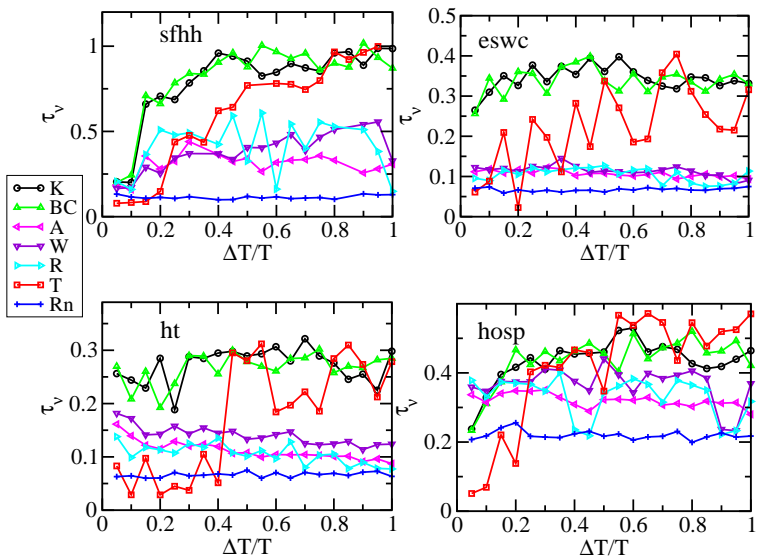}
  \end{center}
  \caption{Infection delay ratio $\tau_{\mathcal{V}}$ as a function of the temporal
    window $\Delta T$ in a probabilistic SI with $\beta=0.2$, for
    different immunization protocols, various datasets, and $f=0.05$. }
  \label{fig:delay02}
\end{figure}

We also verify the robustness of our results using a probabilistic SI
process with $\beta=0.2$. We consider the same immunization strategies
and we compute the same quantities as in the case $\beta=1$.  Given
the probabilistic nature of the spreading process, we now average the
above observables over $10^2$ realizations of the SI process.  Figure
\ref{fig:delay02} shows that our results hold in the case of a
probabilistic spread, although in this case the infection
delay ratio $\tau_{\mathcal{V}}$ presents a noisy behavior, due to the
stochastic fluctuations of the spreading
dynamics. The average outbreak ratio $i_{\mathcal{V}}$, not shown,
behaves in a very similar way.  Thus, also in this more realistic case
with $\beta < 1$, a limited knowledge of the contact sequence is
enough to identify which individuals to immunize.

\section{Summary and Discussion}
\label{sec:epid_concl}

Within the growing body of work concerning temporal networks, few
studies have considered the issue of immunization strategies and
of their relative efficiency. In general terms, the amount of information that
can be extracted from the data at hand about the characteristics of
the nodes and links is a crucial ingredient for the design of optimal
immunization strategies. Understanding how much information is needed
in order to design good (and even optimal) strategies, and how the
efficiency of the strategies depend on the information used, remain
largely open questions whose answers might depend on the precise
dataset under investigation.

We have here leveraged several datasets describing contact patterns
between individuals in various contexts and performed simulations in
order to measure the effect of different immunization strategies on
simple SI spreading processes. We have considered immunization
strategies designed according to different principles, different ways
of using information about the data, and different levels of
randomness.  Strategies range from the completely random $\textbf{Rn}$
to the $\textbf{A}$, $\textbf{W}$ and $\textbf{R}$ strategies that
include a random choice, to the fully deterministic $\textbf{K}$,
$\textbf{BC}$ and $\textbf{T}$ that are based on various node's
characteristics. Moreover, $\textbf{K}$ uses only local information
while $\textbf{BC}$ and $\textbf{T}$ rely on the global knowledge of
the connection patterns.
The most efficient strategies, as measured by the changes in
the velocity of the spread and in the final number of nodes infected,
are the deterministic protocols, namely $\textbf{K}$ and
$\textbf{BC}$.  Strategies based on random choices, even when they are
designed in order to try to immunize ``important" nodes, are less
efficient.

\begin{figure}[tb]
  \begin{center}
      \includegraphics[width=0.8\textwidth]{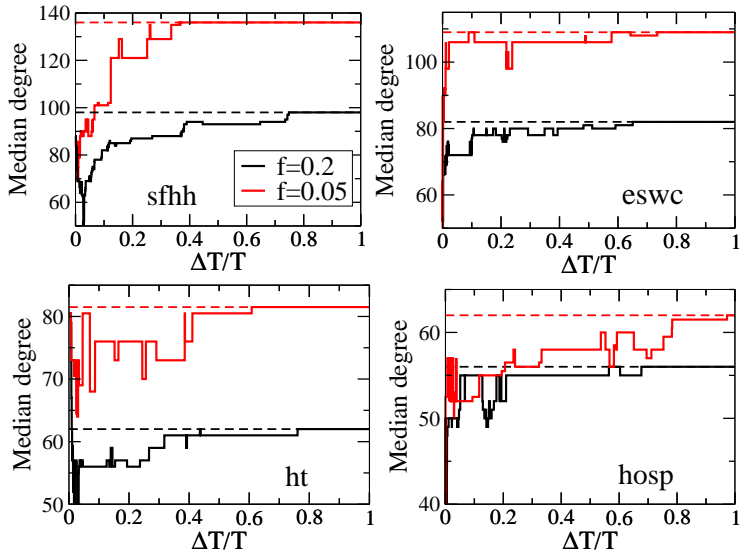}
  \end{center}
  \caption{Median degree in the aggregated network of the
   fraction $f=0.05$ (red) and the $f=0.2$ (black) nodes chosen by
    the $\textbf{K}$ strategy at $\Delta T$, vs $\Delta T/T$. The
    dashed horizontal lines mark the values of the median degrees in
   the network aggregated over the whole temporal window $[0,T]$. }
  \label{fig:fhighestdegnode}
  \end{figure}

We have moreover investigated how the performance of the various
strategies depends on the time window on which the nodes'
characteristics are measured. A longer time window corresponds indeed
a priori to an increase in the available information and hence to the
possibility to better optimize the strategies. We have found, however,
a clear saturation effect in the efficiency increase of the various
strategies as the training window on which they are designed
increases. This is particularly the case when the fraction of
immunized individuals is small (Fig. \ref{fig:delay_dt}), for which a
small $\Delta T$ is enough to reach saturation, while the saturation
is more gradual for larger fractions of immunized
(Fig. \ref{fig:delay_dt_f02}). The emergence of a saturation timescale
is noteworthy as it is unexpected in such a system in which broad
distributions of contacts and inter-contact times are observed and no
characteristic timescale can be defined
\cite{10.1371/journal.pone.0011596}.  Finally, the strategies that
involve a random component yield results that are largely independent
on the amount of information considered.

In order to understand these results in more details, we have
considered the evolution with time of the nodes' properties. In
particular, we compare the nodes with the largest degree in the fully
aggregated network with the set of nodes ${\cal S}_{\textbf K}(\Delta T)$ 
chosen by following the $\textbf{K}$ strategy on the training
window $[0,\Delta T]$. To this aim, we show in
Fig. \ref{fig:fhighestdegnode} the median of the degrees, in the
  fully aggregated network, of the nodes of ${\cal S}_{\textbf K}(\Delta T)$, 
as a function of $\Delta T$. The median rapidly
reaches its final value, showing that, even for short $\Delta T$, the
set of immunized nodes ${\cal S}_{\textbf K}(\Delta T)$ has 
properties (here the degree) similar to the ones of the
set ${\cal S}_{\textbf K}(T)$ that would be obtained by taking
into account the whole dataset of length $T$.

\begin{figure}[t]
  \begin{center}
     \includegraphics[width=0.7\textwidth]{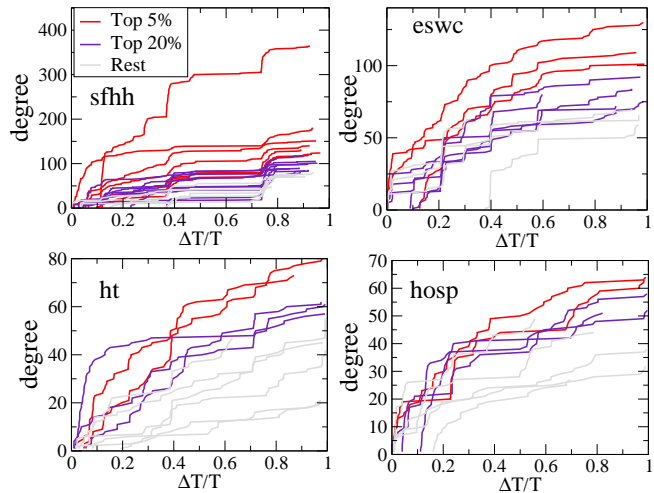}
     \end{center}
  \caption{Degree of nodes on the network aggregated over $[0,\Delta T]$,
    vs. $\Delta T/T$. The top 5\% of nodes with
    highest degree on the fully aggregated network are shown in red
    (only one third is shown for clarity). 
    The nodes ranked between the top 5\% and top 20\% in the fully aggregated
   network are shown in blue (only one sixth is shown). The
   evolution of the aggregated degree of a small number of other
    nodes is shown in grey for comparison. }
  \label{fig:degotime}
\end{figure}

Figure \ref{fig:degotime} moreover displays the evolution of the
degree aggregated over the training window $[0,\Delta T]$, as a
function of $\Delta T/T$, for several nodes. The top 5\% of
nodes with the largest degree in the fully aggregated network are
ranked among the most connected nodes already for small training
windows $\Delta T$.  The top 20\% of the ranking fluctuates more and
takes longer to stabilize.  Overall, while the precise ordering scheme
of the nodes according to their degree is not entirely stable with respect to
increasing values of $\Delta T$, a coarse ordering is rather rapidly
reached: the nodes that reach a large degree at the end of the dataset
are rapidly ranked among the highest degree nodes, and the nodes that
in the end have a low degree are as well rapidly categorized as
such. This confirms the result of Fig. \ref{fig:fhighestdegnode} and
explains why the $\textbf{K}$ strategy reaches its best efficiency
even at short training windows for small $f$, and with a more gradual
saturation for larger fractions of immunized nodes.

The fact that high degree nodes are identified early on in the
information collection process comes here as a surprise: for a
temporal network with Poissonian events, all the information on the
relative importance of links and nodes is present in the data as soon
as the observation time is larger than the typical timescale of the
dynamics; this is however a priori not the case for the bursty
dynamics observed in real-world temporal networks, for which no
characteristic timescale can be defined. Various factors can explain
the observed stability in the ranking of nodes. In first instance, some
nodes can possess some intrinsic properties giving them an important a
priori position in the network (for instance, nurses in a hospital, or
senior scientists in a conference) that ensure them a larger degree
than other nodes even at short times.  Secondly, the
stability of the ranking could in fact be only temporary, and due to
the fact that nodes arriving earlier in the dataset have a larger
probability to gather a large number of contacts. In this case, the
observed stability of the ordering scheme might decrease on longer
timescales.  In order to be able to discriminate between these
possibilities, datasets extending on much longer timescales would
however need to be collected.



The main conclusion of our study is therefore twofold. On the one
hand, a limited amount of information on the contact patterns is
sufficient to design relevant immunization strategies; on the other
hand, the strong variation in the contact patterns significantly
limits the efficiency of any strategy based on importance ranking of
nodes, even if such deterministic strategies still perform much better
than the ``recent'' or ``weight'' protocols that are generalizations
of the ``acquaintance'' strategy. Moreover, strategies based on simple
quantities such as the aggregated degree perform as well as or
better than strategies based on more involved measures such as the
infection delay ratio defined in Sec.~\ref{sec:epid_models}. We
also note that, contrarily to the case investigated by
\cite{Lee:2010fk}, the ``recent'' and ``weight'' strategies, which try
to exploit the temporal structure of the data, do not perform 
better than the simpler ``acquaintance'' strategy.  Such apparent
discrepancy might have various causes. In particular,
\cite{Lee:2010fk} consider spreading processes starting exactly at
$t=\Delta T$ while we average over different possible starting times.
The datasets used are moreover of different nature (\cite{Lee:2010fk}
indeed obtain contrasted results for different datasets) and have
different temporal resolutions. A more detailed comparison of the
different datasets' properties would be needed in order to fully
understand this point, as discussed for instance by \cite{Masuda13}.

\makeatletter
\def\toclevel@chapter{-1}
\makeatother

\chapter{Conclusions and future perspectives}
\label{chap:concl}

Data revolution in social science 
revealed complex patterns of interactions in human dynamics,
such as heterogeneity and burstiness of social contacts.
The recently uncovered temporal dimension of social interactions, indeed,
calls for a renewed effort in analysis and modeling of empirical time-vaying networks.
I contributed to pursue this program with the work presented in this thesis,
 focusing on a twofold track: 
Modeling of dynamical social systems and the study of the impact of temporally evolving substrates 
on dynamical processes running on top of them,
addressed in part \ref{part:modeling} and part \ref{part:dynproc}, respectively.
 Below a general summary of such a work is given, 
 including perspectives for future work,
 for more specific considerations see the concluding sections of each Chapter.

Firstly, in Chapter \ref{chap:intro} 
we introduced some basic concepts and definitions of the time-varying networks formalism, 
 and we presented some empirical data of social dynamics, 
 discussing their main features.
 We showed that the scientific collaboration networks \cite{newmancitations01} 
 present a temporal dimension that can be used to define the concept of activity potential,
 discussed more in detail later in Chapter \ref{chap:activitydriven}.
 We also analyzed the main statistical properties of the face-to-face interactions networks,
 recorded by the the SocioPatterns collaboration,
 such as the bursty dynamics of social interactions,
  revealed by the heavy tailed form of the  
 distributions of contact duration between pairs of individuals and
 gap times between consecutive interactions.
 These empirical data of human contact networks 
 have been largely used in the rest of the thesis,
since they constitute the benchmark for developing 
the modeling effort presented in Chapter \ref{chap:f2f}, 
as well as the substrate on which 
dynamical processes considered in part \ref{part:dynproc} take place.
Recently, a lot of attention has been devoted to the study of \emph{multi-layered} or \emph{multiplex} networks.
Real complex systems, indeed, are often composed of several layers of interrelated networks,
in which the same actors interact between them on different layers. 
Social networks, among the others, seem to be particularly appropriate to be described as multiplex,
since social interactions between the same individuals can occur in different contexts or
mediated by different means, such as face-to-face versus online communication.
Up to date, however, the great majority of works has dealt with the theoretical study of static multiplex,
while little or no attention has been given to the empirical analysis of real multi-layered networks.
This is a promising perspective for future work on temporal networks,
 it would be particularly interesting to consider the burstiness of human dynamics on different layers of interactions,
and the effect of the correlation between the different time series on diffusion processes.

In the first part of the thesis we focused on modeling of social dynamics, 
with a twofold aim: 
reproduction of empirical data properties and analytic treatment of the models considered.
In Chapter \ref{chap:f2f} we presented and discussed the behavior of a simple model
able to replicate the main statistical properties of empirical face-to-face interactions,
at different levels of aggregation, such as individual, group and collective scales.
The model considers individuals involved in a social gathering as 
performing a random walk in space, 
and it is based on the concept of  social ``attractiveness": 
socially attractive people (due to their status or role in the gathering) 
are more likely to make people stop around them, 
 so they start to interact.
Interestingly, the remarkable agreement between the model and experimental data 
is robust with respect to variations of the internal parameters of the model.
The fact that the attractiveness model is able to reproduce the collective dynamics of the individuals,
as shown with the reachability in the random walk case,
 is promising for the forthcoming study of dynamical processes, 
 such as a spreading of information or diseases, 
  taking place on the temporal networks produces by the model.
Moreover, the key role of the concept of social attractiveness in the model definition
naturally encourages further empirical research in this direction,
 in order to deepen the relation between the individual's attractiveness
 and his behavior in the social gathering, 
 revealed by the number and duration of his interactions.

Chapter \ref{chap:activitydriven} has been devoted to
 the analytic study of the activity-driven model,
 aimed to capture the relation between the 
 dynamics of time-varying networks and 
 the topological properties of their corresponding aggregated social networks.
Through a mapping to a hidden variables formalism,
we obtained analytic expressions for several topological properties of the time-integrated networks,
 as a function of the integration time and the form of the activity potential.
 We also explored the connectivity properties of the evolving network, 
 as revealed by the giant connected component size and the percolation threshold,
 for both cases of uncorrelated and correlated networks, finding notable differences.
 Despite the success of the activity driven model in explaining the formation of hubs and 
 the scale free form of the degree distribution, 
 our analysis revealed that several topological properties of real social networks 
 are still missing in the present formulation of the model,
 opening directions for its possible extensions.
 In particular, preliminary studies showed that the model remains analytically tractable 
 if the probability to fire new connections is ruled by a renewal process.
   This non-Markovian model would be able to reproduce the heavy tailed form of the inter-event time distribution,
  one of the most striking features currently missing in the original activity driven model.  
  
In the second part of the thesis we studied the behavior of diffusive processes 
taking place on temporal networks, constituted by empirical face-to-face interactions data.
In Chapter \ref{chap:RW} we considered random walks, 
the simplest diffusion model and a paradigm of dynamical processes,
since its behavior provides fundamental insights on more complex diffusive processes.
Thanks to different randomization strategies we introduced, 
we were able to single out the crucial role of 
temporal correlations in slowing down the random walk exploration.
Our results contribute to shed light on the dynamics of random walks, 
and represent a reference for the study of general dynamical processes unfolding on temporal networks.
 Future work along this line of research will be devoted to the investigation of the behavior of a random walker
moving on top of a multi-layered network, 
such as a social network constituted by a communication layer and a physical layer.
Particular attention will be dedicated to the study of the impact of the temporal correlation between the layers 
on the random walk exploration and the network navigability.

Finally, Chapter \ref{chap:epidemic} dealt with spreading dynamics,
focusing on the case of a simple SI model taking place on temporal networks.
We complemented the study of the epidemic spreading 
by investigating the impact of different immunization strategies,
in order to reduce and delay the infection outbreak.
We addressed in particular the effect of the length of the temporal window 
used to gather information in order to design the immunization strategy,
finding that a limited amount of information of the contact patterns is sufficient to
identify the individuals to immunize so to maximize the effect of the vaccination protocol.
Moreover, our results indicate that strategies based on simple quantities 
such as the individual's aggregated degree perform as well as 
 complex strategies more difficult to implement.
 Spreading dynamics have been recently studied on top of multiplex \cite{Granell2013}.
 In the scenario considered, an epidemic process is taking place on a layer of physical contacts,
 while information awareness regarding the disease is spreading on the communication layer
 involving the same individuals.
 The interplay between the two processes is analyzed under a theoretical point of view,
 revealing the effect of the interrelation with the awareness process on the epidemic threshold.
 These findings prompt for further studies, aimed to compare theoretical predictions with 
 the results of spreading dynamics on empirical multi-layered networks, 
 and to study the impact of the temporal dimension on the phase diagram of the epidemics incidence.




\setcounter{secnumdepth}{-1}
\chapter{Acknowledgements}

I am deeply grateful to Prof. Romualdo Pastor Satorras for 
the effort he put in educate me to critical judgement and thinking, 
and for his passionate and inspiring way of mentoring. 
I am indebted with him also for his expert guidance to approach physics problems, 
and for his constant human support. 
In these years I have had the lucky opportunity to work in close contact with Andrea Baronchelli, 
learning a lot: Andrea, thank you very much, I really enjoyed it. 
I would like to express my gratitude to Universitat Politecnica de Catalunya for its support, 
and for the unique human environment that I found there. 
My collaboration network is one of most valuable asset I gained during my PhD studies:
 Alain, Ciro, Alessandro, Nicola and Claudio, thank you all, 
 and thanks also to your institutions for their kind hospitality.
 Many thanks also to all the people who crossed my path during my PhD studies
 and made these years so nice and stimulating, 
 through many discussions and ideas.
Finally, a special mention for everything that is not directly involved in this thesis, 
but contributed to bring it to light:
thanks to my family and old friends, 
 and thanks to Barcelona and my new life here.

\setcounter{secnumdepth}{-1}
\chapter{List of Publications}

The work exposed in this thesis has been published in the following papers and preprints
(number of citation updated to September 2014, according to Google Scholar):

\begin{itemize}

\item M. Starnini, A. Baronchelli, A. Barrat and R. Pastor-Satorras \\
 \emph{Random walks on temporal networks} \\ 
 Physical Review E, 85, 056115, May 2012, \\
 Times cited: 36, 
 Results exposed in Chapter \ref{chap:intro} and Chapter \ref{chap:RW}.
 
\item M. Starnini, A. Baronchelli, and R. Pastor-Satorras \\
 \emph{Modeling human dynamics of face-to-face interaction networks} \\ 
 Phys. Rev. Lett., 110, 168701, Apr 2013, \\
 Times cited: 10, 
 Results exposed in Chapter \ref{chap:f2f}.

\item M. Starnini, A. Baronchelli, and R. Pastor-Satorras \\
 \emph{Model reproduces individual, group and collective dynamics of \\ human contact networks} \\ 
  preprint, \url{http://arxiv.org/abs/1409.0507}, Sept. 2014,\\
 Results exposed in Chapter \ref{chap:f2f}.
 
 \item M. Starnini and R. Pastor-Satorras \\
 \emph{Topological properties of a time-integrated activity-driven network} \\ 
 Physical Review E, 87, 062807, June 2013,\\
 Times cited: 5,
 Results exposed in Chapter \ref{chap:activitydriven}.
 
 \item  M. Starnini and R. Pastor-Satorras \\
 \emph{Temporal percolation in activity-driven networks} \\ 
 Physical Review E, 89, 032807, Mar 2014, \\
 Times cited: 1, 
 Results exposed in Chapter \ref{chap:activitydriven}. 
 
\item M. Starnini, A. Machens, C. Cattuto, A. Barrat, and R. Pastor-Satorras \\
 \emph{Immunization strategies for epidemic processes in  \\ time-varying contact networks} \\ 
 Journal of Theoretical Biology, 337 p. 89 – 100, 2013,  \\
 Times cited: 5, 
 Results exposed in Chapter \ref{chap:epidemic}. 

\end{itemize}

Other papers and book chapters related with my PhD studies and not included in the present thesis are:

\begin{itemize}

\item M. Starnini, A. Baronchelli, and R. Pastor-Satorras \\
 \emph{Ordering dynamics of the multi-state voter model} \\ 
 Journal of Statistical Mechanics: Theory and Experiment, P10027, 2012, \\
 Times cited: 9.

\item P. Moretti, A. Baronchelli, M. Starnini and R. Pastor-Satorras \\
 \emph{Generalized voter-like models on heterogeneous networks} \\ 
 Chapter in Dynamics On and Of Complex Networks, Volume 2, 285-300.

\end{itemize}

\setcounter{secnumdepth}{1}





\bibliographystyle{abbrv}

\bibliography{Bibliography}

\end{document}